\documentclass{book}
\usepackage[spanish,english]{babel}
\usepackage{courier}
\usepackage{pst-all}
\usepackage{pstcol}
\usepackage{graphicx}
\usepackage{pstricks}
\usepackage{pst-plot}
\usepackage{makeidx}

\makeindex

\definecolor{gris-claro}{cmyk}{0,0,0,0.30}

\psset{subgriddiv=0,gridlabels=7pt,griddots=5}
\newpsobject{showgrid}{psgrid}{subgriddiv=1,griddots=10,gridlabels=6pt}

\begin{document}
\newcounter{num}
\setcounter{num}{1}

\title{Simulation for biology with Excel}

\author{Jos\'e del Carmen Rodr\'\i guez Santamar\'\i a}
\date{}

\maketitle



\section{PREFACE}

One of the implication of the complexity of nature is that we cannot produce analytical formulas for every thing we would like to measure. In other cases, a formula could exists but nobody has found a reason to find it. In other times, mathematicians have hidden their knowledge behind unsurmountable barriers.  The solution to all problems is  simulation, which is the  building  of an artificial world in the memory of a computer with exactly the form we like, to study it and to report the results.

The use of simulation has been traditionally restricted to people with very good knowledge in computing science and surely some problems will continue to demand the skills of experts. Nevertheless, times have changed and  it will be shown in this work that a  biologist could become a good simulator. Our confidence rests on the modern tools that now we have at hand, which are   simple and powerful and that lie   forgotten in everyone's desktop. Our propose is to use the Visual Basic that is attached to Word and Excel to prove our assertion.  We have used Windows XP and  Excel  2000.

We begin with games with words, including a simulation of evolution. Next, we pass to statistical analysis, mathematical models and a visual simulation of evolution.

The code included in this material must be modified as part of the exercises.  To prevent unnecessary  criticisms, the following statements might  be added to the modifications:

'Initial release by Jose.

'Modified by  (type in here your name).

\begin{flushright}

Jos\'e

\end{flushright}

\tableofcontents

\newpage

\mainmatter

\chapter{Getting Started}

\textbf{\thenum. }  \addtocounter{num}{1} A computer is a devise that has the potential ability of executing verbal instructions. A program is a set of verbal instruction, written in  a specific programming language, that can be interpreted and executed by a computer. Code is a set of instructions that eventually contain correctly written programs.  When a machine is managed by a computer,  we differentiate between the software, the set of programs that rule it, and the hardware, the wiring that converts programs into specific actions. A Robot is a machine that has a computer that has been programmed to have certain independence.

A developer is a person that is committed to the design of software, i.e., of programs  with a specific function that is defined before beginning but that maybe modified along the process.  It is absolutely sure that the first trial of a code is inviable.  But if one manage to produce a viable code, most surely it will not do what one was so eagerly expecting. The process of adjusting the code to the predefined purpose is called  debugging. The word bug    arrived to us as follows: when the first computer was made, the Eniac, it was done with tubes and relays. One day,  the computer refused to work. The machinery was examined and there was a bug trapped by a relay. Since that day, a bug is anything that hinders the correct functioning of a program.

It is important to notice that the genome is indeed a set of   verbal instructions to program ribosomes to build specified proteins. So, the genome is for us the natural example of software. A cell is a robotic entity: it is directed by a software and has a lot of independence. In fact, as we go on, many questions of biological interest will spring up and  specially  in relation with the genome.

Our platform of software design is Excel which is provided with the Macro technology. It  is the implementation of a concrete philosophy of software design:

1) Automation: Usual repetitive tasks must be generalized, automated and implemented into a piece of code,  which  is called a Macro.

2) Reusability: Any given Macro must be ready to be incorporated into more larger Macros.

3) Mutability: Any Macro must be modifiable to fit special needs.

4) Personality: Any user shall have the opportunity to work according to his or her own style and needs.

5) Proficiency:  the searching and calling of a Macro must be millions of times easier than designing it from extant material or ab initio.

6)  Social proneness: programming is now a world wide activity. One shall not get apart form the academic community nor from  the whole world.

\bigskip

\textbf{\thenum. }  \addtocounter{num}{1} A  procedure is a portion of code that produces a result when it is run.  A Macro is a portion of code that is automatically produced by Excel. We distinguish two types of procedures: subs, which do  specific tasks, and functions, which  process an input and calculates an output that is to be reported. A set of procedures maybe gathered in a module to execute a great task. Modules  are organized in books that are called projects.  Our default project is personal.XLS.

\bigskip

\textbf{\thenum. }  \addtocounter{num}{1} The personal.XLS project is saved as an inner part of Excel. If you lose Excel, you will also lose your project. You can make a back up of your projects in a word document.

\bigskip
\textbf{\thenum. }  \addtocounter{num}{1} Our two aims in this chapter  are to learn how to use Excel to run an existing  program and how to record and play our own Macros.  Apart form the instructions below, you may use the help of Excel to learn more and clear doubts. Please, look what is there written about macro and module.

\section{Commenting on Excel}

If you already are acquainted with Excel, you will find the exercises below too boring; please, skip over them and just read the comments.

\textbf{\thenum. }  \addtocounter{num}{1} Excel is a powerful and easy to use  open platform for programming. This means that one can easily  accommodate it to  personal needs, choice and style.

\bigskip
\textbf{\thenum. }  \addtocounter{num}{1} Exercise. Excel is platform for programming, i.e., one can run many different programs, even in parallel. To verify  that:

1. Open Excel and open a newsheet (if it did not open automatically).

2. Pick any cell, write \textbf{3}.

3. Pick a second cell just below the first one. Write \textbf{4}.

4. In a third cell, do: a) write \textbf{=}, b) click on the cell that contains the \textbf{3}, (It shall appear something like G5 ) Add \textbf{+} d) click on the cell that contains the \textbf{4} e) apply \textbf{return}. You shall see the result 3+4=7.

5. Replace numbers  \textbf{3} and \textbf{4} by \textbf{6} \textbf{9} and you will see that the result updates automatically.

Warning: In Excel, the multiplication is a star *. The division is a slash /.

\bigskip
\textbf{\thenum. }  \addtocounter{num}{1} Exercise. Find the way to program Excel to calculate the sum  and the multiplication of two given numbers. This means that Excel gives the appearance of running multiple programs at the same time, in parallel.

We have therefore a program to sum and multiply any two numbers, thus this program could be named \textbf{(x+y,xy)}.

Our conclusion is that we have a program that works on restricted inputs. In the same way, we could write programs to multiply two numbers, to find the average of a set of numbers and so on. We notice that it is very easy to program with Excel, this means that Excel adds not overwhelming complexity to the already very difficult task of programming.

\bigskip
\textbf{\thenum. }  \addtocounter{num}{1}  We can chain different programs.

Exercise. Make a program $x+y$ that adds two numbers. Make another program, called  $50\%z$, that takes a number $z$ and calculates its  $50\%$. Chain the two programs to produce one single program, called  $50\%(x+y)$, that calculates the 50\% of the sum of two numbers.

\bigskip
\textbf{\thenum. }  \addtocounter{num}{1} We can apply a program recurrently.

\bigskip
\textbf{\thenum. }  \addtocounter{num}{1}  Example. Let us make a program $x+1$ that adds one to a number $x$.

Let us pick up any cell and write there \textbf{5}. Program the cell just below to add one to the first cell. Copy the  cell with the \textbf{6}. Paste it in the cells below over the same column. You shall get 5,6,7,8,.....

\bigskip

\bigskip
\textbf{\thenum. }  \addtocounter{num}{1}  We can apply the program in parallel many times.

\bigskip
\textbf{\thenum. }  \addtocounter{num}{1} Example. Let us do  seven sums of seven couples of numbers  that are written in stack.

First, we write seven couples of numbers, in two columns. Next, we program Excel  to add the two numbers of the first couple and to report the result just in front of the first couple, on the same row.  Then, we must copy the result and paste  it over the cells where we expect the results of the other sums: all seven sums shall appear .

\bigskip
\textbf{\thenum. }  \addtocounter{num}{1}  Any program is portable.

\bigskip
\textbf{\thenum. }  \addtocounter{num}{1}  Exercise.  Fill in some cells with numbers and operations. Select a region that contains all the cells that were occupied by this task and copy it to the clipboard. Paste the copied information at any other place of the same sheet, or in another sheet or in another Excel book and test whether or not the program kept all its properties.

\bigskip

\bigskip
\textbf{\thenum. }  \addtocounter{num}{1}  Excel allows automatization. For instance, we know how to add two numbers. Using this program, one could  add three numbers. In effect, one   adds the third to the result of adding the first two.  Next, we could add four numbers, and five, and.. ..But this is inefficient: when a task is common, one automatizes it.

\bigskip
\textbf{\thenum. }  \addtocounter{num}{1}  Exercise.  Find a button that sums n selected numbers in a row or in a column. What shall you do to find the sum of all the elements of a matrix?

\bigskip
\textbf{\thenum. }  \addtocounter{num}{1}  One cannot aspire to be prepared with all programs that one eventually could need. Instead, Excel has prepared the way that one could fills in the personal needs in an efficient and powerful way. A fundamental ingredient of this facility is the Macro technology. At this moment, we will not pursue the understanding of the Macro technology, instead we will develop our  own code and run directly because that  is by far easier for a beginner. We will need to return to the Macro technology a bit later.

\bigskip
\textbf{\thenum. }  \addtocounter{num}{1} The key to everything here is to play and  keep   playing.

\section{Running prepared code}

\textbf{\thenum. }  \addtocounter{num}{1} To run code in VBA, please, download the Tex source. You can open it with any word processor. Nevertheless, if you paste the tex source into a Text Processor, you will see the code in a special color. You can scan the web for TeXnicCenter, a Tex Editor, which can be downloaded freely. If you have pasted the Tex source into a document of any word processor and you are ready to run a piece of code, then:

1) Select in the document that contains the Tex source the portion that you want to copy. (If one uses Acrobat, one loses  indentation,  some mathematical characters are not correctly  copied, and some long line are truncated.) Nevertheless, just for the beginning, the acrobat reader  works quite well.

2) Copy the selected code to the clipboard.

3) Open Excel and visit \textbf{Tools},  next the \textbf{Macro} zone, and call the \textbf{VBA editor}. The Visual Basic window will pop up.

4) In the \textbf{Insert} menu, choose \textbf{module}. A window will pop up. Paste there your code.

5) To run the code, when it appears in the VB window, click on the first line of the dominant procedure of the code and then on the advancing button of a recording-like machine that appears in the menu bar. Our dominant procedures are Public and, with rare exceptions, are at the end of each module.

6) The output of our first program   will appear in a special window that is called Instant Window. Please, open it in the VBA menu - See - Instant Window.

7) Play and keep playing: explore here and there. Discover how and where to save your code. If you save your code, you may experience problems with duplicate names. In that case, change the name to anyone of the conflicting Public subroutines. You may consider to keep your code in a word document that you can save apart.

\bigskip
\textbf{\thenum. }  \addtocounter{num}{1} Exercise. Run  the following code, which contains  a program that sums two specified numbers:

\begin{verbatim}

Public Sub MacroSum()
 x = 2
 y = 3
Debug.Print " The sum of x =" & x & " and "
Debug.Print " y = " & y & " is " & x + y
End Sub

\end{verbatim}

Remember that the output appears in the instant window, which is activated from the menu See.

\bigskip
\textbf{\thenum. }  \addtocounter{num}{1} Warranty: if there is a piece of code that is not executed at once by the VBA of Excel, please, express your complaints to Jose at

jorodrig@uniandes.edu.co

If you receive no answer, please look for the next  corrected version of this material.  This is the first version.

\bigskip
\textbf{\thenum. }  \addtocounter{num}{1} The recording-like machine in the menu bar of VBA is the debugging-executing machine. If you activate this machine, you must deactivate it before passing to any other task. This machine interdicts  any further work anywhere.

To stop the VBA executing machine when a program is running, use control + Pause(interruption). Please, look just now where  those key are located on you keypad. This is a secret that you must keep in mind: sometimes an undesirable execution has been started and it may run for hours. To  halt it, use Ctrl + Pause.

To stop the VBA executing machine when a program is not running, use control + Pause(interruption) else click on the appropriate button of the recording-like control.  Learn to distinguish when that machine is operating and when it is not: if all buttons are in black,  it is deactivated else it is running.

\bigskip
\textbf{\thenum. }  \addtocounter{num}{1} Exercise. Copy your MacroSum to the clipboard and paste the copy just below the original one. Modify the copy to make a new macro that multiplies two numbers. To multiple use a star *.

\bigskip
\textbf{\thenum. }  \addtocounter{num}{1} Exercise.  Copy the next code to you personal book and run it with F8. This is structured style of programming was  a great invention of the 1970s: you can build great structures by recursion to independent units that are known to perfectly function. To run a code with various substructures, you shall place the cursor at the beginning of the first line of the dominant substructure. In our case, it is the public procedure.

\begin{verbatim}

Private Function Sum(x, y)
  Sum = x + y
End Function


Public Sub MacroSum()
 x = 2
 y = 3
z = Sum(x, y)
Debug.Print " The sum of x =" & x & " and "
Debug.Print " y = " & y & " is " & z
End Sub

\end{verbatim}

\bigskip
\textbf{\thenum. }  \addtocounter{num}{1} Exercise.   A module is a set of subs and functions that eventually could   serve for a given purpose.  If a sub or function was declared as private, it is accessible only from the module of work.  All along this work, all subs are defined Private with exception of the  sub that directs the module, which is declared Public. This is done for easiness of debugging at the time of design and of deciphering the code at the time of study.

In time of debugging, you can call any sub or function from any other sub or function, but take care of initializations. To run the whole program as a whole, place the cursor at the beginning of the public sub, that for as is the  dominant  sub, and put the execution machine to work by clicking at the go-on button.

\bigskip
\textbf{\thenum. }  \addtocounter{num}{1} Exercise.  Copy the next code to you personal book and play with it. Experiment with F8 to execute this new code. The instruction Dim reserves a place in computer memory for a variable. In VBA this is not mandatory but avoids bugs.  Notice, moreover, that here we introduce a function, i.e., a procedure that calculates an output. Observe also that the function is Private while the lower sub is Public. This will be a rule for us: the main procedure is a Public Sub, and subordinated procedures are Private. At last,   procedures are in ascendent order of dominance: the lower level first, the second level second and so on. This is not mandatory, but it is a rule of levelization that avoids bugs.

\begin{verbatim}


Dim z


Private Function Sum(x, y)
' A function calculates an output
 Sum = x + y
End Function

Private Sub Report()
' A sub or subroutine does a task
Debug.Print " The sum of x =" & x & " and "
Debug.Print " y = " & y & " is " & z
End Sub


Public Sub MacroSum()
 x = 2
 y = 3
z = Sum(x, y)
Call Report
End Sub



\end{verbatim}

\bigskip
\textbf{\thenum. }  \addtocounter{num}{1}  Develop a program to divide two numbers. Prompt that program with a division by zero.

\bigskip
\textbf{\thenum. }  \addtocounter{num}{1} To prevent that a division by zero spoils a running, one can use an if-then control condition as in the following program:

\begin{verbatim}

Private Function Division(x, y)
  Division = x / y
End Function


Public Sub MacroDivision()
'The values of x and y maybe changed.
 x = 2
 y = 0
If y = 0 Then
      Debug.Print " The division  by zero is not possible "
   Else
   z = Division(x, y)
   Debug.Print " The division  of x =" & x & " over "
Debug.Print " y = " & y & " is " & z
End If
End Sub


\end{verbatim}

\bigskip
\textbf{\thenum. }  \addtocounter{num}{1}  Copy the code to your personal book and run it with F8 and test it for various values of the variable y.

\bigskip
\textbf{\thenum. }  \addtocounter{num}{1}  Graduation: develop a structured program that makes the following:

1. Writes your name, birth date,   last employment and the date.

2. Calculates you age in years, in weeks and in  days.

3. If you are younger than 22, writes  your   favorite sport.  If you are older than 22, writes your favorite team.

\chapter{Words}

While Excel was developed to deal with numbers and graphics, Word was developed to deal with verbal texts. Now, the genetic material is naturally described by a string of words, so a word processor could be more suitable to study certain questions about the evolutionary process determined by the replication and further selection of DNA mutants. In consequence, this chapter could be run with the help of Microsoft word and its inbuilt VBA:

\bigskip
\textbf{\thenum. }  \addtocounter{num}{1}  Exercise. a) Copy the next text into a Word document:

AUU-UUA-UUA-UAU-AUA-UAU-AUA-UAU-AUA-UAU-AUU-UAU-AUA-UAU-AUU-UAA

b) Use the Replace function of Word to change everywhere the following codons by its amino acid equivalents.

AUU : Ile (isoleucine)

UUA: Leu (Leucine)

UAU: Tyr (Tyrosine)

AUA :  Arg ( Arginine)

AAU : Asn (Asparagine )

AAA: Lys (Lysine)

UUU: Phe (Phenylalanine)

UAA: Term (chain terminating codon)

The RNA sequence is translated into the following protein string:

Ile-Leu-Leu-Tyr-Arg-Tyr-Arg-Tyr-Arg-Tyr-Ile-Tyr-Arg-Tyr-Ile-Term

\bigskip
\textbf{\thenum. }  \addtocounter{num}{1}   Exercise. Type your own RNA strings with only A and U and   use the replace function  to get the protein equivalents.

Word also has the possibility of recording Macros, but they are so complex that we will do good if we avoid them. Instead, we will learn to produce the code on our own.

\bigskip
\textbf{\thenum. }  \addtocounter{num}{1}  Exercise. Let us learn how to copy a string written in  a  Word  document to a variable inside computer memory. To do that, please,

a) Copy the next code to the clipboard:

\begin{verbatim}

Public Sub MacroCopy()
'
Dim x

     Selection.Copy
     x = Selection.Text
     Debug.Print x
End Sub

 \end{verbatim}

 b) Paste the code into your Macro Book: call Tools, Macro, Editor of Visual basic, Insert a Module (in the menu bar, Insert) and paste there the code.

 c) Save your module.

 c) Open a page in Word. Write your name and select it. Go to Tools, Macro, Macros and look there for your Macro1. Execute it  three times.

 d) In the VB bar, look for the See sub-menu and active the Instant Window: you will see there your name written three times.

 We may use this form of reading strings to input very large RNA sequences for latter exercises.

\section{Manipulation of Strings}

\bigskip
\textbf{\thenum. }  \addtocounter{num}{1}  Visual Basic  for Microsoft Windows has many functions to manipulate strings. To see a complete list, type 'string functions' in the VB help. In the following Macro, we see how one can paste one string to the end of another.

\begin{verbatim}

Public Sub MacroConcatenation()

'This Macro concatenates 3 strings.

 Dim x, y, z
  a = "I"
  b = " "
  c = "work"
  d = "."
  z = a & b & c & d
  Debug.Print z

End Sub

\end{verbatim}

\bigskip
\textbf{\thenum. }  \addtocounter{num}{1}  Exercise. Run  this macro under F8 and disentangle what is done by each instruction. Keep the Instant Window open.

\bigskip
\textbf{\thenum. }  \addtocounter{num}{1}  Exercise. Develop a subroutine that takes three input variables, your first name, your last name and your birth date. As output, the sub produces one single string with all that information, with signs of punctuation and spaces.

\section{Translation}

\bigskip
\textbf{\thenum. }  \addtocounter{num}{1}  To see how other string function operate, let us develop a program that translates a RNA sequence with only U and A into the corresponding protein chain. The code follows:

\begin{verbatim}

Option Explicit

Dim a, b, c, z
Dim GenCode(1 To 8, 1 To 2)
Dim lc, NCodons As Integer



Private Sub Read_RNA_String()
  Selection.Copy
  c = Selection.Text
  Debug.Print c
End Sub

Private Sub Genetic_Code()

'We keep the code in a two column array.

GenCode(1, 1) = "AUU"
GenCode(1, 2) = "Ile"

GenCode(2, 1) = "UUA"
GenCode(2, 2) = "Leu"

GenCode(3, 1) = "UAU"
GenCode(3, 2) = "Tyr"

GenCode(4, 1) = "AUA"
GenCode(4, 2) = "Arg"

GenCode(5, 1) = "AAU"
GenCode(5, 2) = "Asn"

GenCode(6, 1) = "AAA"
GenCode(6, 2) = "Lys"

GenCode(7, 1) = "UUU"
GenCode(7, 2) = "Phe"

GenCode(8, 1) = "UAA"
GenCode(8, 2) = "Term"

End Sub

Private Function Number(a)
Dim i As Integer
 For i = 1 To 8
  If GenCode(i, 1) = a Then Number = i
 Next i
End Function

Public Sub Translation()
Dim i, k, l, m As Integer
Dim Start   As Integer
'We read the source RNA sequence, c,
'from a selected region in a word page.
'We must translate that sequence into
'A protein chain.
Call Read_RNA_String

'The  genetic code is read
Call Genetic_Code

'Translation is prepared.

'We get the lenght  of  c:
  lc = Len(c)
'We get the number of codons:
NCodons = lc / 3

'We translate codon by codon
'The output sequece is initialized
z = " "
For i = 1 To NCodons
 'We copy  the ith codon into a
 Start = 3 * (i - 1) + 1
 a = Mid(c, Start, 3)
 'We revise the Genetic code to know
 'the correspondiong amino acid, b:
 m = Number(a)
 b = GenCode(m, 2)
 'We synthesize z
 z = z & "-" & b
Next i
'We report z
 Debug.Print "The protein chain is " & z

End Sub

\end{verbatim}

\bigskip
\textbf{\thenum. }  \addtocounter{num}{1}  Exercise. a) Copy the code to your clipboard and paste it into  a new module in your VB.

b) Save your module.

a) Open a document  in Word, paste there the next RNA string:

AUUUUAUUAUAUAUAUAUAUAUAUAUAUAUAUUUAUAUAUAUAUUUAA

b) Select the string.

c) In Word, go to Tools, Macro, Macros and open Translation. Execute the code.

d) Read at the Instant Window of VB the output of the translation.

e) The output of this program must coincide with the output that you know from the first exercise.

\bigskip
\textbf{\thenum. }  \addtocounter{num}{1}  Graduation: develop a program to translate any RNA sequence into its amino acid equivalent. Test it over real sequences and compare its output with known results. You can paste your sequences to a word document and read it from there such as we learned in a previous exercise.

\chapter{Point Mutations}

\bigskip
\textbf{\thenum. }  \addtocounter{num}{1}  Evolution is very rich in diversity of mechanism but the simplest one is point mutation, which is the change in a bases by another one. Let us learn how to simulate point mutations. But instead of working with DNA or RNA sequences, we will work with   words and phrases in English, whose mutations  concerns the evolution of languages.

\bigskip
\textbf{\thenum. }  \addtocounter{num}{1}  Languages change: In Papua, one can see that the languages evolve in front  of own eyes and that the linguistic diversity is limited only by the number of villages. The reason is not hard to find: the dwellers have a very strong tradition of enmity with neighbors and so their small villages are isolated linguistic evolutionary laboratories. Why? Because in any house there are always linguistic experiments. In fact, all around the world and in some languages more than in others, children invent words and syntactic gyros that are accepted by the adults but that rarely surpass the limits of the local family. When the family is in isolation, we have automatically a source of evolution.

\bigskip
\textbf{\thenum. }  \addtocounter{num}{1}  Let us implement a program to produce point mutations, very simple changes  in which a character is substituted by another. Our first task is to produce a program to generate characters at random. We will use a function that is directly implemented by VBA: we use the function Chr(i) that  associates a character to a number in within 65 and 90. Let us see the code, in which we introduce the for .. ..next control statement.

\begin{verbatim}

Public Sub MacroChar()
 For i = 65 To 90
  z = Chr(i)
  Debug.Print i & " is  " & z
 Next i
End Sub

\end{verbatim}

\bigskip
\textbf{\thenum. }  \addtocounter{num}{1}  Exercise. Run the code under F8 and follow the effect of the For statement. Modify the  to develop another program that reports all the 256 symbols of the "ASCII CODE", which is the name of this correspondence. The symbols are numbered from 0 to 255.

\bigskip
\textbf{\thenum. }  \addtocounter{num}{1}  Exercise.  Develop a program that calculates the double of the numbers from 10 to 20.

\bigskip
\textbf{\thenum. }  \addtocounter{num}{1}  Exercise.  Develop a program that writes the even numbers from 10 to 20  but jumps over 15.

\bigskip
\textbf{\thenum. }  \addtocounter{num}{1} To generate random characters, we may use the   function that is included in the next module.

\begin{verbatim}

Private Function RandomChr()
 RandomChr = Chr(Int(Rnd * (1 + (90 - 65)) + 65))
End Function

Public Sub MacroRndChar11()
 For i = 1 To 100
  letter = RandomChr()
  Debug.Print letter
 Next i
End Sub

\end{verbatim}

\bigskip
\textbf{\thenum. }  \addtocounter{num}{1} Let us generate now random strings 7 characters long.

\begin{verbatim}

Private Function RandomChr(i)
 RandomChr = Chr(Int(Rnd * (1 + (90 - 65)) + 65))
End Function

Public Sub MacroRndString()
 For i = 1 To 100
 'We initialize the string
  z = ""
  For j = 1 To 7
  'We generate a random char
   letter = RandomChr(i)
   'We glue the letter to the string
   z = z & letter
   Next j
   Debug.Print z
 Next i
End Sub

\end{verbatim}

\bigskip
\textbf{\thenum. }  \addtocounter{num}{1} Exercise. Run the code and count the number of English words generated by this procedure. If needed, you can copy your words to the clipboard and next paste it to  a document  of Word, where you can test your random strings with the orthographic facility to see which words have a meaning in English.

\bigskip
\textbf{\thenum. }  \addtocounter{num}{1} The previous exercise is crucial: it allows us to conclude that most random words are senseless in English.

\textbf{\thenum. }  \addtocounter{num}{1} Exercise. Verify that the same conclusion is valid if you change English by whatever language. If necessary, install orthographic  correctors for French, Spanish, German, Italian or whatever.

\bigskip
\textbf{\thenum. }  \addtocounter{num}{1} Exercise. Do you feel hurt in you personal honor? We are trying to extract general rules from an experiment with strings 7 characters long. And what about the shorter or larger words? Did you notice that many languages have a gap between lengths 8 and 10? How can you verify that? How can you explain that? Please, make the necessary arrangements in the previous code and carry the suitable experiments to answer the questions and specially to strengthen the desired conclusion: most strings are senseless in any language. Bold to explain this fact and include in you explanation considerations in regard with human memory and expressive functionality of the language as a fair representation of a complex reality.

\bigskip
\textbf{\thenum. }  \addtocounter{num}{1} We may make now a prediction: if any word is submitted to a mutation process of random substitution character per character, the resultant series of strings would lose meaning very soon.  To test that prediction, let us develop the appropriate code. We need to develop a function to makes punctual substitutions. It is found in the next code:

\begin{verbatim}

Private Function Subs(place, letter, c)
 'This functions changes the char in the   place place of c
 'by the char letter.  Dim Start, la, lc
 Dim s1, s2, z

  'We copy  the first part of c
  s1 = Left(c, place - 1)
  'Debug.Print "The left part of c is " & s1
  'We    get the lenght  of c
  lc = Len(c)
  'Debug.Print "The string  c has " & lc & " characters"

  'We copy  the right part of c
  s2 = Right(c, lc - place)
  'Debug.Print "The Right part of c is " & s2

  'We concatenate the left part of c  to letter to the rightpart of c
  Subs = s1 & letter & s2
End Function

Public Sub MacroMutation()
 'We define the orginal string
  c = "ORIGINAL"
   Debug.Print "The original word is " & c
  'We get the lenght of the string
  n = Len(c)
  'We make a point mutation:

  'We generate a random char
   letter = RandomChr(i)
   Debug.Print "The new LetterContent is " & letter
   'We search for a place to make the substitution
   place = Int(Rnd * n) + 1
   Debug.Print "The place of substitution is " & place
   'We make the substitution
   c = Subs(place, letter, c)
   Debug.Print "The mutated word is " & c
End Sub

\end{verbatim}

\bigskip
\textbf{\thenum. }  \addtocounter{num}{1} Exercise. Play with the code.

\bigskip
\textbf{\thenum. }  \addtocounter{num}{1} In the next code we  can see how point mutations operate in tandem:

\begin{verbatim}

Private Function RandomChr()
 RandomChr = Chr(Int(Rnd * (1 + (90 - 65)) + 65))
End Function

Private Function Subs(place, letter, c)
 'This functions changes the char in the nth place
 'by the char letter in the string c.
  Dim Start, la, lc
 Dim s1, s2, z

  'We copy  the first part of c
  s1 = Left(c, place - 1)
  'Debug.Print "The left part of c is " & s1
  'We    get the lenght  of c
  lc = Len(c)
  'Debug.Print "The string  c has " & lc & " characters"

  'We copy  the right part of c
  s2 = Right(c, lc - place)
  'Debug.Print "The Right part of c is " & s2

  'We concatenate the left part of c  to letter to the rightpart of c
  Subs = s1 & letter & s2
End Function

Public Sub MacroMMutation()
 'We define the orginal string
  c = "ORIGINAL"
  'We get the lenght of the string
  n = Len(c)
  'We make point mutations in series
  For j = 1 To 20
  'We generate a random char
   letter = RandomChr()
   'We search for a place to make the substitution
   place = Int(Rnd * n) + 1
   'We make the substitution
   c = Subs(place, letter, c)
   Debug.Print c
   Next j
End Sub

\end{verbatim}

\bigskip
\textbf{\thenum. }  \addtocounter{num}{1}   Exercise. Make enough experiments that allow you to conclude that point mutations in tandem rapidly end in nonsense. Is the length of the original string an important parameter for the discussion of this assertion? Please, make experiments changing the length of the original string. Experiment also with phrases and paragraphs, which you can directly copy from any text into the original string.

\bigskip
\textbf{\thenum. }  \addtocounter{num}{1}  Graduation. In a modern languages there are near to one million words.  Instead of words, it is possible to use numbers, and specifically the numbers from 1 to 1000000.  In this new encoding, most  mutations would alter the meaning of a string. Prove that, if you can. But, our languages have gathered words of similar etymology and meaning in groups with similar members and different groups or clusters are separated by a sea of nonsense.  One application of this fact is that when one commits an error of typing, an automatic orthographic facility may suggest a list containing the   correction with almost probability one. Design an orthographic corrector with a dictionary with 20 words and 5 clusters.

\chapter{A Library of Mutations}

To implement evolution we need a set of basic mutations: substitution, deletion, duplication, inversion, insertion, transposition. Let us implement them one by one beginning with substitution. At the root of all these mutations, we find the operation of concatenation that consists in gluing together two strings. This operation is directly encoded by VBA with the ampersand, \&., as follows:

\begin{verbatim}

Private Function Glue(a, b)
  Glue = a & b
End Function


Public Sub Concatenation()
Dim z
 a = "KKKKK"
  b = "QQQQ"
  z = a & b
  Debug.Print "The output string is  " & z
End Sub

\end{verbatim}

\bigskip
\textbf{\thenum. }  \addtocounter{num}{1} Exercise. Run the code and test it under various values of a and b,  clearly state what the ampersand function does.

\bigskip
\textbf{\thenum. }  \addtocounter{num}{1} We also may need the inverse relation, to cut  a string in two parts. We take the string c and divide it into two parts, the first one with n characters:

\begin{verbatim}

Option Explicit
Dim a, b, c
Dim n As Integer


Private Sub Division(c, n)
Dim lc As Integer
 'We take string c and divide it in two parts,
 'the first part with n characters.

 'We copy  the first part of c
  a = Left(c, n)
  Debug.Print "The left part of c is " & a
  'We    get the lenght of  c
   lc = Len(c)
  Debug.Print "The string  c has " & lc & " characters"
  'We copy  the right part of c
  b = Right(c, lc - n)
  Debug.Print "The Right part of c is " & b
End Sub


Public Sub Scission()
 c = "KKKKKQQQQ"
 n = 5
 Call Division(c, n)
End Sub

\end{verbatim}

\bigskip
\textbf{\thenum. }  \addtocounter{num}{1} Exercise. Run the code and test it under various values of c and n. Clearly state what the functions Left and Right do and which are their parameters.  If necessary, help yourself with the VBA help.

\bigskip
\textbf{\thenum. }  \addtocounter{num}{1} To replace a substring by another, one may imitates the next Macro:

\begin{verbatim}

Public Sub Macro3()
'
'This Macro replaces a substring a by b inside c
'
 Dim x, y, z
  a = "AUU"
  b = "-Ile- "
  c = "KKKKKAUUQQQQ"
  'We get where a begins
  Start = InStr(c, a)
  Debug.Print "a begins at position " & Start
  'We copy  the first part of c
  s1 = Left(c, Start - 1)
  Debug.Print "The left part of c is " & s1
  'We    get the lenghts of  the strings
  la = Len(a)
  Debug.Print "The string  a has " & la & " characters"
   lc = Len(c)
  Debug.Print "The string  c has " & lc & " characters"

  'We copy  the right part of c
  s2 = Right(c, lc - Start - la + 1)
  Debug.Print "The Right part of c is " & s2

  'We concatenate the left part of c  to b to the rightpart of c
  z = s1 & b & s2
  Debug.Print "The output string is  " & z

End Sub

\end{verbatim}

\bigskip
\textbf{\thenum. }  \addtocounter{num}{1} Exercise. Open the Instant Window and Run the Macro3 with F8.

\bigskip
\textbf{\thenum. }  \addtocounter{num}{1} Exercise. Design a function, Substitution(a,b,c), that replaces the first occurrence of a by b inside c. Compare your solution with the one found in the next Module.

\begin{verbatim}

Private Function Substitution(a, b, c)
'
'This Function substitutes the substring a by b inside c
'
 Dim Start, la, lc As Integer
 Dim s1, s2, z

  'We get where a begins
  Start = InStr(c, a)
  Debug.Print "a begins at position " & Start
  'We copy  the first part of c
  s1 = Left(c, Start - 1)
  Debug.Print "The left part of c is " & s1
  'We    get the lenghts of  the strings
  la = Len(a)
  Debug.Print "The string  a has " & la & " characters"
   lc = Len(c)
  Debug.Print "The string  c has " & lc & " characters"

  'We copy  the right part of c
  s2 = Right(c, lc - Start - la + 1)
  Debug.Print "The Right part of c is " & s2

  'We concatenate the left part of c  to b to the right part of c
  Substitution = s1 & b & s2
End Function

Public Sub Macro4()
 a = "AUU"
  b = "-Ile- "
  c = "KKKKKAUUQQQQ"
  z = Substitution(a, b, c)
  Debug.Print "The output string is  " & z
End Sub

\end{verbatim}

\bigskip
\textbf{\thenum. }  \addtocounter{num}{1} Exercise. Run the code and verify that it functions just the same way as   Macro3 does.

\bigskip
\textbf{\thenum. }  \addtocounter{num}{1} Let pass now to deletion. One may desire to delete a given substring from a source string or delete certain number of characters from a given string beginning at certain specific place.

To delete the first occurrence of an specific substring from a source string, one may use the Replace function Replace(a,b,c) where a is the substring to be deleted, b is the nil  string "", and c is the source string. The code follows:

\begin{verbatim}

Private Function Substitution(a, b, c)
'
'This Function replaces the substring a by b inside c
'
 Dim Start, la, lc As Integer
 Dim s1, s2, z

  'We get where a begins
  Start = InStr(c, a)
  Debug.Print "a begins at position " & Start
  'We copy  the first part of c
  s1 = Left(c, Start - 1)
  Debug.Print "The left part of c is " & s1
  'We    get the lengths of  the strings
  la = Len(a)
  Debug.Print "The string  a has " & la & " characters"
   lc = Len(c)
  Debug.Print "The string  c has " & lc & " characters"

  'We copy  the right part of c
  s2 = Right(c, lc - Start - la + 1)
  Debug.Print "The Right part of c is " & s2

  'We concatenate the left part of c  to b to the right part of c
  Substitution = s1 & b & s2
End Function

Private Function Delete(a, c)
'This function deletes the first
'occurrence of a from c.
  b = ""
  Delete = Substitution(a, b, c)
  Debug.Print "The output string is  " & z
End Function

Public Sub Macro5()
'We  delete  the first
'occurrence of a from c.

  a = "AUU"
  c = "KKKKKAUUQQQQ"
  z = Delete(a, c)
  Debug.Print "The output string is  " & z
End Sub

\end{verbatim}

\bigskip
\textbf{\thenum. }  \addtocounter{num}{1} Exercise. Run the code with F8 and explain it.

\bigskip
\textbf{\thenum. }  \addtocounter{num}{1} To delete certain number of characters from a given string beginning at certain specific place, we may define the function Delete(n,Start,c), as it is done in the next module.

\begin{verbatim}

Option Explicit
Dim n, Start
Dim c


Private Function Delete(n, Start, c)
'
'This function deletes n characters
'starting from Start out of c

 Dim lc As Integer
 Dim s1, s2



  'We copy  the first part of c
  s1 = Left(c, Start - 1)
  Debug.Print "The left part of c is " & s1

   'We    get the length  of  the string c
  lc = Len(c)
  Debug.Print "The string  c has " & lc & " characters"
  'We copy  the right part of c
  s2 = Right(c, lc - Start - n + 1)
  Debug.Print "The Right part of c is " & s2

  'We concatenate the left part of c  to   the right part of c
  Delete = s1 & s2
End Function


Public Sub Macro6()
Dim z
'We  delete  3 characters
'beginning at place Start from c.

  n = 3
  Start = 6
  c = "KKKKKAUUQQQQ"
  z = Delete(n, Start, c)
  Debug.Print "The output string is  " & z
End Sub

\end{verbatim}

\bigskip
\textbf{\thenum. }  \addtocounter{num}{1} Now, we implement a duplication, which applied in tandem produces gene amplification.

\begin{verbatim}

Private Function Duplicate(a, c)
'
'This Function duplicates the first occurrence
'of substring a  inside c
'
 Dim Start, la, lc As Integer
 Dim s1, s2, z

  'We get where a begins
  Start = InStr(c, a)
  Debug.Print "a begins at position " & Start
  'We copy  the first part of c
  s1 = Left(c, Start - 1)
  Debug.Print "The left part of c is " & s1
  'We    get the lengths of  the strings
  la = Len(a)
  Debug.Print "The string  a has " & la & " characters"
   lc = Len(c)
  Debug.Print "The string  c has " & lc & " characters"

  'We copy  the right part of c
  s2 = Right(c, lc - Start - la + 1)
  Debug.Print "The Right part of c is " & s2

  'We concatenate the left part of c  to a to a to the right part of c
  Duplicate = s1 & a & a & s2
End Function





Public Sub Macro7()
Dim z
'We  duplicate string a inside c.

  a = "AUU"

  c = "KKKKKAUUQQQQ"
  z = Duplicate(a, c)
  Debug.Print "The output string is  " & z
End Sub

\end{verbatim}

\bigskip
\textbf{\thenum. }  \addtocounter{num}{1} Exercise. Run the code with F8 and explain it.

\bigskip
\textbf{\thenum. }  \addtocounter{num}{1} Now we deal with an inversion. Our aim is to invert the subsequence a inside c:

\begin{verbatim}

Private Function Inversion(a)
 'This function inverts a
 Dim la, i, Start As Integer
 Dim d, f

 'We initialize the output
 d = ""

 'We get the length of a
 la = Len(a)

 'We copy the first character to the last place
 'and so on.

 For i = 1 To la
 Start = i
 chari = Mid(a, Start, 1)

  d = chari & d
 Next i
 Inversion = d

End Function


Private Function Invert(a, c)
'
'This Function inverts the first occurrence
'of substring a  inside c
'
 Dim Start, la, lc As Integer
 Dim s1, s2, z

  'We get where a begins
  Start = InStr(c, a)
  Debug.Print "a begins at position " & Start
  'We copy  the first part of c
  s1 = Left(c, Start - 1)
  Debug.Print "The left part of c is " & s1
  'We    get the lengths of  the strings
  la = Len(a)
  Debug.Print "The string  a has " & la & " characters"
   lc = Len(c)
  Debug.Print "The string  c has " & lc & " characters"
  'We invert a properly
  d = Inversion(a)
  'We copy  the right part of c
  s2 = Right(c, lc - Start - la + 1)
  Debug.Print "The Right part of c is " & s2


  'We concatenate the left part of c  to d to the right part of c
  Invert = s1 & d & s2
End Function


Public Sub Macro8()
Dim z
'We  invert string a inside c.

  a = "ABC"

  c = "KKKKKABCQQQQ"
  z = Invert(a, c)
  Debug.Print "The output string is  " & z
End Sub

\end{verbatim}

Exercise. Run the code,  and test it variously.

Let us implement now an insertion of a substring a at place k of string c.

\begin{verbatim}

Private Function Insert(a, Start, c)
'
'This Function takes substring a
'and posits it at place Start inside c
'
 Dim lc As Integer
 Dim s1, s2, z

  'We copy  the left part of c
  s1 = Left(c, Start - 1)
  Debug.Print "The left part of c is " & s1
  'We    get the lengths of  the string c
   lc = Len(c)
  Debug.Print "The string  c has " & lc & " characters"
  'We copy  the right part of c
  s2 = Right(c, lc - Start + 1)
  Debug.Print "The Right part of c is " & s2
  'We concatenate the left part of c  to a to the right part of c
  Insert = s1 & a & s2
End Function


Public Sub Macro9()
Dim z
'We  insert a at place Start of c.

  a = "ABC"
  Start = 6
  c = "KKKKKQQQQ"
  z = Insert(a, Start, c)
  Debug.Print "The output string is  " & z
End Sub

\end{verbatim}

\bigskip
\textbf{\thenum. }  \addtocounter{num}{1} Exercise. Run the code under F8, understand it and  test it variously.

\bigskip
\textbf{\thenum. }  \addtocounter{num}{1} And what about   transpositions? They are very important in biology, say, with the jumping genes first discovered in Maize. A transposition takes a substring a from inside c and moves it to another place. We  will  implement transpositions as follows: they result from a deletion followed by an  insertion.

\begin{verbatim}

Private Function Substitution(a, b, c)
'
'This Function replaces the substring a by b inside c
'
 Dim Start, la, lc As Integer
 Dim s1, s2, z

  'We get where a begins
  Start = InStr(c, a)
  'Debug.Print "a begins at position " & Start
  'We copy  the first part of c
  s1 = Left(c, Start - 1)
  'Debug.Print "The left part of c is " & s1
  'We    get the lenghts of  the strings
  la = Len(a)
  'Debug.Print "The string  a has " & la & " characters"
   lc = Len(c)
  'Debug.Print "The string  c has " & lc & " characters"

  'We copy  the right part of c
  s2 = Right(c, lc - Start - la + 1)
  'Debug.Print "The Right part of c is " & s2

  'We concatenate the left part of c  to b to the right part of c
  Substitution= s1 & b & s2
End Function

Private Function Delete(a, c)
'This function deletes the first
'occurrence of a from c.
  b = ""
  Delete = Substitution(a, b, c)
End Function


Private Function Insert(a, Start, c)
'
'This Function takes substring a
'and posits it at place Start inside c
'
 Dim lc As Integer
 Dim s1, s2, z

  'We copy  the left part of c
  s1 = Left(c, Start - 1)
  'Debug.Print "The left part of c is " & s1
  'We    get the lenghts of  the string c
   lc = Len(c)
  'Debug.Print "The string  c has " & lc & " characters"
  'We copy  the right part of c
  s2 = Right(c, lc - Start + 1)
  'Debug.Print "The Right part of c is " & s2
  'We concatenate the left part of c  to a to the right part of c
  Insert = s1 & a & s2
End Function



Private Function Transposition(a, k, c)
'We cut substring a from c and move  it
'k places in the mutilated c.
'If k>0, movement is forwards else it is backwards.

Dim z
Dim Start1, Start2 As Integer

  'We get where a begins
  Start1 = InStr(c, a)
  'We  remove a from c
  c = Delete(a, c)
  'We insert a at place Start of the new c.
  Start = Start1 + k
  z = Insert(a, Start, c)
  Transposition = z

End Function

Public Sub Macro10()
Dim z
  a = "ABC"
  k = 2
  c = "KKKKKABCQQQQ"
  z = Transposition(a, k, c)
  Debug.Print "The output string is  " & z
End Sub

\end{verbatim}

\bigskip
\textbf{\thenum. }  \addtocounter{num}{1} Exercise. Test the code variously and run it under F8 to disentangle it.

\bigskip
\textbf{\thenum. }  \addtocounter{num}{1} Graduation: We have been working with linear strings. In nature, some genetic sequences are circular. Find the best way to encode circular sequences into a linear one  in order to generalize all defined string  functions to cover both cases, linear and close. Make the generalization.

\chapter{Evolution}

\bigskip
\textbf{\thenum. }  \addtocounter{num}{1} We will develop  a simulation of a population.  If you are not a biologist, feel lucky: today, simulated evolution is a tool to solve every kind of problems. The hypotheses of work is that  if a problem can be solved, it can also  be solved by evolution. One may think of many situations under which this hypotheses could look unsound  but with more care and patience, ideas to convert the hypotheses into a program of work may come to the mind. So, keep working.

\bigskip
\textbf{\thenum. }  \addtocounter{num}{1} Let us simulate a population of strings in which its members need to feed to reproduce and grow but have parasites that cut part of their tails. We also simulate an observer that wants to know what happens to the minimum and maximum length of the strings. Let us  develop the program step by step. To begin with, we make a simple sketch:

\begin{verbatim}

Private Sub Initialization()
Debug.Print "I am initialization"
End Sub


Private Sub Dynamics()
Debug.Print "I am Dynamics "
End Sub

Private Sub Observation()
 Debug.Print "I am Observation "
End Sub

Public Sub Main()
Dim NGen As Integer
'The population has NIndiv members
 NIndiv = 10
 Initialization
 'We run NGen generations
 NGen = 100
 For i = 1 To NGen
   Dynamics
   Observation
  Next i
End Sub

\end{verbatim}

\bigskip
\textbf{\thenum. }  \addtocounter{num}{1} Exercise. Run the code under F8.  If you fall into the black hole of an infinite loop, break the running with Ctrl + Pause and play with the mode of debugging: Run until the position of the cursor. Rationalize each sentence of the code.

\bigskip
\textbf{\thenum. }  \addtocounter{num}{1} Now, we can fill in the Initialization procedure:

\begin{verbatim}


Dim Individual(1 To 100)
Dim NIndiv As Integer

Private Function RandomChr()
 RandomChr = Chr(Int(Rnd * (1 + (90 - 65)) + 65))
End Function



Private Sub Initialization()
Dim i, j As Integer
 'We generate NIndiv individuals (strings)
 'ten characters long.
 'Sequences are completely random
 For i = 1 To NIndiv
  'An individual is assembled char by char
  Individual(i) = ""
  For j = 1 To 10
    Individual(i) = Individual(i) & RandomChr()
  Next j
  Debug.Print "The individual " & i & " is " & Individual(i)
 Next i
 'We generate a parasite pattern
 PatternParasite = "PARASITE"
 'We generate a food pattern
 PatternFood = "ALGAE"
End Sub


Private Sub Dynamics()
Debug.Print "I am Dynamics "
End Sub

Private Sub Observation()
 Debug.Print "I am Observation "
End Sub

Public Sub Main()
Dim NGen As Integer
'The population has NIndiv members
 NIndiv = 10
 Call Initialization
 'We run NGen generations
 NGen = 1
 For i = 1 To NGen
   Call Dynamics
   Call Observation
  Next i
End Sub

\end{verbatim}

\bigskip
\textbf{\thenum. }  \addtocounter{num}{1} Exercise. Run the code and verify that it creates 10 strings generated at random. Modify the code to create 20 strings and to run for 2 generations.

\bigskip
\textbf{\thenum. }  \addtocounter{num}{1} Now, we can fill in the procedure Dynamics with a sketch. Because the code is becoming a little heavy, we take an antibug measure: any variable is explicitly declared, so that if type ever instead of over we could notice it. This is done with the Option Explicit that is always written at the beginning of the code.

\begin{verbatim}

Option Explicit

Dim Individual(1 To 100)
Dim NIndiv As Integer
Dim PatternParasite, PatternFood

Private Function RandomChr()
 RandomChr = Chr(Int(Rnd * (1 + (90 - 65)) + 65))
End Function

Private Sub Initialization()
Dim i, j As Integer
 'We generate NIndiv individuals (strings)
 'ten characters long.
 'Sequences are completly random
 For i = 1 To NIndiv
  'An individual is assembled char by char
  Individual(i) = ""
  For j = 1 To 10
    Individual(i) = Individual(i) & RandomChr()
  Next j
  Debug.Print "The individual " & i & " is " & Individual(i)
 Next i
 'We generate a parasite pattern
 PatternParasite = "PARASITE"
 'We generate a food pattern
 PatternFood = "ALGAE"
End Sub

Private Sub Digestion(i)
  Debug.Print "I am Digestion(i) "
End Sub

Private Sub Corrosion(i)
  Debug.Print "I am Corrosion(i) "
End Sub

Private Sub Reproduction()
  Debug.Print "I am Reproduction"
End Sub

Private Sub Mutation()
  Debug.Print "I am Mutation"
End Sub


Private Sub Dynamics()
Dim i As Integer
 'All individuals feed
  For i = 1 To NIndiv
   'The string Algae is digested and inserted
   'into the individual char by char
  Call Digestion(i)
  Next i
 'All individuals are tested by the parasite,
  'which corrodes any  large substring similar to it located at any end.
  For i = 1 To NIndiv
   Call Corrosion(i)
  Next i
 'The top ten are allowed to  reproduce:
 'a child occupies the place of the shortest individual
  Call Reproduction
  'The new population is subjected to mutation
  Call Mutation
End Sub

Private Sub Observation()
 Debug.Print "I am Observation "
End Sub

Public Sub Main()
Dim i As Integer
Dim NGen As Integer
'The population has NIndiv members
 NIndiv = 10
 Call Initialization
 'We run NGen generations
 NGen = 1
 For i = 1 To NGen
   Call Dynamics
   Call Observation
  Next i
End Sub

\end{verbatim}

\bigskip
\textbf{\thenum. }  \addtocounter{num}{1}  Exercise. Run the code and verify that it works as it is due.

\bigskip
\textbf{\thenum. }  \addtocounter{num}{1}  To Fill in the sub Digestion, we need to develop a code that can be tested apart. The process of feeding is simulated as follows: the program  takes the letters of the string "ALGAE"' and inserts each one of them in a random place in the string representing a given individual. This is the equivalent of taking a protein, breaking to pieces and inserting each piece in a new protein of the feeding organism. The code of Digestion follows and is tested from a Macro Test:

\begin{verbatim}

Dim Individual(1 To 2)

Private Function Insert(a, Start, c)
'
'This Function takes substring a
'and posits it at place Start inside c
'
 Dim lc As Integer
 Dim s1, s2, z

  'We copy  the left part of c
  s1 = Left(c, Start - 1)
  'Debug.Print "The left part of c is " & s1
  'We    get the lengths of  the string c
   lc = Len(c)
  'Debug.Print "The string  c has " & lc & " characters"
  'We copy  the right part of c
  s2 = Right(c, lc - Start + 1)
  'Debug.Print "The Right part of c is " & s2
  'We concatenate the left part of c  to a to the right part of c
  Insert = s1 & a & s2
End Function

Private Sub Digestion(i)
Dim a
Dim l, Start As Integer

  a = "A"
  l = Len(Individual(i))
  Start = Int(Rnd * l) + 1
 Individual(i) = Insert(a, Start, Individual(i))
 a = "L"
  l = Len(Individual(i))
  Start = Int(Rnd * l) + 1
 Individual(i) = Insert(a, Start, Individual(i))
 a = "G"
  l = Len(Individual(i))
  Start = Int(Rnd * l) + 1
 Individual(i) = Insert(a, Start, Individual(i))
 a = "A"
  l = Len(Individual(i))
  Start = Int(Rnd * l) + 1
 Individual(i) = Insert(a, Start, Individual(i))
 a = "E"
  l = Len(Individual(i))
  Start = Int(Rnd * l) + 1
 Individual(i) = Insert(a, Start, Individual(i))
End Sub



Public Sub Test()
 Dim i As Integer
 Individual(1) = "FISH"
 Individual(2) = "TURTLE"
  For i = 1 To 2
   Call Digestion(i)
   Debug.Print "The Individual "; i; "is "; Individual(i)
  Next i
End Sub

\end{verbatim}

\bigskip
\textbf{\thenum. }  \addtocounter{num}{1}  Exercise. Run the code and test it variously to see that it really works. Observe that some sentences that write messages to the Instant Window have been silenced but could be activated by just erasing the symbol that marks a comment.

\bigskip
\textbf{\thenum. }  \addtocounter{num}{1}  Now, we can insert our Digestion(i) procedure into our code:

\begin{verbatim}

Option Explicit

Dim Individual(1 To 100)
Dim NIndiv As Integer
Dim PatternParasite, PatternFood

Private Function RandomChr()
 RandomChr = Chr(Int(Rnd * (1 + (90 - 65)) + 65))
End Function

Private Sub Initialization()
Dim i, j As Integer
 'We generate NIndiv individuals (strings)
 'ten characters long.
 'Sequences are completly random
 For i = 1 To NIndiv
  'An individual is assembled char by char
  Individual(i) = ""
  For j = 1 To 10
    Individual(i) = Individual(i) & RandomChr()
  Next j
  Debug.Print "The individual " & i & " is " & Individual(i)
 Next i
 'We generate a parasite pattern
 PatternParasite = "PARASITE"
 'We generate a food pattern
 PatternFood = "ALGAE"
End Sub


Private Function Insert(a, Start, c)
'
'This Function takes substring a
'and posits it at place Start inside c
'
 Dim lc As Integer
 Dim s1, s2, z

  'We copy  the left part of c
  s1 = Left(c, Start - 1)
  'Debug.Print "The left part of c is " & s1
  'We    get the lengths of  the string c
   lc = Len(c)
  'Debug.Print "The string  c has " & lc & " characters"
  'We copy  the right part of c
  s2 = Right(c, lc - Start + 1)
  'Debug.Print "The Right part of c is " & s2
  'We concatenate the left part of c  to a to the right part of c
  Insert = s1 & a & s2
End Function


Private Sub Digestion(i)
Dim a
Dim l, Start As Integer

  a = "A"
  l = Len(Individual(i))
  Start = Int(Rnd * l) + 1
 Individual(i) = Insert(a, Start, Individual(i))
 a = "L"
  l = Len(Individual(i))
  Start = Int(Rnd * l) + 1
 Individual(i) = Insert(a, Start, Individual(i))
 a = "G"
  l = Len(Individual(i))
  Start = Int(Rnd * l) + 1
 Individual(i) = Insert(a, Start, Individual(i))
 a = "A"
  l = Len(Individual(i))
  Start = Int(Rnd * l) + 1
 Individual(i) = Insert(a, Start, Individual(i))
 a = "E"
  l = Len(Individual(i))
  Start = Int(Rnd * l) + 1
 Individual(i) = Insert(a, Start, Individual(i))
End Sub


Private Sub Corrosion(i)
  Debug.Print "I am Corrosion(i) "
End Sub

Private Sub Reproduction()
  Debug.Print "I am Reproduction"
End Sub

Private Sub Mutation()
  Debug.Print "I am Mutation"
End Sub


Private Sub Dynamics()
Dim i As Integer
 'All individuals feed
  For i = 1 To NIndiv
   'The string Algae is digested and inserted
   'into the individual char by char
  Call Digestion(i)
  Next i
 'All individuals are tested by the parasite,
  'which corrodes any  large substring similar to it located at any end.
  For i = 1 To NIndiv
   Call Corrosion(i)
  Next i
 'The top ten are allowed to  reproduce:
 'a child occupies the place of the shortest individual
  Call Reproduction
  'The new population is subjected to mutation
  Call Mutation
End Sub

Private Sub Observation()
 Debug.Print "I am Observation "
End Sub



Public Sub Main()
Dim i As Integer
Dim NGen As Integer
'The population has NIndiv members
 NIndiv = 10
 Call Initialization
 'We run NGen generations
 NGen = 1
 For i = 1 To NGen
   Call Dynamics
   Call Observation
  Next i
End Sub

\end{verbatim}

\bigskip
\textbf{\thenum. }  \addtocounter{num}{1}  Exercise: Run the code and take your time to test  whether or not the procedure Digestion(i)  works here as softly as in isolation.

\bigskip
\textbf{\thenum. }  \addtocounter{num}{1}  Let us now fill in the procedure Corrosion(i). Why do we need this procedure? Because Digestion causes every individual to grow and so we would like a mechanism to refrain  growth. To that aim, we invented the String PARASITE that interacts with an individual in such a way that if the individual begins with PARA, that portion is deleted, and if the individual ends with SITE, that portion is deleted. We do not expect that our parasitic procedure could be very effective to stop growth because  our only interest is to  see how one can implement a game of two contrary forces in evolution. Our procedure Corrosion(i) can be tested in isolation:

\begin{verbatim}

Dim Individual(1 To 2)

Private Function Substitution(a, b, c)
'
'This Function replaces the substring a by b inside c
'
 Dim Start, la, lc As Integer
 Dim s1, s2, z

  'We get where a begins
  Start = InStr(c, a)
  'Debug.Print "a begins at position " & Start
  'We copy  the first part of c
  s1 = Left(c, Start - 1)
  'Debug.Print "The left part of c is " & s1
  'We    get the lengths of  the strings
  la = Len(a)
  'Debug.Print "The string  a has " & la & " characters"
   lc = Len(c)
  'Debug.Print "The string  c has " & lc & " characters"

  'We copy  the right part of c
  s2 = Right(c, lc - Start - la + 1)
  'Debug.Print "The Right part of c is " & s2

  'We concatenate the left part of c  to b to the right part of c
  Substitution = s1 & b & s2
End Function

Private Function Delete(a, c)
'This function deletes the first
'occurrence of a from c.
  b = ""
  Delete = Substitution(a, b, c)
End Function


Private Sub Corrosion(i)
'Here we see what a PARA-SITE does to individuals.
Dim l
Dim Beginning, Ending
 'The parasite deletes from the beginning of the individual
 'any substring   matching "PARA".
 Beginning = Left(Individual(i), 4)
 If Beginning = "PARA" Then
   Individual(i) = Delete(Beginning, Individual(i))
 End If
  'The parasite deletes from the tail of the individual
 'any substring   matching "SITE".
 Ending = Right(Individual(i), 4)
 l = Len(Individual(i))
 If Ending = "SITE" Then Individual(i) = Right(Individual(i), l - 3)
End Sub

Public Sub Test()
 Dim i As Integer
 Individual(1) = "PARALION"
 Individual(2) = "ELEPHANTSITE"
  For i = 1 To 2
   Call Corrosion(i)
   Debug.Print "The Individual "; i; "is "; Individual(i)
  Next i
End Sub

\end{verbatim}

\bigskip
\textbf{\thenum. }  \addtocounter{num}{1}  Exercise. Test the code to see that it fails to fill its function. It deletes  the left end of an input string if it matches PARA, but it fails to delete the right end of the string if it matches SITE. Fix the bug and compare your solution with that  found just below. Hint: prompt the help of VBA of Word with the input "String functions"' and look there what are exactly the properties of the functions Left and Right.

\bigskip
\textbf{\thenum. }  \addtocounter{num}{1}  A solution to the anti corrosion bug follows:

\begin{verbatim}

Dim Individual(1 To 2)

Private Function Substitution(a, b, c)
'
'This Function replaces the substring a by b inside c
'
 Dim Start, la, lc As Integer
 Dim s1, s2, z

  'We get where a begins
  Start = InStr(c, a)
  'Debug.Print "a begins at position " & Start
  'We copy  the first part of c
  s1 = Left(c, Start - 1)
  'Debug.Print "The left part of c is " & s1
  'We    get the lengths of  the strings
  la = Len(a)
  'Debug.Print "The string  a has " & la & " characters"
   lc = Len(c)
  'Debug.Print "The string  c has " & lc & " characters"

  'We copy  the right part of c
  s2 = Right(c, lc - Start - la + 1)
  'Debug.Print "The Right part of c is " & s2

  'We concatenate the left part of c  to b to the right part of c
  Substitution = s1 & b & s2
End Function

Private Function Delete(a, c)
'This function deletes the first
'occurrence of a from c.
  b = ""
  Delete = Substitution(a, b, c)
End Function


Private Sub Corrosion(i)
'Here we see what a PARA-SITE does to individuals.
Dim l
Dim Beginning, Ending
 'The parasite deletes from the beginning of the individual
 'any substring   matching "PARA".
 Beginning = Left(Individual(i), 4)
 If Beginning = "PARA" Then
    Individual(i) = Delete(Beginning, Individual(i))
  End If
  'The parasite deletes from the tail of the individual
 'any substring   matching "SITE".
 Ending = Right(Individual(i), 4)
 Debug.Print "The Beginning is "; Beginning
 Debug.Print "The End is "; Ending
 l = Len(Individual(i))
 If Ending = "SITE" Then Individual(i) = Left(Individual(i), l - 4)
End Sub

Public Sub Test()
 Dim i As Integer
 Individual(1) = "PARALION"
 Individual(2) = "ELEPHANTSITE"
  For i = 1 To 2
  Debug.Print "The original Individual "; i; "is "; Individual(i)
   Call Corrosion(i)
   Debug.Print "The corroded Individual "; i; "is "; Individual(i)
  Next i
End Sub

\end{verbatim}

Exercise. Run this code and compare it with yours.

We can insert now our procedure Corrosion(i) to our main code:

\begin{verbatim}

Option Explicit

Dim Individual(1 To 100)
Dim NIndiv As Integer
Dim PatternParasite, PatternFood

Private Function RandomChr()
 RandomChr = Chr(Int(Rnd * (1 + (90 - 65)) + 65))
End Function

Private Sub Initialization()
Dim i, j As Integer
 'We generate NIndiv individuals (strings)
 'ten characters long.
 'Sequences are completly random
 For i = 1 To NIndiv
  'An individual is assembled char by char
  Individual(i) = ""
  For j = 1 To 10
    Individual(i) = Individual(i) & RandomChr()
  Next j

 Next i
 'We generate a parasite pattern
 PatternParasite = "PARASITE"
 'We generate a food pattern
 PatternFood = "ALGAE"
End Sub


Private Function Insert(a, Start, c)
'
'This Function takes substring a
'and posits it at place Start inside c
'
 Dim lc As Integer
 Dim s1, s2, z

  'We copy  the left part of c
  s1 = Left(c, Start - 1)
  'Debug.Print "The left part of c is " & s1
  'We    get the lengths of  the string c
   lc = Len(c)
  'Debug.Print "The string  c has " & lc & " characters"
  'We copy  the right part of c
  s2 = Right(c, lc - Start + 1)
  'Debug.Print "The Right part of c is " & s2
  'We concatenate the left part of c  to a to the right part of c
  Insert = s1 & a & s2
End Function


Private Sub Digestion(i)
Dim a
Dim l, Start As Integer

  a = "A"
  l = Len(Individual(i))
  Start = Int(Rnd * l) + 1
 Individual(i) = Insert(a, Start, Individual(i))
 a = "L"
  l = Len(Individual(i))
  Start = Int(Rnd * l) + 1
 Individual(i) = Insert(a, Start, Individual(i))
 a = "G"
  l = Len(Individual(i))
  Start = Int(Rnd * l) + 1
 Individual(i) = Insert(a, Start, Individual(i))
 a = "A"
  l = Len(Individual(i))
  Start = Int(Rnd * l) + 1
 Individual(i) = Insert(a, Start, Individual(i))
 a = "E"
  l = Len(Individual(i))
  Start = Int(Rnd * l) + 1
 Individual(i) = Insert(a, Start, Individual(i))
End Sub




Private Function Substitution(a, b, c)
'
'This Function replaces the substring a by b inside c
'
 Dim Start, la, lc As Integer
 Dim s1, s2, z

  'We get where a begins
  Start = InStr(c, a)
  'Debug.Print "a begins at position " & Start
  'We copy  the first part of c
  s1 = Left(c, Start - 1)
  'Debug.Print "The left part of c is " & s1
  'We    get the lenghts of  the strings
  la = Len(a)
  'Debug.Print "The string  a has " & la & " characters"
   lc = Len(c)
  'Debug.Print "The string  c has " & lc & " characters"

  'We copy  the right part of c
  s2 = Right(c, lc - Start - la + 1)
  'Debug.Print "The Right part of c is " & s2

  'We concatenate the left part of c  to b to the right part of c
  Substitution = s1 & b & s2
End Function

Private Function Delete(a, c)
'This function deletes the first
'occurrence of a from c.
  b = ""
  Delete = Substitution(a, b, c)
End Function


Private Sub Corrosion(i)
'Here we see what a PARA-SITE does to individuals.
Dim l
Dim Beginning, Ending
 'The parasite deletes from the beginning of the individual
 'any substring   matching "PARA".
 Beginning = Left(Individual(i), 4)
 If Beginning = "PARA" Then
   Individual(i) = Delete(Beginning, Individual(i))
  End If
  'The parasite deletes from the tail of the individual
 'any substring   matching "SITE".
 Ending = Right(Individual(i), 4)
 Debug.Print "The Beginning is "; Beginning
 Debug.Print "The End is "; Ending
 l = Len(Individual(i))
 If Ending = "SITE" Then Individual(i) = Left(Individual(i), l - 4)
End Sub

Private Sub Reproduction()
  Debug.Print "I am Reproduction"
End Sub

Private Sub Mutation()
  Debug.Print "I am Mutation"
End Sub


Private Sub Dynamics()
Dim i As Integer
 'All individuals feed

  For i = 1 To NIndiv
   'The string Algae is digested and inserted
   'into the individual char by char
   Debug.Print "The hungry individual " & i & " is " & Individual(i)
   Call Digestion(i)
   Debug.Print "The full individual " & i & " is " & Individual(i)
 'All individuals are tested by the parasite,
  'which corroes any  large substring similar to it located at any end.
   Call Corrosion(i)
   Debug.Print "The corroded individual " & i & " is " & Individual(i)
  Next i
 'The top ten are allowed to  reproduce:
 'a child occupies the place of the shortest individual
  Call Reproduction
  'The new population is subjected to mutation
  Call Mutation
End Sub

Private Sub Observation()
 Debug.Print "I am Observation "
End Sub

Public Sub Main()
Dim i As Integer
Dim NGen As Integer
'The population has NIndiv members
 NIndiv = 10
 Call Initialization
 'We run NGen generations
 NGen = 1
 For i = 1 To NGen
   Call Dynamics
   Call Observation
  Next i
End Sub

\end{verbatim}

\bigskip
\textbf{\thenum. }  \addtocounter{num}{1}  Exercise. Run the code and verify that our anti growth measure is useless because the probability of matching the corroding prescription is almost nil. We have committed an error of design. This means that we are challenged to find a more effective balancing force to the growth tendency.

\bigskip
\textbf{\thenum. }  \addtocounter{num}{1}  We are forced to find a very effective purging measure. Let us delete a part of the string. The new code follows. It contains a new procedure, Purge(i):

\begin{verbatim}

Option Explicit

Dim Individual(1 To 100)
Dim NIndiv As Integer
Dim PatternParasite, PatternFood

Private Function RandomChr()
 RandomChr = Chr(Int(Rnd * (1 + (90 - 65)) + 65))
End Function

Private Sub Initialization()
Dim i, j As Integer
 'We generate NIndiv individuals (strings)
 'ten characters long.
 'Sequences are completely random
 For i = 1 To NIndiv
  'An individual is assembled char by char
  Individual(i) = ""
  For j = 1 To 10
    Individual(i) = Individual(i) & RandomChr()
  Next j

 Next i
 'We generate a parasite pattern
 PatternParasite = "PARASITE"
 'We generate a food pattern
 PatternFood = "ALGAE"
End Sub


Private Function Insert(a, Start, c)
'
'This Function takes substring a
'and posits it at place Start inside c
'
 Dim lc As Integer
 Dim s1, s2, z

  'We copy  the left part of c
  s1 = Left(c, Start - 1)
  'Debug.Print "The left part of c is " & s1
  'We    get the lengths of  the string c
   lc = Len(c)
  'Debug.Print "The string  c has " & lc & " characters"
  'We copy  the right part of c
  s2 = Right(c, lc - Start + 1)
  'Debug.Print "The Right part of c is " & s2
  'We concatenate the left part of c  to a to the right part of c
  Insert = s1 & a & s2
End Function


Private Sub Digestion(i)
Dim a
Dim l, Start As Integer

  a = "A"
  l = Len(Individual(i))
  Start = Int(Rnd * l) + 1
 Individual(i) = Insert(a, Start, Individual(i))
 a = "L"
  l = Len(Individual(i))
  Start = Int(Rnd * l) + 1
 Individual(i) = Insert(a, Start, Individual(i))
 a = "G"
  l = Len(Individual(i))
  Start = Int(Rnd * l) + 1
 Individual(i) = Insert(a, Start, Individual(i))
 a = "A"
  l = Len(Individual(i))
  Start = Int(Rnd * l) + 1
 Individual(i) = Insert(a, Start, Individual(i))
 a = "E"
  l = Len(Individual(i))
  Start = Int(Rnd * l) + 1
 Individual(i) = Insert(a, Start, Individual(i))
End Sub




Private Function Substitution(a, b, c)
'
'This Function replaces the substring a by b inside c
'
 Dim Start, la, lc As Integer
 Dim s1, s2, z

  'We get where a begins
  Start = InStr(c, a)
  'Debug.Print "a begins at position " & Start
  'We copy  the first part of c
  s1 = Left(c, Start - 1)
  'Debug.Print "The left part of c is " & s1
  'We    get the lengths of  the strings
  la = Len(a)
  'Debug.Print "The string  a has " & la & " characters"
   lc = Len(c)
  'Debug.Print "The string  c has " & lc & " characters"

  'We copy  the right part of c
  s2 = Right(c, lc - Start - la + 1)
  'Debug.Print "The Right part of c is " & s2

  'We concatenate the left part of c  to b to the right part of c
  Substitution = s1 & b & s2
End Function

Private Function Delete(a, c)
'This function deletes the first
'occurrence of a from c.
  b = ""
  Delete = Substitution(a, b, c)
End Function


Private Sub Corrosion(i)
'Here we see what a PARA-SITE does to individuals.
Dim l
Dim Beginning, Ending
 'The parasite deletes from the beginning of the individual
 'any substring   matching "PARA".
 Beginning = Left(Individual(i), 4)
 If Beginning = "PARA" Then
    Individual(i) = Delete(Beginning, Individual(i))
 End If
  'The parasite deletes from the tail of the individual
 'any substring   matching "SITE".
 Ending = Right(Individual(i), 4)
 'Debug.Print "The Beginning is "; Beginning
 'Debug.Print "The End is "; Ending
 l = Len(Individual(i))
 If Ending = "SITE" Then Individual(i) = Left(Individual(i), l - 4)
End Sub

Private Sub Purge(i)
Dim c, s1, s2
Dim lc, Start, n As Integer
 'In this procedure individuals are mutilated
 'as a measure to hinder overgrowth.
 'We delete at a random Start
 'a random number of chars from the original string.

 c = Individual(i)
 lc = Len(Individual(i))
 Start = Int(Rnd * lc) + 1
  'We copy  the first part of c
  s1 = Left(c, Start - 1)
  'Debug.Print "The left part of c is " & s1
  n = Int(Rnd * (lc - Start)) + 1

  'We copy  the right part of c
  s2 = Right(c, lc - Start - n + 1)
  'Debug.Print "The Right part of c is " & s2

  'We concatenate the left part of c  to  the right part of c
 Individual(i) = s1 & s2


End Sub

Private Sub Reproduction()
  Debug.Print "I am Reproduction";
End Sub

Private Sub Mutation()
  Debug.Print "I am Mutation"
End Sub


Private Sub Dynamics()
Dim i As Integer
 'All individuals feed

  For i = 1 To NIndiv
   'The string Algae is digested and inserted
   'into the individual char by char
   Debug.Print "The hungry individual " & i & " is " & Individual(i)
   Call Digestion(i)
   Debug.Print "The full individual " & i & " is " & Individual(i)
 'All individuals are tested by the parasite,
  'which corroes any  large substring similar to it located at any end.
   Call Corrosion(i)
   Debug.Print "The corroded individual " & i & " is " & Individual(i)
   Call Purge(i)
   Debug.Print "The purged  individual " & i & " is " & Individual(i)
  Next i
 'The top ten are allowed to  reproduce:
 'a child occupies the place of the shortest individual
  Call Reproduction
  'The new population is subjected to mutation
  Call Mutation
End Sub

Private Sub Observation()
 Debug.Print "I am Observation "
End Sub

Public Sub Main()
Dim i As Integer
Dim NGen As Integer
'The population has NIndiv members
 NIndiv = 10
 Call Initialization
 'We run NGen generations
 NGen = 1
 For i = 1 To NGen
   Call Dynamics
   Call Observation
  Next i
End Sub

\end{verbatim}

\bigskip
\textbf{\thenum. }  \addtocounter{num}{1}  Exercise. Test the code to verify that it is a good weapon against generalized overgrowth and that it produces a very good variability in the length of chains.

\bigskip
\textbf{\thenum. }  \addtocounter{num}{1}  Now that we have managed to produce variability, we at last may implement natural selection. That requires a fitness function. Let us take the fitness of an individual to be the length of the string, i.e., the number of chars that the string contains. Next, we sort individuals by fitness and we allow to reproduce only the top ten. We have a constant number of individuals, so reproduction of someone means the killing of another: a child of a top ten overwrites one of the ten  shortest individuals. Let us prove the reproduction procedure in isolation. Our proposal for the reproduction procedure is the following:

\begin{verbatim}

Dim Individual(1 To 5)
Dim Fitness(1 To 5)
Dim Order(1 To 5)

Private Sub Sorting()
Dim i, j, Champ As Integer
Dim Fitness(1 To 100) As Integer
 'We define a fitness function equal to the length of the string
 For i = 1 To NIndiv
  Fitness(i) = Len(Individual(i))
 Next i
 'We sort individual by fitness
 For i = 1 To NIndiv
 Champ = 1
 For j = 1 To NIndiv
  If Fitness(j) > Fitness(Champ) Then Champ = j
  Next j
  'The array Order keeps a record of fitness by decreasing order.
  Order(i) = Champ
  Fitness(Champ) = 0
Next i
End Sub

Private Sub Copying()
Dim k As Integer
 For i = 1 To NIndiv
  k = Order(NIndiv + 1 - i)
  Individual(k) = Individual(Order(i))
 Next i
End Sub

Private Sub Reproduction()
 'Individuals are sorted by length
 Call Sorting
 For i = 1 To 5
  Debug.Print Individual(Order(i))
 Next i
 'The top ten produce a copy that
 'substitutes the ten shorter ones.
 Call Copying
End Sub


Public Sub ReproTest()
 Individual(1) = "abcde"
 Individual(2) = "a "
 Individual(3) = "ade"
 Individual(4) = "ab de"
 Individual(5) = "ae"
 NIndiv = 5
 Call Reproduction
End Sub

\end{verbatim}

\bigskip
\textbf{\thenum. }  \addtocounter{num}{1}  Exercise. Test the code and verify  that it is inviable, i.e., the debugging machine rejects it.  The debugger automatically colors a sentence with yellow: Debug.Print Individual(Order(i)). PLease, drag the cursor to point to  \textbf{Order(i)}. The debugger will show that  \textbf{Order(i)=empty}. This means that we have problems with the filling of the array \textbf{Order}. Let us include a report of some  different values.

The reason of rejection is that we have an index outside the allowed range. Thus, we compose our own debugging work:

\begin{verbatim}

Option Explicit

Dim Individual(1 To 5)
Dim Order(1 To 5)
Dim NIndiv




Private Sub Sorting()
Dim i, j, Champ As Integer
Dim Fitness(1 To 100) As Integer
 'We define a fitness function equal to the lenght of the string
 For i = 1 To NIndiv
  Fitness(i) = Len(Individual(i))
 Next i
 'We sort individual by fitness
 For i = 1 To NIndiv
 Champ = 1
 For j = 1 To NIndiv
  If Fitness(j) > Fitness(Champ) Then Champ = j
  Next j
  'The array Order keeps a record of fitness by decreasing order.
  Order(i) = Champ
  Fitness(Champ) = 0
  Debug.Print  i & "th ind. is No " & Champ & " " & Individual(Champ)
 Next i

End Sub

Private Sub Copying()
Dim i, k As Integer

 For i = 1 To 2
  k = Order(NIndiv + 1 - i)
  Individual(k) = Individual(Order(i))
 Next i

 Debug.Print "THE NEW POPULATION IS"

 For i = 1 To NIndiv
 Debug.Print Individual(i)
 Next i

End Sub

Private Sub Reproduction()
Dim i As Integer
 'Individuals are sorted by length
 Call Sorting
 For i = 1 To 5
  Debug.Print Individual(Order(i))
 Next i
 'The top two produce a copy that
 'substitutes the two shorter ones.
 Call Copying
End Sub


Public Sub ReproTest()
 Individual(1) = "abcde"
 Individual(2) = "a"
 Individual(3) = "ade"
 Individual(4) = "abde"
 Individual(5) = "ae"
 NIndiv = 5
 Call Reproduction
End Sub

\end{verbatim}

\bigskip
\textbf{\thenum. }  \addtocounter{num}{1}  Exercise. Test the code and verify that it works: It sorts the original population by length and afterwards kills the two shorter individuals and replaces them by the two larger ones.

\bigskip
\textbf{\thenum. }  \addtocounter{num}{1}  Let us insert, with the appropriate number corrections,  the Reproduction procedure into the main code:

\begin{verbatim}

Option Explicit

Dim Individual(1 To 100)
Dim NIndiv As Integer
Dim PatternParasite, PatternFood
Dim Order(1 To 100) As Integer


Private Function RandomChr()
 RandomChr = Chr(Int(Rnd * (1 + (90 - 65)) + 65))
End Function

Private Sub Initialization()
Dim i, j As Integer
 'We generate NIndiv individuals (strings)
 'ten characters long.
 'Sequences are completely random
 For i = 1 To NIndiv
  'An individual is assembled char by char
  Individual(i) = ""
  For j = 1 To 10
    Individual(i) = Individual(i) & RandomChr()
  Next j

 Next i
 'We generate a parasite pattern
 PatternParasite = "PARASITE"
 'We generate a food pattern
 PatternFood = "ALGAE"
End Sub


Private Function Insert(a, Start, c)
'
'This Function takes substring a
'and posits it at place Start inside c
'
 Dim lc As Integer
 Dim s1, s2, z

  'We copy  the left part of c
  s1 = Left(c, Start - 1)
  'Debug.Print "The left part of c is " & s1
  'We    get the lengths of  the string c
   lc = Len(c)
  'Debug.Print "The string  c has " & lc & " characters"
  'We copy  the right part of c
  s2 = Right(c, lc - Start + 1)
  'Debug.Print "The Right part of c is " & s2
  'We concatenate the left part of c  to a to the right part of c
  Insert = s1 & a & s2
End Function


Private Sub Digestion(i)
Dim a
Dim l, Start As Integer

  a = "A"
  l = Len(Individual(i))
  Start = Int(Rnd * l) + 1
 Individual(i) = Insert(a, Start, Individual(i))
 a = "L"
  l = Len(Individual(i))
  Start = Int(Rnd * l) + 1
 Individual(i) = Insert(a, Start, Individual(i))
 a = "G"
  l = Len(Individual(i))
  Start = Int(Rnd * l) + 1
 Individual(i) = Insert(a, Start, Individual(i))
 a = "A"
  l = Len(Individual(i))
  Start = Int(Rnd * l) + 1
 Individual(i) = Insert(a, Start, Individual(i))
 a = "E"
  l = Len(Individual(i))
  Start = Int(Rnd * l) + 1
 Individual(i) = Insert(a, Start, Individual(i))
End Sub




Private Function Substitution(a, b, c)
'
'This Function replaces the substring a by b inside c
'
 Dim Start, la, lc As Integer
 Dim s1, s2, z

  'We get where a begins
  Start = InStr(c, a)
  'Debug.Print "a begins at position " & Start
  'We copy  the first part of c
  s1 = Left(c, Start - 1)
  'Debug.Print "The left part of c is " & s1
  'We    get the lengths of  the strings
  la = Len(a)
  'Debug.Print "The string  a has " & la & " characters"
   lc = Len(c)
  'Debug.Print "The string  c has " & lc & " characters"

  'We copy  the right part of c
  s2 = Right(c, lc - Start - la + 1)
  'Debug.Print "The Right part of c is " & s2

  'We concatenate the left part of c  to b to the right part of c
  Substitution = s1 & b & s2
End Function

Private Function Delete(a, c)
'This function deletes the first
'occurrence of a from c.
  b = ""
  Delete = Substitution(a, b, c)
End Function


Private Sub Corrosion(i)
'Here we see what a PARA-SITE does to individuals.
Dim l
Dim Beginning, Ending
 'The parasite deletes from the beginning of the individual
 'any substring   matching "PARA".
 Beginning = Left(Individual(i), 4)
 If Beginning = "PARA" Then
   Individual(i) = Delete(Beginning, Individual(i))
  End If
  'The parasite deletes from the tail of the individual
 'any substring   matching "SITE".
 Ending = Right(Individual(i), 4)
 'Debug.Print "The Beginning is "; Beginning
 'Debug.Print "The End is "; Ending
 l = Len(Individual(i))
 If Ending = "SITE" Then Individual(i) = Left(Individual(i), l - 4)
End Sub

Private Sub Purge(i)
Dim c, s1, s2
Dim lc, Start, n As Integer
 'In this procedure individuals are mutilated
 'as a measure to hinder overgrowth.
 'We delete at a random Start
 'a random number of chars from the original string.

 c = Individual(i)
 lc = Len(Individual(i))
 Start = Int(Rnd * lc) + 1
  'We copy  the first part of c
  s1 = Left(c, Start - 1)
  'Debug.Print "The left part of c is " & s1
  n = Int(Rnd * (lc - Start)) + 1

  'We copy  the right part of c
  s2 = Right(c, lc - Start - n + 1)
  'Debug.Print "The Right part of c is " & s2

  'We concatenate the left part of c  to  the right part of c
 Individual(i) = s1 & s2


End Sub


Private Sub Sorting()
Dim i, j, Champ As Integer
Dim Fitness(1 To 100) As Integer
 'We define a fitness function equal to the lenght of the string
 For i = 1 To NIndiv
  Fitness(i) = Len(Individual(i))
 Next i
 'We sort individual by fitness
 For i = 1 To NIndiv
 Champ = 1
 For j = 1 To NIndiv
  If Fitness(j) > Fitness(Champ) Then Champ = j
  Next j
  'The array Order keeps a record of fitness by decreasing order.
  Order(i) = Champ
  Fitness(Champ) = 0
  Debug.Print  i & "th ind. is No " & Champ & " " & Individual(Champ)
 Next i

End Sub

Private Sub Copying()
Dim i, k As Integer

 For i = 1 To 10
  k = Order(NIndiv + 1 - i)
  Individual(k) = Individual(Order(i))
 Next i

 Debug.Print "THE NEW POPULATION IS"

 For i = 1 To NIndiv
 Debug.Print Individual(i)
 Next i

End Sub

Private Sub Reproduction()
Dim i As Integer
 'Individuals are sorted by length
 Call Sorting
 For i = 1 To 10
  Debug.Print Individual(Order(i))
 Next i
 'The top ten produce a copy that
 'substitutes the bottom ten.
 Call Copying
End Sub

Private Sub Mutation()
  Debug.Print "I am Mutation"
End Sub


Private Sub Dynamics()
Dim i As Integer
 'All individuals feed

  For i = 1 To NIndiv
   'The string Algae is digested and inserted
   'into the individual char by char
   Debug.Print "The hungry individual " & i & " is " & Individual(i)
   Call Digestion(i)
   Debug.Print "The full individual " & i & " is " & Individual(i)
 'All individuals are tested by the parasite,
  'which corroes any  large substring similar to it located at any end.
   Call Corrosion(i)
   Debug.Print "The corroded individual " & i & " is " & Individual(i)
   Call Purge(i)
   Debug.Print "The purged  individual " & i & " is " & Individual(i)
  Next i
 'The top ten are allowed to  reproduce:
 'a child occupies the place of the shortest individual
  Call Reproduction
  'The new population is subjected to mutation
  Call Mutation
End Sub

Private Sub Observation()
 Debug.Print "I am Observation "
End Sub

Public Sub Main()
Dim i As Integer
Dim NGen As Integer
'The population has NIndiv members, NIndiv >10
 NIndiv = 15
 Call Initialization
 'We run NGen generations
 NGen = 1
 For i = 1 To NGen
   Call Dynamics
   Call Observation
  Next i
End Sub

\end{verbatim}

\bigskip
\textbf{\thenum. }  \addtocounter{num}{1}  Exercise. Run the code and test whether or not it works properly.

\bigskip
\textbf{\thenum. }  \addtocounter{num}{1}  Now, we implement a mutational process. Since we have a very robust reproductive system that tends to enlarge the average length, we could insert a  deleting process together with another process to enhance the possibilities of parasitism. Reversion of the tail could be a good option.

We delete a random number of chars form the beginning such as it is shown in the next module:

\begin{verbatim}


Dim Individual(1 To 100)

Private Function Substitution(a, b, c)
'
'This Function replaces the substring a by b inside c
'
 Dim Start, la, lc As Integer
 Dim s1, s2, z

  'We get where a begins
  Start = InStr(c, a)
  'Debug.Print "a begins at position " & Start
  'We copy  the first part of c
  s1 = Left(c, Start - 1)
  'Debug.Print "The left part of c is " & s1
  'We    get the lengths of  the strings
  la = Len(a)
  'Debug.Print "The string  a has " & la & " characters"
   lc = Len(c)
  'Debug.Print "The string  c has " & lc & " characters"

  'We copy  the right part of c
  s2 = Right(c, lc - Start - la + 1)
  'Debug.Print "The Right part of c is " & s2

  'We concatenate the left part of c  to b to the right part of c
  Substitution = s1 & b & s2
End Function

Private Function Delete(a, c)
'This function deletes the first
'occurrence of a from c.
  b = ""
  Delete = Substitution(a, b, c)
End Function

Private Sub Deletion(i)
Dim Beginning
  l = Len(Individual(i))
  n = Int(Rnd * l) + 1
  Beginning = Left(Individual(i), n)
  Individual(i) = Delete(Beginning, Individual(i))
 End Sub


Public Sub test()
 Individual(1) = "123456789"
 i = 1
 Call Deletion(i)
 Debug.Print Individual(1)
End Sub

\end{verbatim}

\bigskip
\textbf{\thenum. }  \addtocounter{num}{1}  Exercise. Run the code some times in tandem to see that it works properly.

\bigskip
\textbf{\thenum. }  \addtocounter{num}{1}  Let us work now the Inversion procedure:

\begin{verbatim}

Dim Individual(1 To 100)


Private Function Inversion(a)
 'This function inverts a
 Dim la, i, Start As Integer
 Dim d, f

 'We initialize the output
 d = ""

 'We get the length of a
 la = Len(a)

 'We copy the first character to the last place
 'and so on.

 For i = 1 To la
 Start = i
 chari = Mid(a, Start, 1)

  d = chari & d
 Next i
 Inversion = d

End Function


Private Function Invert(a, c)
'
'This Function inverts the first occurrece
'of substring a  inside c
'
 Dim Start, la, lc As Integer
 Dim s1, s2, z

  'We get where we must begin
  Start = Len(c) - 3
  Debug.Print "a begins at position " & Start
  'We copy  the first part of c
  s1 = Left(c, Start - 1)
  Debug.Print "The left part of c is " & s1
  'We    get the lengths of  the strings
  la = Len(a)
  Debug.Print "The string  a has " & la & " characters"
   lc = Len(c)
  Debug.Print "The string  c has " & lc & " characters"
  'We invert a properly
  d = Inversion(a)
  'We concatenate the left part of c  to d
  Invert = s1 & d & s2
End Function


Public Sub InversionTail()
Dim z
'We  invert the last part of each individual.
  i = 1
  Individual(1) = "123456789"
  a = Right(Individual(i), 3)
  c = Individual(i)
  z = Invert(a, c)
  Debug.Print "The output string is  " & z
End Sub

\end{verbatim}

 Exercise. Run the code, detect a bug and fix it. And compare your solution with the one below:

\begin{verbatim}

Dim Individual(1 To 100)


Private Function Inversion(a)
 'This function inverts a
 Dim la, i, Start As Integer
 Dim d, f

 'We initialize the output
 d = ""

 'We get the length of a
 la = Len(a)

 'We copy the first character to the last place
 'and so on.

 For i = 1 To la
 Start = i
 chari = Mid(a, Start, 1)

  d = chari & d
 Next i
 Inversion = d

End Function


Private Function Invert(a, c)
'
'This Function inverts the first occurrence
'of substring a  inside c
'
 Dim Start, la, lc As Integer
 Dim s1, z

  'We get where we must begin
  Start = Len(c) - 3
  Debug.Print "a begins at position " & Start
  'We copy  the first part of c
  s1 = Left(c, Start)
  Debug.Print "The left part of c is " & s1
  'We    get the lengths of  the strings
  la = Len(a)
  Debug.Print "The string  a has " & la & " characters"
   lc = Len(c)
  Debug.Print "The string  c has " & lc & " characters"
  'We invert a properly
  d = Inversion(a)
  'We concatenate the left part of c  to d
  Invert = s1 & d
End Function


Public Sub InversionTail()
Dim z
'We  invert the last part of each individual.
  i = 1
  Individual(1) = "123456789"
  a = Right(Individual(i), 3)
  c = Individual(i)
  z = Invert(a, c)
  Debug.Print "The output string is  " & z
End Sub

\end{verbatim}

We can now assembled our new procedures into the whole program

\begin{verbatim}

Option Explicit

Dim Individual(1 To 100)
Dim NIndiv As Integer
Dim PatternParasite, PatternFood
Dim Order(1 To 100) As Integer
Dim b

Private Function RandomChr()
 RandomChr = Chr(Int(Rnd * (1 + (90 - 65)) + 65))
End Function

Private Sub Initialization()
Dim i, j As Integer
 'We generate NIndiv individuals (strings)
 'ten characters long.
 'Sequences are completely random
 For i = 1 To NIndiv
  'An individual is assembled char by char
  Individual(i) = ""
  For j = 1 To 10
    Individual(i) = Individual(i) & RandomChr()
  Next j

 Next i
 'We generate a parasite pattern
 PatternParasite = "PARASITE"
 'We generate a food pattern
 PatternFood = "ALGAE"
End Sub


Private Function Insert(a, Start, c)
'
'This Function takes substring a
'and posits it at place Start inside c
'
 Dim lc As Integer
 Dim s1, s2, z

  'We copy  the left part of c
  s1 = Left(c, Start - 1)
  'Debug.Print "The left part of c is " & s1
  'We    get the lengths of  the string c
   lc = Len(c)
  'Debug.Print "The string  c has " & lc & " characters"
  'We copy  the right part of c
  s2 = Right(c, lc - Start + 1)
  'Debug.Print "The Right part of c is " & s2
  'We concatenate the left part of c  to a to the right part of c
  Insert = s1 & a & s2
End Function


Private Sub Digestion(i)
Dim a
Dim l, Start As Integer

  a = "A"
  l = Len(Individual(i))
  Start = Int(Rnd * l) + 1
 Individual(i) = Insert(a, Start, Individual(i))
 a = "L"
  l = Len(Individual(i))
  Start = Int(Rnd * l) + 1
 Individual(i) = Insert(a, Start, Individual(i))
 a = "G"
  l = Len(Individual(i))
  Start = Int(Rnd * l) + 1
 Individual(i) = Insert(a, Start, Individual(i))
 a = "A"
  l = Len(Individual(i))
  Start = Int(Rnd * l) + 1
 Individual(i) = Insert(a, Start, Individual(i))
 a = "E"
  l = Len(Individual(i))
  Start = Int(Rnd * l) + 1
 Individual(i) = Insert(a, Start, Individual(i))
End Sub

Private Function Substitution(a, b, c)
'
'This Function replaces the substring a by b inside c
'
 Dim Start, la, lc As Integer
 Dim s1, s2, z

  'We get where a begins
  Start = InStr(c, a)
  'Debug.Print "a begins at position " & Start
  'We copy  the first part of c
  s1 = Left(c, Start - 1)
  'Debug.Print "The left part of c is " & s1
  'We    get the lengths of  the strings
  la = Len(a)
  'Debug.Print "The string  a has " & la & " characters"
   lc = Len(c)
  'Debug.Print "The string  c has " & lc & " characters"

  'We copy  the right part of c
  s2 = Right(c, lc - Start - la + 1)
  'Debug.Print "The Right part of c is " & s2

  'We concatenate the left part of c  to b to the right part of c
  Substitution = s1 & b & s2
End Function

Private Function Delete(a, c)
'This function deletes the first
'occurrence of a from c.
  b = ""
  Delete = Substitution(a, b, c)
End Function


Private Sub Corrosion(i)
'Here we see what a PARA-SITE does to individuals.
Dim l
Dim Beginning, Ending
 'The parasite deletes from the beginning of the individual
 'any substring   matching "PARA".
 Beginning = Left(Individual(i), 4)
 If Beginning = "PARA" Then
  Individual(i) = Delete(Beginning, Individual(i))
 End If
  'The parasite deletes from the tail of the individual
 'any substring   matching "SITE".
 Ending = Right(Individual(i), 4)
 'Debug.Print "The Beginning is "; Beginning
 'Debug.Print "The End is "; Ending
 l = Len(Individual(i))
 If Ending = "SITE" Then Individual(i) = Left(Individual(i), l - 4)
End Sub

Private Sub Purge(i)
Dim c, s1, s2
Dim lc, Start, n As Integer
 'In this procedure individuals are mutilated
 'as a measure to hinder overgrowth.
 'We delete at a random Start
 'a random number of chars from the original string.

 c = Individual(i)
 lc = Len(Individual(i))
 Start = Int(Rnd * lc) + 1
  'We copy  the first part of c
  s1 = Left(c, Start - 1)
  'Debug.Print "The left part of c is " & s1
  n = Int(Rnd * (lc - Start)) + 1

  'We copy  the right part of c
  s2 = Right(c, lc - Start - n + 1)
  'Debug.Print "The Right part of c is " & s2

  'We concatenate the left part of c  to  the right part of c
 Individual(i) = s1 & s2


End Sub


Private Sub Sorting()
Dim i, j, Champ As Integer
Dim Fitness(1 To 100) As Integer
 'We define a fitness function equal to the length of the string
 For i = 1 To NIndiv
  Fitness(i) = Len(Individual(i))
 Next i
 'We sort individual by fitness
 For i = 1 To NIndiv
 Champ = 1
 For j = 1 To NIndiv
  If Fitness(j) > Fitness(Champ) Then Champ = j
  Next j
  'The array Order keeps a record of fitness by decreasing order.
  Order(i) = Champ
  Fitness(Champ) = 0
  Debug.Print  i & "th ind. is No " & Champ & " " & Individual(Champ)
 Next i

End Sub

Private Sub Copying()
Dim i, k As Integer

 For i = 1 To 10
  k = Order(NIndiv + 1 - i)
  Individual(k) = Individual(Order(i))
 Next i

 Debug.Print "THE NEW POPULATION IS"

 For i = 1 To NIndiv
 Debug.Print Individual(i)
 Next i

End Sub

Private Sub Reproduction()
Dim i As Integer
 'Individuals are sorted by length
 Call Sorting
 For i = 1 To 10
  Debug.Print Individual(Order(i))
 Next i
 'The top ten produce a copy that
 'substitutes the bottom ten.
 Call Copying
End Sub


Private Sub Deletion(i)
Dim l, n As Integer
Dim Beginning
  l = Len(Individual(i))
  n = Int(Rnd * l) + 1
  Beginning = Left(Individual(i), n)
  Individual(i) = Delete(Beginning, Individual(i))
 End Sub


Private Function Inversion(a)
 'This function inverts a
 Dim la, i, Start As Integer
 Dim d, f

 'We initialize the output
 d = ""

 'We get the length of a
 la = Len(a)

 'We copy the first character to the last place
 'and so on.

 For i = 1 To la
 Start = i
 chari = Mid(a, Start, 1)

  d = chari & d
 Next i
 Inversion = d

End Function


Private Function Invert(a, c)
'
'This Function inverts the first occurrence
'of substring a  inside c
'
 Dim Start, la, lc As Integer
 Dim s1, z

  'We get where we must begin
  Start = Len(c) - 3
  Debug.Print "a begins at position " & Start
  'We copy  the first part of c
  s1 = Left(c, Start)
  Debug.Print "The left part of c is " & s1
  'We    get the lenghts of  the strings
  la = Len(a)
  Debug.Print "The string  a has " & la & " characters"
   lc = Len(c)
  Debug.Print "The string  c has " & lc & " characters"
  'We invert a properly
  d = Inversion(a)
  'We concatenate the left part of c  to d
  Invert = s1 & d
End Function


Private Sub InversionTail(i)
Dim z
Dim l
'We  invert the last part of each individual.
  l = Len(Individual(i))
  If l > 2 Then
    a = Right(Individual(i), 3)
    c = Individual(i)
    z = Invert(a, c)
  End If
  Debug.Print "The output string is  " & z
End Sub



Private Sub Mutation()
Dim Luck As Integer
Dim i As Integer
 'We consider only two types of mutation:
 'deletion of the beginning
 ' and inversion of the end.

 'Which mutation affects which individual
 'is a matter of gambling.
 'We throw a 0 else a 1
 For i = 1 To NIndiv
  Luck = Int(Rnd + 1)
  If Luck = 1 Then Call Deletion(i) Else Call InversionTail(i)
 Next i
End Sub


Private Sub Dynamics()
Dim i As Integer
 'All individuals feed

  For i = 1 To NIndiv
   'The string Algae is digested and inserted
   'into the individual char by char
   Debug.Print "The hungry individual " & i & " is " & Individual(i)
   Call Digestion(i)
   Debug.Print "The full individual " & i & " is " & Individual(i)
 'All individuals are tested by the parasite,
  'which corroes any  large substring similar to it located at any end.
   Call Corrosion(i)
   Debug.Print "The corroed individual " & i & " is " & Individual(i)
   Call Purge(i)
   Debug.Print "The purged  individual " & i & " is " & Individual(i)
  Next i
 'The top ten are allowed to  reproduce:
 'a child occupies the place of the shortest individual
  Call Reproduction
  'The new population is subjected to mutation
  Call Mutation
End Sub

Private Sub Observation()
 Debug.Print "I am Observation "
End Sub

Public Sub Main()
Dim i As Integer
Dim NGen As Integer
'The population has NIndiv members, NIndiv >10
 NIndiv = 15
 Call Initialization
 'We run NGen generations
 NGen = 1
 For i = 1 To NGen
   Call Dynamics
   Call Observation
  Next i
End Sub

\end{verbatim}

\bigskip
\textbf{\thenum. }  \addtocounter{num}{1}  Exercise. Run the code and take a minute  to understand the bases of your security in the  belief that the code does exactly what it has been planned to do.

\bigskip
\textbf{\thenum. }  \addtocounter{num}{1}  We can inbuilt now an observer that looks for the minimum and maximum lengths at each generation. Since this offers no trouble, let us write it directly into the main code. The final code follows:

\begin{verbatim}

Option Explicit

Dim Individual(1 To 100)
Dim NIndiv As Integer
Dim PatternParasite, PatternFood
Dim Order(1 To 100) As Integer
Dim b
Dim Gen As Integer


Private Function RandomChr()
 RandomChr = Chr(Int(Rnd * (1 + (90 - 65)) + 65))
End Function

Private Sub Initialization()
Dim i, j As Integer
 'We generate NIndiv individuals (strings)
 'ten characters long.
 'Sequences are completely random
 For i = 1 To NIndiv
  'An individual is assembled char by char
  Individual(i) = ""
  For j = 1 To 10
    Individual(i) = Individual(i) & RandomChr()
  Next j

 Next i
 'We generate a parasite pattern
 PatternParasite = "PARASITE"
 'We generate a food pattern
 PatternFood = "ALGAE"
End Sub


Private Function Insert(a, Start, c)
'
'This Function takes substring a
'and posits it at place Start inside c
'
 Dim lc As Integer
 Dim s1, s2, z

  'We copy  the left part of c
  s1 = Left(c, Start - 1)
  'Debug.Print "The left part of c is " & s1
  'We    get the lenghts of  the string c
   lc = Len(c)
  'Debug.Print "The string  c has " & lc & " characters"
  'We copy  the right part of c
  s2 = Right(c, lc - Start + 1)
  'Debug.Print "The Right part of c is " & s2
  'We concatenate the left part of c  to a to the right part of c
  Insert = s1 & a & s2
End Function


Private Sub Digestion(i)
Dim a
Dim l, Start As Integer

  a = "A"
  l = Len(Individual(i))
  Start = Int(Rnd * l) + 1
 Individual(i) = Insert(a, Start, Individual(i))
 a = "L"
  l = Len(Individual(i))
  Start = Int(Rnd * l) + 1
 Individual(i) = Insert(a, Start, Individual(i))
 a = "G"
  l = Len(Individual(i))
  Start = Int(Rnd * l) + 1
 Individual(i) = Insert(a, Start, Individual(i))
 a = "A"
  l = Len(Individual(i))
  Start = Int(Rnd * l) + 1
 Individual(i) = Insert(a, Start, Individual(i))
 a = "E"
  l = Len(Individual(i))
  Start = Int(Rnd * l) + 1
 Individual(i) = Insert(a, Start, Individual(i))
End Sub

Private Function Substitution(a, b, c)
'
'This Function replaces the substring a by b inside c
'
 Dim Start, la, lc As Integer
 Dim s1, s2, z

  'We get where a begins
  Start = InStr(c, a)
  'Debug.Print "a begins at position " & Start
  'We copy  the first part of c
  s1 = Left(c, Start - 1)
  'Debug.Print "The left part of c is " & s1
  'We    get the lengths of  the strings
  la = Len(a)
  'Debug.Print "The string  a has " & la & " characters"
   lc = Len(c)
  'Debug.Print "The string  c has " & lc & " characters"

  'We copy  the right part of c
  s2 = Right(c, lc - Start - la + 1)
  'Debug.Print "The Right part of c is " & s2

  'We concatenate the left part of c  to b to the right part of c
  Substitution = s1 & b & s2
End Function

Private Function Delete(a, c)
'This function deletes the first
'occurrence of a from c.
  b = ""
  Delete = Substitution(a, b, c)
End Function


Private Sub Corrosion(i)
'Here we see what a PARA-SITE does to individuals.
Dim l
Dim Beginning, Ending
 'The parasite deletes from the beginning of the individual
 'any substring   matching "PARA".
 Beginning = Left(Individual(i), 4)
 If Beginning = "PARA" Then
   Individual(i) = Delete(Beginning, Individual(i))
 End If
  'The parasite deletes from the tail of the individual
 'any substring   matching "SITE".
 Ending = Right(Individual(i), 4)
 'Debug.Print "The Beginning is "; Beginning
 'Debug.Print "The End is "; Ending
 l = Len(Individual(i))
 If Ending = "SITE" Then Individual(i) = Left(Individual(i), l - 4)
End Sub

Private Sub Purge(i)
Dim c, s1, s2
Dim lc, Start, n As Integer
 'In this procedure individuals are mutilated
 'as a measure to hinder overgrowth.
 'We delete at a random Start
 'a random number of chars from the original string.

 c = Individual(i)
 lc = Len(Individual(i))
 Start = Int(Rnd * lc) + 1
  'We copy  the first part of c
  s1 = Left(c, Start - 1)
  'Debug.Print "The left part of c is " & s1
  n = Int(Rnd * (lc - Start)) + 1

  'We copy  the right part of c
  s2 = Right(c, lc - Start - n + 1)
  'Debug.Print "The Right part of c is " & s2

  'We concatenate the left part of c  to  the right part of c
 Individual(i) = s1 & s2


End Sub


Private Sub Sorting()
Dim i, j, Champ As Integer
Dim Fitness(1 To 100) As Integer
 'We define a fitness function equal to the length of the string
 For i = 1 To NIndiv
  Fitness(i) = Len(Individual(i))
 Next i
 'We sort individual by fitness
 For i = 1 To NIndiv
 Champ = 1
 For j = 1 To NIndiv
  If Fitness(j) > Fitness(Champ) Then Champ = j
  Next j
  'The array Order keeps a record of fitness by decreasing order.
  Order(i) = Champ
  Fitness(Champ) = 0
  Debug.Print i & "th ind. is No " & Champ & " " & Individual(Champ)
 Next i

End Sub

Private Sub Copying()
Dim i, k As Integer

 For i = 1 To 10
  k = Order(NIndiv + 1 - i)
  Individual(k) = Individual(Order(i))
 Next i

 Debug.Print "THE NEW POPULATION IS"

 For i = 1 To NIndiv
 Debug.Print Individual(i)
 Next i

End Sub

Private Sub Reproduction()
Dim i As Integer
 'Individuals are sorted by length
 Call Sorting
 For i = 1 To 10
  Debug.Print Individual(Order(i))
 Next i
 'The top ten produce a copy that
 'substitutes the bottom ten.
 Call Copying
End Sub


Private Sub Deletion(i)
Dim l, n As Integer
Dim Beginning
  l = Len(Individual(i))
  n = Int(Rnd * l) + 1
  Beginning = Left(Individual(i), n)
  Individual(i) = Delete(Beginning, Individual(i))
 End Sub


Private Function Inversion(a)
 'This function inverts a
 Dim la, i, Start As Integer
 Dim d, f

 'We initialize the output
 d = ""

 'We get the lenght of a
 la = Len(a)

 'We copy the first character to the last place
 'and so on.

 For i = 1 To la
 Start = i
 chari = Mid(a, Start, 1)

  d = chari & d
 Next i
 Inversion = d

End Function


Private Function Invert(a, c)
'
'This Function inverts the first occurrence
'of substring a  inside c
'
 Dim Start, la, lc As Integer
 Dim s1, z

  'We get where we must begin
  Start = Len(c) - 3
  Debug.Print "a begins at position " & Start
  'We copy  the first part of c
  s1 = Left(c, Start)
  Debug.Print "The left part of c is " & s1
  'We    get the lengths of  the strings
  la = Len(a)
  Debug.Print "The string  a has " & la & " characters"
   lc = Len(c)
  Debug.Print "The string  c has " & lc & " characters"
  'We invert a properly
  d = Inversion(a)
  'We concatenate the left part of c  to d
  Invert = s1 & d
End Function


Private Sub InversionTail(i)
Dim z
Dim l
'We  invert the last part of each individual.
  l = Len(Individual(i))
  If l > 2 Then
    a = Right(Individual(i), 3)
    c = Individual(i)
    z = Invert(a, c)
  End If
  Debug.Print "The output string is  " & z
End Sub



Private Sub Mutation()
Dim Luck As Integer
Dim i As Integer
 'We consider only two types of mutation:
 'deletion of the beginning
 ' and inversion of the end.

 'Which mutation affects which individual
 'is a matter of gambling.
 'We throw a 0 else a 1
 For i = 1 To NIndiv
  Luck = Int(Rnd + 1)
  If Luck = 1 Then Call Deletion(i) Else Call InversionTail(i)
 Next i
End Sub


Private Sub Dynamics()
Dim i As Integer
 'All individuals feed

  For i = 1 To NIndiv
   'The string Algae is digested and inserted
   'into the individual char by char
   Debug.Print "The hungry individual " & i & " is " & Individual(i)
   Call Digestion(i)
   Debug.Print "The full individual " & i & " is " & Individual(i)
 'All individuals are tested by the parasite,
  'which corroes any  large substring similar to it located at any end.
   Call Corrosion(i)
   Debug.Print "The corroded individual " & i & " is " & Individual(i)
   Call Purge(i)
   Debug.Print "The purged  individual " & i & " is " & Individual(i)
  Next i
 'The top ten are allowed to  reproduce:
 'a child occupies the place of the shortest individual
  Call Reproduction
  'The new population is subjected to mutation
  Call Mutation
End Sub

Private Sub Observation(Gen)
Dim j As Integer
Dim m, Min, Max As Integer

Max = 0
Min = 32000
For j = 1 To NIndiv
 m = Len(Individual(j))
 Debug.Print "The Individual  " & Individual(j) & " has length " & m
 If m > Max Then Max = m
 If m < Min Then Min = m
Next j
 Debug.Print "Generation " & Gen & " MinLength = " & Min & " MaxLength= " & Max
End Sub

Public Sub Main()
Dim i As Integer
Dim NGen
'The population has NIndiv members, NIndiv >10
 NIndiv = 15
 Call Initialization
 'We run NGen generations
 NGen = 1
 For Gen = 1 To NGen
   Call Dynamics
   Call Observation(Gen)
  Next Gen
End Sub

\end{verbatim}

\bigskip
\textbf{\thenum. }  \addtocounter{num}{1}  Exercise. Run the code and test that the observer does his work accurately. Next, close all debugging helps and keep only  the report of the observer. Change the number of generations from 1 to 10. Run the new code various times to try to catch a tendency in the evolution.

\bigskip
\textbf{\thenum. }  \addtocounter{num}{1}  We have seen that the mutational process has been a good balancing counter force to the process of overgrowth. But, we would like to see a graphic. So, we report apart the minimum values from the maximal ones:

\begin{verbatim}

Option Explicit

Dim Individual(1 To 100)
Dim NIndiv As Integer
Dim PatternParasite, PatternFood
Dim Order(1 To 100) As Integer
Dim b
Dim Gen As Integer
Dim ReportMin(1 To 100), ReportMax(1 To 100) As Integer


Private Function RandomChr()
 RandomChr = Chr(Int(Rnd * (1 + (90 - 65)) + 65))
End Function

Private Sub Initialization()
Dim i, j As Integer
 'We generate NIndiv individuals (strings)
 'ten characters long.
 'Sequences are completely random
 For i = 1 To NIndiv
  'An individual is assembled char by char
  Individual(i) = ""
  For j = 1 To 10
    Individual(i) = Individual(i) & RandomChr()
  Next j

 Next i
 'We generate a parasite pattern
 PatternParasite = "PARASITE"
 'We generate a food pattern
 PatternFood = "ALGAE"
End Sub


Private Function Insert(a, Start, c)
'
'This Function takes substring a
'and posits it at place Start inside c
'
 Dim lc As Integer
 Dim s1, s2, z

  'We copy  the left part of c
  s1 = Left(c, Start - 1)
  'Debug.Print "The left part of c is " & s1
  'We    get the lengths of  the string c
   lc = Len(c)
  'Debug.Print "The string  c has " & lc & " characters"
  'We copy  the right part of c
  s2 = Right(c, lc - Start + 1)
  'Debug.Print "The Right part of c is " & s2
  'We concatenate the left part of c  to a to the right part of c
  Insert = s1 & a & s2
End Function


Private Sub Digestion(i)
Dim a
Dim l, Start As Integer

  a = "A"
  l = Len(Individual(i))
  Start = Int(Rnd * l) + 1
 Individual(i) = Insert(a, Start, Individual(i))
 a = "L"
  l = Len(Individual(i))
  Start = Int(Rnd * l) + 1
 Individual(i) = Insert(a, Start, Individual(i))
 a = "G"
  l = Len(Individual(i))
  Start = Int(Rnd * l) + 1
 Individual(i) = Insert(a, Start, Individual(i))
 a = "A"
  l = Len(Individual(i))
  Start = Int(Rnd * l) + 1
 Individual(i) = Insert(a, Start, Individual(i))
 a = "E"
  l = Len(Individual(i))
  Start = Int(Rnd * l) + 1
 Individual(i) = Insert(a, Start, Individual(i))
End Sub

Private Function Substitution(a, b, c)
'
'This Function replaces the substring a by b inside c
'
 Dim Start, la, lc As Integer
 Dim s1, s2, z

  'We get where a begins
  Start = InStr(c, a)
  'Debug.Print "a begins at position " & Start
  'We copy  the first part of c
  s1 = Left(c, Start - 1)
  'Debug.Print "The left part of c is " & s1
  'We    get the lengths of  the strings
  la = Len(a)
  'Debug.Print "The string  a has " & la & " characters"
   lc = Len(c)
  'Debug.Print "The string  c has " & lc & " characters"

  'We copy  the right part of c
  s2 = Right(c, lc - Start - la + 1)
  'Debug.Print "The Right part of c is " & s2

  'We concatenate the left part of c  to b to the right part of c
  Substitution = s1 & b & s2
End Function

Private Function Delete(a, c)
'This function deletes the first
'occurrence of a from c.
  b = ""
  Delete = Substitution(a, b, c)
End Function


Private Sub Corrosion(i)
'Here we see what a PARA-SITE does to individuals.
Dim l
Dim Beginning, Ending
 'The parasite deletes from the beginning of the individual
 'any substring   matching "PARA".
 Beginning = Left(Individual(i), 4)
 If Beginning = "PARA" Then
   Individual(i) = Delete(Beginning, Individual(i))
 End If
  'The parasite deletes from the tail of the individual
 'any substring   matching "SITE".
 Ending = Right(Individual(i), 4)
 'Debug.Print "The Beginning is "; Beginning
 'Debug.Print "The End is "; Ending
 l = Len(Individual(i))
 If Ending = "SITE" Then Individual(i) = Left(Individual(i), l - 4)
End Sub

Private Sub Purge(i)
Dim c, s1, s2
Dim lc, Start, n As Integer
 'In this procedure individuals are mutilated
 'as a measure to hinder overgrowth.
 'We delete at a random Start
 'a random number of chars from the original string.

 c = Individual(i)
 lc = Len(Individual(i))
 Start = Int(Rnd * lc) + 1
  'We copy  the first part of c
  s1 = Left(c, Start - 1)
  'Debug.Print "The left part of c is " & s1
  n = Int(Rnd * (lc - Start)) + 1

  'We copy  the right part of c
  s2 = Right(c, lc - Start - n + 1)
  'Debug.Print "The Right part of c is " & s2

  'We concatenate the left part of c  to  the right part of c
 Individual(i) = s1 & s2


End Sub


Private Sub Sorting()
Dim i, j, Champ As Integer
Dim Fitness(1 To 100) As Integer
 'We define a fitness function equal to the length of the string
 For i = 1 To NIndiv
  Fitness(i) = Len(Individual(i))
 Next i
 'We sort individual by fitness
 For i = 1 To NIndiv
 Champ = 1
 For j = 1 To NIndiv
  If Fitness(j) > Fitness(Champ) Then Champ = j
  Next j
  'The array Order keeps a record of fitness by decreasing order.
  Order(i) = Champ
  Fitness(Champ) = 0
  'Debug.Print i & "th ind. is  No " & Champ & " " & Individual(Champ)
 Next i

End Sub

Private Sub Copying()
Dim i, k As Integer

 For i = 1 To 10
  k = Order(NIndiv + 1 - i)
  Individual(k) = Individual(Order(i))
 Next i

 'Debug.Print "THE NEW POPULATION IS"

 For i = 1 To NIndiv
 'Debug.Print Individual(i)
 Next i

End Sub

Private Sub Reproduction()
Dim i As Integer
 'Individuals are sorted by length
 Call Sorting
 For i = 1 To 10
  'Debug.Print Individual(Order(i))
 Next i
 'The top ten produce a copy that
 'substitutes the bottom ten.
 Call Copying
End Sub


Private Sub Deletion(i)
Dim l, n As Integer
Dim Beginning
  l = Len(Individual(i))
  n = Int(Rnd * l) + 1
  Beginning = Left(Individual(i), n)
  Individual(i) = Delete(Beginning, Individual(i))
 End Sub


Private Function Inversion(a)
 'This function inverts a
 Dim la, i, Start As Integer
 Dim d, f

 'We initialize the output
 d = ""

 'We get the length of a
 la = Len(a)

 'We copy the first character to the last place
 'and so on.

 For i = 1 To la
 Start = i
 chari = Mid(a, Start, 1)

  d = chari & d
 Next i
 Inversion = d

End Function


Private Function Invert(a, c)
'
'This Function inverts the first occurrence
'of substring a  inside c
'
 Dim Start, la, lc As Integer
 Dim s1, z

  'We get where we must begin
  Start = Len(c) - 3
  'Debug.Print "a begins at position " & Start
  'We copy  the first part of c
  s1 = Left(c, Start)
  'Debug.Print "The left part of c is " & s1
  'We    get the lengths of  the strings
  la = Len(a)
  'Debug.Print "The string  a has " & la & " characters"
   lc = Len(c)
  'Debug.Print "The string  c has " & lc & " characters"
  'We invert a properly
  d = Inversion(a)
  'We concatenate the left part of c  to d
  Invert = s1 & d
End Function


Private Sub InversionTail(i)
Dim z
Dim l
'We  invert the last part of each individual.
  l = Len(Individual(i))
  If l > 2 Then
    a = Right(Individual(i), 3)
    c = Individual(i)
    z = Invert(a, c)
  End If
  'Debug.Print "The output string is  " & z
End Sub



Private Sub Mutation()
Dim Luck As Integer
Dim i As Integer
 'We consider only two types of mutation:
 'deletion of the beginning
 ' and inversion of the end.

 'Which mutation affects which individual
 'is a matter of gambling.
 'We throw a 0 else a 1
 For i = 1 To NIndiv
  Luck = Int(Rnd + 1)
  If Luck = 1 Then Call Deletion(i) Else Call InversionTail(i)
 Next i
End Sub


Private Sub Dynamics()
Dim i As Integer
 'All individuals feed

  For i = 1 To NIndiv
   'The string Algae is digested and inserted
   'into the individual char by char
   'Debug.Print "The hungry individual " & i & " is " & Individual(i)
   Call Digestion(i)
   'Debug.Print "The full individual " & i & " is " & Individual(i)
 'All individuals are tested by the parasite,
  'which corrodes any  large substring similar to it located at any end.
   Call Corrosion(i)
   'Debug.Print "The corroed individual " & i & " is " & Individual(i)
   Call Purge(i)
   'Debug.Print "The purged  individual " & i & " is " & Individual(i)
  Next i
 'The top ten are allowed to  reproduce:
 'a child occupies the place of the shortest individual
  Call Reproduction
  'The new population is subjected to mutation
  Call Mutation
End Sub

Private Sub Observation(Gen)
Dim j As Integer
Dim m, Min, Max As Integer

Max = 0
Min = 32000
For j = 1 To NIndiv
 m = Len(Individual(j))
 'Debug.Print "The Individual  " & Individual(j) & " has length " & m
 If m > Max Then Max = m
 If m < Min Then Min = m
Next j
 'Debug.Print "Generation " & Gen & " MinLenght = " & Min & " MaxLenght = " & Max
 ReportMin(Gen) = Min
 ReportMax(Gen) = Max
End Sub

Public Sub Main()
Dim i As Integer
Dim NGen
'The population has NIndiv members, NIndiv >10
 NIndiv = 15
 Call Initialization
 'We run NGen generations
 NGen = 100
 For Gen = 1 To NGen
   Call Dynamics
   Call Observation(Gen)
  Next Gen
  Debug.Print "MINIMAL VALUES"
 For Gen = 1 To NGen
  Debug.Print ReportMin(Gen)
 Next Gen
 Debug.Print "MAXIMAL VALUES"
 For Gen = 1 To NGen
  Debug.Print ReportMax(Gen)
 Next Gen
End Sub

\end{verbatim}

\bigskip
\textbf{\thenum. }  \addtocounter{num}{1}  Exercise. Run the code and verify that the Instant Window has no enough memory to deal with too  large a report. Verify that 100 lines of report cause no damage at all. Hence, change the number of generations from 100 to 50. Next, copy the whole array of minimal values to the clipboard and paste it into Excel. Call the graphical facility and draw, with Lines, the array. Do the same with the maximal values and deduce whether or not there is a tendency to growth or to fade away.

\bigskip
\textbf{\thenum. }  \addtocounter{num}{1}  We have observed that the Instant window has not been designed as a display for results but only as a debugging facility. So, we could learn to report data into a Word document  to next copy them into the clipboard to paste them into Excel. Or we could learn to edit data directly  into an Excel sheet.

Or we can work directly on  Excel.

That is much more difficult because a sheet of Excel is quite structured and so we must get ready to suffer a bit. Anyway, evolution is a quantitative process and we need statistics, which are better done in Excel. That has its problems but now that we know a lot of programming techniques, those problems will be easily circumvented. Indeed. Excel has many useful possibilities and some of them will be explored in the next chapters.

\bigskip
\textbf{\thenum. }  \addtocounter{num}{1}  Graduation. Develop a simulation of evolution with a population of strings, whose fitness function rewards a middle length, say of 10 chars. The population is affected by viruses that add themselves to any  end  of the strings and there are restriction enzymes that divide the strings in two parts,  the  shorter  of which perishes. Individuals have different aptitude to digest the food.

\chapter{My Excel}

We already have acquired some programming skills, so we may pass now to more sophisticated tasks dealing with statistics and graphics, for which the appropriate tool could be Excel. We will also add various intrigues concerning the genome.

To manipulate Excel from VBA, we shall learn at least two procedures:  to read information from Excel and to print  information into a sheet of Excel. We could understand how that is done if we look at the entrails of the code that is automatically developed by Excel when we record a Macro.

\section{Inside a Macro}

\bigskip
\textbf{\thenum. }  \addtocounter{num}{1}  Every time one invokes a button or  a function in Excel, one is invoking an application, which is a  Macro that has been given independent life. They are in general very complex, so let us design a very simple Macro with the only purpose of looking inside it. A Macro is a recording of a given sequences of actions that one does with Excel. Recording and playing Macros functions just like recording a song in a recording sound machine.

\bigskip
\textbf{\thenum. }  \addtocounter{num}{1}  Example. Let us record a Macro to sum two numbers.

1. Open \textbf{Excel} and close all grids.

2. Select \textbf{Tools} in the Excel menu and then chose \textbf{Macro} and then chose \textbf{Record New Macro}. Give a  name to the Macro, say, \textbf{XplusY} and specify in the Description Box that it is a Macro to add two numbers. The Macro must be saved in the Personal Macro Book. End this operation with the OK button.

3. Open a grid for you.  Program Excel to  calculate  the  sum of two numbers. The   final output of that program looks like this:

\begin{center}
\begin{tabular}{|l|l|c|}\hline
\multicolumn{2}{|c|}{\vphantom{Large Ap} The Macro XplusY }\\ \hline\hline
 & XplusY \\ \hline\hline
Type x & 7 \\ \hline
Type y & 9\\ \hline
Result& 16\\ \hline
\end{tabular}
\end{center}

4. Go to \textbf{Tools}, to \textbf{Macro} and call the \textbf{Stop Recording command}.

5. Leave Excel and save the changes to your Personal Macro Book. In Excel for Windows, that book is called \textbf{Personal.XLS}.

We have our first Macro. It is now available for all times. Let us see how a Macro can be retrieved.

\bigskip
\textbf{\thenum. }  \addtocounter{num}{1}  Exercise. Turn on Excel. Go to Tools, to Macro, to Macros. Look for a Macro with name XplusY. Activate it with a double click or with the appropriate command. When a sheet pops up for you with the information that was recorded, replace the numbers there, 7 and 9, by whatever numbers you desire and test that the XplusY Macro adds them. Next, select the region occupied by the whole Macro and copy it to the clipboard. Paste it wherever you want, verify that it still works. Chain your macro to a new program that calculates $(x+y)/3$. Finally, chain your Macro to another program that calculates $(2x+5y)$.

\bigskip
\textbf{\thenum. }  \addtocounter{num}{1}  Exercise. Create and record  a Macro called \textbf{Square} that takes a number and returns its square. Chain the two Macros, \textbf{XplusY} and \textbf{Square}  to calculate $(a+b)^2$. Write a new program that uses the two given Macros to verify the law that says: $(a+b)^2= a^2 + 2ab+b^2$

\bigskip
\textbf{\thenum. }  \addtocounter{num}{1}  Exercise. Just for fun, Play creating, recoding and chaining diverse Macros. Enrich your Macros with suitable formatting:  while recording a Macro, use  the options given  at the menu Format in the commands for cells, columns or rows.

\bigskip
\textbf{\thenum. }  \addtocounter{num}{1} A Macro has two components: its behavior or phenotype and its entrails or genotype, which is code in VBA. We have been dealing with the phenotype, what a Macro does.

\bigskip
\textbf{\thenum. }  \addtocounter{num}{1}  We can look at the code of an automatically recorded Macro in the same way as we looked at the code that we developed on our own.

\bigskip
\textbf{\thenum. }  \addtocounter{num}{1}  Exercise. Look at the genotype or code of the Macro \textbf{XplusY}.

\bigskip
\textbf{\thenum. }  \addtocounter{num}{1}  Exercise: One could compare a Macro in Excel, and in general in any program,  to a  gene. Thus, genes and Macros are both alphabets with which independent units could be assembled. Compare the properties given to the Macro technology by the philosophy of software design with those given by the philosophy you see behind the Gen technology that operates  in alive beings. Please, notice that the philology behind  Macro technology is a philosophy to guide the design of this technology but the philosophy behind gene technology is given by you a posteriori. To say that an a posteriori philosophy also reflects a philosophy of gene design is to declare that life was created by some intelligent being. By contrast, according to the evolutionary theory, there is no philosophy behind genes apart from surviving of the fittest.

\bigskip
\textbf{\thenum. }  \addtocounter{num}{1}   What a Macro does in front of our eyes, is the result of the code written in VBA: the genotype encodes for the phenotype.    Running a Macro demands to translate the VBA code  into machine code and to execute that translation. This process is known as \textbf{compiling}. A procedure to be compiled or run must be saved before.

\bigskip
\textbf{\thenum. }  \addtocounter{num}{1}   Exercise. Is a Macro like an alive being or is it an alive being? What does a Macro lack to get alive?

\bigskip
\textbf{\thenum. }  \addtocounter{num}{1}   Exercise: Discuss whether or not the Macro technology shall observe the property of continuity. This means that  minor changes in a task must be followed  by minor changes in the Macro, or that similar task must generate similar macros. So one could easily adapt an extant Macro to get a new Macro with a desired  function.

\bigskip
\textbf{\thenum. }  \addtocounter{num}{1}    Now that we have discovered that the entrails of a Macro is a code in VBA, we may face up a  complaint: Why do we need Excel if with VBA, we would have everything?

It is true that VBA has everything one could need, but if we adopt VBA for Excel, many problems could be more easily circumvented. In fact,   programming is a very hard enterprise: it is very difficult to write a code with meaning ( that the VBA debugger machine could process), and  if one writes a code with meaning.  one is nevertheless very far from writing the code that does precisely what it was intended to do. All this implies that, whenever it is possible, one could adopt a system does not overload the problem of designing software. Now, some tasks of software development are made easier with the help of Excel, say, we can couple graphic production to data analysis in a very natural and simple way.  Nevertheless, VBA for windows is not the full fledged VBA, so  one may feel impotent at some times.

\bigskip
\textbf{\thenum. }  \addtocounter{num}{1}  Improve your style. A Macro is a program that records all the actions that you do, be they good, be they bad. The important for us factor is that a program is a concatenation of instructions or moves and the final output depends on the order of the sequence of instructions.

It is very difficult to understand what a Macro says, but our plan is precisely to learn to modify them in   directed ways. Thus, we are looking for a way of implementing directed mutations, as in  genetic engineering. Our counsel is that geneticists know very little about the genome but nevertheless, they can do their work. So us.

To improve our style, let us hold the following simple principle:  one orderly moves over the sheet from left to right and from top to bottom and makes a great effort to look forward and not turn back the head.

\bigskip
\textbf{\thenum. }  \addtocounter{num}{1}  A Macro \textbf{XplusY}  recorded with that style may look like this:

\begin{verbatim}

Public Sub XplusY()
'
' XplusY Macro
' Adds two numbers. Good style.
'

'
    Workbooks.Add
    Range("D4").Select
    ActiveCell.FormulaR1C1 = "x+y"
    Range("C5").Select
    ActiveCell.FormulaR1C1 = "Type x"
    Range("D5").Select
    ActiveCell.FormulaR1C1 = "7"
    Range("C6").Select
    ActiveCell.FormulaR1C1 = "Type y"
    Range("D6").Select
    ActiveCell.FormulaR1C1 = "9"
    Range("C7").Select
    ActiveCell.FormulaR1C1 = "Result"
    Range("D7").Select
    ActiveCell.FormulaR1C1 = "=R[-2]C+R[-1]C"
    Range("D8").Select
End Sub

\end{verbatim}

\bigskip
\textbf{\thenum. }  \addtocounter{num}{1} Exercise. Paste this code to your personal book and run it with F8. Please, divide the screen in such a way  that you could see the VBA window and the Excel window at the same time. Under F8 mode of compiling, one can drag both windows at pleasure.

\bigskip
\textbf{\thenum. }  \addtocounter{num}{1} Add the code to your personal book and run it under F8. Try to get the name of the Macro, its function, to identify which number is x and which is y.

\bigskip
\textbf{\thenum. }  \addtocounter{num}{1} By looking at the code of our Macro, to the things it does when it is compiled step by step, we  learn   that its name is XplusY, that its function is to add two numbers. That one number is called x and that  the other number is  y but we really do not know which is 7 and which is 9.  The result is found in the cell D7, and the halting signal consists in selecting a next cell, that in this case is D8.

\bigskip
\textbf{\thenum. }  \addtocounter{num}{1}  When one records a Macro, it is very easy to observe a bad, undesirable style.  The next  is an example. It is the code recorded by the recording facility, but the operations were done in a very disordered way.

\begin{verbatim}
Public Sub XplusYmessy()
'
' XplusYmessy Macro
' Adds two munbers. Messy style.
'

'
    Workbooks.Add
    Range("D8").Select
    ActiveCell.FormulaR1C1 = "=R[-2]C+R[-1]C"
    Range("D5").Select
    ActiveCell.FormulaR1C1 = "x+y"
    Range("D7").Select
    ActiveCell.FormulaR1C1 = "6"
    Range("C8").Select
    ActiveCell.FormulaR1C1 = "Result"
    Range("D6").Select
    ActiveCell.FormulaR1C1 = "5"
    Range("C7").Select
    ActiveCell.FormulaR1C1 = "y"
    Range("C6").Select
    ActiveCell.FormulaR1C1 = "x"
    Range("C9").Select
End Sub

\end{verbatim}

\bigskip
\textbf{\thenum. }  \addtocounter{num}{1}  Exercise. Verify that the code works. Try to understand what it says. Compare the easiness of understanding of the code of the two versions: the good style and the messy style.

\bigskip
\textbf{\thenum. }  \addtocounter{num}{1}  Exercise. Explicitly define what you mean  by a messy code. Prove that messiness has no upper bound. Would you say that the genome observes a good style of programming or a messy one?  Try to explain why. Make sure to prove that any one answer to the aforementioned question is compatible with the evolutionary theory else express your doubts and fears and try to envisage a form of clearing them out.

\bigskip
\textbf{\thenum. }  \addtocounter{num}{1}  Up to now, a good style of programming is characterized by the easiness of being understood. Now, we add another condition: a good style allows an easy modification of extant Macros or Modules.

\bigskip
\textbf{\thenum. }  \addtocounter{num}{1}  We already know that when one has troubles to produce a procedure that does exactly what one wants, one may use tracers to follow the actions. One usual marker is to order the code  to write specific values of specific variables. When one does not need anymore theses markers, one may convert the corresponding instructions into comments that will left aside by the compiling machine.  VBA has additional facilities,  that you would find in the debugging sub-menu. Play with them. We will use trace all along because they are simple and effective and do not depend on the skills of the reader.

\bigskip
\textbf{\thenum. }  \addtocounter{num}{1}  Example. This is a tracer:

\begin{verbatim}


 Range("D7").Select
    ActiveCell.FormulaR1C1 = "The value of the active cells is "
    Range("D8").Select
    ActiveCell.FormulaR1C1 = "=R[-2]C+R[-1]C"
    Range("D9").Select

\end{verbatim}

\bigskip
\textbf{\thenum. }  \addtocounter{num}{1} Compose a Public sub using this code and run it. Warning: To run this sub you must write a number in the cell D8 and select it.

\bigskip
\textbf{\thenum. }  \addtocounter{num}{1} To modify a Macro, open it in the VBA application. You can modify it as simply as modifying any one text.

\bigskip
\textbf{\thenum. }  \addtocounter{num}{1}  Example. Let us modify our \textbf{XplusY} Macro to produce a new Macro  \textbf{XoperationY} that contains a subtraction x-y, a multiplication x*y, a division x/y and a power $x^y$. Our try is the following: we modify the name and comments. We add a warning that says that division by zero is not allowed. Next, we notice that we do not know how to identify x with "R[-2]C" or with "R[-1]C". For the addition, that was unimportant but not for subtraction or division. Our choice is that x is "R[-2]C" and that y is "R[-1]C". We think that because numbering in Excel grows from top to bottom and from left to right.  Next we add the different operations with the corresponding comments and we include the signal for halting which consists in selecting the next cell. After many trials we find the right code:

\begin{verbatim}

Public Sub XoperationnnY()
'
' XoperationY Macro
' Operations with two numbers. Good style.
'

'
    Workbooks.Add
    Range("D3").Select
    ActiveCell.FormulaR1C1 = "x?y"
    Range("D4").Select
    ActiveCell.FormulaR1C1 = "y not 0"
    Range("C5").Select
    ActiveCell.FormulaR1C1 = "Type x"
    Range("D5").Select
    ActiveCell.FormulaR1C1 = "7"
    Range("C6").Select
    ActiveCell.FormulaR1C1 = "Type y"
    Range("D6").Select
    ActiveCell.FormulaR1C1 = "9"
    Range("C7").Select
    ActiveCell.FormulaR1C1 = "x+y"
    Range("D7").Select
    ActiveCell.FormulaR1C1 = "=R[-2]C+R[-1]C"
    Range("C8").Select
    ActiveCell.FormulaR1C1 = "x-y"
    Range("D8").Select
    ActiveCell.FormulaR1C1 = "=R[-3]C-R[-2]C"
    Range("C9").Select
    ActiveCell.FormulaR1C1 = "x*y"
    Range("D9").Select
    ActiveCell.FormulaR1C1 = "=R[-4]C*R[-3]C"
    Range("C10").Select
    ActiveCell.FormulaR1C1 = "x/y"
    Range("D10").Select
    ActiveCell.FormulaR1C1 = "=R[-5]C/R[-4]C"
    Range("C11").Select
    ActiveCell.FormulaR1C1 = "x^y"
    Range("D11").Select
    ActiveCell.FormulaR1C1 = "=R[-6]C^R[-5]C"
    Range("D12").Select
End Sub

\end{verbatim}

\bigskip
\textbf{\thenum. }  \addtocounter{num}{1}  Exercise. Make a Macro \textbf{XplusminusY}  that takes two numbers and calculates their sum and substraction.   Compare  the code of the two macros \textbf{XplusminusY} and \textbf{Xplus Y}  and and unveil the set of instructions that is the cause of the addition, of the subtraction and of the comments. And also, unveil the instruction  for halting. The reference cell in a sheet is always the selected cell, but determine the system of coordinates in Excel.   Test your theory making a modification in the VBA code of the program \textbf{$XplusY$} to produce a program that moreover subtract the two numbers and then multiplies them.  Next, try to reinvent the $XoperationY$ module.

\bigskip
\textbf{\thenum. }  \addtocounter{num}{1}  Exercise. By looking at your suffering to execute the previous exercise, try to explain why  most random changes in a Module are inviable, i.e., they are not understood by the compiling machine, which is another program that translate your code into actions that the processor unit of your computer  could understand and execute.

\bigskip
\textbf{\thenum. }  \addtocounter{num}{1}  Exercise. Take your $XoperationY$ module and mutate it to produce another Macro that fulfills the same task by produce the desired expressions in different order. This type of mutations are called neutral, i.e., does not affect the fitness of the Macro, which in our case only depends on the calculations of the desired operations without care for the order. When a mutation affects the phenotype, it is called deleterious when it produces a tolerant small damage. It is mortal when the damage is so severe that the organism is not viable, i.e., when it is not understood by the compiler. The mutation could be favorable, when it produces a change in the behavior that makes it more desirable by whatever reason.

\bigskip
\textbf{\thenum. }  \addtocounter{num}{1}  Exercise. Make clear the difference between the value of a mutation in a design project and that of a mutation in nature where there is no objective at hand.

\bigskip
\textbf{\thenum. }  \addtocounter{num}{1}  Define evolution as the change in a program that produce a viable organism, i.e., a phenotype.  Prove that tiny changes in the genotype may cause great changes in the phenotype.   Prove that evolution is just a change of the software . Argue that evolution is  improbable but unavoidable if only the copying   procedure  is subjected to mutations.

\bigskip
\textbf{\thenum. }  \addtocounter{num}{1}  Exercise. Mark a clear difference between evolution in nature and evolution  in the design of software.

\bigskip
\textbf{\thenum. }  \addtocounter{num}{1}  Exercise. Figure out the  software  equivalent  of introns and exons, transcription and ribosomes, protein synthesis and enzymes, biochemical reactions.

\bigskip
\textbf{\thenum. }  \addtocounter{num}{1}  Exercise. Execute the \textbf{XoperationY} Macro with $y=0$, which implies a division by zero. This is the equivalent of a poison.

\bigskip
\textbf{\thenum. }  \addtocounter{num}{1}  Let us modify the \textbf{XoperationY} Macro to make it robust against the poison 'division by zero'. Al we need is to add a checking of the value of $y$.  In our code below, it appears as follows:

\begin{verbatim}

 If y = 0 Then
         ActiveCell.FormulaR1C1 = "Not allowed"
       Else
          ActiveCell.FormulaR1C1 = "=R[-5]C/R[-4]C"
    End If

\end{verbatim}

\bigskip
\textbf{\thenum. }  \addtocounter{num}{1} The complete code follows:

\begin{verbatim}

Public Sub XoperationYrobust()
'
' XoperationYrobust Macro
' Failure with the if condition
'

'
    Workbooks.add
    Range("D3").Select
    ActiveCell.FormulaR1C1 = "x?y"
    Range("D4").Select
    ActiveCell.FormulaR1C1 = "y not 0"
    Range("C5").Select
    ActiveCell.FormulaR1C1 = "Type x"
    Range("D5").Select
    ActiveCell.FormulaR1C1 = "7"
    Range("C6").Select
    ActiveCell.FormulaR1C1 = "Type y"
    Range("D6").Select
    ActiveCell.FormulaR1C1 = "9"
    y = ActiveCell.FormulaR1C1
    Range("C7").Select
    ActiveCell.FormulaR1C1 = "x+y"
    Range("D7").Select
    ActiveCell.FormulaR1C1 = "=R[-2]C+R[-1]C"
    Range("C8").Select
    ActiveCell.FormulaR1C1 = "x-y"
    Range("D8").Select
    ActiveCell.FormulaR1C1 = "=R[-3]C-R[-2]C"
    Range("C9").Select
    ActiveCell.FormulaR1C1 = "x*y"
    Range("D9").Select
    ActiveCell.FormulaR1C1 = "=R[-4]C*R[-3]C"

    Range("C10").Select
    ActiveCell.FormulaR1C1 = "x/y"
    Range("D10").Select
    If y = 0 Then
         ActiveCell.FormulaR1C1 = "Not allowed"
       Else
          ActiveCell.FormulaR1C1 = "=R[-5]C/R[-4]C"
    End If
    Range("C11").Select
    ActiveCell.FormulaR1C1 = "x^y"
    Range("D11").Select
    ActiveCell.FormulaR1C1 = "=R[-6]C^R[-5]C"
    Range("D12").Select
End Sub

\end{verbatim}

\bigskip
\textbf{\thenum. }  \addtocounter{num}{1}   Exercise. Verify that this code is viable and that it perfectly works for those values that are  specified in the code and that it also works when in the sheet the values of  $x$ and $y$ are modified. But when $y=0$ our anti-poisoning measures does not work: modify in the code the value for $y$ : give to it the value zero. Run the code. Test the code with any other value of $y$. Do whatever you want and verify that we have fall into a blind alley.

The reason of our failure is that when the code is compiled, it is executed with the values introduced in the code. When the compiler arrives to the if condition, it takes that branch that fills the condition and the other branch does not appear in the compiled program. So, one fixes the branch once and for all times.

\bigskip
\textbf{\thenum. }  \addtocounter{num}{1}   Exercise. To verify the previous assertion, open your module  \textbf{XoperationYrobust} and compile it step by step, with F8,  when  $y \ne 0$ and then when  $y=0$.

\section{The power of abstraction}

It is really difficult to follow the game of Excel when it deals with cells. Thus, we must deal with this difficulty and moreover we shall overcome the blind alley causing by the disqualification of sub procedures to produce macros with all branches of programming. Both problems are solved if we release VBA from its slavery to the sub procedures, that work with a fixed branch of compilation and get frozen forever. To this aim we introduce the notion of variables and functions that is so natural to the human thought. Everything is clear with the following example, in which we rewrite our previous module  \textbf{sub XoperationYrobust} as a function, in which we deal with $x$ and $y$ as variables that can be directly updated on each call.

\begin{verbatim}

Dim x, y As Integer



Private Sub Capturexy()
 Workbooks.add

  'Captures x
  x = InputBox( _
    prompt:="Please, enter the value of x.", _
    Default:=7)

 ' Captures y
  y = InputBox( _
    prompt:="Please, enter the value of y.", _
    Default:=9)
End Sub



Private Function add(x, y)
    add = x + y
End Function

Private Function subs(x, y)
    subs = x - y
End Function

Private Function mult(x, y)
    mult = x * y
End Function

Private Function div(x, y)
    div = x / y
End Function

Private Function power(x, y)
    power = x ^ y
End Function

Public  Sub Foperationsxy()
    Capturexy
    'Reports x and y
    Range("C5").Select
    ActiveCell.FormulaR1C1 = "x"
    Range("D5").Select
    ActiveCell.FormulaR1C1 = x
    Range("C6").Select
    ActiveCell.FormulaR1C1 = "y"
    Range("D6").Select
    ActiveCell.FormulaR1C1 = y

    'Calculates and reports results
    Range("C7").Select
    ActiveCell.FormulaR1C1 = "x+y"
    Range("D7").Select
    ActiveCell.FormulaR1C1 = add(x, y)
    Range("C8").Select
    ActiveCell.FormulaR1C1 = "x-y"
    Range("D8").Select
    ActiveCell.FormulaR1C1 = subs(x, y)
    Range("C9").Select
    ActiveCell.FormulaR1C1 = "x*y"
    Range("D9").Select
    ActiveCell.FormulaR1C1 = mult(x, y)
    Range("C10").Select
    ActiveCell.FormulaR1C1 = "x/y"
    Range("D10").Select

    'Antipoisoning measure
    If y = 0 Then
         ActiveCell.FormulaR1C1 = "Not allowed"
       Else
          ActiveCell.FormulaR1C1 = div(x, y)
    End If

    Range("C11").Select
    ActiveCell.FormulaR1C1 = "x^y"
    Range("D11").Select
    ActiveCell.FormulaR1C1 = power(x, y)
    Range("D12").Select
End Sub

\end{verbatim}

\bigskip
\textbf{\thenum. }  \addtocounter{num}{1}   Exercise. Copy and paste this code into your personal.XLS and run it. Test it against the zero poison. Poison the code with a letter instead of a number to see what happens.

\bigskip
\textbf{\thenum. }  \addtocounter{num}{1}   Exercise. Study the code above. Copy and adapt suitable parts of its code to compose the code for a function that calculates and reports $(x-y)^2$. Compile, debug, run and test the resultant code.

\bigskip
\textbf{\thenum. }  \addtocounter{num}{1}   Exercise.  We were unable of making a Sub  procedure to be compiled with all branches of commutation. What looked as a tiny problem demanded   a revolution, which was solved by the power of abstraction. Discuss the property of continuity of software in view of this revolution.

\bigskip
\textbf{\thenum. }  \addtocounter{num}{1}   Exercise. Our solution to a blind alley involved a revolution, one of whose ingredients was the appearing of encapsulation of code: small functional pieces of code must now be packed into an individual, personal box. Research about the procaryote vs eucaryote revolution and compare it with our revolution. Pay attention both to the organization of the genome and to the compartmentalization of biochemical tasks.

\bigskip
\textbf{\thenum. }  \addtocounter{num}{1}    Let us learn now how to read the content of a cell.

To select a cell in a sheet of Excel, click it with the mouse. The VBA code for reading the content of a selected and filled cell is the following:

\begin{verbatim}

Public Sub ReadCell()

CellValue = Selection.Value
MsgBox "The cell value is " & CellValue

End Sub

\end{verbatim}

Exercise. Run the \textbf{ReadCell} procedure and test it against numbers and \textbf{chars} (letters and symbols) . Run it step by step with F8.

\bigskip
\textbf{\thenum. }  \addtocounter{num}{1}   Let us learn by example  how to read  a selected region. Copy and paste into your personal.XLS  the next code for reading the content of a selected and filled matrix :

\begin{verbatim}

Public Sub ReadMatrix()

MatrixAddress = Selection.Address
MsgBox "The region has a range " & MatrixAddress
MatrixValue = Selection.Value
MsgBox "The value of cell (1,1) is  " & MatrixValue(1, 1)
MsgBox "The value of cell (4,1) is  " & MatrixValue(4, 1)
MatrixFormula = Selection.Formula
MsgBox "The formula of cell (2,2) is  " & MatrixFormula(2, 2)
MsgBox "The formula of cell (4,2) is  " & MatrixFormula(4, 2)
MsgBox "The formula of cell (3,1) is  " & MatrixFormula(3, 1)
MsgBox "The value of cell (10, 10) is  " & MatrixValue(10, 10)

End Sub

\end{verbatim}

To feed the code, type in any region of a sheet of Excel a matrix with 2 columns and 6 rows. The cells of a matrix are addressed using the couple (i,j). The first index is for the row and the second for the column. Fill the first column with numbers and the second with formulas.   Next, select the matrix and  run the procedure step by step.

\bigskip
\textbf{\thenum. }  \addtocounter{num}{1}   Exercise. Run the \textbf{ReadMatrix} procedure and test it. Run it step by step with F8. Find a bug and debug the code. Test your new code against blanks or void cells.

\section{Repetitive tasks}

\bigskip
\textbf{\thenum. }  \addtocounter{num}{1} When one must go over a repetitive task, one may consider the use of the for.. ..next statement. The following is a For loop:

\begin{verbatim}

Public Sub OneToFive()
 For Counter = 1 To 5
    MsgBox "The value of Counter is " & Counter
 Next Counter
End Sub

\end{verbatim}

Exercise. Run the last code.

 \bigskip
\textbf{\thenum. }  \addtocounter{num}{1 }  The following is another option to execute some repetitive tasks. It is a do.. .. while loop:

\begin{verbatim}

 Public Sub Positive()

x = 4
Do
    x = x - 1
    MsgBox "The value of x is " & x
Loop While x > 0
End Sub

\end{verbatim}

\section{Integers vs reals}

\bigskip
\textbf{\thenum. }  \addtocounter{num}{1} The Dim instruction selects memory to store a variable. Some variables demand too much memory but others not. To optimize the use of memory, variables are classified by its type. Integer  is used for numbers 1,2,3,.... , Single is used for for real numbers of the form 3.14159. When one dimensions a variable but does not write its type, it is a variant variable, which could keep anything.

\bigskip
\textbf{\thenum. }  \addtocounter{num}{1}  Exercise. Run the next program to learn something about the integer type:

\begin{verbatim}

Public Sub Integers()
 'Difference between integers and reals.
 'Integers are between -32000 and 32000
 'If you need to deal with decimal  and larger numbers,
 ' use the Single type.
 'Integers round real numbers.

Dim k As Integer
Dim l  As Integer
Dim r, s  As Single

For i = 1 To 12
    MsgBox " i = " & i
    k = i / 7
    MsgBox " i/7 as Integer is " & k
    r = i / 7
    MsgBox " i/7 as real is " & r
    l = 5000 * i
    MsgBox " the integer 5000 x i is " & l
    s = 5000 * i
    MsgBox " the real 5000 x i  is " & s
Next i

End Sub
\end{verbatim}

\bigskip
\textbf{\thenum. }  \addtocounter{num}{1}  Warning: One usually makes many developing errors. When they are incorporated in within a \textbf{for loop}, one would desire to abort immediately. This can be done with Ctrl + Pause. Another idea to prevent looping when it is not desired is to incorporate a code specifically target to halt the program manually, as in the following code:

\begin{verbatim}

Public Sub Finish()
 For i = 1 To 1000
 n = MsgBox("Do you want to continue looping?", vbYesNo)
 If n = vbNo Then
   Exit For
End If
Next i
End Sub

\end{verbatim}

Exercise. Run the last code in step by step mode. Find the instructions that exit the loop and the condition to do so.

\bigskip
\textbf{\thenum. }  \addtocounter{num}{1}  Example. Let us use a For statement to calculate a mean of a row data.

\begin{verbatim}

Public Sub MeanRawData()
 Dim X(1 To 10)
 X(1) = 1
 X(2) = 3
 X(3) = 5
 X(4) = 3
 X(5) = 4
 X(6) = 2
 X(7) = 6
 X(8) = 4
 X(9) = 3
 X(10) = 5
 X(11) = 3
 Sum = 0
 For i = 1 To 11
  Sum = Sum + X(i)
 Next i
 Mean = Sum / 11
 MsgBox " the mean   is " & Mean
End Sub

\end{verbatim}

\bigskip
\textbf{\thenum. }  \addtocounter{num}{1}  Exercise: Test this code with F8 under VBA.   Verify the answer with a direct calculation.

\bigskip
\textbf{\thenum. }  \addtocounter{num}{1}  Exercise. Develop a code to calculate the variance and standard deviation of the data that appear in the previous code.

\bigskip
\textbf{\thenum. }  \addtocounter{num}{1}  Intrigue: Did you noted that we have produced too much garbage and that the code stacks itself in a very disordered fashion?  Some garbage is automatically dealt with by Excel and VBA but other garbage must be deal with directly by you. In some occasions, garbage remains alive in the code.

\bigskip
\textbf{\thenum. }  \addtocounter{num}{1}  Exercise. The genome is filled with many pieces with unknown function or that are corrupted copies of functional material or of material that function in another species. In our terms, this is garbage. Officially, it  is called junk material.

Let us pass now to a real and important though simple task: that of calculating the basic statistics  of a given histograms.

\section{Histograms}

When the data are listed one by one, one can use Excel functions to calculate their mean, variance and standard deviation.

\bigskip
\textbf{\thenum. }  \addtocounter{num}{1}  Exercise.  Find the mean of the integer numbers 3, 5, 4, 6, 4, 7,  3, 5,  4,  6, 3, 8  by hand and by a function in Excel that is called the mean or the average. That function is located in the menu Insert and submenu function (it is possibly hidden under the subsubmenu statistical functions). There is also a button that calls the menu that contains all functions. Develop a program in VBA that finds the same statistics and compare your answers with those of Excel.

\bigskip
\textbf{\thenum. }  \addtocounter{num}{1}   Excel can deal with very long lists of numbers. But one usually likes  to compress the information, when that is possible. For instance, a list of numbers like that given in the previous exercise can be compressed in a table of frequencies, as follows: The number 3 appears trice. Number 4 appears trice. Number 5 appears twice. Number 6 appears 2 times. Number 7 appears once. Number 8 appears once. We can write this in a table:

\begin{center}
\begin{tabular}{|l|l|l|c|}\hline
\multicolumn{4}{|c|}{\vphantom{Large Ap} X=Number of offspring of 12 different hens}\\ \hline\hline
Number& Absolute  frequency &Relative frequency & Cumulative frequency \\ \hline\hline
3 & 3 & 3/12 & 3/12\\ \hline
4 & 3&3/12 & 6/12\\ \hline
5 & 2&2/12 & 8/12\\ \hline
6 & 2&2/12 & 10/12\\ \hline
7 & 1&1/12 & 11/12\\ \hline
8 & 1&1/12 & 12/12 \\ \hline
\end{tabular}
\end{center}

\bigskip
\textbf{\thenum. }  \addtocounter{num}{1}   An histogram is a drawing that represents a frequency table. Since the correspondence among frequency tables and histograms is one to one, we use histograms and frequency tables as equivalent terms.

\begin{center}
 \begin{pspicture}(2,-0.5)(9,4)\showgrid
 \psline[linewidth=1pt](2 ,0)(9,0)
 \psline[linewidth=1pt](2.5,0)(2.5,3)
 \psline[linewidth=1pt](2.5,3)(3.5,3)
 \psline[linewidth=1pt](3.5,0)(3.5,3)
 \psline[linewidth=1pt](3.5,3)(4.5,3)
 \psline[linewidth=1pt](4.5,0)(4.5,3)
 \psline[linewidth=1pt](4.5,2)(5.5,2)
 \psline[linewidth=1pt](5.5,0)(5.5,2)
 \psline[linewidth=1pt](5.5,2)(6.5,2)
 \psline[linewidth=1pt](6.5,0)(6.5,2)
 \psline[linewidth=1pt](6.5,1)(7.5,1)
 \psline[linewidth=1pt](7.5,0)(7.5,1)
 \psline[linewidth=1pt](7.5,1)(8.5,1)
 \psline[linewidth=1pt](8.5,0)(8.5,1)
\end{pspicture}

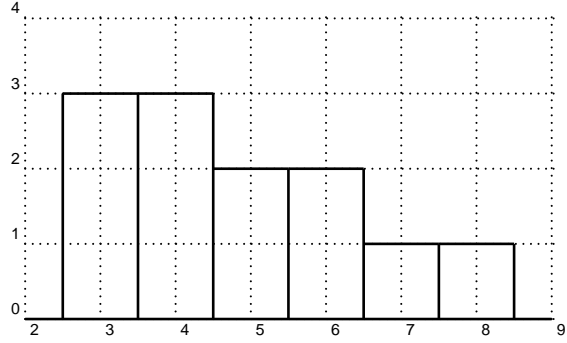
Figure \thefigure. Histogram that represents the previous  frequency table.
\stepcounter{figure}
\end{center}

\bigskip
\textbf{\thenum. }  \addtocounter{num}{1}   To calculate the mean of a list $x$ of $n$ data, one adds all numbers and divide by the number of data. What shall we do to find the mean of a frequency table?

Example. The mean of data 3, 5, 4, 6, 4, 7,  3, 5,  4,  6, 3, 8 is

$\bar x= \frac{3+ 5+ 4+ 6+ 4+ 7+  3+ 5+  4+  6+ 3+ 8}{12} = \frac{3+ 3+3+4+4+4+ 5+ 5+ 6+ 6+  7+       8}{12}$

$ = \frac{3(3)+ 3(4)+ 2(5)+ 2(6)+1( 7)+     1(  8)}{12}$

Therefore, to find the mean of a frequency table, one must multiply the columns of the $x_i$ and $n_i$ entry by entry and then add altogether and the result must be divided by the total number of events that is the sum of all the $n_i$. This is written   as follows: let $x_i$ be   datum numner $i$ and $n_i$ its frequency. Then, the mean of all data is

$\bar x = \frac{\sum n_i xi}{\sum n_i}$

\bigskip
\textbf{\thenum. }  \addtocounter{num}{1}   Exercise. Program Excel to find the mean of the table 1.

\bigskip
\textbf{\thenum. }  \addtocounter{num}{1}   While the mean is the best representative  of a given data, the variance measures the divergence of the population with respect to the mean. Let $X$ be a data and $V_x$ its variance. When the $n$ data are listed, the variance is:

$V_x= \frac{\sum (x_i- \bar x)^2}{n}$

When the data are compressed into a frequency table $(x_i, n_i)$, the variance is

$V_x= \frac{\sum n_i(x_i- \bar x)^2}{\sum n_i}$

\bigskip
\textbf{\thenum. }  \addtocounter{num}{1}   Exercise: take the data 3, 5, 4, 6, 4, 7,  3, 5,  4,  6, 3, 8, write the expression for its variance according to the definition and arrange that expression to recover the formula for the variance that corresponds to a table of frequencies.

\bigskip
\textbf{\thenum. }  \addtocounter{num}{1}   The standard deviation is the positive square root of the variance. Therefore the standard deviation $s_x$ of a frequency  table $(x_i, n_i)$ is

$s_x= \sqrt{\frac{\sum n_i(x_i- \bar x)^2}{\sum n_i}}$

\bigskip
\textbf{\thenum. }  \addtocounter{num}{1}   Exercise. Find in Excel a mathematical function that calculates the square of a number. Find also the square root function. Program Excel to calculate the mean, variance and standard deviation of data 3, 5, 4, 6, 4, 7,  3, 5,  4,  6, 3, 8 , Verify the answer of Excel with your  own calculations.

\bigskip
\textbf{\thenum. }  \addtocounter{num}{1}  We already know what we mean by a frequency table and how to calculate its mean, variance and standard deviation. We eventually could program Excel to make that job for us for any given frequency table. But if we change of frequency table, we must adjust our program. This could generate errors and lost of time. So, let us learn now how to design a program that accepts any frequency table, of whatever length, and produces the mean, variance and standard deviation. To achieve our goal, we need a good deal of generalizations and know-how, so let us advance step by step.

\section{Stadigraphs of a  histogram }

Our goal is to devise a program  that captures a table of frequencies,   that   calculates the mean, the variance, the standard deviation.

\begin{verbatim}

Dim k, n, average, variance, stdeviation As Integer
Dim Table(50, 2)
Dim TableValue


Private Sub Capturen()
'Captures the number of pairs of data
  k = InputBox( _
    prompt:="Please, enter the number of pairs of data", _
    Default:=7)

End Sub

Private Sub CaptureTable()

 TableAddress = Selection.Address
 MsgBox "The region has a range " & TableAddress
 TableValue = Selection.Value

 End Sub




 Private Sub Stadigraphs()
 n = 0
 average = 0
 variance = 0
 stdeviation = 0

 For i = 1 To k
    n = n + TableValue(i, 2)
    average = average + TableValue(i, 1) * TableValue(i, 2)
 Next i
    average = average / n

  For i = 1 To k
variance = variance + TableValue(i, 2) * (TableValue(i, 1) - average) ^ 2
 Next i

 variance = variance / n

 stdeviation = variance ^ 0.5

 MsgBox "The average is  " & average
 MsgBox "The variance is  " & variance
 MsgBox "The stdeviation is  " & stdeviation
End Sub


Public Sub Histogram()
 Call Capturen
 Call CaptureTable
 Call Stadigraphs
End Sub


\end{verbatim}

\bigskip
\textbf{\thenum. }  \addtocounter{num}{1}  Exercise.  Work this code and study its organization.

\bigskip
\textbf{\thenum. }  \addtocounter{num}{1}  Our code functions but it is not very elegant.  So, let us make and additional effort to  polish some  details. The next version gets:

\begin{verbatim}

Dim k, n, average, variance, stdeviation As Integer
Dim Table(1000, 2) As Integer

Dim TableValue



Private Sub Capturen()
'
'Captures the number of pairs of data
'
  k = InputBox( _
    prompt:="Please, enter the number of pairs of data", _
    Default:=7)

End Sub

Private Sub CaptureTable()

 TableAddress = Selection.Address
 MsgBox "The region has a range " & TableAddress
 TableValue = Selection.Value

 End Sub

Private Function Averageh(Table)
 n = 0
 average = 0
 For i = 1 To k
    n = n + TableValue(i, 2)
    average = average + TableValue(i, 1) * TableValue(i, 2)
 Next i
    Averageh = average / n
End Function

Private Function Varianceh(Table)
Dim v
  v  = 0
   For i = 1 To k
    v  = v  + TableValue(i, 2) * (TableValue(i, 1) - average) ^ 2
   Next i

variance = v  / n

End Function

Private Function Stdeviationh(Table)
 stdeviation = variance ^ 0.5
End Function


 Sub Stadigraphsh()

 MsgBox "The average is  " & Averageh(Table)
 MsgBox "The variance is  " & Varianceh(Table)
 MsgBox "The stdeviation is  " & Stdeviationh(Table)

End Sub


Public Sub Histogramh()
 Call Capturen
 Call CaptureTable
 Call Stadigraphs
End Sub



\end{verbatim}

\bigskip
\textbf{\thenum. }  \addtocounter{num}{1}  Here we have good encapsulation but we still shall try to improve the quality of the design: we require from the user to write the number of pairs of data. But that is unnecessary: the alone selection of the  table  shall be enough for all purposes.

\begin{verbatim}

Dim k, n, average, stdeviation As Integer
Dim variance
Dim TableValue


Private Sub CaptureTableh()

 TableAddress = Selection.Address
 MsgBox "The region has a range " & TableAddress
 TableValue = Selection.Value
 TableWidth = Selection.Rows.Count
 MsgBox "Number of rows " & TableWidth
 TableHeight = Selection.Columns.Count
 MsgBox "Number of columns " & TableHeight
 k = TableWidth
 End Sub

Private Function Averageh(TableValue)
 n = 0
 average = 0
 For i = 1 To k
    n = n + TableValue(i, 2)
    average = average + TableValue(i, 1) * TableValue(i, 2)
 Next i
    Averageh = average / n
End Function

Private Function Varianceh(TableValue)
 Dim v
  v  = 0
   For i = 1 To k
v=v+TableValue(i, 2)*(TableValue(i, 1)-Averageh(TableValue)) ^ 2
 Next i
 Varianceh = v  / n
End Function

Private Function Stdeviationh(TableValue)
 Stdeviationh = Varianceh(TableValue) ^ 0.5
End Function

Private Sub Stadigraphsh()
 MsgBox "The average is  " & Averageh(TableValue)
 MsgBox "The variance is  " & Varianceh(TableValue)
 MsgBox "The stdeviation is  " & Stdeviationh(TableValue)
End Sub

Public Sub Histogramh()
 CaptureTableh
 Stadigraphsh
End Sub

\end{verbatim}

\bigskip
\textbf{\thenum. }  \addtocounter{num}{1}  Learn to use this code.

\bigskip
\textbf{\thenum. }  \addtocounter{num}{1}  Let do our code   transparent to a person that knows our formulas for the stadigraphs. We could also change the form of doing the final presentation of results  that they could be automatically reusable  in other tasks. We also could hide all  that information that helped to circumvent some  developing problems but that is unnecessary as for the user as for us.

\bigskip
\textbf{\thenum. }  \addtocounter{num}{1}  Example. A Sub procedure to calculate the mean, the variance and the standard deviation of a frequency table with good, transparent style.

\begin{verbatim}



Private Sub CaptureTableh()

 TableAddress = Selection.Address
 'MsgBox "The region has a range " & TableAddress
 TableValue = Selection.Value
 TableWidth = Selection.Rows.Count
 'MsgBox "Number of rows " & TableWidth
 TableHeight = Selection.Columns.Count
 'MsgBox "Number of columns " & TableHeight
 k = TableWidth
 c1 = Selection.Column
 'MsgBox "First column " & c1
 r1 = Selection.Row
 'MsgBox "First row " & r1

 End Sub

Private  Function Averageh(TableValue)
 n = 0
 average = 0
 For i = 1 To k
    n = n + TableValue(i, 2)
    average = average + TableValue(i, 1) * TableValue(i, 2)
 Next i
    Averageh = average / n
End Function

Private  Function Varianceh(TableValue)
 Dim v , xi, ni
   v  = 0
   For i = 1 To k
    xi = TableValue(i, 1)
    ni = TableValue(i, 2)
    v  = v  + ni * (xi - Averageh(TableValue)) ^ 2
  Next i
  Varianceh = v  / n
End Function

Private Function Stdeviationh(TableValue)
 Stdeviationh = Varianceh(TableValue) ^ 0.5
End Function

 Private Sub Stadigraphsh()
 'MsgBox "The average is  " & Averageh(TableValue)
 'MsgBox "The variance is  " & Varianceh(TableValue)
 'MsgBox "The stdeviation is  " & Stdeviationh(TableValue)

  Range(Cells(r1, c1 + 3), Cells(r1, c1 + 3)).Select
  ActiveCell.FormulaR1C1 = "Average"
  Range(Cells(r1, c1 + 4), Cells(r1, c1 + 4)).Select
  ActiveCell.FormulaR1C1 = Averageh(TableValue)

  Range(Cells(r1 + 1, c1 + 3), Cells(r1 + 1, c1 + 3)).Select
  ActiveCell.FormulaR1C1 = "Variance"
  Range(Cells(r1 + 1, c1 + 4), Cells(r1 + 1, c1 + 4)).Select
  ActiveCell.FormulaR1C1 = Varianceh(TableValue)

  Range(Cells(r1 + 2, c1 + 3), Cells(r1 + 2, c1 + 3)).Select
  ActiveCell.FormulaR1C1 = "Stdeviation"
  Range(Cells(r1 + 2, c1 + 4), Cells(r1 + 2, c1 + 4)).Select
  ActiveCell.FormulaR1C1 = Stdeviationh(TableValue)

  Range(Cells(r1 + 3, c1 + 3), Cells(r1 + 3, c1 + 3)).Select
End Sub


Public Sub Histogramf()
 CaptureTableh
 Stadigraphsh
End Sub

\end{verbatim}

\bigskip
\textbf{\thenum. }  \addtocounter{num}{1}  Exercise. Work over the Histogramf module. To run it, you must type a histogram table (2 columns and, say, 7 rows) into a sheet. Next, select the region and call the Public Sub Histogramf.  Pay attention to  some tracer instructions that appear in the code but that were silenced.

\bigskip
\textbf{\thenum. }  \addtocounter{num}{1}   As we can see, our personal experience shows that the debugging process is a very hard enterprise. You must correct the code, the module must be saved, the compiling  machine must be stopped. The data on Excel must be prepared, the Macro must be invoked and run. False products must be detected and corrected. And one must do these operations time after time an  almost endless number of times.

\bigskip
\textbf{\thenum. }  \addtocounter{num}{1}  Exercise. Made  perfectly clear the difference between the developer debugging process and that of natural selection, if any.

\bigskip
\textbf{\thenum. }  \addtocounter{num}{1}  Please, keep guard against the strong temptation of claiming that your personal experience clearly shows that evolution is not possible. In fact, your personal experience only says that to design software with a purpose is very difficult. But as yet we know nothing about designing software by mere gambling.

\section{Graduation}

To graduate from this chapter, you shall incorporate into Excel a Macro to run a Kolmogorov Smirnov test, which  is used to  decide when a frequency table fits a normal distribution.  Help: Excel has the function \textbf{Standard Normal Distribution} that for a given value of $z$ returns the distribution function at $z$, i.e., $\int ^{z}_{-\infty} \phi(s)ds $, where $phi$ is the density of the normal distribution.

Let us look at an example of how the K-S test works:

\bigskip
\textbf{\thenum. }  \addtocounter{num}{1}  Let us   decide whether or not a distribution of tallness fits a normal distribution. The mean height is 1.65 and the standard deviation is 9.5743. So the $z$ associated to a  tallness $x$ is $z=\frac{x-1.65}{9.5743}$. Tallness is represented by $x$, the absolute frequency of $x$ is $N(x)$, the cumulated absolute frequency is $F(x)$, the relative cumulated frequency of $x$ is $\Phi(x) $, the distribution function of $z$ is $\Phi(z)$ and the absolute value of the difference between the theoretical expected distribution and the experimental one is  $|\Phi(x)-\Phi(z)|$. The null hypothesis in this test is that these   differences  are due to mere randomness. The test is calculated with the help of the next table:

\begin{center}
\begin{tabular}{|l|l|l|l|l|l|l|c|}\hline
\multicolumn{7}{|c|}{\vphantom{Large Ap} Fitting of tallness to a normal distribution  }\\ \hline\hline
 $x$& $N(x)$ &$F(x)$&  $\Phi(x)$ & z & $\Phi(z)$ & $|\Phi(x)-\Phi(z)|$ \\ \hline\hline
150& 12  &12& 0.167  &-1.57     & 0.059 &0.108  \\ \hline
160& 24  &36& 0.5    &-0.522    & 0.301 &0.199  \\ \hline
170& 24  &60& 0.833  &0.522     & 0.699 &0.134  \\ \hline
180 & 12 &72& 1      & 1.57   & 0.941 & 0.058\\ \hline

\end{tabular}
\end{center}

The Maximum value of $|\Phi(x)-\Phi(z)|=0.199$  but, according to the Kolmogorov Smirnov theory and if the null hypothesis is correct, the maximum allowed value would be 0.192. Since the observed difference is greater than the allowed one, we reject the null hypothesis: our data does not fit a normal distribution.  The value 0.192 was calculated as $1.63/\sqrt{n} 1.63/\sqrt{72}= 1.63/8.48=0.192$ and approximation that is valid when the number of events is greater than 35. The constant 1.63 is the value associated to the significance $\alpha=0.01$ and $n$ is the total number of data. For $\alpha=0.05$, that constant is 1.36 and the corresponding maximum allowed difference would be $1.36/\sqrt{72}= 1.36/8.48=0.16$. For this value of $\alpha$, the nul hypothesis also would be rejected (of course).

\chapter{The RAND Generator}

\bigskip
\textbf{\thenum. }  \addtocounter{num}{1} Exercise.  We have used the Random generator without much inquiring, but in this chapter we will look at it with the eyes of a scientists, say,  we would like to know the distribution followed by  the  generated numbers.

\bigskip
\textbf{\thenum. }  \addtocounter{num}{1} Exercise.   Every programming language has a generator of random numbers, which is a  program  that produces  a deterministic output   that  simulates random numbers, i.e., if you have an initial sequence of generated numbers, it is difficult for you to predict which number will be the next. But because we have a deterministic program, numbers get repeated or auto-correlated in the long run. For this reason, the generator of random numbers cannot be used for very long sequences. That is why these numbers all called pseudo-random numbers. VBA and Excel use a recursive function that could be initialized by the  time given by the clock of the system.

\bigskip
\textbf{\thenum. }  \addtocounter{num}{1} Example. 480 random numbers were generated and an histogram with 10 classes was made. To study the null hypotheses that the random numbers have a uniform distribution in within zero and one, a chi square test was run.  The error is defined as $(Observed_i-Expected_i)^2/Expected_i$.

\begin{center}
\begin{tabular}{|l|l|l|l|c|}\hline
\multicolumn{4}{|c|}{\vphantom{Large Ap} Random numbers  }\\ \hline\hline
 Class& Observ.  &Expect. & Error  \\ \hline\hline
 0,1    &44&    48 &    0,333333333 \\ \hline
0,2 &49&    48& 0,020833333\\ \hline
0,3 &40&    48& 1,333333333\\ \hline
0,4 &42&    48& 0,75\\ \hline
0,5 &40&    48& 1,333333333\\ \hline
0,6 &54&    48& 0,75\\ \hline
0,7 &59&    48& 2,520833333\\ \hline
0,8 &48&    48& 0\\ \hline
0,9 &56&    48& 1,333333333\\ \hline
1   &48 &48 &0\\ \hline
$\Sigma$&480&   480&    8,375\\ \hline
\end{tabular}
\end{center}

The total error is compared with the theoretically allowed error, which is given by the Chi-square distribution. We have 9 d.f. For $\alpha=0.20$ the corresponding Chi-square is 12.242. For $\alpha=0.10$ the corresponding Chi-square is 14.684. For   $\alpha=0.05$ the corresponding Chi-square is 16.919 and with $\alpha=0.05$ the corresponding Chi-square is  21.666. Our experimental error is 8.375 which is less than all those values. Therefore, the discrepancy between our distribution and a uniform one is too small to be consider as significant. Hence, we conclude that the generator produces random numbers with a uniform distribution in the interval (0,1).

\bigskip
\textbf{\thenum. }  \addtocounter{num}{1} Exercise. Run your own test over your own generator of random numbers.   Hints: In Excel, select a cell, any one, and look at the menu \textbf{Insert} and call the submenu \textbf{Function}. There, look  in the mathematical section for a function that produces a random number and select it. A random number in within zero and one will appear in your selected cell. Read the formula beneath it and pay attention to its exact syntax. For instance, in Spanish it reads \textbf{=aleatorio()}. Select a region in an open sheet and fill it in random numbers. Try with a column, with a row, with a rectangle. Select a rectangular region with some 500 or 1000 cells and fill it in random numbers. Keep the region selected and call the histogram facility. This can be found in Tools, Data analysis, Histogram. The address of the selected region will appear in the dialog box. The other cell of that box shall be filled with a column with the numbers 0, 0.1, 0.2, .. .., 0.9, 1. (Recall that Excel requires colon instead of dot for decimal notation). Next, program Excel to calculate the error and the total error. Compare the total error with the allowed error by a Chi-square table and decide whether or not the distribution is a uniform one.  If your distribution is not uniform, get a new generator of random numbers and learn to use it because we need one to continue.

\bigskip
\textbf{\thenum. }  \addtocounter{num}{1}. Run  a test for the uniformity of the generator for 100 classes, i.e., when the (0,1) interval is divided in 100 sub intervals of equal length.

\bigskip
\textbf{\thenum. }  \addtocounter{num}{1} Example. In VBA, the generator of random numbers is invoked by \textbf{rnd}.

\begin{verbatim}

Public Function RAND() As Single
 RAND = Rnd
 MsgBox "This is a random number " & RAND
End Function

\end{verbatim}

\bigskip
\textbf{\thenum. }  \addtocounter{num}{1} Exercise. test the code.

\bigskip
\textbf{\thenum. }  \addtocounter{num}{1} Example. The generator produces new numbers each time.

\begin{verbatim}

Public Sub LoopRand()
 For i = 1 To 3
  MsgBox "This is a random number " & Rnd
 Next i
End Sub

\end{verbatim}

\bigskip
\textbf{\thenum. }  \addtocounter{num}{1} Exercise. Test the code.

\bigskip
\textbf{\thenum. }  \addtocounter{num}{1} Example. Let us simulate a coin, whose faces are named $m$ and $f$ as in male and female.

\begin{verbatim}

Public Sub Coin()
Dim s As String
 r = Rnd
 If r > 0.5 Then s = "m" Else s = "f"
  MsgBox "The coin landed " & s
End Sub

\end{verbatim}

\bigskip
\textbf{\thenum. }  \addtocounter{num}{1} Exercise. Test the code with \textbf{F8}. Modify  the previous code to simulate sex ratio.  For instance,  modify  the probability of male and pose it equal to 0.7 and that of female equal to 0.3.

Example. Let us simulate a population of 1000 individuals, each one \textbf{f} else \textbf{m}. Let us define the probability of \textbf{m } to be 0.2 and that of \textbf{ f} to be 0.8.

\begin{verbatim}

Public Sub Coins1000()
 Dim X(1 To 1000) As Integer

 'Generate 1000 individuals
 'Probability of males 0.2
 p = 0.2
 For i = 1 To 1000
  r = Rnd
  If r < p Then X(i) = 1 Else X(i) = 0
 Next i
  'Count the number of males
  Males = 0
  For i = 1 To 1000
  If X(i) = 1 Then Males = Males + 1
 Next i
 MsgBox "The number of Males is  " & Males
 'Statistical test to see the fairness of the simulation
 sigmap = (p * (1 - p) / 1000) ^ 0.5
 zexp = (Males / 1000 - p) / sigmap
  MsgBox "The z of the experiment is " & zexp
  s = 1.96
  MsgBox "The critical z is   " & s
  A1 = "Yes"
  A2 = "No"
  If Abs(zexp) < s Then l = A1 Else l = A2
     MsgBox "Is the generator fair?  " & l
End Sub

\end{verbatim}

\bigskip
\textbf{\thenum. }  \addtocounter{num}{1} Exercise. Run the code under  F8.

\bigskip
\textbf{\thenum. }  \addtocounter{num}{1} Graduation.  Modify the previous code  to produce a population of random numbers of arbitrary size (less than 30000). In the code, the significance 0.05 is inbuilt. Modify the code to accept any significance.

\chapter{Markov chains}

\bigskip
\textbf{\thenum. }  \addtocounter{num}{1} We will deal with a theme that is  nice and powerful: Markov Chains, which studies a system that is best described probabilistically and that has transitions from an allowed state to another. We suppose that the number of allowed states is finite in number.

\bigskip
\textbf{\thenum. }  \addtocounter{num}{1} Let us consider a system that may occupy one of two states. State one is A, state two is B. The transitions probabilities are

$p(A\rightarrow B) = 0.2$

$p(A\rightarrow A) = 0.8$

$p(B\rightarrow A) = 0.3$

$p(B\rightarrow B) = 0.7$

Our aim is to understand  the behavior of the system in the long run for a given initial condition.

Let us consider a total population of 3000 individuals. The  initial distribution is given by   $A_o= 1000$ and $B_o= 2000$. In probabilistic terms, this corresponds to an initial probability of A equal to 1/3 and that of B equal to 2/3.

\begin{verbatim}

Public Sub MarkovChain()
 Dim X(1 To 3000) As Integer
 Dim pa As Single
 'Define probabilities of transition
 paa = 0.8
 pbb = 0.7
 'Ask for the initial conditions
 pa = InputBox("Enter the initial p(A)", "M.C.: Init.  condition for A")
 'Generate 3000 individuals
  MsgBox "The   initial probability of state A is " & pa
 na = 0
 nb = 0
 For i = 1 To 3000
  r = Rnd
  If r < pa Then na = na + 1 Else nb = nb + 1
 Next i
  'Report of  the number of individuals in each state
    MsgBox "The number of individuals in state A is  " & na
    MsgBox "The number of individuals in state B is  " & nb
  'Execute 1000 transitions
  For i = 1 To 1000
    'Decide whether or not an individual in A must transite
    For j = 1 To na
     s = Rnd
     If s > paa Then
       na = na - 1
       nb = nb + 1
      End If
    Next j
    'Decide whether or not an individual in B must transite
    For j = 1 To nb
     s = Rnd
     If s > pbb Then
       nb = nb - 1
       na = na + 1
      End If
    Next j
  Next i
 MsgBox "The number of individuals in state A is  " & na
    MsgBox "The number of individuals in state B is  " & nb
End Sub


\end{verbatim}

\bigskip
\textbf{\thenum. }  \addtocounter{num}{1} Exercise. Run the code.  Warning: In Excel a decimal number must appear with colon as 0,2222 but in VBA with dot as 0.2222. This is a Bug of intercommunication of developer teams. This is typical. The same bug appears in the InputBox: decimal numbers must be typed with colon but in the code they must be written with a dot.

\bigskip
\textbf{\thenum. }  \addtocounter{num}{1} Example. The code was run with diverse initial probabilities, $p_a$ of state A.

\begin{center}
\begin{tabular}{|l |l|l|c|}\hline
\multicolumn{3}{|c|}{\vphantom{Large Ap} Markov Chains  }\\ \hline\hline
$p_a$ & $n_a$ & $f_a$\\ \hline\hline
0.10 &  2088 & 0.696\\ \hline
0.20 & 2067 & 0.689\\ \hline
0.30 & 2084 &  0.695\\ \hline
0.40 & 2023 & 0.674\\ \hline
0.50 & 1975 & 0.658\\ \hline
0.60 & 2089 & 0.696\\ \hline
0.70 & 2084 & 0.694\\ \hline
0.80 & 2093 & 0.698\\ \hline
0.90 & 2068 & 0.689\\ \hline

\end{tabular}
\end{center}

The mean value for $n_a$ is 2063, which corresponds to a proportion of 2063/3000= 0.688.

The Anova to test whether or not the slope of the relation ($p_a$,  $n_a$ ) is different than zero was calculated by Excel and gave the following table:

\begin{center}
\begin{tabular}{|l|l|l|l|l|l|c|}\hline
\multicolumn{6}{|c|}{\vphantom{Large Ap} Anova for regression }\\ \hline\hline
Variation& SS&DF& Mean Square & Fisher &  Critical F \\ \hline\hline
Regression&1    & 68,26666667   &68,26666667    & 0,038543989   & 0,849932894 \\ \hline
Error & 7   & 12397,95556   & 1771,136508    & &     \\ \hline
Total  & 8  & 12466,22222       & & &  \\ \hline
\end{tabular}
\end{center}

\bigskip
\textbf{\thenum. }  \addtocounter{num}{1} We clearly see that the initial condition does not influence the final state. We may postulate that a Markov chain has a particular asymptotic state to which the system tends no matter form which initial condition it is   turned  on.

\bigskip
\textbf{\thenum. }  \addtocounter{num}{1} Complaint: How do we know that after 1000 generations we already achieved the final state?

\bigskip
\textbf{\thenum. }  \addtocounter{num}{1} Exercise. Run the next code. Find the instructions that cause the program to make a report each 100 generations. Verify that by generation 100 we already have attained the asymptotic state. Use Excel to make a graphic of the dynamics.

\begin{verbatim}

Public Sub MarkovChain()
 Dim X(1 To 3000) As Integer
 Dim l As Integer
 Dim pa As Single
 'Define probabilities of transition
 paa = 0.8
 pbb = 0.7
 'Ask for the initial conditions
 pa = InputBox("Enter the initial p(A)", "M. C.: Init. condition for A")
 'Generate 3000 individuals
  'MsgBox "The   initial probability of state A is " & pa
 na = 0
 nb = 0
 For i = 1 To 3000
  r = Rnd
  If r < pa Then na = na + 1 Else nb = nb + 1
 Next i
  'Report of  the number of individuals in each state
    'MsgBox "The number of individuals in state A is  " & na
    'MsgBox "The number of individuals in state B is  " & nb
  'Execute 1000 transitions
  For i = 1 To 1000
    'Decide whether or not an individual in A must  pass to B
    For j = 1 To na
     s = Rnd
     If s > paa Then
       na = na - 1
       nb = nb + 1
      End If
    Next j
    'Decide whether or not an individual in B must pass to A
    For j = 1 To nb
     s = Rnd
     If s > pbb Then
       nb = nb - 1
       na = na + 1
      End If
    Next j
    l = i / 100
    If i = l * 100 Then
     MsgBox "i  is  " & i
     MsgBox "The number of individuals in state A is  " & na
    End If
  Next i
 MsgBox "The number of individuals in state A is  " & na
    'MsgBox "The number of individuals in state B is  " & nb
End Sub

\end{verbatim}

\bigskip
\textbf{\thenum. }  \addtocounter{num}{1} Since we know that the system tends to the equilibrium before the 100th generation, we instruct VBA to report the first 100 generations and to show then in Excel. This is done thanks to the next code:

\begin{verbatim}

Public Sub MarkovChain()
 Dim X(1 To 3000) As Integer
 Dim l As Integer
 Dim pa As Single
 'Define probabilities of transition
 paa = 0.8
 pbb = 0.7
 'Ask for the initial conditions
 pa=InputBox("Enter the init. p(A)", "M.C.: Init. condition for A")

  'MsgBox "The   initial probability of state A is " & pa
 na = 0
 nb = 0
 'Open a new sheet
 Workbooks.add
 'Initializes the sheet
 Range(Cells(1, 2), Cells(1, 2)).Select
  ActiveCell.FormulaR1C1 = "Generation"
  Range(Cells(1, 3), Cells(1, 3)).Select
  ActiveCell.FormulaR1C1 = "Population A"

 'Generates 3000 individuals
 For i = 1 To 3000
  r = Rnd
  If r < pa Then na = na + 1 Else nb = nb + 1
 Next i
  'Report of  the number of individuals in each state
    'MsgBox "The number of individuals in state A is  " & na
    'MsgBox "The number of individuals in state B is  " & nb
    'Reports initial generation
       Range(Cells(2, 2), Cells(2, 2)).Select
  ActiveCell.FormulaR1C1 = 0
  Range(Cells(2, 3), Cells(2, 3)).Select
  ActiveCell.FormulaR1C1 = na
  'Execute 100 transitions
  For i = 1 To 100
    'Decide whether or not an individual in A must transite
    For j = 1 To na
     s = Rnd
     If s > paa Then
       na = na - 1
       nb = nb + 1
      End If
    Next j
    'Decide whether or not an individual in B must transite
    For j = 1 To nb
     s = Rnd
     If s > pbb Then
       nb = nb - 1
       na = na + 1
      End If
    Next j
   Range(Cells(i + 2, 2), Cells(i + 2, 2)).Select
  ActiveCell.FormulaR1C1 = i
  Range(Cells(i + 2, 3), Cells(i + 2, 3)).Select
  ActiveCell.FormulaR1C1 = na
  Next i
 End Sub

\end{verbatim}

\bigskip
\textbf{\thenum. }  \addtocounter{num}{1} Exercise. Run the code. Verify that it produces   a report of the first 100 generations over an Excel sheet. Use Excel to make a graphic of population number in state A against generation number. Hint : use the scatter mode of the drawing facility in Excel. Verify that for this particular Markov  chain the equilibrium has been attained by generation 20. Try various runs to get a complete scenario for the evolution of the system beginning from different initial conditions.

\bigskip
\textbf{\thenum. }  \addtocounter{num}{1} Our simulation allows us to postulate that any Markov chain tends towards an equilibrium and that the system arrives near the equilibrium   in the very first generations. This implies that if  a natural system is described by a Markov chain, it is probably near equilibrium. Otherwise there shall exist forces that maintain the system far from equilibrium.

\bigskip
\textbf{\thenum. }  \addtocounter{num}{1} Now that we have made sure that the Markov chain attends its equilibrium rather soon, we can compare our asymptotic state with the mathematical prediction.

Let us use the following conventions

$p_{aa}$= transition probability from state A to state A in one generation to the next.

$p_{ab}$= transition probability from state A to state B in one generation to the next.

$p_{ba}$= transition probability from state B to state A in one generation to the next.

$p_{bb}$= transition probability from state B to state B in one generation to the next.

This implies that the proportions evolve according to the following equations:

$x_a (t+1) =p_{aa} x_a (t) + p_{ba} x_b(t)$

$x_b (t+1) =p_{ab} x_a (t) + p_{bb} x_b(t)$

For our example, $p_{aa}= 0.8$ so, $p_{ab}=0.2$ and $p_{bb}=0.7$ so $p_{ba}=0.3$ Then, we get

$x_a (t+1) =0.8 x_a (t) + 0.3 x_b(t)$

$x_b (t+1) =0.2 x_a (t) + 0.7 x_b(t)$

In equilibrium, the occupied state at $t+1$ coincides with the state at $t$. Hence, the proportions at equilibrium obey the equations

$x_a   =0.8 x_a   + 0.3 x_b $

$x_b   =0.2 x_a   + 0.7 x_b $

Replacing $x_b=1-x_a$ in the first equation, we get

$x_a   =0.8 x_a   + 0.3 (1-x_a) = 0.8 x_a   + 0.3 1-0.3x_a$

$x_a   =0.5 x_a   + 0.3 $

$0.5x_a   =  0.3 $

$x_a   =  0.3/0.5 = 3/5=0.6$

\bigskip
\textbf{\thenum. }  \addtocounter{num}{1} Thus, if we have 3000 individuals, the expected number of individuals in state A when the system is in equilibrium shall be 9000/5= 1800. Our experimental results render values near 2063, with a proportion of 0.688. Is there a sensible difference?

Let use a $z$ test to  compare the experimental and the expected frequencies. We have a binomial distribution with n=3000, p=0.6, 1-p=0.4, so

$\sigma_p= \sqrt{ p(1-p)/n}=\sqrt{ 0.6(0.4)/3000} = 0.009$:

 Hence $z = (0.6-0.688)/0.009 = -9.7$ which is very improbable by mere randomness. How can we explain the discrepancy? We could speak of error multiplications from generation from generation but on the other hand, the experimental results are very close one to another, so that explanation seems to be unjustified. We conclude that our generator of random numbers is not very good for predicting accurate values but the mathematical theory beneath all this assures that our simulation is qualitatively good.

\bigskip
\textbf{\thenum. }  \addtocounter{num}{1} Graduation. Simulate a Markov Chain with 3 states and study its evolution. Write a presentation in Power point  including 5 slices.

\section{Improving the style}

\bigskip
\textbf{\thenum. }  \addtocounter{num}{1} No one is able to produce at once a program that makes what it must. Every developer tries many times before arriving to a correct solution. Experienced developers may write some 60 lines of debugged  code per day. Usually, one tends to refine  the style of the software after one sees that it works. But experience has shown that a good style of programming is very useful to diminishing  the suffering of software development. Experts write advices of diverse flavor. In fact, the modern languages of programming come with environments specially suited to ease the debugging process.

A good style includes    encapsulating separate task into different capsules. For the last program, this could be done as follows:

\begin{verbatim}



 Private  Sub Initialization()
 'Opens a new sheet
 Workbooks.add
 'Initializes the sheet with the titles
 Range(Cells(1, 2), Cells(1, 2)).Select
  ActiveCell.FormulaR1C1 = "Generation"
  Range(Cells(1, 3), Cells(1, 3)).Select
  ActiveCell.FormulaR1C1 = "Population A"
  End Sub

  Private Function Seed(pa)
 'Generates 3000 individuals
 na = 0
 nb = 0
 'MsgBox "The   initial probability of state A is " & pa
 For i = 1 To 3000
  r = Rnd
  If r < pa Then na = na + 1 Else nb = nb + 1
 Next i
 Seed = na
  'Reports of  the number of individuals in each state
    'MsgBox "The number of individuals in state A is  " & na
    'MsgBox "The number of individuals in state B is  " & nb
    'Reports initial generation
       Range(Cells(2, 2), Cells(2, 2)).Select
  ActiveCell.FormulaR1C1 = 0
  Range(Cells(2, 3), Cells(2, 3)).Select
  ActiveCell.FormulaR1C1 = na
  End Function

  Private Function Dynamics(na, nb, paa, pbb)
  'Executes 100 transitions
  For i = 1 To 100
    'Decides whether or not an individual in A must pass to B
    For j = 1 To na
     s = Rnd
     If s > paa Then
       na = na - 1
       nb = nb + 1
      End If
    Next j
    'Decides whether or not an individual in B must pass to A
    For j = 1 To nb
     s = Rnd
     If s > pbb Then
       nb = nb - 1
       na = na + 1
      End If
    Next j
   Range(Cells(i + 2, 2), Cells(i + 2, 2)).Select
  ActiveCell.FormulaR1C1 = i
  Range(Cells(i + 2, 3), Cells(i + 2, 3)).Select
  ActiveCell.FormulaR1C1 = na
  Next i
   End Function

Public Sub MarkovChaind()
 Dim na As Integer, nb As Integer, l As Integer
 Static pa As Single, paa As Single, pbb As Single
 Dim r As Single

 'Asks for the initial conditions
 pa=InputBox("Enter the init. p(A)", "M. C.: Init. condition for A")
 'MsgBox "The   initial probability of state A is " & pa
 'Defines probabilities of transition
 paa = 0.8
 pbb = 0.7
 Initialization
 na = Seed(pa)
 nb = 3000 - na
 na = Dynamics(na, nb, paa, pbb)
End Sub

\end{verbatim}

\bigskip
\textbf{\thenum. }  \addtocounter{num}{1} Exercise. Observe the changes that were needed to just achieve encapsulation. Encapsulation creates many troubles for the developer of software. Why do developers prefer to face up these problems? It is because this generates  a style that easies debugging, the unavoidable process of correcting errors in the design of software. The same problem is seemingly faced in the manage of software in the cell: there are concrete signals to determine when a DNA sequence is to be used in within  a given organelle or if the product must go to the main stream. Prove that the fact that the structure of the genome   follows an encapsulation style does not imply that the design was necessarily generated by an intelligent being.

\bigskip
\textbf{\thenum. }  \addtocounter{num}{1} It is very unpleasant that the mathematical predictions of the Markov chain theory does not coincide with the results of the simulation. Let us consider the possibility that the generator of random numbers is a little strayed. This is usual among those generators. So, let us develop a program to produce two events, one with probability 0.7 and the other with probability 0.3, so that we could  compare expectation and experiment. We make a simple experiment to see how the experimental frequency evolves, as follows:

\begin{verbatim}

Public Sub tracking()
 'Traces the experimental frequency
 Dim l As Integer

 pa = 0.8
 na = 0
 nb = 0

 For i = 1 To 500
  r = Rnd
  If r < pa Then na = na + 1 Else nb = nb + 1
  exppa = na / i
  l = i / 100
  If i = l * 100 Then
   MsgBox "Generation  " & i
   MsgBox "The experimental p(a) is  " & exppa
   End If
 Next i

End Sub

\end{verbatim}

Now, we automatize everything for 30000 generations.

\begin{verbatim}

Private Sub Titles()
 'Opens a new sheet
 Workbooks.add
 'Initializes the sheet with the titles
 Range(Cells(1, 2), Cells(1, 2)).Select
  ActiveCell.FormulaR1C1 = "Generation"
  Range(Cells(1, 3), Cells(1, 3)).Select
  ActiveCell.FormulaR1C1 = "Frequency of A"
  End Sub

Public Sub TrackinProb()
 '
 'Traces the experimental frequency
 Dim l As Integer
 '
 pa = 0.8
 na = 0
 nb = 0
 Count = Count + 1
 Titles
 For i = 1 To 30000
  r = Rnd
  If r < pa Then na = na + 1 Else nb = nb + 1
  exppa = na / i
  l = i / 300
  If i = l * 300 Then
    Count = Count + 1
    Range(Cells(Count, 2), Cells(Count, 2)).Select
    ActiveCell.FormulaR1C1 = i
    Range(Cells(Count, 3), Cells(Count, 3)).Select
    ActiveCell.FormulaR1C1 = exppa
   End If
 Next i
End Sub


\end{verbatim}

The results appeared in a sheet, a graphic was drawn using the Excel graphic facility  and a tendency to overestimate the probabilities were evidenced: there is a surplus of 0.003. Let us now make the correction in our Markow Chain simulation. Let us took the module \textbf{Public Sub MarkovChain()} and run it with $paa= 0.797$ and $pbb =0.697 $ instead of 0.8 and 0.7 to see what happens. The result, after a long time of running, was that after 30000 generations the final occupation number of state A was 2020 with a relative frequency of 2020/3000 = 0.673, which is too far afield from the expected 0.6.

\bigskip
\textbf{\thenum. }  \addtocounter{num}{1} Exercise. Now, you: re-engineer the  module \textbf{Public Sub MarkovChain()} to execute 30000 generations and to report only the final result. Impose the conditions $paa= 0.797$ and $pbb =0.697 $. Compare expected and experimental results.

\bigskip
\textbf{\thenum. }  \addtocounter{num}{1} Graduation: If you find like we that our corrections produces no improving in the fitting of the theoretical expectations, then you are challenged to produce something better than us and to find the remedy to our problem. Maybe there is none, in that case, explain why.

\chapter{Lotka Volterra}

Let us simulate a pray-predator system. Very simple and very instructive is the model proposed by Lotka and  Volterra. We will solve the deterministic model in first place to pass then to consider an stochastic one.

The deterministic model is as follows: we have a pray, say rabbits, that feed on grass at a constant rate. The number of offspring of rabbits that is born per unit time is proportional to the number of rabbits (all rabbits can reproduce and sex is forgiven). Predation is the only cause of a dead of a rabbit. The number of rabbit's deaths  is proportional to the number of encounters among rabbits and wolves, which in its turn is proportional to the population numbers of both species. Moreover, wolves die at natural death. The number of  dead wolves  per unit time is proportional to the population.
 
\begin{verbatim}

Public Sub LotkaVolterra()

'Study of predation of a pray-predator system.
'Deterministic model

'Generates 500 rabbits,  50 wolves
 nr = 500
 nw = 50
 'Rate of rabbit's reproduction
 r = 0.1
 'Rate of predation
 p = 0.01
 'Rate of wolves' death
 k = 0.05
 'Titles
 'Opens a new sheet
  Workbooks.add
  Range(Cells(1, 2), Cells(1, 2)).Select
  ActiveCell.FormulaR1C1 = "Rabbits"
  Range(Cells(1, 3), Cells(1, 3)).Select
  ActiveCell.FormulaR1C1 = "Wolves"
'We run N generations
 n = 1000
 For i = 1 To n
'Rabbits reproduce
 nr = nr + r * nr
 'Rabbits are prayed
 nr = nr - p * nr * nw
 'Wolves reproduce
 nw = nw + 0.1 * p * nr * nw
 'Wolves die
 nw = nw - k * nw
'Report
  Range(Cells(i + 1, 2), Cells(i + 1, 2)).Select
  ActiveCell.FormulaR1C1 = nr
  Range(Cells(i + 1, 3), Cells(i + 1, 3)).Select
  ActiveCell.FormulaR1C1 = nw
Next i
End Sub

\end{verbatim}

\bigskip
\textbf{\thenum. }  \addtocounter{num}{1} Exercise. Run the code. Use the drawing facility of Excel with Lines mode to make a drawing and verify that a cyclic dynamics is observed. Make a  study of periodicity in terms of the different parameters.  Determine the value of the  parameters to get a soft sinusoidal dynamics.

\bigskip
\textbf{\thenum. }  \addtocounter{num}{1} Exercise. Write down the system of ordinary differential equations that is solved by this program.

\bigskip
\textbf{\thenum. }  \addtocounter{num}{1} Let us consider now the stochastic version of this model. We first consider a very simple code that includes the following loop, where nr is the number of rabbits and r is a threshold:

\begin{verbatim}

 For j = 1 To nr
   s = Rnd
   If s < r Then nr = nr + 1
 Next j

 \end{verbatim}

 Let us observe that the parameter nr enters  both as a control statement, in the first line,  and as a modifying property in within the code. This causes a self reference looping that in general is very difficult to manage. Nevertheless, we use small values of $r$ and so children per generation are not too many and the ensuing perturbation is tolerable. Let us see why: the intention of this code is that rabbits have a son if a random number happens to be less than the threshold r. Thus, the number of children is proportional to nr and to r. But, the code does not match the intention: the number of rabbits augments with each child, or in other words, each child is qualified to  have children just from the very instant of birth. Anyway, as we will see somewhere below, a correction to this problem is possible although it generates more work and more code.

\begin{verbatim}

Public Sub StochastLVolterra()

'Study of predation of a pray-predator system.
'Stochastic model

'Generates 500 rabbits,  50 wolves
 nr = 500
 nw = 50
 'Probability of rabbit's reproduction
 r = 0.1
 'probability of predation
 p = 0.002
 'Probability of wolf's death
 wd = 0.05
 'Titles
 'Opens a new sheet
  Workbooks.add
  Range(Cells(1, 2), Cells(1, 2)).Select
  ActiveCell.FormulaR1C1 = "Rabbits"
  Range(Cells(1, 3), Cells(1, 3)).Select
  ActiveCell.FormulaR1C1 = "Wolves"
'We run n generations
 n = 1000
 For i = 1 To n
'Rabbits reproduce
  For j = 1 To nr
   s = Rnd
   If s < r Then nr = nr + 1
 Next j
 'Rabbits are prayed, wolves reproduce
  For l = 1 To nw
   For j = 1 To nr
    s = Rnd
    If s < p Then nr = nr - 1
    If s < 0.1 * p Then nw = nw + 1
   Next j
  Next l
 'Wolves die
 For j = 1 To nw
   s = Rnd
   If s < wd Then nw = nw - 1
 Next j
'Report
  Range(Cells(i + 1, 2), Cells(i + 1, 2)).Select
  ActiveCell.FormulaR1C1 = nr
  Range(Cells(i + 1, 3), Cells(i + 1, 3)).Select
  ActiveCell.FormulaR1C1 = nw
Next i
End Sub

\end{verbatim}

\bigskip
\textbf{\thenum. }  \addtocounter{num}{1} Exercise. Run the code and make a graphic of the dynamics.

\bigskip
\textbf{\thenum. }  \addtocounter{num}{1} Exercise. Try to figure out the form of the stochastic differential equations that are defined by a continuous approximation to this simulation.

\bigskip
\textbf{\thenum. }  \addtocounter{num}{1} Exercise. Make an experimental program of investigation of this virtual world. Execute it. Compare the deterministic and the stochastic model.

\bigskip
\textbf{\thenum. }  \addtocounter{num}{1} One observes that the virtual world that we have built is unstable: a spur of rabbits always appear and after that wolves reproduce too much and then rabbits disappear. To make that world stable, we could cause rabbits to reproduce only in the measure that reproduction is allowed by   finite resources. Our new setting  is the logistic Lotta Volterra model.

\begin{verbatim}

Public Sub LogisticLVolterra()

'Study of predation of a pray-predator system.
'Stochastic model.
'Reproduction of rabbits obeys a logistic model

'Generates 500 rabbits,  50 wolves
 nr = 500
 nw = 50
 'Probability of rabbit's reproduction
 r = 0.1
 'Maximum number of Rabbits
 Max = 2000
 'probability of predation
 p = 0.002
 'Probability of wolf's death
 wd = 0.05
 'Titles
 'Opens a new sheet
  Workbooks.add
  Range(Cells(1, 2), Cells(1, 2)).Select
  ActiveCell.FormulaR1C1 = "Rabbits"
  Range(Cells(1, 3), Cells(1, 3)).Select
  ActiveCell.FormulaR1C1 = "Wolves"
'We run n generations
 n = 1000
 For i = 1 To n
'Rabbits reproduce
  For j = 1 To nr
   s = Rnd
   If s < r * (1 - nr / Max) Then nr = nr + 1
 Next j
 'Rabbits are prayed, wolves reproduce
  For l = 1 To nw
   For j = 1 To nr
    s = Rnd
    If s < p Then nr = nr - 1
    If s < 0.1 * p Then nw = nw + 1
   Next j
  Next l
 'Wolves die
 For j = 1 To nw
   s = Rnd
   If s < wd Then nw = nw - 1
 Next j
'Report
  Range(Cells(i + 1, 2), Cells(i + 1, 2)).Select
  ActiveCell.FormulaR1C1 = nr
  Range(Cells(i + 1, 3), Cells(i + 1, 3)).Select
  ActiveCell.FormulaR1C1 = nw
Next i
End Sub

\end{verbatim}

\bigskip
\textbf{\thenum. }  \addtocounter{num}{1} Exercise. Run the code and verify that it gets stable. Is it stable against any perturbation allowed by the other parameters? Would we need to purge wolves when they get too many? How would be do that?

\bigskip
\textbf{\thenum. }  \addtocounter{num}{1} Exercise. Rewrite the code with an encapsulating style. Make sure that the new program produces the same results as the former one.

\bigskip
\textbf{\thenum. }  \addtocounter{num}{1} Graduation. Add sex to the model: both populations are composed of males and females.

\chapter{Surviving of the fittest}

\bigskip
\textbf{\thenum. }  \addtocounter{num}{1} In this chapter we will understand evolution as the differential surviving of the members of a population in response to environment pressures. Everything gets clear with a single example. To that purpose, let us simulate a field observation that played a historical role in the supporting of the evolutionary theory: white moths survive predation by birds better than black ones in a clean environment but worse in a blacked one.

\section{Moths}

\bigskip
\textbf{\thenum. }  \addtocounter{num}{1} We suppose that the number of encounters among birds and moths is proportional to their population numbers. Next, the constant of proportionality is type moth dependent. In a clean environment, the constant of proportionality is greater for blacks moths than for white one and in a contaminated environment, things happen the other way around. We suppose that  moths reproduce in equal proportions independent of the type but that they obey a logistic model, which takes into account the role of ecological restrictions. We fix the maximum  number of moths  in 1000, counting both types. Birds are predators and we will suppose that they feed on moths only. For stability reasons, we also add a logistic grow. Their rate of reproduction depends on their capacity to get food. Our initial setting is a clean environment with white moths abounding and black ones depleted.

All that we need is to modify a bit our previous pray-predator model. This code runs substantially slowly than the previous ones. Please, divide the screen in two portions, one for Excel and the other for VBA: that would allow you to follow the dynamics.

\begin{verbatim}

Public Sub Moths()

'Study of predation of moths by birds
'Moths could be white or brown.
'Reproduction  obeys a logistic model
'Two environments: clean and dirty.
'Stochastic model.


'Generates 500 moths,
 nw = 450
 nbr = 450
'Generates 10 birds
 nb = 10
 'Probability of moth's reproduction
 mr = 0.1
 'Maximum number of Moths
 Maxm = 2000
 'Maximum number of Birds
 Maxb = 20
 'probability of predation on white moths
 pw = 0.002
 'probability of predation on brown moths
 pbr = 0.003
 'Probability of bird's death
 bd = 0.05
 'Titles
 'Opens a new sheet
  Workbooks.add
  Range(Cells(1, 2), Cells(1, 2)).Select
  ActiveCell.FormulaR1C1 = "White moths"
  Range(Cells(1, 3), Cells(1, 3)).Select
  ActiveCell.FormulaR1C1 = "Brown moths"
  Range(Cells(1, 4), Cells(1, 4)).Select
  ActiveCell.FormulaR1C1 = "Birds"
'We run n generations
 n = 1000
 For i = 1 To n
'White moths reproduce
  For j = 1 To nw
   s = Rnd
   If s < mr * (1 - (nw + nbr) / Maxm) Then nw = nw + 1
 Next j
 'Brown moths reproduce
  For j = 1 To nbr
   s = Rnd
   If s < mr * (1 - (nw + nbr) / Maxm) Then nbr = nbr + 1
 Next j
 'Moths are prayed, birds reproduce
  For l = 1 To nb
  'White moths
   For j = 1 To nw
    s = Rnd
    If s < pw Then nw = nw - 1
    If s < 0.1 * pw * (1 - nb / Maxb) Then nb = nb + 1
   Next j
   'brown moths
   For j = 1 To nbr
    s = Rnd
    If s < pbr Then nbr = nbr - 1
    If s < 0.1 * pbr * (1 - nb / Maxb) Then nb = nb + 1
   Next j
  Next l
 'Birds die
 For j = 1 To nb
   s = Rnd
   If s < bd Then nb = nb - 1
 Next j
'Report
  Range(Cells(i + 1, 2), Cells(i + 1, 2)).Select
  ActiveCell.FormulaR1C1 = nw
  Range(Cells(i + 1, 3), Cells(i + 1, 3)).Select
  ActiveCell.FormulaR1C1 = nbr
  Range(Cells(i + 1, 4), Cells(i + 1, 4)).Select
  ActiveCell.FormulaR1C1 = nb
Next i
End Sub

\end{verbatim}

\bigskip
\textbf{\thenum. }  \addtocounter{num}{1} Exercise. Play with parameters to prevent the extinction of the unfavored type of moths. What does  happen to oscillations in the population numbers?

\bigskip
\textbf{\thenum. }  \addtocounter{num}{1} Exercise. Do we have here a  methodological error? In fact, in the code, white moths reproduce first than brown ones? Does this generate a selection in favor of one type and in detriment of the other? How could you know that? Please, test your ideas.

\bigskip
\textbf{\thenum. }  \addtocounter{num}{1} In principle, any tiny difference between the predation parameters should cause the disappearance of the unfavored type of moths. But, it is known that the unfavored type of moths always have representatives in the population. The only explanation is mutation: the offspring of a type of moth not necessarily belongs to the same type. The implementation follows. We see for the first time the appearance of nested ifs. To understand the mode of compilation, we apply a simple rule: when an If conditions is written in a single line, there is no need of End If statement. But if the If is written in various lines, we need to end with the End If statement.



\bigskip
\textbf{\thenum. }  \addtocounter{num}{1} Exercise. Verify that with this setting, brown moths never disappear.  Can you restore oscillations somehow?

\bigskip
\textbf{\thenum. }  \addtocounter{num}{1} Exercise. Develop a code to repeat the following field observation. Initial conditions: clean environment, prevalence of white moths. After equilibrium has been achieved, external  contamination shifts the selection parameters in favor of brown moths. An ensuing change in the population  shows that brown moths prevail. Hint: make a copy of a suitable portion of the code and paste it just before the end of the original one and update the values of the parameters for the second part of the simulation. Delete from the second part the portion corresponding to titles and to the opening of a new sheet and to the parameters that remain the same. In the reporting section update the row: the new stadium shall begin at row 1001. To know that you have succeed, draw the dynamics for 2000 generations: at generation 1000, a flip flop of white and brown numbers shall be observed.

\bigskip
\textbf{\thenum. }  \addtocounter{num}{1} Exercise. Compare your code with the following one, which also solves the previous exercise.



\section{In hindsight}

\bigskip
\textbf{\thenum. }  \addtocounter{num}{1} By this moment one knows by experience that  it is easily to commit many errors in a developing   software. But the most important thing is to understand why can we know that we commit an error, a bug. In first place, it is because we have an aim that we follow. In second place, it is because   any tiny change in the code causes a possible abrupt change in the resulting dynamics. In technical terms we say that software executing is at the edge of chaos: this means that we still know what we do and what we want, but if we deviate a bit, we pay an exorbitant price in the results.

\bigskip
\textbf{\thenum. }  \addtocounter{num}{1} The same applies to genetic software: it is very good to produce mutations, if they are viable. In effect, minor changes in the genetic messages could cause exorbitant changes in the morphology or physiology of the carrier organism. But in nature, we have no aim, so the word bug has no meaning in nature. It seems that the only point of contact among human developers and the blind developer is that both have problems with viability: it is not easy to concatenate symbols that could produce a code that could be executed, be it it in software development or in genetics.

\bigskip
\textbf{\thenum. }  \addtocounter{num}{1} Research: investigate the rate of inviability or natural abortion in flies, mammals and humans. How could we relate those rates with our problem of developing viable code?

\bigskip
\textbf{\thenum. }  \addtocounter{num}{1} Let us dig a bit on another trait that is very important for humans: they can produce viable software in very different styles. To understand this assertion better, let us rewrite the last code in two other styles. The code that was reported as a sample solution to the last exercise, already has an style: it solves a task in the simplest way using what we have at hand. The result is known as a duplication plus some minor but fundamental adaptations. This is very reminiscent of the genetic code. Let us become acquainted with two other styles.

\section{The GOTO statement}

\bigskip
\textbf{\thenum. }  \addtocounter{num}{1} A  duplication of code into a code could be considered naive by most developers. The following solution, based on the GOTO statement  is more recursive but is is considered entangled. Next we show a less entangled style. The GOTO statement means that one may label a line with a number or a name  and return to it when needed. That is  clear in the next code that is another solution to the same previous exercise:

\begin{verbatim}

Public Sub MothsFlipFlop()

'Study of predation of moths by birds
'Moths could be white or black.
'Reproduction  obeys a logistic model
'Two environments: clean and dirty.
'Stochastic model.

'Clean environment
'Generates 500 moths,
 nw = 450
 nbr = 450
'Generates 10 birds
 nb = 10
 'Probability of moth's reproduction
 mr = 0.1
 'Maximum number of Moths
 Maxm = 2000
 'Maximum number of Birds
 Maxb = 20
'Probability of bird's death
 bd = 0.05
 'Mutation rate of moths
 mut = 0.01
 'probability of predation on white moths
 pw = 0.002
 'probability of predation on brown moths
 pbr = 0.003
 'Titles
 'Opens a new sheet
  Workbooks.add
  Range(Cells(1, 2), Cells(1, 2)).Select
  ActiveCell.FormulaR1C1 = "White moths"
  Range(Cells(1, 3), Cells(1, 3)).Select
  ActiveCell.FormulaR1C1 = "Brown moths"
  Range(Cells(1, 4), Cells(1, 4)).Select
  ActiveCell.FormulaR1C1 = "Birds"
  'We need a counter of cycles
  Cycle = 1
  'In the first cycle the report begins on line 2
  Delay = 1
  'We jump over the initialization of the second cycle
  GoTo Always
  'The second cycle is initialized here
  Secondcycle:
'This is executed only in the second time
'Probabilities of predation are inverted
'probability of predation on white moths
 pw = 0.003
 'probability of predation on brown moths
 pbr = 0.002
 The report in the second cycle begins on line 1002
 Delay = 1001
 'This is sub routine that is always executed
Always:
 'We run n generations
 n = 1000
 For i = 1 To n
'White moths reproduce and mutate
  For j = 1 To nw
   s = Rnd
   If s < mr * (1 - (nw + nbr) / Maxm) Then
     s = Rnd
     If s < mut Then nbr = nbr + 1 Else nw = nw + 1
    End If
 Next j
 'Brown moths reproduce and mutate
  For j = 1 To nbr
   s = Rnd
   If s < mr * (1 - (nw + nbr) / Maxm) Then
    s = Rnd
    If s < mut Then nw = nw + 1 Else nbr = nbr + 1
   End If
 Next j
 'Moths are prayed, birds reproduce
  For l = 1 To nb
  'White moths
   For j = 1 To nw
    s = Rnd
    If s < pw Then nw = nw - 1
    If s < 0.1 * pw * (1 - nb / Maxb) Then nb = nb + 1
   Next j
   'brown moths
   For j = 1 To nbr
    s = Rnd
    If s < pbr Then nbr = nbr - 1
    If s < 0.1 * pbr * (1 - nb / Maxb) Then nb = nb + 1
   Next j
  Next l
 'Birds die
 For j = 1 To nb
   s = Rnd
   If s < bd Then nb = nb - 1
 Next j
'Report
  Range(Cells(i + Delay, 2), Cells(i + Delay, 2)).Select
  ActiveCell.FormulaR1C1 = nw
  Range(Cells(i + Delay, 3), Cells(i + Delay, 3)).Select
  ActiveCell.FormulaR1C1 = nbr
  Range(Cells(i + Delay, 4), Cells(i + Delay, 4)).Select
  ActiveCell.FormulaR1C1 = nb
Next i
Cycle = Cycle + 1
If Cycle = 2 Then GoTo Secondcycle
End Sub

\end{verbatim}

\bigskip
\textbf{\thenum. }  \addtocounter{num}{1} We have used the GoTo instruction to solve a problem in the simplest form.  In nature, that would be all to it. But not always: if resources are limited and are important for genetic software synthesis and execution, shorter programs will prevail. But in software engineering, the GoTo style could create really intricate programs that after a while you cannot understand. That is why one prefers encapsulating style.

\bigskip
\textbf{\thenum. }  \addtocounter{num}{1} Example. Let us  rewrite the last program in an encapsulating style. The level of encapsulation is a bit higher than  the required one to leave away the  GoTo instructions.

\begin{verbatim}


Private Sub Titling()
 'Titles
 'Opens a new sheet
  Workbooks.add
  Range(Cells(1, 2), Cells(1, 2)).Select
  ActiveCell.FormulaR1C1 = "White moths"
  Range(Cells(1, 3), Cells(1, 3)).Select
  ActiveCell.FormulaR1C1 = "Brown moths"
  Range(Cells(1, 4), Cells(1, 4)).Select
  ActiveCell.FormulaR1C1 = "Birds"
End Sub

Private Sub Report(i, delay, nw, nbr, nb)
 'Report
  Range(Cells(i + delay, 2), Cells(i + delay, 2)).Select
  ActiveCell.FormulaR1C1 = nw
  Range(Cells(i + delay, 3), Cells(i + delay, 3)).Select
  ActiveCell.FormulaR1C1 = nbr
  Range(Cells(i + delay, 4), Cells(i + delay, 4)).Select
  ActiveCell.FormulaR1C1 = nb
End Function


Private Sub Dynamics(n, nw, nbr, nb, mr, Maxm, Maxb, bd, mut, pw, pbr, delay)
  For i = 1 To n
'White moths reproduce and mutate
  For j = 1 To nw
   s = Rnd
   If s < mr * (1 - (nw + nbr) / Maxm) Then
     s = Rnd
     If s < mut Then nbr = nbr + 1 Else nw = nw + 1
    End If
 Next j
 'Brown moths reproduce and mutate
  For j = 1 To nbr
   s = Rnd
   If s < mr * (1 - (nw + nbr) / Maxm) Then
    s = Rnd
    If s < mut Then nw = nw + 1 Else nbr = nbr + 1
   End If
 Next j
 'Moths are prayed, birds reproduce
  For l = 1 To nb
  'White moths
   For j = 1 To nw
    s = Rnd
    If s < pw Then nw = nw - 1
    If s < 0.1 * pw * (1 - nb / Maxb) Then nb = nb + 1
   Next j
   'brown moths
   For j = 1 To nbr
    s = Rnd
    If s < pbr Then nbr = nbr - 1
    If s < 0.1 * pbr * (1 - nb / Maxb) Then nb = nb + 1
   Next j
  Next l
 'Birds die
 For j = 1 To nb
   s = Rnd
   If s < bd Then nb = nb - 1
 Next j
 Call Report(i, delay, nw, nbr, nb)
Next i
End Sub


Private Sub MothsFlipFlopEncap()

'Study of predation of moths by birds
'Moths could be white or black.
'Reproduction  obeys a logistic model
'Two environments: clean and dirty.
'Stochastic model.
'Clean environment
'Generates 500 moths,
 nw = 450
 nbr = 450
'Generates 10 birds
 nb = 10
 'Probability of moth's reproduction
 mr = 0.1
 'Maximum number of Moths
 Maxm = 2000
 'Maximum number of Birds
 Maxb = 20
'Probability of bird's death
 bd = 0.05
 'Mutation rate of moths
 mut = 0.01
 'probability of predation on white moths
 pw = 0.002
 'probability of predation on brown moths
 pbr = 0.003
 Titling
  'In the first cycle the report begins on line 2
  delay = 1
 'We run n generations on each cyle
 n = 1000
 Call Dynamics(n, nw, nbr, nb, mr, Maxm, Maxb, bd, mut, pw, pbr, delay)
'The second cycle is initialized here
'Secondcycle:
'This is executed only in the second time
'The environment changes from clean to dirty
'Probabilities of predation are inverted
'Probability of predation on white moths
 pw = 0.003
 'probability of predation on brown moths
 pbr = 0.002
 'The report in the second cycle begins on line 1002
 delay = 1001
Call Dynamics(n, nw, nbr, nb, mr, Maxm, Maxb, bd, mut, pw, pbr, delay)
End Sub

\end{verbatim}

\bigskip
\textbf{\thenum. }  \addtocounter{num}{1} Exercise. Run the code and verify that it produces the same dynamics as the previous programs. Pay attention to the form as very complex procedures are called. Observe that the dynamics procedure is too complex to be understood by a beginner, therefore, cut it down to  various digestible pieces. You can do that because you ceased to be a  beginner long ago.

\bigskip
\textbf{\thenum. }  \addtocounter{num}{1} There is moreover a bunch of styles that aim at   perfectionism because there are too many divergent forms of getting more perfect: In fact, VBA provides with exuberance many specific facilities to develop embellished and friendly interfaces. Our only purpose is to facilitate the use of the code as a laboratory tool, so that it could be run time after time with the minimum effort to change parameters and to record results.

Our first problem is this: when one modifies a code time after time to update parameters, one always commit undesirable modifications. So, our first duty is to prevent putting hands on the code. To do this, we capture the modifications directly from the Excel sheet. In second place, we facilitate recording by positing  the values of the parameters just in front of the output of the program. The user is meant to use Excel to do a graphic of the results that would be attached to the same sheet, which could be edited and saved for future use.

Our solutions is as follows: we design a template for the parameters, the user modifies it to call next   that module that executes the simulation.

The template is built in the  subroutine TemplateInputData of the next module. The simulation is then invoked by calling the subroutine MothsBirdsVirtualLab

\begin{verbatim}

Public Sub TemplateInputData()
 'Titles
 'Opens a new sheet
  Workbooks.add
  Range(Cells(1, 4), Cells(1, 4)).Select
  ActiveCell.FormulaR1C1 = "White and brown moths vs. Birds"
  Range(Cells(3, 4), Cells(3, 4)).Select
  ActiveCell.FormulaR1C1 = " Please, Modify Parameters"
  Range(Cells(5, 4), Cells(5, 4)).Select
  ActiveCell.FormulaR1C1 = "Highlight the range of   parameters"
  Range(Cells(6, 4), Cells(6, 4)).Select
  ActiveCell.FormulaR1C1 = "Call the WhiteBrownMothsBirds macro"
  Selection.EntireColumn.AutoFit
  Range(Cells(8, 4), Cells(8, 4)).Select
  ActiveCell.FormulaR1C1 = " Initial number of white moths"
  Range(Cells(8, 5), Cells(8, 5)).Select
  ActiveCell.FormulaR1C1 = " 450"
  Range(Cells(9, 4), Cells(9, 4)).Select
  ActiveCell.FormulaR1C1 = " Initial number of brown moths"
  Range(Cells(9, 5), Cells(9, 5)).Select
  ActiveCell.FormulaR1C1 = " 450"
  Range(Cells(10, 4), Cells(10, 4)).Select
  ActiveCell.FormulaR1C1 = " Initial number of birds  "
  Range(Cells(10, 5), Cells(10, 5)).Select
  ActiveCell.FormulaR1C1 = " 10"
  Range(Cells(11, 4), Cells(11, 4)).Select
  ActiveCell.FormulaR1C1 = "Probability of moth's reproduction "
  Range(Cells(11, 5), Cells(11, 5)).Select
  ActiveCell.FormulaR1C1 = " 0.1"
  Range(Cells(12, 4), Cells(12, 4)).Select
  ActiveCell.FormulaR1C1 = " Maximum number of moths  "
  Range(Cells(12, 5), Cells(12, 5)).Select
  ActiveCell.FormulaR1C1 = " 2000"
  Range(Cells(13, 4), Cells(13, 4)).Select
  ActiveCell.FormulaR1C1 = " Maximum number of birds "
  Range(Cells(13, 5), Cells(13, 5)).Select
  ActiveCell.FormulaR1C1 = " 20"
  Range(Cells(14, 4), Cells(14, 4)).Select
  ActiveCell.FormulaR1C1 = " Probability of bird's death"
  Range(Cells(14, 5), Cells(14, 5)).Select
  ActiveCell.FormulaR1C1 = " 0.05"
  Range(Cells(15, 4), Cells(15, 4)).Select
  ActiveCell.FormulaR1C1 = " Mutation rate of moths"
  Range(Cells(15, 5), Cells(15, 5)).Select
  ActiveCell.FormulaR1C1 = " 0.01"
  Range(Cells(16, 4), Cells(16, 4)).Select
  ActiveCell.FormulaR1C1 = "Probability of predation on white moths"
  Range(Cells(16, 5), Cells(16, 5)).Select
  ActiveCell.FormulaR1C1 = " 0.002"
  Range(Cells(17, 4), Cells(17, 4)).Select
  ActiveCell.FormulaR1C1 = "Probability of predation on brown moths"
  Range(Cells(17, 5), Cells(17, 5)).Select
  ActiveCell.FormulaR1C1 = " 0.003"
  Range(Cells(18, 4), Cells(18, 4)).Select
  ActiveCell.FormulaR1C1 = " Number of generations per cycle"
  Range(Cells(18, 5), Cells(18, 5)).Select
  ActiveCell.FormulaR1C1 = " 1000"
End Sub


Private Sub Titling()
 'Clears old tiltes

  Range(Cells(1, 4), Cells(6, 4)).ClearContents

 'Titles
  Range(Cells(4, 4), Cells(4, 4)).Select
  ActiveCell.FormulaR1C1 = "White and brown moths vs birds"
  Range(Cells(5, 4), Cells(5, 4)).Select
  ActiveCell.FormulaR1C1 = "Experiment Number"
  Range(Cells(1, 1), Cells(1, 1)).Select
  ActiveCell.FormulaR1C1 = "White moths"
  Range(Cells(1, 2), Cells(1, 2)).Select
  ActiveCell.FormulaR1C1 = "Brown moths"
  Range(Cells(1, 3), Cells(1, 3)).Select
  ActiveCell.FormulaR1C1 = "Birds"
End Sub

Private Function Report(i, delay, nw, nbr, nb)
 'Reports results
  Range(Cells(i + delay, 1), Cells(i + delay, 1)).Select
  ActiveCell.FormulaR1C1 = nw
  Range(Cells(i + delay, 2), Cells(i + delay, 2)).Select
  ActiveCell.FormulaR1C1 = nbr
  Range(Cells(i + delay, 3), Cells(i + delay, 3)).Select
  ActiveCell.FormulaR1C1 = nb
End Function

Private Sub Dynamics(n,nw,nbr,nb,mr,Maxm,Maxb,bd,mut,pw,pbr,delay)
  For i = 1 To n
'White moths reproduce and mutate
  For j = 1 To nw
   s = Rnd
   If s < mr * (1 - (nw + nbr) / Maxm) Then
     s = Rnd
     If s < mut Then nbr = nbr + 1 Else nw = nw + 1
    End If
 Next j
 'Brown moths reproduce and mutate
  For j = 1 To nbr
   s = Rnd
   If s < mr * (1 - (nw + nbr) / Maxm) Then
    s = Rnd
    If s < mut Then nw = nw + 1 Else nbr = nbr + 1
   End If
 Next j
 'Moths are prayed, birds reproduce
  For l = 1 To nb
  'White moths
   For j = 1 To nw
    s = Rnd
    If s < pw Then nw = nw - 1
    If s < 0.1 * pw * (1 - nb / Maxb) Then nb = nb + 1
   Next j
   'brown moths
   For j = 1 To nbr
    s = Rnd
    If s < pbr Then nbr = nbr - 1
    If s < 0.1 * pbr * (1 - nb / Maxb) Then nb = nb + 1
   Next j
  Next l
 'Birds die
 For j = 1 To nb
   s = Rnd
   If s < bd Then nb = nb - 1
 Next j
 Dummy2 = Report(i, delay, nw, nbr, nb)
Next i
End Sub


Public Static Sub MothsBirdsVirtualLab()
 Dim MatrixValue
 Dim RowNum As Integer
 Dim Column As Range

 MatrixAddress = Selection.Address
 Set Column = ActiveCell.CurrentRegion
 MsgBox "The region has a range " & MatrixAddress
 MatrixValue = Selection.Value
 'Reads the number of moths
 nw = MatrixValue(1, 1)
 nbr = MatrixValue(2, 1)
'Reads the number of birds
 nb = MatrixValue(3, 1)
 'Reads the probability of moth's reproduction
 mr = MatrixValue(4, 1)
 'Reads the maximum number of Moths
 Maxm = MatrixValue(5, 1)
 'Reads the maximum number of Birds
 Maxb = MatrixValue(6, 1)
'Reads the probability of bird's death
 bd = MatrixValue(7, 1)
 'Reads the mutation rate of moths
 mut = MatrixValue(8, 1)
 'Reads the probability of predation on white moths
 pw = MatrixValue(9, 1)
 'Reads the probability of predation on brown moths
 pbr = MatrixValue(10, 1)
  'Reads the number of generation per each cyle
 n = MatrixValue(11, 1)
 Call Titling
 delay = 1
Call Dynamics(n,nw,nbr,nb,mr,Maxm,Maxb,bd,mut,pw,pbr,delay)
'The second cycle is initialized here
'Secondcycle:
'This is executed only in the second time
'The environment changes from clean to dirty
'Probabilities of predation are inverted
'Probability of predation on white moths
 pw = 0.003
 'probability of predation on brown moths
 pbr = 0.002
 'The report in the second cycle begins on line 1002
 delay = 1002
Call Dynamics(n,nw,nbr,nb,mr,Maxm,Maxb,bd,mut,pw,pbr,delay)
End Sub

\end{verbatim}

\bigskip
\textbf{\thenum. }  \addtocounter{num}{1} Exercise. Run the code as follows:  Posit the cursor at the beginning of the subroutine TemplateInputData and run it . A sheet will pop up, in which you can read some default parameters. Modify then as  you want. Select the region that contains the numerical values, return to VBA, select the subroutine MothsBirdsVirtualLab and run it. Make a graphic of the resultant output data and write the number of the experiment. You have the option of saving this sheet for further study.

\bigskip
\textbf{\thenum. }  \addtocounter{num}{1} At this stage, we are already mature to pursue some level of sophistication. We have two fronts: to have the code robust against methodological errors and to embellish the interface with the user. So, we will deal with the following problems: 1) the self referencing loop of reproduction of individuals  2) nature functions in parallel, rabbits and wolves reproduce at the same time,  but our programs are executed serially, instruction after instruction. This causes an asymmetry whose effect we need to discard. 3) We will use some of the tremendous facilities given by VBA to design nice and friendly interfaces. All this demands some effort, so let us arm ourselves with patience. The first two problems will be solved in the next code but the building of the interface is more difficult. Let us proceed.

1. In Excel, call tools, then Macros and then VBA.

2. In the menu bar, choose Insert and there click   Userform. In response, two boxes pop up. One is a box that serves as a table for design of our interface and the second box contains some patterns that could be clicked on and posited in the work table. Play with this setting to learn the name of each pattern, the form to posit each one on the table, the form of modifying it to suit your special needs: you can modify both the titles and the size of each pattern. Learn to convert the design of the work table into code and run it.

3. Play a bit more: did you noted that every element that you posited in the work table have a lot of properties that do not appear in the code? That is why, you need to follow the instructions below to make an interface such as we need it. An interface contains the following types of boxes: 1) labels, where we write instructions or the name of required parameters 2) TextBoxes where the user types specific values of parameters 3) Control buttons to cancel and to accept the reading of the parameters by the VBA executing machine.

4. Call a userform and pay attention to its name, say, userform12, which has been given by the system. Remember this number: we are not working with a new module but with a new form. In consequence, you shall search for it under the index FORMS, Userforms  that one can see after calling the project explorer ( in the menu See).

5. In strict order, insert in the work table the following items. In first place we need 13 labels to specify the parameters and the titles. The order is as follows: 1)Initial number of white moths, 2)Initial number of brown moths, 3) Initial number of birds, 4)Probability of moth's  reproduction, 5) Maximum number of moths, 6) Maximum number of birds, 7) Probability of bird's death, 8) Mutation rate of moths, 9) Probability of predation on white moths, 10) Probability of predation on brown moths, 11) Number of generations per cycle 12) Title: White and Brown Moths vs Birds 13 ) Subtitle: Modify parameters and click OK. Notice that Title and subtitle are labels 12 and 13 but they must be at the top of the form.

6. Insert 11 TextBoxes for the 11 parameters to be read. TextBox\textbf{1} one must be just in from of Label\textbf{1} and so on. The default values of the parameters shall be posited just in the work table. For decimal figures, try dot else colon to see which does it. The default values are:  Moths: nw = 450, nbr = 450, Birds:  nb = 10; Probability of moth's reproduction: mr = 0.1, Maximum number of Moths : Maxm = 2000, 'Maximum number of Birds: Maxb = 20, Probability of bird's death: bd = 0.05, Mutation rate of moths: mut = 0.01, Probability of predation on white moths: pw = 0.002, Probability of predation on brown moths: pbr = 0.003, Generations on each cyle: n = 1000. Type on each TextBox only the numerical values.

6. Insert two command buttons: one for CANCEL and the other for OK. Rewrite their names with the corresponding values. The first is Cancel, the second is OK.

7. Read the code that is associated to the table. It must look something like this. In some part of this code does not appear, it may be called by a control bar at the top of the code window.

\begin{verbatim}


Private Sub CommandButton1_Click()

End Sub

Private Sub CommandButton2_Click()

End Sub

Private Sub Label1_Click()

End Sub

Private Sub Label10_Click()

End Sub

Private Sub Label11_Click()

End Sub

Private Sub Label12_Click()

End Sub

Private Sub Label13_Click()

End Sub

Private Sub Label2_Click()

End Sub

Private Sub Label3_Click()

End Sub

Private Sub Label4_Click()

End Sub

Private Sub Label5_Click()

End Sub

Private Sub Label6_Click()

End Sub

Private Sub Label7_Click()

End Sub

Private Sub Label8_Click()

End Sub

Private Sub Label9_Click()

End Sub

Private Sub TextBox1_Change()

End Sub

Private Sub TextBox10_Change()

End Sub

Private Sub TextBox11_Change()

End Sub

Private Sub TextBox2_Change()

End Sub

Private Sub TextBox3_Change()

End Sub

Private Sub TextBox4_Change()

End Sub

Private Sub TextBox5_Change()

End Sub

Private Sub TextBox6_Change()

End Sub

Private Sub TextBox7_Change()

End Sub

Private Sub TextBox8_Change()

End Sub

Private Sub TextBox9_Change()

End Sub

Private Sub UserForm_Click()

End Sub

\end{verbatim}

8) Search the code corresponding to the first command, that for Cancel. It looks like this:

\begin{verbatim}

Private Sub CommandButton1_Click()

End Sub

\end{verbatim}

9) Insert End as the body of this subroutine. The result looks as follows:

\begin{verbatim}

Private Sub CommandButton1_Click()
  'Cancel Buttom
  End
End Sub

\end{verbatim}

11) Run the code and check the effect of your command button CANCEL.

12) Copy at the very beginning of the code of your form the following text

\begin{verbatim}

Option Explicit
Dim i, n, l, nw, nbr, nb, Maxm, Maxb, delay, dlay

Dim mr As Double
Dim bd As Double
Dim mut As Double
Dim bw As Double
Dim pw As Double
Dim pbr As Double

\end{verbatim}

Let us observe that in our pursuing of sophistication we are taking a better control of our variables: the Option Explicit says that any variable shall be declared before it could be used. That would avoid bugs originated in typing errors in which one changes   nbr by  nrb and so on.

13) Search for the code corresponding to the command button whose number is  two: in the work table it was marked with the OK flag but here in the code we just know that it is the command button number 2 because we created it in second order after the cancel button. Insert in the corresponding sub the next code:

\begin{verbatim}

'OK Buttom

'Reads Data

 nw = TextBox1.Text
 nbr = TextBox2.Text
 nb = TextBox3.Text
 mr = TextBox4.Text
 Maxm = TextBox5.Text
 Maxb = TextBox6.Text
 bd = TextBox7.Text
 mut = TextBox8.Text
 pw = TextBox9.Text
 pbr = TextBox10.Text
 n = TextBox11.Text
 Call Action

\end{verbatim}

 14) Paste the following code below the last line:



15) To run the code, keep in mind  that the execution of a code can be interrupted at will with the keys CTRL + PAUSE (INTERRUPT). Also keep in mind that if the execution has aborted by whatever reason, the control of your machine is settled by VBA, so return to it and stop the executing  machine in the recording controls. If the code runs at once, it produces a  n Excel sheet filled with the numbers from 1 to 2000. That takes a lot of time, some two to three minutes in fast machines. At the end, close the interface window by clicking in the red X at  the upper right corner of the interface. Make a graphic of the output: it shall look just as you are used to see the output in the previous programs. If if does not run or if it produces a different output to what we are used to, consider the next possible cause:

16) There is an interface bug in the EXCEL-VBA application: Excel uses colon for decimal numbers while VBA uses dot. The effect of the this bug is that one could write a decimal number say, 0.002 and the system reads it as 2. For that reason, there are some tracer boxes inside the code that can be turned on by just erasing the initial marker of commentary, which is a vertical mark, ', at the beginning of the line. Play with those tracers. Else, learn to use the Instant Message Box used by the Debugging facility. Also play with the cursor when the program is run under F8.

17) If you fail anyway, please, consider that one is greatly helped by youngsters that know how to click buttons.

\bigskip
\textbf{\thenum. }  \addtocounter{num}{1} Exercise. Digest the code.

\bigskip
\textbf{\thenum. }  \addtocounter{num}{1} Intrigue: even sophisticated programs have bugs. Does the genome has bugs or is Bug a term reserved for intentional creators of software? Actually, why are we so perfect? Are we indeed perfect or are we late to visit hospitals and   grandfathers?

\bigskip
\textbf{\thenum. }  \addtocounter{num}{1} Exercise.   Learn to use this module  as a  tool for your ecological studies. As a first study, make experiments to disentangle the following question: the system oscillates under certain parameters: which and in  which range? Under other parameters, the system is stable: which and in which range? Some oscillations may lead to the annihilation of the species: which and in which range? Does we need the division of the pray species in two subtypes, white and brown, to assure the stability of the pray-predator system? Elaborate your laboratory inform as an article.

\bigskip
\textbf{\thenum. }  \addtocounter{num}{1} Intrigue: The perfectionism displayed in the last code is justified here because the program could be considered as a specific virtual laboratory of ecology. How do we justify perfectionism in nature? If we use an economic directive, we could say:   perfectionism pays when the cost of its implementation is overcome by  the fruits it produces. Is that correct? Does perfectionism have a cost in nature? Which are the components of that cost? Would that balance pose a limit to the development of perfectionism?

\bigskip
\textbf{\thenum. }  \addtocounter{num}{1}  Exercise. We have learned that under some setting of parameters, a pray-predator stochastic system necessarily oscillates and that an outburst of the pray is followed by an outburst of predators that kill the prays and that causes predators to die at starving. In short: a stochastic pray - predator system cannot survive for ever. The remedy was to pose an artificially maximum number of predators which governs the predator reproduction through a logistic model. When this is considered, the whole system becomes stable for appropriate values of the  parameters. Please, make an estimation of the range of stability of the system.

 \bigskip
\textbf{\thenum. }  \addtocounter{num}{1}  Notice that in nature, our artificially imposed maximum number of predators is also artificially imposed: it corresponds to territoriality. We have proved that territoriality is necessary for the stability of a pray-predator system. Equivalently: without territoriality there cannot exist pray-predator systems. Now, develop the simplest model that proves that a  system of the form grass-game must be enhanced into a system of the form grass-game-predator - high level predator with territoriality. This is certainly true because there are always fluctuations of the low level links of the feeding chain and some of them favor the multiplication of high level populations. This would cause a complete exhaustion of low level components. A consequence is the starvation to death of high level members. Net result: the system disappears. Would you explain why territoriality could be tied to a single sex or even to the alpha member of the hierarchy? What about feminine participation in the control of picks of natality?

\bigskip
\textbf{\thenum. }  \addtocounter{num}{1}  Graduation.  Develop a virtual laboratory to study a food chain. Keep a record of all your attempts and try to understand the  evolution of your style and of the complexity and realism of your codes.

\chapter{Simulation in  statistics}

In this chapter we will show how one can use simulation to run a test for statistics. We begin with running a test for a one factor anova. Next, we run an anova with blockade. Because these tests are well known, we could calibrate our method.  Next, we pass to study a block design with incomplete data. In that way, we open a new world of possibilities. As usually, we have assumed in all these studies a normal distribution in the null hypotheses. So, we must learn how to generate random numbers with a normal distribution and that is why we begin with the simulation of a binomial distribution.

\section{The binomial distribution}

The binomial distribution could be considered as the abstraction of the counting procedure of the number of males among the spring of a family with n children. Thus, a binomial distribution is characterized by two parameters, let us denote them $n$ and $p$. The first, $n$ is the number of children per family and the second $p$ is the probability of a male, where we mean that $1-p$ is the probability of female. Although natural populations have a value of $p$ close to 1/2, we allow it to take any value in within 0 and 1. In similar way, $n$  could take on any integer value.

 \bigskip
\textbf{\thenum. }  \addtocounter{num}{1}  Example. The following Macro simulates 1000 families, each one with 7 children and the probability of male is 0.2.

\begin{verbatim}

Option Explicit
Public Sub Binomial()

 Dim n As Integer
 Dim p As Double
 Dim s As Double
 Dim i As Integer
 Dim j As Integer
 Dim Males As Integer



 'Opens a new sheet
  Workbooks.add
  'Title
  Range(Cells(1, 4), Cells(1, 4)).Select
  ActiveCell.FormulaR1C1 = "Numbers of males per family"

 'Initialize the random generator with the clock.
 'To repeat exactly the same experiment,
 'randomize must be silenced.
 Randomize

 p = 0.2
 n = 7
 'MsgBox "p= " & p

 For i = 1 To 1000
  Males = 0
  For j = 1 To n
  s = Rnd
   'MsgBox "p-s= " & p - s
 If s < p Then Males = Males + 1
  Next j
  'Report
  Range(Cells(i, 1), Cells(i, 1)).Select
  ActiveCell.FormulaR1C1 = Males
 Next i

End Sub

\end{verbatim}

 \bigskip
\textbf{\thenum. }  \addtocounter{num}{1}  Exercise. Run the Macro. Add a column with the numbers from 0 to 8 one after the other. Call, from Data Analysis, the Histogram facility. Feed that sub with the thousand output numbers in the first entry of the dialog box and with the numbers from 0 to 8 in the second. Draw the histogram with the scatter mode of the graphic facility. Use a Macro developed by us to calculate the mean and the variance of our data. Compare our data with those predicted by the mathematical theory. Calculate the total number of male children and the relative observed frequency of   males. Run a test to compare the observed frequency with the expected one.

\section{The normal distribution}

 \bigskip
\textbf{\thenum. }  \addtocounter{num}{1}  Since the normal distribution is the continuous fitting to the binomial distribution, we could use the generator of the binomial distribution of the previous section to generate numbers with a normal distribution. Of course, we must, in first place, rescale data to be fitted in an interval from zero to certain positive range.

 \bigskip
\textbf{\thenum. }  \addtocounter{num}{1}  Exercise. Work out the details of the previous idea and embody them into a Macro. Hint: You must be given the variance $V>6$ and the mean $M>3$ of the normal distribution. Equate $V$ to $np(1-p)= n(1/2)(1/2) =n/4$ and clear for $n$. Use a previous Macro to generate random numbers with the binomial distribution with  $n$ and $p=1/2$. To each output number add the term $M-n/2$. These new numbers obey a normal distribution with the required mean and variance. Explain why all this functions.

 \bigskip
\textbf{\thenum. }  \addtocounter{num}{1}  Let us learn now how to develop another generator of numbers with a normal distribution, with mean $M=10$ and standard deviation $s=2$. In contrast with the previous one, this new generator will output numbers with decimal figures.

\begin{verbatim}

Option Explicit
Public Sub NormalGenerator()

 Dim n As Integer
 Dim i As Integer

 'Opens a new sheet
  Workbooks.add
  'Title
  Range(Cells(1, 4), Cells(1, 4)).Select
  ActiveCell.FormulaR1C1 = "Normal distribution"

 'The instruction Randomize
 'initializes the random generator with the clock.
 'To repeat exactly the same experiment,
 'Randomize must be silenced else
 'the generator must be
 'initialized with the same value,
 'for example:

 Randomize ([3])

'Generates 1000 random numbers
'with a normal distribution
'with mean m and
'standard deviation s

'Generates 1000 random numbers with
'a normal distribution
'with mean 10
'and standard deviation 2

 For i = 1 To 1000

  'Generates a random number
  'with uniform distribution in [0,1]
  Range(Cells(i, 1), Cells(i, 1)).Select
  ActiveCell.FormulaR1C1 = "=RAND()"

  'The Excel function "=NORMINV(RC[-1],10,2)"
  'generates a random number with normal distribution
  'The first entry is a r.n. in [0,1]
  'The second is the mean
  'The third is the standard deviation

  Range(Cells(i, 2), Cells(i, 2)).Select
    ActiveCell.FormulaR1C1 = "=NORMINV(RC[-1],10,2)"
 Next i

End Sub

\end{verbatim}

 \bigskip
\textbf{\thenum. }  \addtocounter{num}{1}  Exercise. Test this code: prepare the screen to see  Excel and  VBA at the same time. The program runs during  various minutes. If this Macro does not run in the version of Excel that you have, record a macro that uses the function Inverse Normal distribution and look the exact form as that function is referenced in your Excel version. Make the corresponding change in the code, where that function has been called as "=NORMINV(RC[-1],10,2)". After execution, you will get 1000 random numbers with a normal distribution with mean 10 and standard deviation 2. To get convinced of that, please, fill a column in the open Excel Sheet with numbers from 0 to 20 and call the Macro Histogram, from Data Analysis from Tools in Excel. Feed the first entry of the popping dialog box  with the address of the region that contains the random numbers with normal distribution and the second entry with the numbers from 0 to 20. Verify that you have 1000 numbers indeed: to calculate the inverse normal distribution, it is used a numerical algorithm that sometimes does not converge. If a given number is lacked, fill in the corresponding cell  by hand. Next, draw the histogram with the drawing facility in the mode scatter. Run a Kolmogorov Smironov test to verify that the resultant distribution is indeed normal.

 \bigskip
\textbf{\thenum. }  \addtocounter{num}{1}  Exercise. Generate 1000 random numbers with a normal distribution with mean 40 and standard deviation 5. Execute all the required ritual, as in the previous exercise.

 \bigskip
\textbf{\thenum. }  \addtocounter{num}{1}   It may happens to you the same as to me, that I was  unable to find the way to generate the normal distribution from VBA directly.  The reason is that Excel does not uses a full fledged version of VBA but a special adaptation to Excel, which is truncated and that is called VBE. This is annoying, but  let us return to study of an anova by means of a simulation.

\section{One factor anova}

A one factor anova is the analysis of variance with the purpose of studying differences in means among diverse samples obtained  by a random sampling procedure. The diverse samples come from diverse treatments.

 \bigskip
\textbf{\thenum. }  \addtocounter{num}{1}  Example. We study strawberries under three diverse temperatures and measure the crop in kg per square meter. The results and the anova test by Excel follow:

\begin{center}
\begin{tabular}{ |l|l|l|c|}\hline
9?C & 18?C &25?C\\ \hline\hline
1   &3  &1\\ \hline\hline
2   &4& 1\\ \hline\hline
1   &5& 2\\ \hline\hline
\end{tabular}
\end{center}

The One Factor anova analysis carried out by Excel renders the following table:

\begin{center}
\begin{tabular}{|l|l|l|l|l|l|l|c|}\hline
\multicolumn{7}{|c|}{\vphantom{Large Ap} Anova for three temperatures }\\ \hline\hline
Origin of variation& SS&DF& Mean Square & Fisher & Probab & Critical F \\ \hline\hline
Among groups & 14,22222222  & 2 & 7,111111111 &     12,8 &  0,0068453   & 5,143249382 \\ \hline
Within groups &     3,333333333 & 6 &   0,555555556      & & & \\ \hline
Total  & 17,55555556    & 8 &&&&  \\ \hline
\end{tabular}
\end{center}

As we see, the anova test authorizes us to claim that some differences in temperature is matched  by a difference in average production.

 \bigskip
\textbf{\thenum. }  \addtocounter{num}{1}  Let us learn now how to carry the same test with the help of   a simulation. The central idea is simple and is the same for all problems no matter how complex they could be. Let us see it:

1) One examines exactly what one did in an experiment.

2) One declares exactly what is the null hypothesis that one wants to test.

3) One builds an artificial, virtual world, with the same characteristics as those proposed by the null hypotheses.

4) One runs the virtual world under the same conditions as those met by the experiment.

5) For each run in the virtual world one measures exactly the same stadigraph  as in the real experiment.

6) One compares the experimental result found in the real world with the distribution of results found in the virtual world.

7) If the stadigraph associated to the experimental result gets a value that is an outlier  with respect to the distribution of the virtual world, one declares that the experiment does not support the null hypotheses. Otherwise, one declares that the null hypotheses explains our experiment.

The implementation of this directive for our temperature experiment is as follows.

1) We arranged three levels of temperature with three replicas per temperature.

2) The null hypothesis is that temperature plays no special role in the amount of crop of the strawberry. Instead, all observed differences are due to uncontrolled variables, such as the intrinsic genetic variability that causes that different plants produce different amounts of fruit.

3) Let us build an artificial, virtual world, with the same characteristics as those proposed by the null hypotheses.

We must, in first place, refine our null hypotheses. We said that all observed differences are effect of uncontrolled variables. But we have the observed differences in front of our eyes, so we are allowed to characterize them. The first supposition is that the fluctuations were created by many independent factors, so they follow a normal distribution: our simplest model is that all observed values dance around its average with steps that fluctuate with  a normal distribution. Hence, we must find unbiased estimators of the mean and of the standard deviation of the fluctuations.

Now, if the null hypotheses is correct, the division of data into three columns is unjustified: all data belong to the same sample. In that case, the unbiased estimator $\widehat{\mu}$ of the mean is the ordinary mean:

$\widehat{\mu}= \frac{1+2+1+3+4+5+1+1+2}{9}= 20/9$

If the null hypothesis is correct, the unbiased estimator $\widehat{\sigma}$ of the standard deviation is

$\widehat{\sigma} = \sqrt{\frac{\sum(x_i- \widehat{\mu})^2} {n-1} }= $

$ = \sqrt{ \frac{\sum x^2 } {n-1}- \frac{n \widehat{\mu}^2} {n-1 }}$

$ =  \sqrt{\frac{1^2+2^2+1^2+3^2+4^2+5^2+1^2+1^2+2^2}{8} - \frac{9 (20/9)^2} {8 }} $

$ =  \sqrt{\frac{62}{8} - \frac{400/9 } {8 }} $

 $ =  \sqrt{7.75 - 5.55} = \sqrt{2.2}=1.483$

 Thus, an observed value is the sum of the mean value, $20/9=2.222$ plus a fluctuation that follows a normal distribution with mean zero and standard deviation 1.483. This conditions define our virtual world: random numbers with a normal distribution with mean 2,2222 and standard deviation 1.483.

4) Let us  run now the virtual world under the same conditions as those met by the experiment.

The  module just below produces all we need. The subroutine Anova1F produces a  matrix that  is filled in random numbers with a uniform distribution in the interval [0,1]. The subroutine F3x3 produces a   matrix that uses the random numbers of the first matrix to feed the inverse normal distribution with mean 2,2222 and standard deviation 1.483. The first subroutine was automatically recorded by the Macro Facility. The second sub was recorded upon the output of the first Macro. Next, the main sub, virtualF, was developed.



5) For each run in the virtual world one measures exactly the same stadigraph  as in the real experiment.

For a given matrix of data with K columns and N entries = number of rows times number of columns, the stadigraph F is calculated as follows:

$SSW= \sum^k_{i=1} \sum^{n_i}_{j=1} (X_{ij} -\bar{X_i})^2$

$SSA= \sum^k_{i=1} 3(\bar{X_i }  -\bar{X}) ^2 $

Then

$F= \frac{SSA/(k-1)}{SSW/(N-k)}$

This stadigraph is calculated in the second sub and is called X. The code was fed with the matrix of data that we are analyzing and produced the same F as Excel: the F for our experimental data was 12.8.   This testing was necessary to see whether or not our program was well encoded. It was not on the first trial. The correction process showed that   debugging A Macro that has been automatically recorded  is excessively difficult. It becomes obvious that Programming is easier in VBA while we could use Excel for displaying, memorizing and graphic elaboration of results.

6) One compares the experimental result found in the real world with the distribution of results found in the virtual world.

To do this, we must repeat the experiment in the virtual world a significant number of times. We choose 1000 times, but someone would prefer 10000 times.

The program outputs 1000 values of the stadigraph F. They were compressed into an histogram and then drawn and the distribution resulted to be a typical F.  The histogram is reported in the next table:

\begin{center}
\begin{tabular}{ |l|l|l|c|}\hline
Class   &Frequency  &Cumulated Freq. \\ \hline\hline
0   &0& 0\\ \hline\hline
1   &577&   577\\ \hline\hline
2   &196&   773\\ \hline\hline
3   &104&   877\\ \hline\hline
4   &46&    923\\ \hline\hline
5   &27&    950\\ \hline\hline
6   &13 &963\\ \hline\hline
7   &9& 972\\ \hline\hline
8   &6& 978\\ \hline\hline
9   &5& 983\\ \hline\hline
10& 4   &987\\ \hline\hline
11& 2   &989\\ \hline\hline
12& 1   &990\\ \hline\hline
13& 2   &992\\ \hline\hline
14& 1   &993\\ \hline\hline
15& 1&  994\\ \hline\hline
$\geq15$& 6&    1000\\ \hline\hline
\end{tabular}
\end{center}

Now, we are ready to take a verdict.

7) If the stadigraph associated to the experimental result gets a value that is an outlier  with respect to the distribution of the virtual world, one declares that the experiment does not support the null hypotheses. Otherwise, one declares that the null hypotheses explains our experiment.

The experimental stadigraph takes the value 12.8. There were 990 data among 1000 below that value. This means that our experimental result renders an outliers value whose probability is less than 0.01. (The corresponding value predicted by Excel, and calculated on theoretical grounds,  was   0.007. This close correspondence is necessary to have confidence in  the exactitude of our simulation methods to test statistical hypothesis.)  As a consequence, we declare that, with a 0.01 significance, temperature is a  factor that influences the average production of strawberries.

 \bigskip
\textbf{\thenum. }  \addtocounter{num}{1}  Exercise. Work out the simulation of a one  factor anova test  for a matrix of data with  4 treatments, 3 data in the first treatment, 4 in the second, 5 in the third, and 6 in the fourth.  Hint: try not to adapt the previous program to your needs. It is easier to begin anew from scratch to record the appropriate Macro.

\section{A block design}

 \bigskip
\textbf{\thenum. }  \addtocounter{num}{1}  The receipt to calculate a Randomized block anova is a slight modification of the one way  anova, which is exemplified with our mitochondria case study. We have raw data as follows:

\begin{center}
\begin{tabular}{|l|l|l|l|c|}\hline
\multicolumn{4}{|c|}{\vphantom{Large Ap}  Mitochondria  }\\ \hline
\multicolumn{4}{|c|}{\vphantom{Large Ap}  ATP production }\\ \hline\hline
&pH 4& pH 6 & pH 8\\ \hline
$30^{o}$&5 & 8 & 6 \\ \hline
$36^{o}$&7 & 10 &8\\ \hline
$42^{o}$& 4 & 7&3  \\ \hline
\end{tabular}
\end{center}

The statistical analysis is as follows:

We have $k=3$ treatments and $l=3$ levels of blockade,  $N= kl=9$.

$GS_i$= Great  Square under treatment $i$ = $(\sum^{n_i}_{j=1}X_{ij})^2$

$GS_1=(5+7+4)^2= 16^2=256$

$GS_2=(8+10+7)^2= 25^2= 625$

$GS_3= (6+8+3)^2= 17^2= 289$

$T= \sum^k_{i=1} \sum^{n_i}_{j=1} X_{ij}= 5+7+4+ 8+10+7+ 6+8+3= 58$

$GS$= Great  Square = $(\sum^k_{i=1} \sum^{n_i}_{j=1} X_{ij})^2 $

$GS= T^2= (5+7+4+ 8+10+7+ 6+8+3)^2= 58^2= 3364$

\bigskip

 Among Columns SS =  $\sum^k_{i=1}\frac{GS_i}{ l}-\frac{GS}{N}$
=(256+ 625+ 289)/3 - 3364/9= 390  -373.7 = 16.3

Among columns DF= Among columns degrees of freedom= k-1= 3-1=2.

Among columns Mean square= $\frac{Among Columns SS}{Among columns DF}= 16.3/2=8.1$

We repeat the procedure with blocks or rows or levels of blockade:

$BGS_1=(5+8+6)^2= 19^2=361$

$BGS_2=(7+10+8)^2= 25^2= 625$

$BGS_3= (4+7+3)^2= 14^2= 196$

$T= \sum^k_{i=1} \sum^{n_i}_{j=1} X_{ij}= 5+7+4+ 8+10+7+ 6+8+3= 58$

$GS$= Great  Square = $(\sum^k_{i=1} \sum^{n_i}_{j=1} X_{ij})^2 $

$GS= T^2= (5+7+4+ 8+10+7+ 6+8+3)^2= 58^2= 3364$

\bigskip

 Among Rows SS =  $\sum^k_{i=1}\frac{GS_i}{ k}-\frac{GS}{N}$
=(361+ 625+ 196)/3 - 3364/9= 394  -373.7 = 20.3

Among rows DF= Among rows degrees of freedom= l-1= 3-1=2.

Among rows Mean square= $\frac{Among Rows SS}{Among rows DF}= 20.3/2=10.1$

Now we calculate the square of each entry.

\begin{center}
\begin{tabular}{|l|l|c|}\hline
\multicolumn{3}{|c|}{\vphantom{Large Ap} $X^2_{ij}$  }\\ \hline\hline
Diet 1& Diet 2&Diet 3\\ \hline\hline
25 & 64 & 36 \\ \hline
49 & 100 &64\\ \hline
16 & 49&9  \\ \hline
\end{tabular}
\end{center}

\bigskip

Total SS= Total Sum of Squares=   $\sum^k_{i=1} \sum^{n_i}_{j=1}(X_{ij}^2)-\frac{GS}{N}$

25 + 64 +36 +49 + 100 +64+16 + 49+9

=$(25 + 64 +36 +49 + 100 +64+16 + 49+9 )-\frac{3364}{9} = 412 -373.7 = 38.3$

Error SS=error sum of squares=  Total SS - Among columns SS -Among row SS= 38.3-16.3-20.3= 1.7

Error DF= Total DF - Among groups DF - Among rows DF=  $ 8-2-2=4$.

Error Mean square= $\frac{Error SS}{error DF}=1.7/4=0.43$

$Fisher for columns= \frac{Among \  columns \  mean  \ square}{Error \  Mean  \ square}= \frac{8.1 }{0.43}=18.8$

$Fisher for rows= \frac{Among \  rows \  mean  \ square}{Error \  Mean  \ square}= \frac{10.1 }{0.43}= 23.4$

Next, we compare this experimental stadigraphs with the critical ones given by a table or by Excel. For columns and rows, that value was the same: 6.94. We conclude that pH and Temperature are factors that influence the average production of ATP by the mitochondria.

 \bigskip
\textbf{\thenum. }  \addtocounter{num}{1}  Let us see how we can run the same test but based on a simulation. Let us pay attention to the fact that our methodology is  exactly the same for every design.

The following code makes the complete task.



Let us explain now in which form this code answers the questions posed by  our code of science:

1) One examines exactly what one did in an experiment. We arranged an experiment in which diverse levels of two factors were combined as input to a mitochondria system. The output measured was the production of ATP.

2) One declares exactly what is the null hypothesis that one wants to test: in our experiment, the null hypotheses is that the two factors do not influence the average response of our system and that all observed differences were due to the effect of uncontrolled variables.

3) One builds an artificial, virtual world, with the same characteristics as those proposed by the null hypotheses. So, let us characterize the answers of our system. To that effect we use all the information available and organize it with a theoretical idea: random fluctuations around the mean value appeared as a result of many independent factors, so a normal distribution is expected. Let us evaluate the mean and the standard deviation of our experimental data using  unbiased estimators. The information of the experimental data was consigned in the data matrix, but according to the null hypotheses, the matrix is an artificial unimportant arrangement.

Thus, we have just data, which are 5, 8,  6, 7,  10, 8, 4,  7,3. The mean of these numbers is 58/9=6.444. The standard deviation is:

$\widehat{\sigma} = \sqrt{\frac{\sum(x_i- \widehat{\mu})^2} {n-1} }= $

$ = \sqrt{ \frac{\sum x^2 } {n-1}- \frac{n \widehat{\mu}^2} {n-1 }}$

$ =  \sqrt{\frac{5^2+8^2+6^2+7^2+10^2+8^2+4^2+7^2+3^2}{8} - \frac{9 (58/9)^2} {8 }} $

$ =  \sqrt{\frac{412}{8} - \frac{400/9 } {8 }} $

 $ =  \sqrt{51.5 - 46.722} = \sqrt{4.778}= 2.1858$

4) One runs the virtual world under the same conditions as those met by the experiment. Thus, with the sub \textbf{Rand3x3} we filled a 3x3 matrices  with random numbers with a normal distribution with mean 6.444 and standard deviation 2.1858. Let us notice that we reused the software that was developed for a previous program. \textbf{Reusability} is considered as an important aspect of software development skills.

5) For each run in the virtual world one measures exactly the same stadigraph  as in the real experiment. We measures two F stadigraphs, one for rows and the other for columns, which  were calculated by the sub Blocks3x3. The result was tabulated and the output was:

\bigskip
\begin{center}
\begin{tabular}{|l|l|l|c|}\hline
\multicolumn{3}{|c|}{\vphantom{Large Ap} $F for a Block design $  }\\ \hline\hline
Class& Columns 2&Rows\\ \hline\hline
0   &0 & 0\\ \hline
1   &533&   578\\ \hline
2   &199&   194\\ \hline
3   &81 &92\\ \hline
4   &47 &38\\ \hline
5   &35 &21\\ \hline
6&  25& 22\\ \hline
7   &20 &14\\ \hline
8   &11&    4\\ \hline
9   &13&    8\\ \hline
10& 7   &2\\ \hline
11& 4&  1\\ \hline
12& 7&  7\\ \hline
13& 4   &2\\ \hline
14& 2   &2\\ \hline
15& 2&  0\\ \hline
16& 0   &1\\ \hline
17& 0   &2\\ \hline
18& 1   &1\\ \hline
19 &    3   & 0\\ \hline
20& 2&  1\\ \hline
21  &1& 1\\ \hline
22& 1   &0\\ \hline
23& 0   &0\\ \hline
24& 0   &1\\ \hline
25& 0   &1\\ \hline
26& 0&  0\\ \hline
27& 0&  1\\ \hline
28& 0   &0\\ \hline
29  &1& 1\\ \hline
30  &0& 2\\ \hline
And greater &1& 3\\ \hline
\end{tabular}
\end{center}

6) One compares the experimental result found in the real world with the distribution of results found in the virtual world. The F stadigraphs of the experiment, calculated up to two significant figures,  were 18.8 for columns and 23.4 for rows. (Calculated by Excel they were 18,25 for columns and 22,75 for rows.) There were 10 values among 1000 that were higher than 18.8 and 6 that were higher than 22.75.

7) If the stadigraph associated to the experimental result gets a value that is an outlier  with respect to the distribution of the virtual world, one declares that the experiment does not support the null hypotheses. Otherwise, one declares that the null hypotheses explains our experiment. This directive reads in our case as follows:   with a significance of 0.01,  we may conclude that temperature and pH influence the average production of ATP by mitochondria.

 \bigskip
\textbf{\thenum. }  \addtocounter{num}{1}  Exercise. Make a simulation study of a block design 3x4.

 \bigskip
\textbf{\thenum. }  \addtocounter{num}{1}  Exercise. We have produced some histograms of the distribution of certain stadigraphs. These histograms are estimations of well known theoretical distributions. Use an appropriate  statistical test to compare our estimations with theoretical ones.

\section{An incomplete block design}

 \bigskip
\textbf{\thenum. }  \addtocounter{num}{1}  In an experiment one varies the form of the input to the system and records the output. Then, one submits the data to a statistical analysis. The problem is that the mathematics associated to real circumstances is in general very complex or even unknown. That is why some few designs have been worked out completely and the experimenter tries to box science in within them.

But it may happen that under the best wishes, the final result does not fit the statistical box planned for the experiment. For instance, one may lose a datum in a block design and an entry in the corresponding rectangular data matrix gets empty. The consequence is that one loses the possibility of making a block anova to test the hypotheses concerning the insensibility of the mean to changes in the treatments. Mathematicians have solved the problem for an incomplete block design and they know what must be done. But we cannot expect the same for every design.

Our proposal for further study is simulation:  it is transparent and with some work it can be adapted to the most terrible cases and to the most extravagant design of  experiments. More to the point, you can execute your experiment box-freely. But more wisely, you can simulate a lot of  circumstances before your experiment: after all, a simulation is hundred or thousands  times cheaper than a real experiment.

 \bigskip
\textbf{\thenum. }  \addtocounter{num}{1}  To fix ideas, let us analyze a block design with one lost datum. Our matrix is the following:

\begin{center}
\begin{tabular}{|l|l|l|l|c|}\hline
\multicolumn{4}{|c|}{\vphantom{Large Ap}  Mitochondria  }\\ \hline
\multicolumn{4}{|c|}{\vphantom{Large Ap}  ATP production }\\ \hline\hline
&pH 4& pH 6 & pH 8\\ \hline
$30^{o}$&5 & 8 & 6 \\ \hline
$36^{o}$&7 &   &8\\ \hline
$42^{o}$& 4 & 7&3  \\ \hline
\end{tabular}
\end{center}

Were the matrix complete, we would calculate an F stadigraph which is expected to have an F distribution. But the design is not complete, so we adapt the calculation of the F stadigraph to our circumstances but causing the minimum damage. In our case that is straightforward. The statistical analysis is a slight modification of  the known one for a complete design:

We have $k=3$ treatments and $3$ levels of blockade  for the first column, 2 for the second and 3 for the third.

$GS_i$= Great  Square under treatment $i$ = $(\sum^{n_i}_{j=1}X_{ij})^2$

$GS_1=(5+7+4)^2= 16^2=256$

$GS_2=(8+7)^2= 15^2= 225$

$GS_3= (6+8+3)^2= 17^2= 289$

$T= \sum^k_{i=1} \sum^{n_i}_{j=1} X_{ij}= 5+7+4+ 8+7+ 6+8+3= 48$

$GS$= Great  Square = $(\sum^k_{i=1} \sum^{n_i}_{j=1} X_{ij})^2 $

$GS= T^2= (5+7+4+ 8+ 7+ 6+8+3)^2= 48^2= 2304$

\bigskip

 Among Columns SS =  $\sum^k_{i=1}\frac{GS_i}{ l_i}-\frac{GS}{N}$. In our case,   the total number of data is $N= 8$.

=(256/3+ 225/2+ 289/3)  - 2304/8= 294,16666666666666666666666666667  - 288= 6,1666

Among columns DF= Among columns degrees of freedom= k-1= 3-1=2.

Among columns Mean square= $\frac{Among Columns SS}{Among columns DF}= 6.1666/2=3.083$

We repeat the procedure with blocks or rows or levels of blockade:

$BGS_1=(5+8+6)^2= 19^2=361$

$BGS_2=(7+8)^2= 15^2= 225$

$BGS_3= (4+7+3)^2= 14^2= 196$

$T= \sum^k_{i=1} \sum^{n_i}_{j=1} X_{ij}= 5+7+4+ 8+ 7+ 6+8+3= 48$

\bigskip

 Among Rows SS =  $\sum^k_{i=1}\frac{GS_i}{ k}-\frac{GS}{N}$
=(361/3+ 225/2+ 196/3) - 2304/8=  298,16666666666666666666666666667  -288 = 10.166

Among rows DF= Among rows degrees of freedom= l-1= 3-1=2.

Among rows Mean square= $\frac{Among Rows SS}{Among rows DF}= 10.166/2=5.083$

Now we calculate the square of each entry.

\begin{center}
\begin{tabular}{|l|l|c|}\hline
\multicolumn{3}{|c|}{\vphantom{Large Ap} $X^2_{ij}$  }\\ \hline\hline
Diet 1& Diet 2&Diet 3\\ \hline\hline
25 & 64 & 36 \\ \hline
49 &   &64\\ \hline
16 & 49&9  \\ \hline
\end{tabular}
\end{center}

\bigskip

Total SS= Total Sum of Squares=   $\sum^k_{i=1} \sum^{n_i}_{j=1}(X_{ij}^2)-\frac{GS}{N}$

=$(25 + 64 +36 +49  +64+16 + 49+9 )-\frac{2304}{8} = 312 -288 = 24$

Error SS=error sum of squares=  Total SS - Among columns SS -Among row SS= 24-6.1666-10.1666=7.667

Error DF= Total DF - Among groups DF - Among rows DF=  $ 7-2-2=3$.

Error Mean square= $\frac{Error SS}{error DF}=7.667/3= 2.555$

$Fisher for columns= \frac{Among \  columns \  mean  \ square}{Error \  Mean  \ square}= \frac{3.083 }{2.555}= 1.2$

$Fisher for rows= \frac{Among \  rows \  mean  \ square}{Error \  Mean  \ square}= \frac{5.083 }{2.555}= 1.99$

Now, we shall compare these values with the corresponding distributions. The problem is that nobody knows how are they. Nevertheless, since our methodology is a slight modification of the usual one, the small values of the stadigraphs condition us to believe that our data have nothing special and that are explained by the null hypotheses: we have gather as yet no proofs that the studied factors could influence the output of the experiment. Anyway, let us  carry out our simulation that is framed in within our  code for science:

1) One examines exactly what one did in the  experiment. Done.

2) One declares exactly what is the null hypothesis that one wants to test: we want to test the null hypotheses that under a change of treatment there is no change in the average output.

3) One builds   an artificial, virtual world, with the same characteristics as those proposed by the null hypotheses. To that aim, we declare the null hypotheses: all data come from the same source with a normal distribution and their arrangement in 3 columns and 3 rows is artificial without any bases. Thus we gather all data, calculate the mean and the standard deviation:

The mean of 5+7+4+ 8+ 7+ 6+8+3 is 48/8= 6. The standard deviation is $\sqrt{24/7}= \sqrt{3.4285714}=1.8516401$. Our supposition is that the data source is described by a mean of 6 and a fluctuation that follows a normal distribution with mean zero and standard deviation  1.8516401.

4) One runs the virtual world under the same conditions as those met by the experiment. Thus, we fabricate a generator of random numbers with mean 6 and standard deviation 1.8516401. To make that, we use our  random generator attuned to the prescribed parameters.

5) For each run in the virtual world one measures exactly the same stadigraph  as in the real experiment. A program that makes all that and that is easier to debug is the following:



The program was tested against our experimental data and it produced the same results as we. So, we thought that all was O.K. But when a simulation with 1000 rounds was attempted, it was found that some F values were negative, due to the negativity of the error mean square. Hence, we must recognized that our slight modification of the F stadigraph produced a nightmare of results.

 \bigskip
\textbf{\thenum. }  \addtocounter{num}{1} The solution is to compute the stadigraphs from fundamental definitions and begin all anew. Our definitions are as follows:

$r_i =$ number of data in the row $ith$.

$c_i = $number of data in the column $jth$.

$\bar{x_{.i}}$ = mean of column $i$

$\bar{x_{i.}}$ = mean of row $i$

$\bar{x}$ = general mean

TotalSS= $\sum (x_{ij}-\bar{x})^2$

ACSS = $\sum c_i (\bar{x_{.i}}-\bar{x})^2$, where

ARSS = $\sum r_i (\bar{x_{i.}}-\bar{x})^2$

Error SS = $\sum ( x{ij}-x_{i.}-x_{.i}-\bar{x})^2 $

ACMS = ACSS / 2

ARMS = ARSS / 2

EMS = ERRORSS / 3

Fcolumns = ACMS / EMS
Frows = ARMS / EMS

For our matrix these data produces:

$r_i =$ number of data in the row $ith$: $r_1 =3$, $r_2 =2$, $r_3 =3$

$c_i = $number of data in the column $jth$:$c_1 =3$, $c_2 =2$, $c_3 =3$

$x_{.i}$ = mean of column $i$: $x_{.1}=5.33$, $x_{.2}=7.5$, $x_{.3}=5.66$

$x_{i.}$ = mean of row $i$:$x_{1.}=6.33$, $x_{2.}=7.5$, $x_{3.}=4.66$

$\bar{x}$ = general mean = 6

TotalSS= $\sum (x_{ij}-\bar{x})^2=24$

ACSS = $\sum c_i (\bar{x_{.i}}-\bar{x})^2=6,166666667$,

ARSS = $\sum r_i (\bar{x_{i.}}-\bar{x})^2=10,16666667$

These values were those obtained by our previous work. With them, we calculated the error SS by the formula:

ErrorSS = TotalSS-ACSS-ARSS which rendered negative values in many cases.

Therefore, we must conclude that the total variance cannot be distributed in the present  case in three terms: that caused by rows, that by columns and that by uncontrolled factors. We need to take into account the diverse covariances, which could be positive or negative. This implies that we are anymore studying differences among means; instead, we are estimating  complex combinations of means, variances and covariances. Such uncertainty is proper of non parametric analysis.

The fundamental definition of Error SS is the following

Error SS = $\sum ( x{ij}-x_{i.}-x_{.i}-\bar{x})^2 $

This renders:

Error SS = 1155,666667

ACMS = ACSS / 2 = 6,166666667/2 = 3,08

ARMS = ARSS / 2 = 10,16/2 = 5,08

EMS = ERRORSS / 3 = 1155,666667/3 = 385,22

Fcolumns = ACMS / EMS =3,08/385,22= 0.007

Frows = ARMS / EMS = 0,013

Seemingly, we need no simulation to know that these values have no statistical  significance. Instead, let us waste our energies working out another idea.

\section{The duplex method}

 \bigskip
\textbf{\thenum. }  \addtocounter{num}{1}  We have tried to adapt the anova methodology to the case of incomplete data. To that aim, we perturbed the methodology to fit our circumstances and we kept data intact. Results were disappointing.

 \bigskip
\textbf{\thenum. }  \addtocounter{num}{1}  Let us try now another  strategy: let us perturb data and keep the methodology intact. In this method we compare two distributions. The first begins with the filling in the empty cell of  data matrix  with a random number generated according to the normal distribution inferred form the other real data. We apply the known  methodology on this matrix. Varying the random number we get varying stadigraphs, which define a distribution. This is our first distribution. The second one comes form  matrices completely filled in random numbers all around. We get in that way pairs of distributions to be compared.

 \bigskip
\textbf{\thenum. }  \addtocounter{num}{1}  Let us test the null hypotheses that our system is completely insensible in average to the external row and column factors. Thus, under this null hypotheses, data are completely characterized by a  global mean and the corresponding variance around it.

Example. Let analyze the following matrix

\begin{center}
\begin{tabular}{|l|l|l|c|}\hline
\multicolumn{4}{|c|}{\vphantom{Large Ap} $X_{ij}$  }\\ \hline\hline
& Treat 1& Treat 2& Treat 3\\ \hline\hline
B1&1 & 3 & 3 \\ \hline
B2&2 &   &4\\ \hline
B3&1 & 3&4  \\ \hline
B4&2 & 4 &3  \\ \hline
\end{tabular}
\end{center}

The sub that does that is the following:



 \bigskip
\textbf{\thenum. }  \addtocounter{num}{1}  Exercise. Digest the code. Help yourself activating the tracer instructions and using the different modes of execution.

 \bigskip
\textbf{\thenum. }  \addtocounter{num}{1}  Exercise. Run the code. It produces 4 distributions. In a first book, it presents the distributions of F stadigraphs for Columns and Rows of the original matrix that has been filled at its empty cell with a random number with a normal distribution. In a second book, the program presents the  stadigraphs but for a matrix that has been completely filled in random numbers. Draw the 4 distributions. Make a K-S test to compare the distributions for columns under both treatments and then do the same but for rows.

 \bigskip
\textbf{\thenum. }  \addtocounter{num}{1}  The output of the previous module may be translated to a graphic   and at once one perceives that the distributions corresponding to the experimental setting (that was completed with a random number  in the empty place) is very different than that corresponding to a matrix completely filled in random numbers with the same distribution as the original augmented matrix. In consequence, we see a great future for the duplex method and we proceed to make it more expedite.

 \bigskip
\textbf{\thenum. }  \addtocounter{num}{1}  The module below contains a code that directly produces the tables corresponding to the histograms of the different distributions.



 \bigskip
\textbf{\thenum. }  \addtocounter{num}{1}  Exercise. Run the code. Pay attention to the form as the histograms are generated. Make a graphic of the histograms and make a prognostic of the significance of the differences among experimental and control (generated by a random procedure) series. Run a K-S test to verify your suspicions.

 \bigskip
\textbf{\thenum. }  \addtocounter{num}{1}  Exercise. The spirit of an anova design  with blocks is that when one control a background variable, the experiment turns to be more sensitive: a change of treatment  needs not to  produce extremely large effects to be detected. Our simulations does not consider this: we just consider that the studied system was insensitive to a change of both row and column treatments. In that vein, we considered the model: datum = average value  + fluctuation. Thus, one could desire to implement that directive. Let us suppose that treatments are listed in the columns and blocks in the rows. We could consider the model

datum in row r= great average + average effect of row r + fluctuation.

Please, implement this new model under the duplex method.

\chapter{My first drawing }

 \bigskip
\textbf{\thenum. }  \addtocounter{num}{1}  Excel is certainly not designed for drawing, but one can remake it into a very powerful drawing platform. Let us see the know-how to do that.

 \bigskip
\textbf{\thenum. }  \addtocounter{num}{1}  Exercise. Record a Macro that color cells B2 to B7 in various distinct colors. To that aim, choose the pallet from your see menu, from  the ToolBar. Compare the resultant code with the following.

\begin{verbatim}

Public  Sub Gradient()
'
' Gradient Macro
'
'Opens a new sheet
  Workbooks.add
'
    Range("C2").Select
    With Selection.Interior
        .ColorIndex = 1
        .Pattern = xlSolid
    End With
    Range("C3").Select
    With Selection.Interior
        .ColorIndex = 9
        .Pattern = xlSolid
    End With
    Range("C4").Select
    With Selection.Interior
        .ColorIndex = 55
        .Pattern = xlSolid
    End With
    Range("C5").Select
    With Selection.Interior
        .ColorIndex = 12
        .Pattern = xlSolid
    End With
    Range("C6").Select
    With Selection.Interior
        .ColorIndex = 50
        .Pattern = xlSolid
    End With
    Range("C7").Select
    With Selection.Interior
        .ColorIndex = 43
        .Pattern = xlSolid
    End With
    Range("C8").Select
    With Selection.Interior
        .ColorIndex = 33
        .Pattern = xlSolid
    End With
    Range("C9").Select
    With Selection.Interior
        .ColorIndex = 8
        .Pattern = xlSolid
    End With
    Range("C10").Select
    With Selection.Interior
        .ColorIndex = 44
        .Pattern = xlSolid
    End With
    Range("C11").Select
    With Selection.Interior
        .ColorIndex = 36
        .Pattern = xlSolid
    End With
    Range("C12").Select
End Sub

\end{verbatim}

 \bigskip
\textbf{\thenum. }  \addtocounter{num}{1}  Exercise. Run the code. The displayed pattern has been  designed to try to give the impression of a gradient. Say, at the top we have a cool region while at the bottom we have a warmer one. Or, at the top we have a region with almost no population, while at the bottom the regions are densely populated. Or, at the top we have a low altitude region while at the bottom a high land. And so on. Employed colors do not contain the red one, so one could say that we use a cold mood.

 \bigskip
\textbf{\thenum. }  \addtocounter{num}{1}  Exercise. Run the following code and reinvent a Macro that does approximately the same.



Other drawing possibilities of Excel are considered in the next Macro

\begin{verbatim}

Public Sub graphic()
'Opens a new sheet
  Workbooks.add
 Dim triArray(1 To 4, 1 To 2) As Single
 Dim pts(1 To 7, 1 To 2) As Single
pts(1, 1) = 0
pts(1, 2) = 0
pts(2, 1) = 72
pts(2, 2) = 72
pts(3, 1) = 100
pts(3, 2) = 40
pts(4, 1) = 20
pts(4, 2) = 50
pts(5, 1) = 90
pts(5, 2) = 120
pts(6, 1) = 60
pts(6, 2) = 30
pts(7, 1) = 150
pts(7, 2) = 90
Set myDocument = Worksheets(1)
myDocument.Shapes.AddCurve pts

triArray(1, 1) = 25
triArray(1, 2) = 100
triArray(2, 1) = 100
triArray(2, 2) = 150
triArray(3, 1) = 150
triArray(3, 2) = 50
triArray(4, 1) = 25
 ' The last point has same coordinates as the first
triArray(4, 2) = 100
Set myDocument = Worksheets(1)
myDocument.Shapes.AddPolyline triArray

End Sub

\end{verbatim}

 \bigskip
\textbf{\thenum. }  \addtocounter{num}{1}  We have seen a method  to color a cell of a sheet. Let us see another one that is more suitable for our programming techniques. Let us draw a filled rectangle.

\begin{verbatim}

Public Sub rectangle()

'Opens a new sheet
  Workbooks.add


'Fills a rectangle cell by cell
For i = 1 To 5
 For j = 1 To 7
  Range(Cells(i, j), Cells(i, j)).Select
  With Selection.Interior
        .ColorIndex = 8
        .Pattern = xlSolid
    End With
Next j
Next i

'Fills a rectangle at once
 Range(Cells(8, 8), Cells(15, 12)).Select
  With Selection.Interior
        .ColorIndex = 18
        .Pattern = xlSolid
    End With
 Range(Cells(1, 9), Cells(1, 9)).Select
End Sub

\end{verbatim}

 \bigskip
\textbf{\thenum. }  \addtocounter{num}{1}  Graduation: draw an artwork and hang it at your office.

\chapter{The Asp}

 \bigskip
\textbf{\thenum. }  \addtocounter{num}{1}  Let us use our drawing possibilities to run a simulation of evolution: we design a poisonous  snake, an asp,  an our target is to devise an evolutionary algorithm to teach the snake to crawl. We will proceed step by step. Remember that asps are poisonous even from birth.  So, be careful.

We begin with a  design, with our own intelligence, of  the algorithm that makes the snake to crawl. Next, we forget our own ideas and implement an evolutionary algorithm whose objective is to learn to crawl. As an exercise, you could  compare our algorithm and that devised by evolution.

\section{The Robotic Asp}

 \bigskip
\textbf{\thenum. }  \addtocounter{num}{1}  Our snake has a head, 10 links and a tail. All twelve parts or cells have one unit length. The 12 parts can move independently  but   the structure of an organism must be kept: movements cannot separate the organism which must remain in one piece. We are allowed to move over the corners of a reticulate. In principle, every cell has 8 possibilities of movement plus another one of making nothing.

One shall check whether or not a given movement disrupts the organisms. For instance, if the snake is totally stretched, a forward movement of the head would disrupt the organism and so it is not allowed. If the snake is stretched, allowed movements are sidewards.

Every part can move autonomously but the organism must be kept in one piece. We posit the snake at the bottom of our screen and its task is to arrive at the upper edge in the minimum number of moves.

The next sub presents  a solution devised by the author, by a human. It would be interesting to compare a human solution with those produced by evolution.

\begin{verbatim}

Option Explicit
Dim VectMoves(1 To 9, 1 To 2)
Dim BodyAddress(1 To 12, 1 To 2)
Dim InitialRow




Private Sub Display()
'Prepares the screen
 Workbooks.add
    Range("A1:IH100").Select
    Selection.RowHeight = 3
    Selection.ColumnWidth = 0.3
    ActiveWindow.WindowState = xlMaximized
    Range("G9:T9").Select
End Sub

Private Sub InitialPosition()
Dim l, k
 'The asp has 12 parts,
 'each part has its own coordinates on the screen,
 'which are kept in a 12x2 array.
 'This is the initial position.
 'The head is at row InitialRow
 InitialRow = 80
For l = 1 To 12
 k = InitialRow - 12
 BodyAddress(l, 1) = 50
 BodyAddress(l, 2) = k + l
 Next l
End Sub

Private Sub DrawBody()
Dim l, i, j
 'Call Display
 'Call InitialPosition

 'Paints the head
  j = BodyAddress(1, 1)
  i = BodyAddress(1, 2)
  Range(Cells(i, j), Cells(i, j)).Select
  With Selection.Interior
        .ColorIndex = 9
        .Pattern = xlSolid
    End With

 'Paints the body
  For l = 2 To 12
  j = BodyAddress(l, 1)
  i = BodyAddress(l, 2)
  'MsgBox "l " & l
  'MsgBox "i " & i
  'MsgBox "j " & j
   Range(Cells(i, j), Cells(i, j)).Select
   With Selection.Interior
        .ColorIndex = 14
        .Pattern = xlSolid
     End With
  Next l
End Sub



Public Sub asp20()
Dim i, j, l, m, n
Dim Delay, h

 Call Display
 Call InitialPosition
 Call DrawBody
 'To move a link:
 'the cell occupied by  the link is cleared,
 'the cell to be occupied is painted.


 For l = 1 To 68
  For m = 1 To 12
   n = 13 - m
  j = BodyAddress(n, 1)
  i = BodyAddress(n, 2)
  Range(Cells(i, j), Cells(i, j)).Select
  Selection.Clear

  'MsgBox "i " & i
  'MsgBox "j " & j

  'Moves in zig-zag
  j = j + (-1) ^ l
  i = i - 1
  BodyAddress(n, 1) = j
  BodyAddress(n, 2) = i

  'MsgBox "i " & i
  'MsgBox "j " & j
   Range(Cells(i, j), Cells(i, j)).Select


   'If movement is too fast, nothing can be seen.
   'To attune the velocity of the as
   'Augment or diminish the delay as   pleased

   Delay = 250
   For h = 1 To Delay
      With Selection.Interior
        .ColorIndex = 14
        .Pattern = xlSolid
     End With
   Next h
  Next m
 Next l

End Sub

\end{verbatim}

 \bigskip
\textbf{\thenum. }  \addtocounter{num}{1}  Exercise. Please, run the code. To do that, assign the left part of the screen to Excel and keep 30\% of the screen at the right for VBE. In that way, you could look at your code with F8 and at its execution directly on the sheet of Excel. Notice that the asp has a distinct head at the beginning of the process but later on  the head turns to get the same color as the rest of the body. Make the necessary emendation.

 \bigskip
\textbf{\thenum. }  \addtocounter{num}{1}  Exercise. Would you continue to enjoy life while, by contrast, the asp cannot take out its tongue?

 \bigskip
\textbf{\thenum. }  \addtocounter{num}{1}  Let us move now into evolution, which is today a method to solve any type of problems. Ours is to teach the asp to crawl as fast as possible. We already know how to do that but we will search a way to begin from nothing and, with the help of evolution, to end with a very wise snake.

Our first  task is to translate our problem  into the language of evolution, which is understood here as composed of three items: 1) the change in the information content of a set of instructions which fulfill a given task. 2) The assignation of a reward to any information set according to its performance. 3) The reward is the right to get copies of the same information that pass into the next generation.

 \bigskip
\textbf{\thenum. }  \addtocounter{num}{1}  Example. A set of information contains instructions to make  our asp move.  In our original setting, we pose a population of asps at the bottom of the screen, whose information set has been assembled at random.  Asps may move and reproduce but the number of offspring of each asp is proportional to the velocity of the mother  to arrive to the top of the screen. The   offspring of each asp may have its set of instructions different than that of its mother, with mutations.

To implement this program, let us take some steps. The first is to rewrite the code for our asp in the required form, in an evolutionary language. Our set of instructions is as follows:

Select a link of the asp and have it move one unit upwards and one unit sidewards; do this for each link from tail to head, in that order. Repeat this over and over until finishing.

Our set of instruction is highly informative because 1) it does the task, 2) it does the task in the least possible time and 3) there are a lot of information sets that has nothing to do with the task or with its optimization. In fact, each move of each link is the selection of a given option among a list.

 \bigskip
\textbf{\thenum. }  \addtocounter{num}{1}  Thus, if we use cartesian coordinates and we imagine that the initial position of the link is at (0,0), then the nine possibilities of movement are:

(0,0); (0,1);(1,1);(1,0); (1,-1); (0,-1);(-1,-1);(-1,0);(-1,1);

 \bigskip
\textbf{\thenum. }  \addtocounter{num}{1}  Exercise. Run the next Macro to figure out these nine possibilities. These moves have a number, which is the order as they appear here.

\begin{verbatim}

Public Sub Nine()
'
' Nine Macro
'
    Workbooks.add

    Range("D6").Select
    ActiveCell.FormulaR1C1 = """(0,1)"""
    Range("F6").Select
    ActiveCell.FormulaR1C1 = """(1,1)"""
    Range("F9").Select
    ActiveCell.FormulaR1C1 = """(1,0)"""
    Range("F12").Select
    ActiveCell.FormulaR1C1 = """(1,1)"""
    Range("F12").Select
    ActiveCell.FormulaR1C1 = """(1,-1)"""
    Range("D12").Select
    ActiveCell.FormulaR1C1 = """(0,-1)"""
    Range("B12").Select
    ActiveCell.FormulaR1C1 = """(-1,-1)"""
    Range("B9").Select
    ActiveCell.FormulaR1C1 = """(-1,0)"""
    Range("B6").Select
    ActiveCell.FormulaR1C1 = """(-1,1)"""
    Range("D9").Select
    ActiveCell.FormulaR1C1 = """(0,0)"""
    Range("D9").Select
End Sub

\end{verbatim}

 \bigskip
\textbf{\thenum. }  \addtocounter{num}{1}  Exercise. Prove that the 9 listed  allowed movements do not exhaust all possibilities: in some occasions one can move two units apart and nevertheless the organism does not get disrupted. Anyway, a movement of two units can be achieved using  two moves in tandem, so we will ignore movements two units long.

 \bigskip
\textbf{\thenum. }  \addtocounter{num}{1}  Any one instruction of our instruction set says which link to move and in which direction. So, we may posit all instructions one after the other in an arrangement that is called a chromosome: it is vector with 2 columns and 13 rows. In each row we keep an instruction: which link to move is kept in the first column and which move among the nine in the second column.  The information contained in the chromosome is decoded, interpreted and executed by a unit that we call ribosome. At present time we need no more complexity. The following sub contains the code:

\begin{verbatim}

Option Explicit
Dim VectMoves(1 To 9, 1 To 2)
Dim BodyAddress(1 To 12, 1 To 2)
Dim InitialRow
Dim Chromosome(1 To 13, 1 To 2)
Dim Side

Private Sub Moves()

'The 9 moves are kept in an array
'with 9 rows and 2 columns.
'The first column keeps the x
'The second column for y
'The first column contains the number of the move.

VectMoves(1, 1) = 0
VectMoves(1, 2) = 1
VectMoves(2, 1) = 1
VectMoves(2, 2) = 1
VectMoves(3, 1) = 1
VectMoves(3, 2) = 0
VectMoves(4, 1) = 1
VectMoves(4, 2) = -1
VectMoves(5, 1) = 0
VectMoves(5, 2) = -1
VectMoves(6, 1) = -1
VectMoves(6, 2) = -1
VectMoves(7, 1) = -1
VectMoves(7, 2) = 0
VectMoves(8, 1) = -1
VectMoves(8, 2) = 1
VectMoves(9, 1) = 0
VectMoves(9, 2) = 0

End Sub


Private Sub Display()
'Prepares the screen
 Workbooks.add
    Range("A1:IH100").Select
    Selection.RowHeight = 3
    Selection.ColumnWidth = 0.3
    ActiveWindow.WindowState = xlMaximized
    Range("G9:T9").Select
End Sub

Private Sub InitialPosition()
Dim l, k
 'The asp has 12 parts,
 'each part has its own coordinates on the screen,
 'which are kept in a 12x2 array.
 'This is the initial position.
 'The head is at row InitialRow
 InitialRow = 80
For l = 1 To 12
 k = InitialRow - 12
 BodyAddress(l, 1) = 50
 BodyAddress(l, 2) = k + l
 Next l
End Sub

Private Sub DrawBody()
Dim l, i, j, h, Delay
 'Call Display
 'Call InitialPosition

 'Paints the head
  j = BodyAddress(1, 1)
  i = BodyAddress(1, 2)
  Range(Cells(i, j), Cells(i, j)).Select
  With Selection.Interior
        .ColorIndex = 9
        .Pattern = xlSolid
    End With

 'Paints the body
  For l = 2 To 12
  j = BodyAddress(l, 1)
  i = BodyAddress(l, 2)
  'MsgBox "l " & l
  'MsgBox "i " & i
  'MsgBox "j " & j
   Range(Cells(i, j), Cells(i, j)).Select
   With Selection.Interior
        .ColorIndex = 14
        .Pattern = xlSolid
     End With
  Next l
End Sub

Private Sub InitialChromosome()
Dim m, n
 For m = 1 To 12
   n = 13 - m
 'Move the tail first and then the others links
  Chromosome(m, 1) = n
 'Our move for all links is the same:
 'Move northeast or northwest
 'Northeast: (1,1), whose number is 2,
 'Northwest: (-1,1), whose number is 8.
 'Which side is decided by the variable Side,
 '0 means  northwest,
 '1 is  northeast.
'Initial direction: Northeast=2.
' because Side = 0
  Chromosome(m, 2) = 2 + 6 * Side
 Next m
End Sub

Private Sub ActualChromosome()
'Advance in zig-zag
 If Side = 0 Then Side = 1 Else Side = 0
   Call InitialChromosome
End Sub
Private Sub Ribosome()
Dim m, n, i, j
Dim Delay, h
Dim Move

 For m = 1 To 12
   'Read the link to be moved
   n = Chromosome(m, 1)
   'The occupied cell is cleared
  j = BodyAddress(n, 1)
  i = BodyAddress(n, 2)
  Range(Cells(i, j), Cells(i, j)).Select
  Selection.Clear

  'MsgBox "i " & i
  'MsgBox "j " & j

  'The move to be done is read
  Move = Chromosome(m, 2)
  'MsgBox "Move " & Move
  'MsgBox "+i " & VectMoves(Move, 1)
  'MsgBox "+j " & VectMoves(Move, 2)
  j = j + VectMoves(Move, 1)
  i = i - VectMoves(Move, 2)
  BodyAddress(n, 1) = j
  BodyAddress(n, 2) = i

  'The new cell is painted
  'MsgBox "New i " & i
  'MsgBox " New j " & j
   Range(Cells(i, j), Cells(i, j)).Select


   'If movement is too fast, nothing can be seen.
   'To attune the velocity of the asp
   'Augment or diminish the delay as   pleased

   Delay = 250
   For h = 1 To Delay
      With Selection.Interior
        .ColorIndex = 14
        .Pattern = xlSolid
     End With
   Next h
  Next m

End Sub

Private Sub Reproduction()
 'Our asp needs no reproduction
 'because it is perfect.
End Sub

Public Sub asp19()
Dim i, j, l, m, n
 Call Moves
 Call Display
 Call InitialPosition
 Call DrawBody
 Side = 0
 Call InitialChromosome
 'To move a link:
 'the cell occupied by  the link is cleared,
 'the cell to be occupied is painted.
 For l = 1 To 68
  Call ActualChromosome
  Call Ribosome
  Call Reproduction
 Next l
End Sub

\end{verbatim}

 \bigskip
\textbf{\thenum. }  \addtocounter{num}{1}  Exercise. Run the code and try to understand it.

 \bigskip
\textbf{\thenum. }  \addtocounter{num}{1}  The debugging of this particular code resulted to be   more difficult than usual. Thus, it could be now  the right occasion to introduce a technique of programming that was invented long a go  to deal with the complexities associated with debugging problems. The    philosophical directive underneath these advances   could be expressed as follows:

1)Developing code is very difficult. Hence, you shall not expect to develop code the right way on your first trial. To be sure, a code 40 lines long could be corrected some 50 times before one could say that one has what one wanted.

2) Divide and conquer. This means that complex problems become easier if one divides them in subproblems. Or, if one is studying a system, that becomes easier if one divides the system into subsystems.

3) Divide by the natural divisions: distort not the natural structures in your models and simulations.  If we are studying a vertebrate, then one could divide it into head, trunk and limbs.

4) Your programming language and style of encoding shall reflect this correspondence: there shall be a transparent correspondence among observations, data, algorithm of analysis, simulations and displayed results.

 \bigskip
\textbf{\thenum. }  \addtocounter{num}{1}  This philosophy has led today to very efficient  platforms for complex programming, such as the OOP, or object oriented programming ideology. Let us see the most ancient  implementation of this philosophy.  To fix ideas, let us think of a chromosome of our asp. It is an ordered set of instructions, each with two types of information: which link to move and  in which direction. Because the chromosome is ordered, it is a vector. But in its entries we do not keep numbers but a set of instructions consistent in  which link to move and in which direction. In its turn, a direction has two components, x and y. Therefore, we are dealing with structures of the type

Exon = (Link, Move)

Move = (Movex, MoveY)

Please, pay attention  to the form as this is implemented in VBA for our code for the asp. In all modern languages the form is very similar if not equal.

\begin{verbatim}

Option Explicit

Type LinkAddresses
 X As Integer
 Y As Integer
End Type

Type Moves
 X As Integer
 Y As Integer
End Type


Type Exons
 Link As Integer
 Move As Integer
End Type



Dim VectMoves(1 To 9) As Moves
Dim LinkAddress(1 To 12) As LinkAddresses
Dim Chromosome(1 To 13) As Exons

Dim InitialRow
Dim Side

Private Sub Moves()
VectMoves(1).X = 0
VectMoves(1).Y = 1
VectMoves(2).X = 1
VectMoves(2).Y = 1
VectMoves(3).X = 1
VectMoves(3).Y = 0
VectMoves(4).X = 1
VectMoves(4).Y = -1
VectMoves(5).X = 0
VectMoves(5).Y = -1
VectMoves(6).X = -1
VectMoves(6).Y = -1
VectMoves(7).X = -1
VectMoves(7).Y = 0
VectMoves(8).X = -1
VectMoves(8).Y = 1
VectMoves(9).X = 0
VectMoves(9).Y = 0
End Sub


Private Sub Display()
'Prepares the screen
'Excel already works with this "new" style!
'To be sure, Excel works with the OOP style!

 Workbooks.add
    Range("A1:IH100").Select
    Selection.RowHeight = 3
    Selection.ColumnWidth = 0.3
    ActiveWindow.WindowState = xlMaximized
    Range("G9:T9").Select
End Sub

Private Sub InitialPosition()
Dim l, k
 'The asp has 12 links,
 'each part has its own coordinates on the screen,
 'which are kept in the LinkAddress   array.
 'This is the initial position.
 'The head is at row InitialRow
 InitialRow = 80
For l = 1 To 12
 k = InitialRow - 12
 LinkAddress(l).X = 50
 LinkAddress(l).Y = k + l
 Next l
End Sub

Private Sub DrawBody()
Dim l, i, j, h, Delay
 'Call Display
 'Call InitialPosition

 'Paints the head
  j = LinkAddress(1).X
  i = LinkAddress(1).Y
  Range(Cells(i, j), Cells(i, j)).Select
  With Selection.Interior
        .ColorIndex = 9
        .Pattern = xlSolid
    End With

 'Paints the body
  For l = 2 To 12
  j = LinkAddress(l).X
  i = LinkAddress(l).Y
  'MsgBox "l " & l
  'MsgBox "i " & i
  'MsgBox "j " & j
   Range(Cells(i, j), Cells(i, j)).Select
   With Selection.Interior
        .ColorIndex = 14
        .Pattern = xlSolid
     End With
  Next l
End Sub

Private Sub InitialChromosome()
Dim m, n
 For m = 1 To 12
   n = 13 - m
 'Move the tail first and then the others links
  Chromosome(m).Link = n
 'We have two types of moves:
 'Northeast or northwest
 'Northeast: (1,1), whose number is 2,
 'Northwest: (-1,1), whose number is 8.
 'Which move  is decided by the variable Side,
 '0 means  northwest,
 '1 is  northeast.
'Initial direction: Northeast=2.
' because Side = 0
  Chromosome(m).Move = 2 + 6 * Side
 Next m
End Sub

Private Sub ActualChromosome()
'Advance in zig-zag
 If Side = 0 Then Side = 1 Else Side = 0
   Call InitialChromosome
End Sub

Private Sub Ribosome()
Dim m, n, i, j
Dim Delay, h
Dim Move


 For m = 1 To 12
 'Reads the link to be moved
 n = Chromosome(m).Link
  'To move a link:
 'the cell occupied by  the link is cleared,
 'the cell to be occupied is painted.

 'The occupied cell is cleared
  j = LinkAddress(n).X
  i = LinkAddress(n).Y
  Range(Cells(i, j), Cells(i, j)).Select
  Selection.Clear

  'MsgBox "i " & i
  'MsgBox "j " & j

  'The move to be done is read
  Move = Chromosome(m).Move
  'MsgBox "Move " & Move
  'MsgBox "+i " & VectMoves(Move).X
  'MsgBox "+j " & VectMoves(Move).Y
  j = j + VectMoves(Move).X
  i = i - VectMoves(Move).Y
  LinkAddress(n).X = j
  LinkAddress(n).Y = i

  'The new cell is painted
  'MsgBox "New i " & i
  'MsgBox " New j " & j
   Range(Cells(i, j), Cells(i, j)).Select


   'If movement is too fast, nothing can be seen.
   'To attune the velocity of the asp
   'Augment or diminish the delay as   pleased

   Delay = 250
   For h = 1 To Delay
      With Selection.Interior
        .ColorIndex = 14
        .Pattern = xlSolid
     End With
   Next h
  Next m
End Sub

Public Sub Reproduction()
 'Our asp needs no reproduction
 'because it is perfect.
End Sub
Public Sub asp18()
Dim i, j, l, m, n


 Call Moves
 Call Display
 Call InitialPosition
 Call DrawBody
 Side = 0
 Call InitialChromosome
 For l = 1 To 68
  Call ActualChromosome
  Call Ribosome
  Call Reproduction
 Next l
End Sub


\end{verbatim}

 \bigskip
\textbf{\thenum. }  \addtocounter{num}{1}  Exercise. Digest the code. Please, notice that in spite of the difficulty originated by the newness of the new flavor of the language, it is easier to understand. We observe that the more complex is a code created by humans,  the more structured it becomes. In fact, that is the only way as a human can deal with complexity -when that is possible.

 \bigskip
\textbf{\thenum. }  \addtocounter{num}{1}  Let us continue preparing the path to evolution. The next step is to make 50 clones of our asp. Let us see:

\begin{verbatim}

Option Explicit

Type LinkAddresses
 X As Integer
 Y As Integer
End Type

Type Moves
 X As Integer
 Y As Integer
End Type

Type Exons
 Link As Integer
 Move As Integer
End Type

Type asps
 Chromosome(1 To 13) As Exons
 LinkAddress(1 To 12) As LinkAddresses
 Color As Integer
 Side As Integer
End Type

Dim asp(1 To 20) As asps
Dim VectMoves(1 To 9) As Moves
Dim Side As Integer
Dim InitialRow



Private Sub Moves()
VectMoves(1).X = 0
VectMoves(1).Y = 1
VectMoves(2).X = 1
VectMoves(2).Y = 1
VectMoves(3).X = 1
VectMoves(3).Y = 0
VectMoves(4).X = 1
VectMoves(4).Y = -1
VectMoves(5).X = 0
VectMoves(5).Y = -1
VectMoves(6).X = -1
VectMoves(6).Y = -1
VectMoves(7).X = -1
VectMoves(7).Y = 0
VectMoves(8).X = -1
VectMoves(8).Y = 1
VectMoves(9).X = 0
VectMoves(9).Y = 0
End Sub


Private Sub Display()
'Prepares the screen
'Excel already works with this "new" style!
'To be sure, Excel works with the OOP style!

 Workbooks.add
    Range("A1:IH100").Select
    Selection.RowHeight = 3
    Selection.ColumnWidth = 0.3
    ActiveWindow.WindowState = xlMaximized
    Range("G9:T9").Select
End Sub

Private Sub InitialPosition(A As Integer)
Dim l, k
 'The index A distinguishes each asp.
 'Each asp has 12 links,
 'each part has its own coordinates on the screen,
 'which are kept in the LinkAddress   array.
 'The asps are posited at the bottom of the screen.
 'Head are at row InitialRow
 InitialRow = 80
For l = 1 To 12
 k = InitialRow - 12
 asp(A).LinkAddress(l).X = 8 * A
 asp(A).LinkAddress(l).Y = k + l
 Next l
End Sub

Private Sub DrawBody(A As Integer)
Dim l, i, j, h, Delay
 'Call Display
 'Call InitialPosition

 'Paints the body
 'Color is a number less than 40

  asp(A).Color = Int(50 * Rnd)
  For l = 1 To 12
  j = asp(A).LinkAddress(l).X
  i = asp(A).LinkAddress(l).Y
  'MsgBox "l " & l
  'MsgBox "i " & i
  'MsgBox "j " & j
   Range(Cells(i, j), Cells(i, j)).Select
   With Selection.Interior
        .ColorIndex = asp(A).Color
        .Pattern = xlSolid
     End With
  Next l
End Sub

Private Sub InitialChromosome(A As Integer)
Dim m, n
 For m = 1 To 12
   n = 13 - m
 'Move the tail first and then the others links
  asp(A).Chromosome(m).Link = n
 'We have two types of moves:
 'Northeast or northwest
 'Northeast: (1,1), whose number is 2,
 'Northwest: (-1,1), whose number is 8.
 'Which move  is decided by the variable Side,
 '0 means  northwest,
 '1 is  northeast.
'Initial direction: random


  asp(A).Chromosome(m).Move = 2 + 6 * asp(A).Side
 Next m
End Sub
Private Sub ActualChromosome(A As Integer)
'Advance in zig-zag
 If asp(A).Side = 0 Then asp(A).Side = 1 Else asp(A).Side = 0
   Call InitialChromosome(A)
End Sub

Private Sub Ribosome(A As Integer)
Dim m, n, i, j
Dim Delay, h
Dim Move

 For m = 1 To 12
 'Reads the link to be moved
 n = asp(A).Chromosome(m).Link
  'To move a link:
 'the cell occupied by  the link is cleared,
 'the cell to be occupied is painted.

 'The occupied cell is cleared
  j = asp(A).LinkAddress(n).X
  i = asp(A).LinkAddress(n).Y
  Range(Cells(i, j), Cells(i, j)).Select
  Selection.Clear

  'MsgBox "i " & i
  'MsgBox "j " & j

  'The move to be done is read
  Move = asp(A).Chromosome(m).Move
  'MsgBox "Move " & Move
  'MsgBox "+i " & VectMoves(Move).X
  'MsgBox "+j " & VectMoves(Move).Y
  j = j + VectMoves(Move).X
  i = i - VectMoves(Move).Y
  asp(A).LinkAddress(n).X = j
  asp(A).LinkAddress(n).Y = i

  'The new cell is painted
  'MsgBox "New i " & i
  'MsgBox " New j " & j
   Range(Cells(i, j), Cells(i, j)).Select


   'If movement is too fast, nothing can be seen.
   'To attune the velocity of the asp
   'Augment or diminish the delay as   pleased

   Delay = 10
   For h = 1 To Delay
      With Selection.Interior
        .ColorIndex = asp(A).Color
        .Pattern = xlSolid
     End With
   Next h
  Next m

End Sub

Private Sub Reproduction(A As Integer)
 'Our asp needs no reproduction
 'because it is perfect.
End Sub

Public Sub asp17()
Dim i, j, l, m, n
Dim A As Integer
Dim NClones As Integer
 Call Moves
 Call Display
 'Initialization
 NClones = 20
  For A = 1 To NClones
   asp(A).Side = Int(Rnd+0.5)
   Call InitialPosition(A)
   Call DrawBody(A)
   Call InitialChromosome(A)
  Next A
'Race
 For l = 1 To 68
  For A = 1 To NClones
  Call ActualChromosome(A)
  Call Ribosome(A)
  Call Reproduction(A)
  Next A
 Next l
End Sub

\end{verbatim}

 \bigskip
\textbf{\thenum. }  \addtocounter{num}{1}  Exercise. Run the code. Notice that all clones judiciously advance in perfect order. Is that not too boring? Make the suitable changes to have them advancing in random order with respect to its position on the screen.

 \bigskip
\textbf{\thenum. }  \addtocounter{num}{1}  To come closer to evolution, we must consider that the asps  would move over the screen  not necessarily upwards. Most probably they would wander around. So, we must foresee that no part of an asp can occupy the same part of any other or of itself. To solve the ensuing problem, we have a compromise: we pay with memory lost else with time wasting. We could use memory to keep track of the occupied cells of the screen else we could look at the color of each cell to inquire whether or not it is occupied by some asp. Because memory is, at the present level of complexity, more scare than time, we choose to waste time instead of memory.

 \bigskip
\textbf{\thenum. }  \addtocounter{num}{1}  Exercise. Make some experiments with Macro recording to know which is the default color of cells in an Excel Sheet. You could verify the answer found by the author in the code below because that color is a forbidden   for a snake. This new condition is the only change made to the previous code.

\begin{verbatim}

Option Explicit

Type LinkAddresses
 X As Integer
 Y As Integer
End Type

Type Moves
 X As Integer
 Y As Integer
End Type

Type Exons
 Link As Integer
 Move As Integer
End Type

Type asps
 Chromosome(1 To 13) As Exons
 LinkAddress(1 To 12) As LinkAddresses
 Color As Integer
 Side As Integer
End Type

Dim asp(1 To 20) As asps
Dim VectMoves(1 To 9) As Moves
Dim Side As Integer
Dim InitialRow



Private Sub Moves()
VectMoves(1).X = 0
VectMoves(1).Y = 1
VectMoves(2).X = 1
VectMoves(2).Y = 1
VectMoves(3).X = 1
VectMoves(3).Y = 0
VectMoves(4).X = 1
VectMoves(4).Y = -1
VectMoves(5).X = 0
VectMoves(5).Y = -1
VectMoves(6).X = -1
VectMoves(6).Y = -1
VectMoves(7).X = -1
VectMoves(7).Y = 0
VectMoves(8).X = -1
VectMoves(8).Y = 1
VectMoves(9).X = 0
VectMoves(9).Y = 0
End Sub


Private Sub Display()
'Prepares the screen
'Excel already works with this "new" style!
'To be sure, Excel works with the OOP style!

 Workbooks.add
    Range("A1:IH100").Select
    Selection.RowHeight = 3
    Selection.ColumnWidth = 0.3
    ActiveWindow.WindowState = xlMaximized
    Range("G9:T9").Select
End Sub

Private Sub InitialPosition(A As Integer)
Dim l, k
 'The index A distinguishes each asp.
 'Each asp has 12 links,
 'each part has its own coordinates on the screen,
 'which are kept in the LinkAddress   array.
 'The asps are posited at the bottom of the screen.
 'Head are at row InitialRow
 InitialRow = 80
For l = 1 To 12
 k = InitialRow - 12
 asp(A).LinkAddress(l).X = 8 * A
 asp(A).LinkAddress(l).Y = k + l
 Next l
End Sub

Private Function Fcolor(A As Integer)
 'Color is a number less than 40
 'but different than 2.
Rebegin:  Fcolor = Int(40 * Rnd)
 If Fcolor = 2 Then GoTo Rebegin
 'MsgBox "Color  " & Fcolor
End Function

Private Sub DrawBody(A As Integer)
Dim l, i, j, h, Delay
 'Call Display
 'Call InitialPosition
 'Paints the body
  asp(A).Color = Fcolor(A)
  For l = 1 To 12
  j = asp(A).LinkAddress(l).X
  i = asp(A).LinkAddress(l).Y
  'MsgBox "l " & l
  'MsgBox "i " & i
  'MsgBox "j " & j
   Range(Cells(i, j), Cells(i, j)).Select
   With Selection.Interior
        .ColorIndex = asp(A).Color
        .Pattern = xlSolid
     End With
  Next l
End Sub

Private Sub InitialChromosome(A As Integer)
Dim m, n
 For m = 1 To 12
   n = 13 - m
 'Move the tail first and then the others links
  asp(A).Chromosome(m).Link = n
 'We have two types of moves:
 'Northeast or northwest
 'Northeast: (1,1), whose number is 2,
 'Northwest: (-1,1), whose number is 8.
 'Which move  is decided by the variable Side,
 '0 means  northwest,
 '1 is  northeast.
'Initial direction: random
  asp(A).Chromosome(m).Move = 2 + 6 * asp(A).Side
 Next m
End Sub

Private Sub ActualChromosome(A As Integer)
'Advance in zig-zag
 If asp(A).Side = 0 Then asp(A).Side = 1 Else asp(A).Side = 0
   Call InitialChromosome(A)
End Sub

Private Sub Ribosome(A As Integer)
Dim m, n, i, j
Dim Delay, h
Dim Move

 For m = 1 To 12
 'Reads the link to be moved
 n = asp(A).Chromosome(m).Link
  'To move a link:
 'the cell occupied by  the link is cleared,
 'the cell to be occupied is painted.

 'The occupied cell is cleared
  j = asp(A).LinkAddress(n).X
  i = asp(A).LinkAddress(n).Y
  Range(Cells(i, j), Cells(i, j)).Select
  Selection.Clear

  'MsgBox "i " & i
  'MsgBox "j " & j

  'The move to be done is read
  Move = asp(A).Chromosome(m).Move
  'MsgBox "Move " & Move
  'MsgBox "+i " & VectMoves(Move).X
  'MsgBox "+j " & VectMoves(Move).Y
  j = j + VectMoves(Move).X
  i = i - VectMoves(Move).Y
  asp(A).LinkAddress(n).X = j
  asp(A).LinkAddress(n).Y = i

  'The new cell is painted
  'MsgBox "New i " & i
  'MsgBox " New j " & j
   Range(Cells(i, j), Cells(i, j)).Select


   'If movement is too fast, nothing can be seen.
   'To attune the velocity of the asp
   'Augment or diminish the delay as   pleased

   Delay = 10
   For h = 1 To Delay
      With Selection.Interior
        .ColorIndex = asp(A).Color
        .Pattern = xlSolid
     End With
   Next h
  Next m
End Sub

Private Sub Reproduction(A As Integer)
 'Our asp needs no reproduction
 'because it is perfect.
End Sub

Public Sub asp16()
Dim i, j, l, m, n
Dim A As Integer
Dim NClones As Integer
 Call Moves
 Call Display
 'Initialization
 NClones = 20
  For A = 1 To NClones
   asp(A).Side = Int(Rnd+0.5)
   Call InitialPosition(A)
   Call DrawBody(A)
   Call InitialChromosome(A)
  Next A
'Race
 For l = 1 To 68
  For A = 1 To NClones
  Call ActualChromosome(A)
  Call Ribosome(A)
  Call Reproduction(A)
  Next A
 Next l
End Sub

\end{verbatim}

 \bigskip
\textbf{\thenum. }  \addtocounter{num}{1}  Exercise. Run the code various times and notice that it always repeats the same colors in the same positions. Modify the code to generate different colors at the same position at differing executions. To interrupt execution pulse Ctrl + Pause (Interruption). Let us keep in mind that one shall be able to recover repeatability because one would like sometimes to disentangle what exactly occurred in some occasions.

 \bigskip
\textbf{\thenum. }  \addtocounter{num}{1}  In the next code we rewrite the instructions of the chromosomes with the following change: instead of witting an instruction saying that one time the asp shall move all its body  to the Northeast and the following time all its body to the northwest, we write the corresponding instruction for each link in tandem.   With that new correction, we have chromosomes with very simple structures, without regulatory sequences. The code follows:

\begin{verbatim}

Option Explicit

Type LinkAddresses
 X As Integer
 Y As Integer
End Type

Type Moves
 X As Integer
 Y As Integer
End Type

Type Exons
 Link As Integer
 Move As Integer
End Type

Type asps
'Each chromosome has 24 instructions
 Chromosome(1 To 25) As Exons
 LinkAddress(1 To 12) As LinkAddresses
 Color As Integer
 Side As Integer
End Type

Dim asp(1 To 20) As asps
Dim VectMoves(1 To 9) As Moves
Dim Side As Integer
Dim InitialRow



Private Sub Moves()
VectMoves(1).X = 0
VectMoves(1).Y = 1
VectMoves(2).X = 1
VectMoves(2).Y = 1
VectMoves(3).X = 1
VectMoves(3).Y = 0
VectMoves(4).X = 1
VectMoves(4).Y = -1
VectMoves(5).X = 0
VectMoves(5).Y = -1
VectMoves(6).X = -1
VectMoves(6).Y = -1
VectMoves(7).X = -1
VectMoves(7).Y = 0
VectMoves(8).X = -1
VectMoves(8).Y = 1
VectMoves(9).X = 0
VectMoves(9).Y = 0
End Sub


Private Sub Display()
'Prepares the screen
'Excel already works with this "new" style!
'To be sure, Excel works with the OOP style!

 Workbooks.add
    Range("A1:IH100").Select
    Selection.RowHeight = 3
    Selection.ColumnWidth = 0.3
    ActiveWindow.WindowState = xlMaximized
    Range("G9:T9").Select
End Sub

Private Sub InitialPosition(A As Integer)
Dim l, k
 'The index A distinguishes each asp.
 'Each asp has 12 links,
 'each part has its own coordinates on the screen,
 'which are kept in the LinkAddress   array.
 'The asps  are posited at the bottom of the screen.
 'Head are at row InitialRow
 InitialRow = 80
For l = 1 To 12
 k = InitialRow - 12
 asp(A).LinkAddress(l).X = 8 * A
 asp(A).LinkAddress(l).Y = k + l
 Next l
End Sub

Private Function Fcolor(A As Integer)
 'Color is a number less than 40
 'but different than 2.
Rebegin:  Fcolor = Int(40 * Rnd)
 If Fcolor = 2 Then GoTo Rebegin
 'MsgBox "Color  " & Fcolor
End Function

Private Sub DrawBody(A As Integer)
Dim l, i, j, h, Delay
 'Call Display
 'Call InitialPosition
 'Paints the body
  asp(A).Color = Fcolor(A)
  For l = 1 To 12
  j = asp(A).LinkAddress(l).X
  i = asp(A).LinkAddress(l).Y
  'MsgBox "l " & l
  'MsgBox "i " & i
  'MsgBox "j " & j
   Range(Cells(i, j), Cells(i, j)).Select
   With Selection.Interior
        .ColorIndex = asp(A).Color
        .Pattern = xlSolid
     End With
  Next l
End Sub

Private Sub InitialChromosome(A As Integer)
Dim m, n
'Move in Zig-Zag:
'Move the whole body Northeast
 For m = 1 To 13
   n = 13 - m
 'Move the tail first and then the others links
  asp(A).Chromosome(m).Link = n
 'We have two types of moves:
 'Northeast or northwest
 'Northeast: (1,1), whose number is 2,
 'Northwest: (-1,1), whose number is 8.
  asp(A).Chromosome(m).Move = 2
 Next m

 'Move the whole body Northwest
 For m = 13 To 25
   n = 25 - m
 'Move the tail first and then the others links

  asp(A).Chromosome(m).Link = n
 'We have two types of moves:
 'Northeast or northwest
 'Northeast: (1,1), whose number is 2,
 'Northwest: (-1,1), whose number is 8.
  asp(A).Chromosome(m).Move = 8
 Next m
End Sub

Private Sub ActualChromosome(A As Integer)
 'We have no evolution as yet
Call InitialChromosome(A)
End Sub

Private Sub Ribosome(A As Integer)
Dim m, n, i, j
Dim Delay, h
Dim Move

 For m = 1 To 24
 'Reads the link to be moved
 n = asp(A).Chromosome(m).Link
  'To move a link:
 'the cell occupied by  the link is cleared,
 'the cell to be occupied is painted.

 'The occupied cell is cleared

  j = asp(A).LinkAddress(n).X
  i = asp(A).LinkAddress(n).Y
  Range(Cells(i, j), Cells(i, j)).Select
  Selection.Clear

  'MsgBox "i " & i
  'MsgBox "j " & j

  'The move to be done is read
  Move = asp(A).Chromosome(m).Move
  'MsgBox "Move " & Move
  'MsgBox "+i " & VectMoves(Move).X
  'MsgBox "+j " & VectMoves(Move).Y
  j = j + VectMoves(Move).X
  i = i - VectMoves(Move).Y
  asp(A).LinkAddress(n).X = j
  asp(A).LinkAddress(n).Y = i

  'The new cell is painted
  'MsgBox "New i " & i
  'MsgBox " New j " & j
   Range(Cells(i, j), Cells(i, j)).Select


   'If movement is too fast, nothing can be seen.
   'To attune the velocity of the asp
   'Augment or diminish the delay as   pleased

   Delay = 10
   For h = 1 To Delay
      With Selection.Interior
        .ColorIndex = asp(A).Color
        .Pattern = xlSolid
     End With
   Next h
  Next m

End Sub

Private Sub Reproduction(A As Integer)
 'Our asp needs no reproduction
 'because it is perfect.
End Sub

Public Sub asp15()
Dim i, j, l, m, n
Dim A As Integer
Dim NClones As Integer
 Call Moves
 Call Display
 'Initialization
 NClones = 20

 'To generate new random sequences
 'change the number below.

 Randomize (3)

  For A = 1 To NClones

   Call InitialPosition(A)
   Call DrawBody(A)
   Call InitialChromosome(A)
  Next A
'Race
 For l = 1 To 68
  For A = 1 To NClones
  Call ActualChromosome(A)
  Call Ribosome(A)
  Call Reproduction(A)
  Next A
 Next l
End Sub

\end{verbatim}

 \bigskip
\textbf{\thenum. }  \addtocounter{num}{1}  Exercise. Study the code. Let us observe that an apparently tiny change in the objectives caused many correlated changes in the code. Can you explain why we need to dimension a chromosome with 25 exons instead of 24? If it happens to be a bug, please correct it.

 \bigskip
\textbf{\thenum. }  \addtocounter{num}{1}  In the previous code all instructions of a chromosome were executed in tandem for a given asp. That would cause a lack of symmetry in regard with opportunities. Let us rewrite the code in such a way that an asp can execute only one move at a time.

\begin{verbatim}

Option Explicit

Type LinkAddresses
 X As Integer
 Y As Integer
End Type

Type Moves
 X As Integer
 Y As Integer
End Type

Type Exons
 Link As Integer
 Move As Integer
End Type

Type asps
'Each chromosome has 24 instructions
 Chromosome(1 To 25) As Exons
 LinkAddress(1 To 12) As LinkAddresses
 Color As Integer
 Side As Integer
End Type

Dim asp(1 To 20) As asps
Dim VectMoves(1 To 9) As Moves
Dim Side As Integer
Dim InitialRow



Private Sub Moves()
VectMoves(1).X = 0
VectMoves(1).Y = 1
VectMoves(2).X = 1
VectMoves(2).Y = 1
VectMoves(3).X = 1
VectMoves(3).Y = 0
VectMoves(4).X = 1
VectMoves(4).Y = -1
VectMoves(5).X = 0
VectMoves(5).Y = -1
VectMoves(6).X = -1
VectMoves(6).Y = -1
VectMoves(7).X = -1
VectMoves(7).Y = 0
VectMoves(8).X = -1
VectMoves(8).Y = 1
VectMoves(9).X = 0
VectMoves(9).Y = 0
End Sub


Private Sub Display()
'Prepares the screen
'Excel already works with this "new" style!
'To be sure, Excel works with the OOP style!

 Workbooks.add
    Range("A1:IH100").Select
    Selection.RowHeight = 3
    Selection.ColumnWidth = 0.3
    ActiveWindow.WindowState = xlMaximized
    Range("G9:T9").Select
End Sub

Private Sub InitialPosition(A As Integer)
Dim l, k
 'The index A distinguishes each asp.
 'Each asp has 12 links,
 'each part has its own coordinates on the screen,
 'which are kept in the LinkAddress   array.
 'The asps are posited at the bottom of the screen.
 'Head are at row InitialRow
 InitialRow = 80
For l = 1 To 12
 k = InitialRow - 12
 asp(A).LinkAddress(l).X = 8 * A
 asp(A).LinkAddress(l).Y = k + l
 Next l
End Sub

Private Function Fcolor(A As Integer)
 'Color is a number less than 40
 'but different than 2.
Rebegin:  Fcolor = Int(40 * Rnd)
 If Fcolor = 2 Then GoTo Rebegin
  If Fcolor = 21 Then GoTo Rebegin
  If Fcolor = 0 Then GoTo Rebegin
 'MsgBox "Color  " & Fcolor
End Function

Private Sub DrawBody(A As Integer)
Dim l, i, j, h, Delay
 'Call Display
 'Call InitialPosition
 'Paints the body
  asp(A).Color = Fcolor(A)
  For l = 1 To 12
  j = asp(A).LinkAddress(l).X
  i = asp(A).LinkAddress(l).Y
  'MsgBox "l " & l
  'MsgBox "i " & i
  'MsgBox "j " & j
   Range(Cells(i, j), Cells(i, j)).Select
   With Selection.Interior
        .ColorIndex = asp(A).Color
        .Pattern = xlSolid
     End With
  Next l
  'If A = 15 Then MsgBox "Color " & asp(A).Color
End Sub

Private Sub InitialChromosome(m As Integer, A As Integer)
Dim n
'Move in Zig-Zag:

'We consider two cases:
'the first is moving the whole body Northeast,
'the second is moving the whole body Northeast

Select Case m
 Case 1, 2, 3, 4, 5, 6, 7, 8, 9, 10, 11, 12

 n = 13 - m
 'Move the tail first and then the others links
  asp(A).Chromosome(m).Link = n
 'We have two types of moves:
 'Northeast or northwest
 'Northeast: (1,1), whose number is 2,
 'Northwest: (-1,1), whose number is 8.
  asp(A).Chromosome(m).Move = 2

  Case 13, 14, 15, 16, 17, 18, 19, 20, 21, 22, 23, 24

 'Move the whole body Northwest
   n = 25 - m
 'Move the tail first and then the others links

  asp(A).Chromosome(m).Link = n
 'We have two types of moves:
 'Northeast or northwest
 'Northeast: (1,1), whose number is 2,
 'Northwest: (-1,1), whose number is 8.
  asp(A).Chromosome(m).Move = 8
 End Select
End Sub

Private Sub ActualChromosome(m As Integer, A As Integer)
 'We have no evolution as yet
Call InitialChromosome(m, A)
End Sub

Private Sub Ribosome(m As Integer, A As Integer)
Dim n, i, j
Dim Delay, h
Dim Move


 'Reads the link to be moved
 n = asp(A).Chromosome(m).Link
  'To move a link:
 'the cell occupied by  the link is cleared,
 'the cell to be occupied is painted.

 'The occupied cell is cleared

  j = asp(A).LinkAddress(n).X
  i = asp(A).LinkAddress(n).Y
  Range(Cells(i, j), Cells(i, j)).Select
  Selection.Clear

  'MsgBox "i " & i
  'MsgBox "j " & j

  'The move to be done is read
  Move = asp(A).Chromosome(m).Move
  'MsgBox "Move " & Move
  'MsgBox "+i " & VectMoves(Move).X
  'MsgBox "+j " & VectMoves(Move).Y
  j = j + VectMoves(Move).X
  i = i - VectMoves(Move).Y
  asp(A).LinkAddress(n).X = j
  asp(A).LinkAddress(n).Y = i

  'The new cell is painted
  'MsgBox "New i " & i
  'MsgBox " New j " & j
   Range(Cells(i, j), Cells(i, j)).Select


   'If movement is too fast, nothing can be seen.
   'To attune the velocity of the asp
   'Augment or diminish the delay as   pleased

   Delay = 10
   For h = 1 To Delay
      With Selection.Interior
        .ColorIndex = asp(A).Color
        .Pattern = xlSolid
     End With
   Next h
End Sub

Private Sub Reproduction(A As Integer)
 'Our asp needs no reproduction
 'because it is perfect.
End Sub

Public Sub asp14()

'One move at a time.
Dim m As Integer
Dim i, j, l, n
Dim A As Integer
Dim NClones As Integer
 Call Moves
 Call Display
 'Initialization
 NClones = 20

 'To generate new random sequences
 'change the number below.

 Randomize (3)

'A is the index of the asp
'm is the number of the move


  For A = 1 To NClones

   Call InitialPosition(A)
   Call DrawBody(A)
   For m = 1 To 24
    Call InitialChromosome(m, A)
   Next m
  Next A

'Race
'l is the counter of days of life
 For l = 1 To 34
  For m = 1 To 24
   For A = 1 To NClones
    Call ActualChromosome(m, A)
    Call Ribosome(m, A)
   Next A
  Next m
 Next l

End Sub

\end{verbatim}

 \bigskip
\textbf{\thenum. }  \addtocounter{num}{1}  Exercise. Run the code.  Please, pay attention to the new \textbf{Select case statement. } This  is an instruction to select a curse of action depending on the value of certain variable. This statement is equivalent to various if's in parallel that test the value of  the same variable.

 \bigskip
\textbf{\thenum. }  \addtocounter{num}{1} Exercise. Notice that all asps coherently turn to the left else to the right. Please,  correct the code to give individual freedom to each asp.

\section{Stepping into evolution}

 \bigskip
\textbf{\thenum. }  \addtocounter{num}{1} We already know how robotic all knowing snakes behave. Our first task is to clean robotization by providing ways of being free. The price for freedom is ignorance: from now on, the asps of the first generation  only know how to move but they do not know in which direction they shall move. Those asps that by mere randomness happen to observe a tendency to rise along the display  will have the right to reproduce and to get offspring in proportion to its efficacy. In that way, the knowledge of the population will increase from generation to generation. Latter on, we will have the opportunity to compare wise evolved snakes with our robotic version.

In the next code we will have the opportunity to see in which form we provide asps with freedom (and ignorance). All we must do is to initialize them with chromosomes completely generated at random. Those snakes that tend to go down and that are outliers will left no children.

In our first trial, we will see only the asp under the effect of chromosomes assembled with random information.

\begin{verbatim}

Option Explicit

Type LinkAddresses
 X As Integer
 Y As Integer
End Type

Type Moves
 X As Integer
 Y As Integer
End Type

Type Exons
 Link As Integer
 Move As Integer
End Type

Type asps
'Each chromosome has 24 instructions
 Chromosome(1 To 25) As Exons
 LinkAddress(1 To 12) As LinkAddresses
 Color As Integer
 Side As Integer
End Type

Dim asp(1 To 20) As asps
Dim VectMoves(1 To 9) As Moves
Dim Side As Integer
Dim InitialRow



Private Sub Moves()
VectMoves(1).X = 0
VectMoves(1).Y = 1
VectMoves(2).X = 1
VectMoves(2).Y = 1
VectMoves(3).X = 1
VectMoves(3).Y = 0
VectMoves(4).X = 1
VectMoves(4).Y = -1
VectMoves(5).X = 0
VectMoves(5).Y = -1
VectMoves(6).X = -1
VectMoves(6).Y = -1
VectMoves(7).X = -1
VectMoves(7).Y = 0
VectMoves(8).X = -1
VectMoves(8).Y = 1
VectMoves(9).X = 0
VectMoves(9).Y = 0
End Sub


Private Sub Display()
'Prepares the screen
'Excel already works with this "new" style!
'To be sure, Excel works with the OOP style!

 Workbooks.add
    Range("A1:IH100").Select
    Selection.RowHeight = 3
    Selection.ColumnWidth = 0.3
    ActiveWindow.WindowState = xlMaximized
    Range("G9:T9").Select
End Sub

Private Sub InitialPosition(A As Integer)
Dim l, k
 'The index A distinguishes each asp.
 'Each asp has 12 links,
 'each part has its own coordinates on the screen,
 'which are kept in the LinkAddress   array.
 'The asps are posited at the bottom of the screen.
 'Head are at row InitialRow
 InitialRow = 80
For l = 1 To 12
 k = InitialRow - 12
 asp(A).LinkAddress(l).X = 8 * A
 asp(A).LinkAddress(l).Y = k + l
 Next l
End Sub

Private Function Fcolor(A As Integer)
 'Color is a number less than 40
 'but different than 2.
Rebegin:  Fcolor = Int(40 * Rnd)
 If Fcolor = 2 Then GoTo Rebegin
  If Fcolor = 21 Then GoTo Rebegin
  If Fcolor = 0 Then GoTo Rebegin
 'MsgBox "Color  " & Fcolor
End Function

Private Sub DrawBody(A As Integer)
Dim l, i, j, h, Delay
 'Call Display
 'Call InitialPosition
 'Paints the body
  asp(A).Color = Fcolor(A)
  For l = 1 To 12
  j = asp(A).LinkAddress(l).X
  i = asp(A).LinkAddress(l).Y
  'MsgBox "l " & l
  'MsgBox "i " & i
  'MsgBox "j " & j
   Range(Cells(i, j), Cells(i, j)).Select
   With Selection.Interior
        .ColorIndex = asp(A).Color
        .Pattern = xlSolid
     End With
  Next l
  'If A = 15 Then MsgBox "Color " & asp(A).Color
End Sub

Private Sub InitialChromosome(m As Integer, A As Integer)
Dim n
'Move at random,
'without any purpose,
'without any restriction.

\textbf{'Select  which link to move
'Throw a number from 1 to 12

 n = Int(12 * Rnd) + 1
  asp(A).Chromosome(m).Link = n

 'Select  which direction to move on
'Throw a number from 1 to 9
  asp(A).Chromosome(m).Move = Int(9 * Rnd) + 1}
End Sub

Private Sub ActualChromosome(m As Integer, A As Integer)
 'We have no evolution as yet
Call InitialChromosome(m, A)
End Sub

Private Sub Ribosome(m As Integer, A As Integer)
Dim n, i, j
Dim Delay, h
Dim Move


 'Reads the link to be moved
 n = asp(A).Chromosome(m).Link
  'To move a link:
 'the cell occupied by  the link is cleared,
 'the cell to be occupied is painted.

 'The occupied cell is cleared

  j = asp(A).LinkAddress(n).X
  i = asp(A).LinkAddress(n).Y
  Range(Cells(i, j), Cells(i, j)).Select
  Selection.Clear

  'MsgBox "i " & i
  'MsgBox "j " & j

  'The move to be done is read
  Move = asp(A).Chromosome(m).Move
  'MsgBox "Move " & Move
  'MsgBox "+i " & VectMoves(Move).X
  'MsgBox "+j " & VectMoves(Move).Y
  j = j + VectMoves(Move).X
  i = i - VectMoves(Move).Y
  asp(A).LinkAddress(n).X = j
  asp(A).LinkAddress(n).Y = i

  'The new cell is painted
  'MsgBox "New i " & i
  'MsgBox " New j " & j
   Range(Cells(i, j), Cells(i, j)).Select


   'If movement is too fast, nothing can be seen.
   'To attune the velocity of the asp
   'Augment or diminish the delay as   pleased

   Delay = 10
   For h = 1 To Delay
      With Selection.Interior
        .ColorIndex = asp(A).Color
        .Pattern = xlSolid
     End With
   Next h


End Sub

Private Sub Reproduction(A As Integer)
 'Our asp needs no reproduction
 'because it is perfect.
End Sub

Public Sub asp13()

 'Evolution on its first trial

Dim m As Integer
Dim i, j, l, n
Dim A As Integer
Dim NClones As Integer
 Call Moves
 Call Display
 'Initialization
 NClones = 20

 'To generate new random sequences
 'change the number below.

 Randomize (3)

'A is the index of the asp
'm is the number of the move


  For A = 1 To NClones

   Call InitialPosition(A)
   Call DrawBody(A)
   For m = 1 To 24
    Call InitialChromosome(m, A)
   Next m
  Next A

'Race
'l is the counter of days of life
 For l = 1 To 34
  For m = 1 To 24
   For A = 1 To NClones
    Call ActualChromosome(m, A)
    Call Ribosome(m, A)
   Next A
  Next m
 Next l
End Sub

\end{verbatim}

 \bigskip
\textbf{\thenum. }  \addtocounter{num}{1} Exercise. Run the code. We observe that random information has no idea of what an organism is. We see also that random information does not recognize a preferred direction.

 \bigskip
\textbf{\thenum. }  \addtocounter{num}{1} Our job is to devise an evolutionary environment to teach these unwise asps how to behave as an organism and how to advance towards a goal. They do not  know as yet that there are restrictions that one must observe else one risks to spoil everything.

The arisen of anatomical organization in an evolutionary environment is a very important problem but let us save it for a future.  We can assure that  the  movements of the asps cause no rupture to their organisms if disrupting instructions are   ignored.

 \bigskip
\textbf{\thenum. }  \addtocounter{num}{1} Exercise. The next code is viable, i.e., it is understood by the execution machine  but, please, run the code to verify that it is yet disruptive.   Next, try to figure out the cause of the failure and pass to the next code to verify your intuition.

\begin{verbatim}

Option Explicit

Type LinkAddresses
 x As Integer
 y As Integer
End Type

Type Moves
 x As Integer
 y As Integer
End Type

Type Exons
 Link As Integer
 Move As Integer
End Type

Type asps
'Each chromosome has 24 instructions
 Chromosome(1 To 25) As Exons
 LinkAddress(1 To 12) As LinkAddresses
 Color As Integer
End Type

Dim asp(1 To 20) As asps
Dim VectMoves(1 To 9) As Moves
Dim LifeSpan As Integer
Dim InitialRow



Private Sub Moves()
VectMoves(1).x = 0
VectMoves(1).y = 1
VectMoves(2).x = 1
VectMoves(2).y = 1
VectMoves(3).x = 1
VectMoves(3).y = 0
VectMoves(4).x = 1
VectMoves(4).y = -1
VectMoves(5).x = 0
VectMoves(5).y = -1
VectMoves(6).x = -1
VectMoves(6).y = -1
VectMoves(7).x = -1
VectMoves(7).y = 0
VectMoves(8).x = -1
VectMoves(8).y = 1
VectMoves(9).x = 0
VectMoves(9).y = 0
End Sub


Private Sub Display()
'Prepares the screen
'Excel already works with this "new" style!
'To be sure, Excel works with the OOP style!

 Workbooks.add
    Range("A1:IH100").Select
    Selection.RowHeight = 3
    Selection.ColumnWidth = 0.3
    ActiveWindow.WindowState = xlMaximized
    Range("G9:T9").Select
End Sub

Private Sub InitialPosition(A As Integer)
Dim l, k
 'The index A distinguishes each asp.
 'Each asp has 12 links,
 'each part has its own coordinates on the screen,
 'which are kept in the LinkAddress   array.
 'The asps are posited at the bottom of the screen.
 'Head are at row InitialRow
 InitialRow = 50
For l = 1 To 12
 k = InitialRow - 12
 asp(A).LinkAddress(l).x = 8 * A
 asp(A).LinkAddress(l).y = k + l
 Next l
End Sub

Private Function Fcolor(A As Integer)
 'Color is a number less than 40
 'but different than 2.
Rebegin:  Fcolor = Int(40 * Rnd)
 If Fcolor = 2 Then GoTo Rebegin
  If Fcolor = 21 Then GoTo Rebegin
  If Fcolor = 0 Then GoTo Rebegin
 'MsgBox "Color  " & Fcolor
End Function

Private Sub DrawBody(A As Integer)
Dim l, i, j, h, Delay
 'Call Display
 'Call InitialPosition
 'Paints the body
  asp(A).Color = Fcolor(A)
  For l = 1 To 12
  j = asp(A).LinkAddress(l).x
  i = asp(A).LinkAddress(l).y
  'MsgBox "l " & l
  'MsgBox "i " & i
  'MsgBox "j " & j
   Range(Cells(i, j), Cells(i, j)).Select
   With Selection.Interior
        .ColorIndex = asp(A).Color
        .Pattern = xlSolid
     End With
  Next l
  'If A = 15 Then MsgBox "Color " & asp(A).Color
End Sub

Private Sub InitialChromosome(A As Integer)
Dim m, n, t
Dim x, y As Integer
Dim x1, y1 As Integer

  'Select at random  which link to move
'Throw a number from 1 to 12
For m = 1 To 24
 n = Int(12 * Rnd) + 1
  asp(A).Chromosome(m).Link = n

 'Select at random which direction to move on
 'Throw a number from 1 to 9
  asp(A).Chromosome(m).Move = Int(9 * Rnd) + 1
Next m
End Sub

Private Sub MoveOn(m As Integer, A As Integer)
Dim n, i, j As Integer
Dim Delay, h As Integer
Dim Move As Integer
 'To move a link:
 'the cell occupied by  the link is cleared,
 'the cell to be occupied is painted.

 'The occupied cell is cleared
  n = asp(A).Chromosome(m).Link
  j = asp(A).LinkAddress(n).x
  i = asp(A).LinkAddress(n).y
  Range(Cells(i, j), Cells(i, j)).Select
  Selection.Clear

  'The target addres is memorized
  Move = asp(A).Chromosome(m).Move
  j = j + VectMoves(Move).x
  i = i - VectMoves(Move).y
  asp(A).LinkAddress(n).x = j
  asp(A).LinkAddress(n).y = i

  'The target cell is painted
   Range(Cells(i, j), Cells(i, j)).Select


   'If movement is too fast, nothing can be seen.
   'To attune the velocity of the asp,
   'augment or diminish the delay as   pleased

   Delay = 500
   For h = 1 To Delay
      With Selection.Interior
        .ColorIndex = asp(A).Color
        .Pattern = xlSolid
     End With
   Next h
End Sub

Private Sub ActualChromosome(m As Integer, A As Integer)
 'We have no evolution as yet
Call InitialChromosome(A)
End Sub

Private Function Disruptive(m As Integer, A As Integer)
'A move is disruptive if the target address
'is farther than 1.5 units from the neighboring links.
Dim n As Integer
Dim Move As Integer
Dim x, y, x1, y1, x2, x3, y2, y3 As Integer
Dim d1, d2 As Single

  'Reads the link to be moved
  n = asp(A).Chromosome(m).Link
  'Reads the   address of the link
  y = asp(A).LinkAddress(n).x
  x = asp(A).LinkAddress(n).y
  'The (integer) move to be done is read
  Move = asp(A).Chromosome(m).Move

 'The target address is calculated
  y1 = y + VectMoves(Move).x
  x1 = x - VectMoves(Move).y
  'The addresses of the two neighboring links is read
  'And disruption is calculated
    'MsgBox "n " & n
  If ((1 < n) And (n < 12)) Then
    y2 = asp(A).LinkAddress(n - 1).x
    x2 = asp(A).LinkAddress(n - 1).y
    d1 = ((x2 - x1) ^ 2 + (y2 - y1) ^ 2) ^ 0.5
If d1>1.5 Then Disruptive=True Else Disruptive=False
    y3 = asp(A).LinkAddress(n + 1).x
    x3 = asp(A).LinkAddress(n + 1).y
    d2 = ((x3 - x1) ^ 2 + (y3 - y1) ^ 2) ^ 0.5
If d2>1.5 Then Disruptive=True Else Disruptive = False
  End If

  If n = 1 Then
    y2 = asp(A).LinkAddress(2).x
    x2 = asp(A).LinkAddress(2).y
    d1 = ((x2 - x1) ^ 2 + (y2 - y1) ^ 2) ^ 0.5
If d1>1.1 Then Disruptive=True Else Disruptive=False
   End If

  If n = 12 Then
    y2 = asp(A).LinkAddress(11).x
    x2 = asp(A).LinkAddress(11).y
    d2 = ((x2 - x1) ^ 2 + (y2 - y1) ^ 2) ^ 0.5
If d2>1.1 Then Disruptive=True Else Disruptive = False
  End If
End Function



Private Sub Ribosome(m As Integer, A As Integer)
Dim n, i, j As Integer
Dim Delay, h As Integer
Dim Move As Integer
Dim x, y, x1, y1 As Integer
Dim Color As Integer
  'Reads the link to be moved
  n = asp(A).Chromosome(m).Link
  'Reads the   address of the link
  y = asp(A).LinkAddress(n).x
  x = asp(A).LinkAddress(n).y
  'The (integer) move to be done is read
  Move = asp(A).Chromosome(m).Move

 'The target address is calculated
  y1 = y + VectMoves(Move).x
  x1 = x - VectMoves(Move).y

  'There are forbidden borders
  If ((x1 = 0) Or (y1 = 0)) Then Exit Sub

  'Is the target address occupied?
  Range(Cells(x1, y1), Cells(x1, y1)).Select
      With Selection.Interior
       Color = .ColorIndex
      End With

  Select Case Color
    Case 0, 2, 21 'do nothing: the cell is occupied.
    Case Else 'the cell is empty, is it disruptive?
      If Not (Disruptive(m, A)) Then Call MoveOn(m, A)
   End Select
End Sub
Private Sub Reproduction()
 'No reproduction as yet.

End Sub

Private Sub Death()
'No death.
End Sub


Public Sub asp12()

 'Evolution on its first trial.
 'Body structure is automatically conserved.
 'The task is to climb over the display.

Dim m As Integer
Dim i, j, l, n
Dim A As Integer
Dim NClones As Integer
Dim Gen, NGen As Integer
 Call Moves
 Call Display
 'Initialization
 NClones = 20

 'To generate new random sequences
 'change the number below.

 Randomize (3)

'A is the index of the asp.

'Initialization
  For A = 1 To NClones
   Call InitialPosition(A)
   Call DrawBody(A)
    Call InitialChromosome(A)
  Next A


'l is the counter of days of life.
'Life ends with reproduction and dead.
'NGen is the number of generations
'of the whole experiment.
'm is the number of the link


NGen = 1
LifeSpan = 30
For Gen = 1 To NGen
 For l = 1 To LifeSpan
  For m = 1 To 24
   For A = 1 To NClones
    Call Ribosome(m, A)
    'Call ActualChromosome(m, A)
   Next A
  Next m
 Next l
 Reproduction
 Death
 Next Gen
End Sub

\end{verbatim}

 \bigskip
\textbf{\thenum. }  \addtocounter{num}{1}Exercise. An amend was done, as presented in the next code. It solves most problems with one exception. Run the   code to find it.

\begin{verbatim}

Option Explicit

Type LinkAddresses
 x As Integer
 y As Integer
End Type

Type Moves
 x As Integer
 y As Integer
End Type

Type Exons
 Link As Integer
 Move As Integer
End Type

Type asps
'Each chromosome has 24 instructions
 Chromosome(1 To 25) As Exons
 LinkAddress(1 To 12) As LinkAddresses
 Color As Integer
End Type

Dim asp(1 To 20) As asps
Dim VectMoves(1 To 9) As Moves
Dim LifeSpan As Integer
Dim InitialRow



Private Sub Moves()
VectMoves(1).x = 0
VectMoves(1).y = 1
VectMoves(2).x = 1
VectMoves(2).y = 1
VectMoves(3).x = 1
VectMoves(3).y = 0
VectMoves(4).x = 1
VectMoves(4).y = -1
VectMoves(5).x = 0
VectMoves(5).y = -1
VectMoves(6).x = -1
VectMoves(6).y = -1
VectMoves(7).x = -1
VectMoves(7).y = 0
VectMoves(8).x = -1
VectMoves(8).y = 1
VectMoves(9).x = 0
VectMoves(9).y = 0
End Sub


Private Sub Display()
'Prepares the screen
'Excel already works with this "new" style!
'To be sure, Excel works with the OOP style!

 Workbooks.add
    Range("A1:IH100").Select
    Selection.RowHeight = 3
    Selection.ColumnWidth = 0.3
    ActiveWindow.WindowState = xlMaximized
    Range("G9:T9").Select
End Sub

Private Sub InitialPosition(A As Integer)
Dim l, k
 'The index A distinguishes each asp.
 'Each asp has 12 links,
 'each part has its own coordinates on the screen,
 'which are kept in the LinkAddress   array.
 'The asps  are posited at the bottom of the screen.
 'Head are at row InitialRow
 InitialRow = 50
For l = 1 To 12
 k = InitialRow - 12
 asp(A).LinkAddress(l).x = 8 * A
 asp(A).LinkAddress(l).y = k + l
 Next l
End Sub

Private Function Fcolor(A As Integer)
 'Color is a number less than 40
 'but different than 2.
Rebegin:  Fcolor = Int(40 * Rnd)
 If Fcolor = 2 Then GoTo Rebegin
  If Fcolor = 21 Then GoTo Rebegin
  If Fcolor = 0 Then GoTo Rebegin
 'MsgBox "Color  " & Fcolor
End Function

Private Sub DrawBody(A As Integer)
Dim l, i, j, h, Delay
 'Call Display
 'Call InitialPosition
 'Paints the body
  asp(A).Color = Fcolor(A)
  For l = 1 To 12
  j = asp(A).LinkAddress(l).x
  i = asp(A).LinkAddress(l).y
  'MsgBox "l " & l
  'MsgBox "i " & i
  'MsgBox "j " & j
   Range(Cells(i, j), Cells(i, j)).Select
   With Selection.Interior
        .ColorIndex = asp(A).Color
        .Pattern = xlSolid
     End With
  Next l
  'If A = 15 Then MsgBox "Color " & asp(A).Color
End Sub

Private Sub InitialChromosome(A As Integer)
Dim m, n, t
Dim x, y As Integer
Dim x1, y1 As Integer

  'Select at random  which link to move
'Throw a number from 1 to 12
For m = 1 To 24
 n = Int(12 * Rnd) + 1
  asp(A).Chromosome(m).Link = n

 'Select at random which direction to move on
 'Throw a number from 1 to 9
  asp(A).Chromosome(m).Move = Int(9 * Rnd) + 1
Next m
End Sub

Private Sub MoveOn(m As Integer, A As Integer)
Dim n, i, j As Integer
Dim Delay, h As Integer
Dim Move As Integer
 'To move a link:
 'the cell occupied by  the link is cleared,
 'the cell to be occupied is painted.

 'The occupied cell is cleared
  n = asp(A).Chromosome(m).Link
  j = asp(A).LinkAddress(n).x
  i = asp(A).LinkAddress(n).y
  Range(Cells(i, j), Cells(i, j)).Select
  Selection.Clear

  'The target addres is memorized
  Move = asp(A).Chromosome(m).Move
  j = j + VectMoves(Move).x
  i = i - VectMoves(Move).y
  asp(A).LinkAddress(n).x = j
  asp(A).LinkAddress(n).y = i

  'The target cell is painted
   Range(Cells(i, j), Cells(i, j)).Select


   'If movement is too fast, nothing can be seen.
   'To attune the velocity of the asp,
   'augment or diminish the delay as   pleased

   Delay = 500
   For h = 1 To Delay
      With Selection.Interior
        .ColorIndex = asp(A).Color
        .Pattern = xlSolid
     End With
   Next h
End Sub

Private Sub ActualChromosome(m As Integer, A As Integer)
 'We have no evolution as yet
Call InitialChromosome(A)
End Sub

 Private Function Disruptive(m As Integer, A As Integer)
'A move is disruptive if the target address
'is farther than 1.5 units from the neighboring links.
Dim n As Integer
Dim Move As Integer
Dim x, y, x1, y1, x2, x3, y2, y3 As Integer
Dim d1, d2 As Single

  'Reads the link to be moved
  n = asp(A).Chromosome(m).Link
  'Reads the   address of the link
  y = asp(A).LinkAddress(n).x
  x = asp(A).LinkAddress(n).y
  'The (integer) move to be done is read
  Move = asp(A).Chromosome(m).Move

 'The target address is calculated
  y1 = y + VectMoves(Move).x
  x1 = x - VectMoves(Move).y
  'The addresses of the two neighboring links is read
  'And disruption is calculated
    'MsgBox "n " & n
  If ((1 < n) And (n < 12)) Then
    y2 = asp(A).LinkAddress(n - 1).x
    x2 = asp(A).LinkAddress(n - 1).y
    d1 = ((x2 - x1) ^ 2 + (y2 - y1) ^ 2) ^ 0.5
    y3 = asp(A).LinkAddress(n + 1).x
    x3 = asp(A).LinkAddress(n + 1).y
    d2 = ((x3 - x1) ^ 2 + (y3 - y1) ^ 2) ^ 0.5

    'The disruption bug is fixed here.

     If ((d1 > 1.5) Or (d2 > 1.5)) Then
      Disruptive = True
      Else
      Disruptive = False
     End If
  End If

    'Debugging messages
    'MsgBox "d1 " & d1
    'MsgBox "d2 " & d2

  If n = 1 Then
    y2 = asp(A).LinkAddress(2).x
    x2 = asp(A).LinkAddress(2).y
    d1 = ((x2 - x1) ^ 2 + (y2 - y1) ^ 2) ^ 0.5
 If d1 > 1.1 Then Disruptive = True Else Disruptive = False
   End If

  If n = 12 Then
    y2 = asp(A).LinkAddress(11).x
    x2 = asp(A).LinkAddress(11).y
    d2 = ((x2 - x1) ^ 2 + (y2 - y1) ^ 2) ^ 0.5
 If d2 > 1.1 Then Disruptive = True Else Disruptive = False
  End If

 End Function



Private Sub Ribosome(m As Integer, A As Integer)
Dim n, i, j As Integer
Dim Delay, h As Integer
Dim Move As Integer
Dim x, y, x1, y1 As Integer
Dim Color As Integer
  'Reads the link to be moved
  n = asp(A).Chromosome(m).Link
  'Reads the   address of the link
  y = asp(A).LinkAddress(n).x
  x = asp(A).LinkAddress(n).y
  'The (integer) move to be done is read
  Move = asp(A).Chromosome(m).Move

 'The target address is calculated
  y1 = y + VectMoves(Move).x
  x1 = x - VectMoves(Move).y

  'There are forbidden borders
  If ((x1 = 0) Or (y1 = 0)) Then Exit Sub

  'Is the target address occupied?
  Range(Cells(x1, y1), Cells(x1, y1)).Select
      With Selection.Interior
       Color = .ColorIndex
      End With

  Select Case Color
    Case 0, 2, 21 'do nothing: the cell is occupied.
    Case Else 'the cell is empty, is it disruptive?
      If Not (Disruptive(m, A)) Then Call MoveOn(m, A)
   End Select
End Sub

Private Sub Reproduction()
 'No reproduction as yet.

End Sub

Private Sub Death()
'No death.
End Sub

Public Sub asp11()

 'Evolution on its first trial.
 'Body structure is automatically conserved.
 'The task is to climb over the display.

Dim m As Integer
Dim i, j, l, n
Dim A As Integer
Dim NClones As Integer
Dim Gen, NGen As Integer
 Call Moves
 Call Display
 'Initialization
 NClones = 20

 'To generate new random sequences
 'change the number below.

 Randomize (3)

'A is the index of the asp.

'Initialization
  For A = 1 To NClones
   Call InitialPosition(A)
   Call DrawBody(A)
    Call InitialChromosome(A)
  Next A


'l is the counter of days of life.
'Life ends with reproduction and dead.
'NGen is the number of generations
'of the whole experiment.
'm is the number of the link


NGen = 1
LifeSpan = 30
For Gen = 1 To NGen
 For l = 1 To LifeSpan
  For m = 1 To 24
   For A = 1 To NClones
    Call Ribosome(m, A)
    'Call ActualChromosome(m, A)
   Next A
  Next m
 Next l
 Reproduction
 Death
 Next Gen
End Sub

\end{verbatim}

\section{The evolvable asp}

 \bigskip
\textbf{\thenum. }  \addtocounter{num}{1} The code that is viable and is not disruptive is the following:

\begin{verbatim}

Option Explicit

Type LinkAddresses
 x As Integer
 y As Integer
End Type

Type Moves
 x As Integer
 y As Integer
End Type

Type Exons
 Link As Integer
 Move As Integer
End Type

Type asps
'Each chromosome has 24 instructions
 Chromosome(1 To 25) As Exons
 LinkAddress(1 To 12) As LinkAddresses
 Color As Integer
End Type

Dim asp(1 To 20) As asps
Dim VectMoves(1 To 9) As Moves
Dim LifeSpan As Integer
Dim InitialRow



Private Sub Moves()
VectMoves(1).x = 0
VectMoves(1).y = 1
VectMoves(2).x = 1
VectMoves(2).y = 1
VectMoves(3).x = 1
VectMoves(3).y = 0
VectMoves(4).x = 1
VectMoves(4).y = -1
VectMoves(5).x = 0
VectMoves(5).y = -1
VectMoves(6).x = -1
VectMoves(6).y = -1
VectMoves(7).x = -1
VectMoves(7).y = 0
VectMoves(8).x = -1
VectMoves(8).y = 1
VectMoves(9).x = 0
VectMoves(9).y = 0
End Sub


Private Sub Display()
Dim i, j As Integer
'Prepares the screen
'Excel already works with this "new" style!
'To be sure, Excel works with the OOP style!


 Workbooks.Add
    Range("A1:IH100").Select
    Selection.RowHeight = 3
    Selection.ColumnWidth = 0.3
    ActiveWindow.WindowState = xlMaximized

  'Paint the background in white
  'This is necessary because It is not at all clear
  'what is the default background color.
  'The color of a cell will be taken as
  'the indicative of occupation.

  For i = 1 To 110
   For j = 1 To 200
    Range(Cells(i, j), Cells(i, j)).Select
      With Selection.Interior
       .ColorIndex = 2
      End With
   Next j
  Next i
    Range("G9:T9").Select
End Sub

Private Sub InitialPosition(A As Integer)
Dim l, k
 'The index A distinguishes each asp.
 'Each asp has 12 links,
 'each part has its own coordinates on the screen,
 'which are kept in the LinkAddress   array.
 'The asps are posited at the bottom of the screen.
 'Head are at row InitialRow
 InitialRow = 50
For l = 1 To 12
 k = InitialRow - 12
 asp(A).LinkAddress(l).x = 8 * A
 asp(A).LinkAddress(l).y = k + l
 Next l
End Sub

Private Function Fcolor(A As Integer)
 'Color is a number less than 40
 'but different than 2.
Rebegin:  Fcolor = Int(40 * Rnd)
 If Fcolor = 2 Then GoTo Rebegin
  If Fcolor = 21 Then GoTo Rebegin
  If Fcolor = 0 Then GoTo Rebegin
 'MsgBox "Color  " & Fcolor
End Function

Private Sub DrawBody(A As Integer)
Dim l, i, j, h, Delay
 'Call Display
 'Call InitialPosition
 'Paints the body
  asp(A).Color = Fcolor(A)
  For l = 1 To 12
  j = asp(A).LinkAddress(l).x
  i = asp(A).LinkAddress(l).y
  'MsgBox "l " & l
  'MsgBox "i " & i
  'MsgBox "j " & j
   Range(Cells(i, j), Cells(i, j)).Select
   With Selection.Interior
        .ColorIndex = asp(A).Color
        .Pattern = xlSolid
     End With
  Next l
  'If A = 15 Then MsgBox "Color " & asp(A).Color
End Sub

Private Sub InitialChromosome(A As Integer)
Dim m, n, t
Dim x, y As Integer
Dim x1, y1 As Integer

  'Select at random  which link to move
'Throw a number from 1 to 12
For m = 1 To 24
 n = Int(12 * Rnd) + 1
  asp(A).Chromosome(m).Link = n

 'Select at random which direction to move on
 'Throw a number from 1 to 9
  asp(A).Chromosome(m).Move = Int(9 * Rnd) + 1
Next m

End Sub

Private Sub MoveOn(m As Integer, A As Integer)
Dim n, i, j As Integer
Dim Delay, h As Integer
Dim Move As Integer
 'To move a link:
 'the cell occupied by  the link is cleared,
 'the cell to be occupied is painted.

 'The occupied cell is cleared
  n = asp(A).Chromosome(m).Link
  j = asp(A).LinkAddress(n).x
  i = asp(A).LinkAddress(n).y
  Range(Cells(i, j), Cells(i, j)).Select
  Selection.Clear

  'The target addres is memorized
  Move = asp(A).Chromosome(m).Move
  j = j + VectMoves(Move).x
  i = i - VectMoves(Move).y
  asp(A).LinkAddress(n).x = j
  asp(A).LinkAddress(n).y = i

  'The target cell is painted
   Range(Cells(i, j), Cells(i, j)).Select


   'If movement is too fast, nothing can be seen.
   'To attune the velocity of the asp,
   'augment or diminish the delay as   pleased

   Delay = 500
   For h = 1 To Delay
      With Selection.Interior
        .ColorIndex = asp(A).Color
        .Pattern = xlSolid
     End With
   Next h
End Sub

Private Sub ActualChromosome(m As Integer, A As Integer)
 'We have no evolution as yet
Call InitialChromosome(A)
End Sub

 Private Function Disruptive(m As Integer, A As Integer)
'A move is disruptive if the target address
'is farther than 1.5 units from the neighboring links.
Dim n As Integer
Dim Move As Integer
Dim x, y, x1, y1, x2, x3, y2, y3 As Integer
Dim d1, d2 As Single

  'Reads the link to be moved
  n = asp(A).Chromosome(m).Link
  'Reads the   address of the link
  y = asp(A).LinkAddress(n).x
  x = asp(A).LinkAddress(n).y
  'The (integer) move to be done is read
  Move = asp(A).Chromosome(m).Move

 'The target address is calculated
  y1 = y + VectMoves(Move).x
  x1 = x - VectMoves(Move).y
  'The addresses of the two neighboring links is read
  'And disruption is calculated
    'MsgBox "n " & n
  If ((1 < n) And (n < 12)) Then
    y2 = asp(A).LinkAddress(n - 1).x
    x2 = asp(A).LinkAddress(n - 1).y
    d1 = ((x2 - x1) ^ 2 + (y2 - y1) ^ 2) ^ 0.5
    y3 = asp(A).LinkAddress(n + 1).x
    x3 = asp(A).LinkAddress(n + 1).y
    d2 = ((x3 - x1) ^ 2 + (y3 - y1) ^ 2) ^ 0.5

    'The disruption bug is fixed here.

     If ((d1 > 1.5) Or (d2 > 1.5)) Then
       Disruptive = True
      Else
       Disruptive = False
     End If
  End If

    'Debugging messages
    'MsgBox "d1 " & d1
    'MsgBox "d2 " & d2

  If n = 1 Then
    y2 = asp(A).LinkAddress(2).x
    x2 = asp(A).LinkAddress(2).y
    d1 = ((x2 - x1) ^ 2 + (y2 - y1) ^ 2) ^ 0.5
If d1 > 1.1 Then Disruptive = True Else Disruptive = False
   End If

  If n = 12 Then
    y2 = asp(A).LinkAddress(11).x
    x2 = asp(A).LinkAddress(11).y
    d2 = ((x2 - x1) ^ 2 + (y2 - y1) ^ 2) ^ 0.5
If d2 > 1.1 Then Disruptive = True Else Disruptive = False
    End If


 End Function



Private Sub Ribosome(m As Integer, A As Integer)
Dim n, i, j As Integer
Dim Delay, h As Integer
Dim Move As Integer
Dim x, y, x1, y1 As Integer
Dim Color As Integer
  'Reads the link to be moved
  n = asp(A).Chromosome(m).Link
  'Reads the   address of the link
  y = asp(A).LinkAddress(n).x
  x = asp(A).LinkAddress(n).y
  'The (integer) move to be done is read
  Move = asp(A).Chromosome(m).Move

 'The target address is calculated
  y1 = y + VectMoves(Move).x
  x1 = x - VectMoves(Move).y

  'There are forbidden borders
  If ((x1 = 0) Or (y1 = 0)) Then Exit Sub

  'Get the  color of the target address
  Range(Cells(x1, y1), Cells(x1, y1)).Select
      With Selection.Interior
       Color = .ColorIndex
      End With

  'Is the target address occupied?
  'MsgBox "Color " & Color
  'In previous codes, there was here a bug.

  Select Case Color
    Case 2   'the cell is empty, is it disruptive?
         If Not (Disruptive(m, A)) Then Call MoveOn(m, A)
    Case Else 'do nothing: the cell is occupied.

   End Select
End Sub

Private Sub Reproduction()
 'No reproduction as yet.
End Sub

Private Sub Death()
'No death.
End Sub

Public Sub asp10()

 'Evolution on its first trial.
 'Body structure is automatically conserved.
 'The task is to climb over the display.

Dim m As Integer
Dim i, j, l, n
Dim A As Integer
Dim NClones As Integer
Dim Gen, NGen As Integer
 Call Moves
 Call Display
 'Initialization
 NClones = 20

 'To generate new random sequences
 'change the number below.

 Randomize (3)

'A is the index of the asp.

'Initialization
  For A = 1 To NClones
   Call InitialPosition(A)
   Call DrawBody(A)
    Call InitialChromosome(A)
  Next A


'l is the counter of days of life.
'Life ends with reproduction and dead.
'NGen is the number of generations
'of the whole experiment.
'm is the number of the link


NGen = 1
LifeSpan = 30
For Gen = 1 To NGen
 For l = 1 To LifeSpan
  For m = 1 To 24
   For A = 1 To NClones
    Call Ribosome(m, A)
    'Call ActualChromosome(m, A)
   Next A
  Next m
 Next l
 Reproduction
 Death
 Next Gen
End Sub



\end{verbatim}

 \bigskip
\textbf{\thenum. }  \addtocounter{num}{1} Exercise. Run the code and verify that is not disruptive. Witness that by mere randomness some asps began to rise up their tails.

 \bigskip
\textbf{\thenum. }  \addtocounter{num}{1} To raise up the tail is the first task that is necessary to advance upwards: that  is the beginning of an upward movement. In official  jargon, it is said that the potentialities of self organization  of the evolutionary environment caused a fluctuation that eventually could be used to trigger evolution. What is then evolution? it  is the linking of the potentialities of self organization to the reproductive cycle in such a way that those individuals that have certain property with more emphasis are allowed to be reproduced more abundantly than the rest. In that way, the property is extended over the population and is carried over to the extreme.

 \bigskip
\textbf{\thenum. }  \addtocounter{num}{1} Let us implement the evolutionary race: we have already observed that a population of 20 asps is large enough for the appearance of incipient self organizing properties, which in our case is the organization of genomes      causing   their carrier asps to raise up the tail.

We have two tasks: to determine which asps have begun to rise up the tail and to reproduce them more than the rest. We need to include a mechanism of mutation to guarantee that the property of climbing could extend in the population. Let us see that in the next code.

\begin{verbatim}

Option Explicit

Type LinkAddresses
 x As Integer
 y As Integer
End Type

Type Moves
 x As Integer
 y As Integer
End Type

Type Exons
 Link As Integer
 Move As Integer
End Type

Type asps
'Each chromosome has 24 instructions
 Chromosome(1 To 25) As Exons
 LinkAddress(1 To 12) As LinkAddresses
 Color As Integer
End Type

Dim asp(1 To 20) As asps
Dim asp2(1 To 20) As asps
Dim VectMoves(1 To 9) As Moves
Dim LifeSpan As Integer
Dim InitialRow
Dim Order(1 To 20) As Integer






Private Sub Moves()
VectMoves(1).x = 0
VectMoves(1).y = 1
VectMoves(2).x = 1
VectMoves(2).y = 1
VectMoves(3).x = 1
VectMoves(3).y = 0
VectMoves(4).x = 1
VectMoves(4).y = -1
VectMoves(5).x = 0
VectMoves(5).y = -1
VectMoves(6).x = -1
VectMoves(6).y = -1
VectMoves(7).x = -1
VectMoves(7).y = 0
VectMoves(8).x = -1
VectMoves(8).y = 1
VectMoves(9).x = 0
VectMoves(9).y = 0
End Sub


Private Sub Display()
Dim i, j As Integer
'Prepares the screen
'Excel already works with this "new" style!
'To be sure, Excel works with the OOP style!


 Workbooks.add
    Range("A1:IH100").Select
    Selection.RowHeight = 3
    Selection.ColumnWidth = 0.3
    ActiveWindow.WindowState = xlMaximized

  'Paint the background in white
  'This is necessary because It is not at all clear
  'what is the default background color.
  'The color of a cell will be taken as
  'the indicative of occupation.

  For i = 1 To 103
   For j = 1 To 200
    Range(Cells(i, j), Cells(i, j)).Select
      With Selection.Interior
       .ColorIndex = 2
      End With
   Next j
  Next i
    Range("G9:T9").Select
End Sub

Private Sub InitialPosition(A As Integer)
Dim l, k
 'The index A distinguishes each asp.
 'Each asp has 12 links,
 'each part has its own coordinates on the screen,
 'which are kept in the LinkAddress   array.
 'The asps are posited at the bottom of the screen.
 'Head are at row InitialRow
 InitialRow = 100
For l = 1 To 12
 k = InitialRow - 12
 asp(A).LinkAddress(l).x = 8 * A
 asp(A).LinkAddress(l).y = k + l
 Next l
End Sub

Private Function Fcolor(A As Integer)
 'Color is a number less than 40
 'but different than 2.
Rebegin:  Fcolor = Int(40 * Rnd)
 If Fcolor = 2 Then GoTo Rebegin
  If Fcolor = 21 Then GoTo Rebegin
  If Fcolor = 0 Then GoTo Rebegin
 'MsgBox "Color  " & Fcolor
End Function

Private Sub DrawBody(A As Integer)
Dim l, i, j, h, Delay
 'Call Display
 'Call InitialPosition
 'Paints the body
  asp(A).Color = Fcolor(A)
  For l = 1 To 12
  j = asp(A).LinkAddress(l).x
  i = asp(A).LinkAddress(l).y
  'MsgBox "l " & l
  'MsgBox "i " & i
  'MsgBox "j " & j
   Range(Cells(i, j), Cells(i, j)).Select
   With Selection.Interior
        .ColorIndex = asp(A).Color
        .Pattern = xlSolid
     End With
  Next l
  'If A = 15 Then MsgBox "Color " & asp(A).Color
End Sub

Private Sub InitialChromosome(A As Integer)
Dim m, n, t
Dim x, y As Integer
Dim x1, y1 As Integer

  'Select at random  which link to move
'Throw a number from 1 to 12
For m = 1 To 24
 n = Int(12 * Rnd) + 1
  asp(A).Chromosome(m).Link = n

 'Select at random which direction to move on
 'Throw a number from 1 to 9
  asp(A).Chromosome(m).Move = Int(9 * Rnd) + 1
Next m

End Sub

Private Sub MoveOn(m As Integer, A As Integer)
Dim n, i, j As Integer
Dim Delay, h As Integer
Dim Move As Integer
 'To move a link:
 'the cell occupied by  the link is cleared,
 'the cell to be occupied is painted.

 'The occupied cell is cleared
  n = asp(A).Chromosome(m).Link
  j = asp(A).LinkAddress(n).x
  i = asp(A).LinkAddress(n).y
  Range(Cells(i, j), Cells(i, j)).Select
  Selection.Clear

  'The target addres is memorized
  Move = asp(A).Chromosome(m).Move
  j = j + VectMoves(Move).x
  i = i - VectMoves(Move).y
  asp(A).LinkAddress(n).x = j
  asp(A).LinkAddress(n).y = i

  'The target cell is painted
   Range(Cells(i, j), Cells(i, j)).Select


   'If movement is too fast, nothing can be seen.
   'To attune the velocity of the asp,
   'augment or diminish the delay as   pleased

   Delay = 1
   For h = 1 To Delay
      With Selection.Interior
        .ColorIndex = asp(A).Color
        .Pattern = xlSolid
     End With
   Next h
End Sub

Private Sub ActualChromosome(m As Integer, A As Integer)
 'We have no evolution as yet
Call InitialChromosome(A)
End Sub

 Private Function Disruptive(m As Integer, A As Integer)
'A move is disruptive if the target address
'is farther than 1.5 units from the neighboring links.
Dim n As Integer
Dim Move As Integer
Dim x, y, x1, y1, x2, x3, y2, y3 As Integer
Dim d1, d2 As Single

  'Reads the link to be moved
  n = asp(A).Chromosome(m).Link
  'Reads the   address of the link
  y = asp(A).LinkAddress(n).x
  x = asp(A).LinkAddress(n).y
  'The (integer) move to be done is read
  Move = asp(A).Chromosome(m).Move

 'The target address is calculated
  y1 = y + VectMoves(Move).x
  x1 = x - VectMoves(Move).y
  'The addresses of the two neighboring links is read
  'And disruption is calculated
    'MsgBox "n " & n
  If ((1 < n) And (n < 12)) Then
    y2 = asp(A).LinkAddress(n - 1).x
    x2 = asp(A).LinkAddress(n - 1).y
    d1 = ((x2 - x1) ^ 2 + (y2 - y1) ^ 2) ^ 0.5
    y3 = asp(A).LinkAddress(n + 1).x
    x3 = asp(A).LinkAddress(n + 1).y
    d2 = ((x3 - x1) ^ 2 + (y3 - y1) ^ 2) ^ 0.5

    'The disruption bug is fixed here.

    If ((d1 > 1.5) Or (d2 > 1.5)) Then
       Disruptive = True
      Else
       Disruptive = False
     End If
  End If

    'Debugging messages
    'MsgBox "d1 " & d1
    'MsgBox "d2 " & d2

  If n = 1 Then
    y2 = asp(A).LinkAddress(2).x
    x2 = asp(A).LinkAddress(2).y
    d1 = ((x2 - x1) ^ 2 + (y2 - y1) ^ 2) ^ 0.5
 If d1 > 1.1 Then Disruptive = True Else Disruptive = False
   End If

  If n = 12 Then
    y2 = asp(A).LinkAddress(11).x
    x2 = asp(A).LinkAddress(11).y
    d2 = ((x2 - x1) ^ 2 + (y2 - y1) ^ 2) ^ 0.5
If d2 > 1.1 Then Disruptive = True Else Disruptive = False
  End If

 End Function



Private Sub Ribosome(m As Integer, A As Integer)
Dim n, i, j As Integer
Dim Delay, h As Integer
Dim Move As Integer
Dim x, y, x1, y1 As Integer
Dim Color As Integer
  'Reads the link to be moved
  n = asp(A).Chromosome(m).Link
  'Reads the   address of the link
  y = asp(A).LinkAddress(n).x
  x = asp(A).LinkAddress(n).y
  'The (integer) move to be done is read
  Move = asp(A).Chromosome(m).Move

 'The target address is calculated
  y1 = y + VectMoves(Move).x
  x1 = x - VectMoves(Move).y

  'There are forbidden borders
  If ((x1 = 0) Or (y1 = 0)) Then Exit Sub

  'Get the  color of the target address
  Range(Cells(x1, y1), Cells(x1, y1)).Select
      With Selection.Interior
       Color = .ColorIndex
      End With

  'Is the target address occupied?
  'MsgBox "Color " & Color
  'In previous codes, there was  a bug here.

  Select Case Color
    Case 2   'the cell is empty, is it disruptive?
         If Not (Disruptive(m, A)) Then Call MoveOn(m, A)
    Case Else 'do nothing: the cell is occupied.

   End Select
End Sub

Private Sub AssessingFitness()
Dim A, B, m, Champ As Integer
Dim Max, Man As Integer
Dim Fitness(1 To 20) As Integer

'The fitness measures the vertical distance from
'actual position to the 100th row.

 For A = 1 To 20
 Max = 1
  For m = 1 To 12
   Man = asp(A).LinkAddress(m).y
   'MsgBox "vertical" & Man
   If Man > Max Then Max = Man
  Next m
   Fitness(A) = 100 - Max
   'MsgBox "Fitness " & Fitness(A)
 Next A

'Now, we sort asps by fitness

For A = 1 To 4
 Champ = 1
 For B = 1 To 20
  If Fitness(B) > Fitness(Champ) Then Champ = B
  Next B
  Order(A) = Champ
  Fitness(Champ) = 0
   'MsgBox "Next champ " & Champ
Next A
End Sub

Private Sub Mutation()
 Dim A, m, n, l As Integer
 'Let us try out blind point mutations
 'in specific exons
 'in the order of links to be moved
 'and in the direction of movement.
 For A = 1 To 20
 'Select at random  which exon to mutate:
'throw a number from 1 to 24
 m = Int(24 * Rnd) + 1
'Select at random which link to move:
'throw a number from 1 to 12
 n = Int(24 * Rnd) + 1
  asp(A).Chromosome(m).Link = n
 'Select at random which direction to move on
 'Throw a number from 1 to 9
  asp(A).Chromosome(m).Move = Int(9 * Rnd) + 1
 Next A

End Sub

Private Sub Reproduction()
Dim A, m, l As Integer
'There are always 20 asps.
'Our reproductive scheme is as follows:
'10 offspring for the champion,
'5 offspring for the second,
'3 for the third,
'2 for the fourth.
'We place children on order.
'Children are clones of the fathers
'that are copied to the asp2 array.

'Call InitialPosition(8)
'MsgBox "var1 " & asp(8).LinkAddress(7).x
'MsgBox "var2 " & asp2(8).LinkAddress(7).x
 'asp2(8) = asp(8)
 'MsgBox "var2n " & asp2(8).LinkAddress(7).x

 For A = 1 To 20
 Select Case A
   Case 1, 2, 3, 4, 5, 6, 7, 8, 9, 10
     asp2(A) = asp(Order(1))
   Case 11, 12, 13, 14, 15
     asp2(A) = asp(Order(2))
   Case 16, 17, 18
     asp2(A) = asp(Order(3))
  Case 19, 20
     asp2(A) = asp(Order(4))
 End Select
Next A

'Clones are subjected to mutation
Call Mutation
End Sub

Private Sub Death()
Dim A As Integer
'The asp array is overwrited with
'the asp2 array.
For A = 1 To 20
 asp(A) = asp2(A)
Next A
End Sub

Public Sub asp9()
 'Evolution on its first trial.
 'Body structure is automatically conserved.
 'The task is to climb over the display.

Dim m As Integer
Dim i, j, l, n
Dim A As Integer
Dim Nasps As Integer
Dim Gen, NGen As Integer
 Call Moves
 Call Display
 'Initialization

'Nasps is the number of asps,
'Default value = 20.
'NGen is the number of generations
'of the whole generation.
'LifeSpan is the number of   days
'in a life.
'Life ends with reproduction and dead.

 Nasps = 20
 NGen = 10
LifeSpan = 15

 'To generate new random sequences
 'change the number below.

 Randomize (3)

'A is the index of the asp.

'Initialization
  For A = 1 To Nasps
   Call InitialPosition(A)
   Call DrawBody(A)
    Call InitialChromosome(A)
  Next A


'm is the number of the link
'l is the counter of days of life.


For Gen = 1 To NGen
 For l = 1 To LifeSpan
  For m = 1 To 24
   For A = 1 To Nasps
    Call Ribosome(m, A)
   Next A
  Next m
 Next l
 AssessingFitness
 Reproduction
 Mutation
 Death
 Next Gen
End Sub


\end{verbatim}

 \bigskip
\textbf{\thenum. }  \addtocounter{num}{1} Exercise. Run the code and verify that it is a failure: it was designed to execute reproduction in proportion to fitness, so, one expects that the color of the asps remain the same for ever and ever. Thus, something happens that we need to disentangle.

 \bigskip
\textbf{\thenum. }  \addtocounter{num}{1} At the main time, it has become clear that we are already dealing with complex programs. This is evidenced because the debugging problems have become very heavy. An additional problem is that the VBE is executed in interpretation mode, so some errors are evidenced only when they are produced late in the running process. Let us clear this complaint  a bit: there are many forms as one can execute a program. One is to read and execute it line per line: this is the interpreter mode. The other is to execute the program as a whole only after compilation, which is the translation of the code, written in a   humanlike language, into machine language, which is specially suited to deal directly with the electronic processor. A compiled program runs faster than the original code because translation of the code to machine language takes time which is by no means insignificant  for code with many loops.

The interpretation mode of running a program allows one to make corrections them just on the way. In compilation mode that is not possible. Instead: 1)  One cuts the code in many parts by its natural divisions, subs and functions 2) One compiles the code, if it has errors of syntax the compiler express them in a list, one corrects the code and so on until the code produces an executable file in machine language. 3) If one decides to change the code, the compiler only recompiles the portion of the code that has been changed. The rest is untouched and in this way no lack of precious time occurs.

Because it is not always clear which mode to choose, some languages allow one to run in any of the modes. Anyway, language $C^{++}$, which is the default language for very complex programming, runs only in compilation mode. Nevertheless, $C^{++}$ is surrounded by an impenetrable  mysticism. Strange as it could be, Java, the beloved son of $C{++}$ seems to be more popular, specially among youngsters whose life has danced  among  Internet communications.

For the present level of complexity, there is no substantial difference among our Excel-VBE, C and Java. This means that one can get requalified very easily if it is needed. Nevertheless, our approach to programming with Excel-VBE has dispensed us from drawing and disk management problems, which are not simple at all. So, it is important to answer to the question: Is it necessary for us to get requilified to $C^{++}$ or to Java? The reality is simple: all experts in computer science are multilingual, they know VBA, C,  $C{++}$ in various dialects, Fortran 77, Pascal, ADA, Java, List, Small Talk, Pear  and, moreover, they can learn a language from one day to the next. Thus, if you are planning to get an expert, you have not lost the time: you already have a good knowledge of a very modern and powerful language: VBE for Excel. In general, it could be a good idea to be ready to acquire more understanding of other languages, specially because there are many wonderful programs of interest to a bio-simulator  that are written in other languages, specially C.

Let us try to learn, at the meantime,  more about VBE. So,  let us return back to the debugging of our last code.

 \bigskip
\textbf{\thenum. }  \addtocounter{num}{1} We devised a program to simulate evolution, including reproduction with mutation and in proportion to fitness. We debugged  the program from syntactic errors until it run, but it did not worked. How did we know that the program failed? Because, the color of the offspring was programmed to be the very same color of the champion fathers and so, after the first generation a change of color should had happened but it did not.

We do not know where the error could be, so let us begin by adding the following changes: 1) Let us arrange a bit the borders of our reticulate because in the present state the eyes suffer when the program is executed. This implies  that our world has barriers in all its four directions. 2) We need to report the number of the generation. 3) Each new generation of asps shall begin to climb the screen right from the bottom. 3)The procedures responding for evolution must have some effect no matter which.   The corresponding  code, after debugging of syntactic errors,  follows:

\begin{verbatim}

Option Explicit

Type LinkAddresses
 x As Integer
 y As Integer
End Type

Type Moves
 x As Integer
 y As Integer
End Type

Type Exons
 Link As Integer
 Move As Integer
End Type

Type asps
'Each chromosome has 24 instructions
 Chromosome(1 To 25) As Exons
 LinkAddress(1 To 12) As LinkAddresses
 Color As Integer
End Type

Dim Nasps As Integer 'Less than 20
Dim asp(1 To 20) As asps
Dim asp2(1 To 20) As asps
Dim VectMoves(1 To 9) As Moves
Dim LifeSpan As Integer
Dim InitialRow
Dim Order(1 To 20) As Integer
Dim Flag As Integer







Private Sub Moves()
VectMoves(1).x = 0
VectMoves(1).y = 1
VectMoves(2).x = 1
VectMoves(2).y = 1
VectMoves(3).x = 1
VectMoves(3).y = 0
VectMoves(4).x = 1
VectMoves(4).y = -1
VectMoves(5).x = 0
VectMoves(5).y = -1
VectMoves(6).x = -1
VectMoves(6).y = -1
VectMoves(7).x = -1
VectMoves(7).y = 0
VectMoves(8).x = -1
VectMoves(8).y = 1
VectMoves(9).x = 0
VectMoves(9).y = 0
End Sub


Private Sub Display()
Dim i, j As Integer
'Prepares the screen
'Excel already works with this "new" style!
'To be sure, Excel works with the OOP style!


 Workbooks.add
    Range(Cells(1, 1), Cells(103, 170)).Select
    Selection.RowHeight = 3
    Selection.ColumnWidth = 0.3
    ActiveWindow.WindowState = xlMaximized

  'Paint the background in white
  'This is necessary because It is not at all clear
  'what is the default background color.
  'The color of a cell will be taken as
  'the indicative of occupation.

  For i = 1 To 103
   For j = 1 To 170
    Range(Cells(i, j), Cells(i, j)).Select
      With Selection.Interior
       .ColorIndex = 2
      End With
   Next j
  Next i
    Range("G9:T9").Select
End Sub

Private Sub InitialPosition(A As Integer)
Dim l, k
 'The index A distinguishes each asp.
 'Each asp has 12 links,
 'each part has its own coordinates on the screen,
 'which are kept in the LinkAddress   array.
 'The asps are posited at the bottom of the screen.
 'Head are at row InitialRow
 InitialRow = 100
For l = 1 To 12
 k = InitialRow - 12
 asp(A).LinkAddress(l).x = 8 * A
 asp(A).LinkAddress(l).y = k + l
 Next l
End Sub

Private Function Fcolor(A As Integer)
 'Color is a number less than 40
 'but different than 2.
Rebegin:  Fcolor = Int(40 * Rnd)
 If Fcolor = 2 Then GoTo Rebegin
 'MsgBox "Color  " & Fcolor
End Function

Private Sub DrawBody(A As Integer, Flag As Integer)
Dim l, i, j, h, Delay
 'Call Display
 'Call InitialPosition
 'Paints the body
 'Flag = 1 for initialization at random
 'Flag = 0 for reinitialization by inheritance.
  If Flag = 1 Then asp(A).Color = Fcolor(A)
  For l = 1 To 12
  j = asp(A).LinkAddress(l).x
  i = asp(A).LinkAddress(l).y
  'MsgBox "l " & l
  'MsgBox "i " & i
  'MsgBox "j " & j
   Range(Cells(i, j), Cells(i, j)).Select
   With Selection.Interior
        .ColorIndex = asp(A).Color
        .Pattern = xlSolid
     End With
  Next l
  'If A = 15 Then MsgBox "Color " & asp(A).Color
End Sub

Private Sub InitialChromosome(A As Integer)
Dim m, n, t
Dim x, y As Integer
Dim x1, y1 As Integer

  'Select at random  which link to move
'Throw a number from 1 to 12
For m = 1 To 24
 n = Int(12 * Rnd) + 1
  asp(A).Chromosome(m).Link = n

 'Select at random which direction to move on
 'Throw a number from 1 to 9
  asp(A).Chromosome(m).Move = Int(9 * Rnd) + 1
Next m

End Sub

Private Sub MoveOn(m As Integer, A As Integer)
Dim n, i, j As Integer
Dim Delay, h As Integer
Dim Move As Integer
 'To move a link:
 'the cell occupied by  the link is cleared,
 'the cell to be occupied is painted.

 'The occupied cell is cleared
  n = asp(A).Chromosome(m).Link
  j = asp(A).LinkAddress(n).x
  i = asp(A).LinkAddress(n).y
  Range(Cells(i, j), Cells(i, j)).Select
  Selection.Clear

  'The target addres is memorized
  Move = asp(A).Chromosome(m).Move
  j = j + VectMoves(Move).x
  i = i - VectMoves(Move).y
  asp(A).LinkAddress(n).x = j
  asp(A).LinkAddress(n).y = i

  'The target cell is painted
   Range(Cells(i, j), Cells(i, j)).Select


   'If movement is too fast, nothing can be seen.
   'To attune the velocity of the asp,
   'augment or diminish the delay as   pleased

   Delay = 1
   For h = 1 To Delay
      With Selection.Interior
        .ColorIndex = asp(A).Color
        .Pattern = xlSolid
     End With
   Next h
End Sub

Private Sub ActualChromosome(m As Integer, A As Integer)
 'We have no evolution as yet
Call InitialChromosome(A)
End Sub

 Private Function Disruptive(m As Integer, A As Integer)
'A move is disruptive if the target address
'is farther than 1.5 units from the neighboring links.
Dim n As Integer
Dim Move As Integer
Dim x, y, x1, y1, x2, x3, y2, y3 As Integer
Dim d1, d2 As Single

  'Reads the link to be moved
  n = asp(A).Chromosome(m).Link
  'Reads the   address of the link
  y = asp(A).LinkAddress(n).x
  x = asp(A).LinkAddress(n).y
  'The (integer) move to be done is read
  Move = asp(A).Chromosome(m).Move

 'The target address is calculated
  y1 = y + VectMoves(Move).x
  x1 = x - VectMoves(Move).y
  'The addresses of the two neighboring links is read
  'And disruption is calculated
    'MsgBox "n " & n
  If ((1 < n) And (n < 12)) Then
    y2 = asp(A).LinkAddress(n - 1).x
    x2 = asp(A).LinkAddress(n - 1).y
    d1 = ((x2 - x1) ^ 2 + (y2 - y1) ^ 2) ^ 0.5
    y3 = asp(A).LinkAddress(n + 1).x
    x3 = asp(A).LinkAddress(n + 1).y
    d2 = ((x3 - x1) ^ 2 + (y3 - y1) ^ 2) ^ 0.5

    'The disruption bug is fixed here.

    If ((d1 > 1.5) Or (d2 > 1.5)) Then
       Disruptive = True
      Else
       Disruptive = False
     End If
  End If

    'Debugging messages
    'MsgBox "d1 " & d1
    'MsgBox "d2 " & d2

  If n = 1 Then
    y2 = asp(A).LinkAddress(2).x
    x2 = asp(A).LinkAddress(2).y
    d1 = ((x2 - x1) ^ 2 + (y2 - y1) ^ 2) ^ 0.5
If d1 > 1.1 Then Disruptive = True Else Disruptive = False
   End If

  If n = 12 Then
    y2 = asp(A).LinkAddress(11).x
    x2 = asp(A).LinkAddress(11).y
    d2 = ((x2 - x1) ^ 2 + (y2 - y1) ^ 2) ^ 0.5
If d2 > 1.1 Then Disruptive = True Else Disruptive = False
  End If

 End Function



Private Sub Ribosome(m As Integer, A As Integer)
Dim n, i, j As Integer
Dim Delay, h As Integer
Dim Move As Integer
Dim x, y, x1, y1 As Integer
Dim Color As Integer
  'Reads the link to be moved
  n = asp(A).Chromosome(m).Link
  'Reads the   address of the link
  y = asp(A).LinkAddress(n).x
  x = asp(A).LinkAddress(n).y
  'The (integer) move to be done is read
  Move = asp(A).Chromosome(m).Move

 'The target address is calculated
  y1 = y + VectMoves(Move).x
  x1 = x - VectMoves(Move).y

  'There are forbidden borders on all directions
  If ((x1 < 1) Or (y1 < 1) Or (x1 > 103) Or (y1 > 168)) Then
   Exit Sub
  End If
  'Get the  color of the target address
  Range(Cells(x1, y1), Cells(x1, y1)).Select
      With Selection.Interior
       Color = .ColorIndex
      End With

  'Is the target address occupied?
  'MsgBox "Color " & Color
  'In previous codes, there was  a bug here.

  Select Case Color
    Case 2   'the cell is empty, is it disruptive?
         If Not (Disruptive(m, A)) Then Call MoveOn(m, A)
    Case Else 'do nothing: the cell is occupied.

   End Select
End Sub

Private Sub AssessingFitness()
Dim A, B, m, Champ As Integer
Dim Max, Man As Integer
Dim Fitness(1 To 20) As Integer

'The fitness measures the vertical distance from
'actual position to the 100th row.

 For A = 1 To Nasps
 Max = 1
  For m = 1 To 12
   Man = asp(A).LinkAddress(m).y
   'MsgBox "vertical" & Man
   If Man > Max Then Max = Man
  Next m
   Fitness(A) = 100 - Max
   'MsgBox "Fitness " & Fitness(A)
 Next A

'Now, we sort asps by fitness
'and select the best four,
'whose numbers are recorded in Order.
For A = 1 To 4
 Champ = 1
 For B = 1 To 20
  If Fitness(B) > Fitness(Champ) Then Champ = B
  Next B
  Order(A) = Champ
  Fitness(Champ) = 0
   'MsgBox "Next champ " & Champ
Next A
End Sub

Private Sub Mutation()
 Dim A, m, n, l As Integer
 'Let us try out blind point mutations
 'in specific exons,
 'in the order of links to be moved
 'and in the direction of movement.
 For A = 1 To Nasps
 'Select at random  which exon to mutate:
'throw a number from 1 to 24
 m = Int(24 * Rnd) + 1
'Select at random which link to move:
'throw a number from 1 to 12
 n = Int(24 * Rnd) + 1
  asp(A).Chromosome(m).Link = n
 'Select at random which direction to move on
 'Throw a number from 1 to 9
  asp(A).Chromosome(m).Move = Int(9 * Rnd) + 1
 Next A

End Sub

Private Sub Reproduction()
Dim A, m, l As Integer
'There are always 20 asps.
'Our reproductive scheme is as follows:
'10 offspring for the champion,
'5 offspring for the second,
'3 for the third,
'2 for the fourth.
'We place children on order.
'Children are clones of the fathers
'that are copied to the asp2 array.

'Call InitialPosition(8)
'MsgBox "var1 " & asp(8).LinkAddress(7).x
'MsgBox "var2 " & asp2(8).LinkAddress(7).x
 'asp2(8) = asp(8)
 'MsgBox "var2n " & asp2(8).LinkAddress(7).x

 For A = 1 To Nasps
 Select Case A
   Case 1, 2, 3, 4, 5, 6, 7, 8, 9, 10
     asp2(A) = asp(Order(1))
   Case 11, 12, 13, 14, 15
     asp2(A) = asp(Order(2))
   Case 16, 17, 18
     asp2(A) = asp(Order(3))
  Case 19, 20
     asp2(A) = asp(Order(4))
 End Select
Next A

'Clones are subjected to mutation
Call Mutation
End Sub


Private Sub Death()
Dim A As Integer
'The asp array is overwritten with
'the asp2 array.
For A = 1 To Nasps
 asp(A) = asp2(A)
Next A
End Sub

Private Sub Initialization()
Dim A As Integer

  For A = 1 To Nasps
   Call InitialPosition(A)
   Call DrawBody(A, Flag)
    Call InitialChromosome(A)
  Next A

End Sub

Private Sub Reinitialization()
Dim A As Integer

'Beginning from the bottom
'with new genes
  For A = 1 To Nasps
   Call InitialPosition(A)
   Call DrawBody(A, Flag)
  Next A
End Sub



Public Sub asp8()

 'Evolution on its first trial.
 'Body structure is automatically conserved.
 'The task is to climb over the display.

Dim m As Integer
Dim i, j, l, n
Dim A As Integer
Dim Gen, NGen As Integer


 Call Moves
 Call Display
 Call Initialization

'Nasps is the number of asps,
'Default value = 20.
'NGen is the number of generations
'of the whole generation.
'LifeSpan is the number of   days
'in a life.
'Life ends with reproduction and dead.

 Nasps = 20
 NGen = 10
 LifeSpan = 15

 'To generate new random sequences
 'change the number below.

 Randomize (3)

'Flag = 1 for initialization at random
 'Flag = 0 for reinitialization by inheritance.
 Flag = 1
 Call Initialization
 Flag = 0
'm is the number of the link
'l is the counter of days of life.
'A is the index of the asp.

For Gen = 1 To NGen
'Reports the Generation
MsgBox "Generation " & Gen

 For l = 1 To LifeSpan
  For m = 1 To 24
   For A = 1 To Nasps
    Call Ribosome(m, A)
   Next A
  Next m
 Next l
 Call AssessingFitness
 Call Reproduction
 Call Mutation
 Call Death
 Call Reinitialization
 Next Gen
End Sub

\end{verbatim}

 \bigskip
\textbf{\thenum. }  \addtocounter{num}{1} Exercise. Run the code and verify that it works much better but still seems to mix old and new generations. Seemingly, it has problems to forget the old generation, which shall disappear without a trace apart from the offspring.

 \bigskip
\textbf{\thenum. }  \addtocounter{num}{1} The next code can reinitialize the population from bottom and moreover  with winner genes of the last generation.

\begin{verbatim}

Option Explicit

Type LinkAddresses
 x As Integer
 y As Integer
End Type

Type Moves
 x As Integer
 y As Integer
End Type

Type Exons
 Link As Integer
 Move As Integer
End Type

Type asps
'Each chromosome has 24 instructions
 Chromosome(1 To 25) As Exons
 LinkAddress(1 To 12) As LinkAddresses
 Color As Integer
End Type

Dim Nasps As Integer 'Less than 20
Dim asp(1 To 20) As asps
Dim asp2(1 To 20) As asps
Dim VectMoves(1 To 9) As Moves
Dim LifeSpan As Integer
Dim InitialRow
Dim Order(1 To 20) As Integer
Dim Flag As Integer







Private Sub Moves()
VectMoves(1).x = 0
VectMoves(1).y = 1
VectMoves(2).x = 1
VectMoves(2).y = 1
VectMoves(3).x = 1
VectMoves(3).y = 0
VectMoves(4).x = 1
VectMoves(4).y = -1
VectMoves(5).x = 0
VectMoves(5).y = -1
VectMoves(6).x = -1
VectMoves(6).y = -1
VectMoves(7).x = -1
VectMoves(7).y = 0
VectMoves(8).x = -1
VectMoves(8).y = 1
VectMoves(9).x = 0
VectMoves(9).y = 0
End Sub


Private Sub Display()
Dim i, j As Integer
'Prepares the screen
'Excel already works with this "new" style!
'To be sure, Excel works with the OOP style!


 Workbooks.add
    Range(Cells(1, 1), Cells(103, 170)).Select
    Selection.RowHeight = 3
    Selection.ColumnWidth = 0.3
    ActiveWindow.WindowState = xlMaximized

  'Paint the background in white
  'This is necessary because It is not at all clear
  'what is the default background color.
  'The color of a cell will be taken as
  'the indicative of occupation.

  For i = 1 To 103
   For j = 1 To 170
    Range(Cells(i, j), Cells(i, j)).Select
      With Selection.Interior
       .ColorIndex = 2
      End With
   Next j
  Next i
    Range("G9:T9").Select
End Sub

Private Sub InitialPosition(A As Integer)
Dim l, k
 'The index A distinguishes each asp.
 'Each asp has 12 links,
 'each part has its own coordinates on the screen,
 'which are kept in the LinkAddress   array.
 'The asps are posited at the bottom of the screen.
 'Head are at row InitialRow
 InitialRow = 100
For l = 1 To 12
 k = InitialRow - 12
 asp(A).LinkAddress(l).x = 8 * A
 asp(A).LinkAddress(l).y = k + l
 Next l
End Sub

Private Function Fcolor(A As Integer)
 'Color is a number less than 40
 'but different than 2.
Rebegin:  Fcolor = Int(40 * Rnd)
 If Fcolor = 2 Then GoTo Rebegin
 'MsgBox "Color  " & Fcolor
End Function

Private Sub DrawBody(A As Integer, Flag As Integer)
Dim l, i, j, h, Delay
 'Call Display
 'Call InitialPosition
 'Paints the body
 'Flag = 1 for initialization at random
 'Flag = 0 for reinitialization by inheritance.
  If Flag = 1 Then asp(A).Color = Fcolor(A)
  For l = 1 To 12
  j = asp(A).LinkAddress(l).x
  i = asp(A).LinkAddress(l).y
  'MsgBox "l " & l
  'MsgBox "i " & i
  'MsgBox "j " & j
   Range(Cells(i, j), Cells(i, j)).Select
   With Selection.Interior
        .ColorIndex = asp(A).Color
        .Pattern = xlSolid
     End With
  Next l
  'If A = 15 Then MsgBox "Color " & asp(A).Color
End Sub

Private Sub InitialChromosome(A As Integer)
Dim m, n, t
Dim x, y As Integer
Dim x1, y1 As Integer

  'Select at random  which link to move
'Throw a number from 1 to 12
For m = 1 To 24
 n = Int(12 * Rnd) + 1
  asp(A).Chromosome(m).Link = n

 'Select at random which direction to move on
 'Throw a number from 1 to 9
  asp(A).Chromosome(m).Move = Int(9 * Rnd) + 1
Next m

End Sub

Private Sub MoveOn(m As Integer, A As Integer)
Dim n, i, j As Integer
Dim Delay, h As Integer
Dim Move As Integer
 'To move a link:
 'the cell occupied by  the link is cleared,
 'the cell to be occupied is painted.

 'The occupied cell is cleared
  n = asp(A).Chromosome(m).Link
  j = asp(A).LinkAddress(n).x
  i = asp(A).LinkAddress(n).y
  Range(Cells(i, j), Cells(i, j)).Select
  Selection.Clear

  'The target addres is memorized
  Move = asp(A).Chromosome(m).Move
  j = j + VectMoves(Move).x
  i = i - VectMoves(Move).y
  asp(A).LinkAddress(n).x = j
  asp(A).LinkAddress(n).y = i

  'The target cell is painted
   Range(Cells(i, j), Cells(i, j)).Select


   'If movement is too fast, nothing can be seen.
   'To attune the velocity of the asp,
   'augment or diminish the delay as   pleased

   Delay = 1
   For h = 1 To Delay
      With Selection.Interior
        .ColorIndex = asp(A).Color
        .Pattern = xlSolid
     End With
   Next h
End Sub


 Private Function Disruptive(m As Integer, A As Integer)
'A move is disruptive if the target address
'is farther than 1.5 units from the neighboring links.
Dim n As Integer
Dim Move As Integer
Dim x, y, x1, y1, x2, x3, y2, y3 As Integer
Dim d1, d2 As Single

  'Reads the link to be moved
  n = asp(A).Chromosome(m).Link
  'Reads the   address of the link
  y = asp(A).LinkAddress(n).x
  x = asp(A).LinkAddress(n).y
  'The (integer) move to be done is read
  Move = asp(A).Chromosome(m).Move

 'The target address is calculated
  y1 = y + VectMoves(Move).x
  x1 = x - VectMoves(Move).y
  'The addresses of the two neighboring links is read
  'And disruption is calculated
    'MsgBox "n " & n
  If ((1 < n) And (n < 12)) Then
    y2 = asp(A).LinkAddress(n - 1).x
    x2 = asp(A).LinkAddress(n - 1).y
    d1 = ((x2 - x1) ^ 2 + (y2 - y1) ^ 2) ^ 0.5
    y3 = asp(A).LinkAddress(n + 1).x
    x3 = asp(A).LinkAddress(n + 1).y
    d2 = ((x3 - x1) ^ 2 + (y3 - y1) ^ 2) ^ 0.5

    'The disruption bug is fixed here.

    If ((d1 > 1.5) Or (d2 > 1.5)) Then
       Disruptive = True
      Else
       Disruptive = False
     End If
  End If

    'Debugging messages
    'MsgBox "d1 " & d1
    'MsgBox "d2 " & d2

  If n = 1 Then
    y2 = asp(A).LinkAddress(2).x
    x2 = asp(A).LinkAddress(2).y
    d1 = ((x2 - x1) ^ 2 + (y2 - y1) ^ 2) ^ 0.5
If d1 > 1.1 Then Disruptive = True Else Disruptive = False
   End If

  If n = 12 Then
    y2 = asp(A).LinkAddress(11).x
    x2 = asp(A).LinkAddress(11).y
    d2 = ((x2 - x1) ^ 2 + (y2 - y1) ^ 2) ^ 0.5
If d2 > 1.1 Then Disruptive = True Else Disruptive = False
  End If

 End Function



Private Sub Ribosome(m As Integer, A As Integer)
Dim n, i, j As Integer
Dim Delay, h As Integer
Dim Move As Integer
Dim x, y, x1, y1 As Integer
Dim Color As Integer
  'Reads the link to be moved
  n = asp(A).Chromosome(m).Link
  'Reads the   address of the link
  y = asp(A).LinkAddress(n).x
  x = asp(A).LinkAddress(n).y
  'The (integer) move to be done is read
  Move = asp(A).Chromosome(m).Move

 'The target address is calculated
  y1 = y + VectMoves(Move).x
  x1 = x - VectMoves(Move).y

  'There are forbidden borders on all directions
  If ((x1 < 1) Or (y1 < 1) Or (x1 > 103) Or (y1 > 168)) Then
    Exit Sub
  End If

  'Get the  color of the target address
  Range(Cells(x1, y1), Cells(x1, y1)).Select
      With Selection.Interior
       Color = .ColorIndex
      End With

  'Is the target address occupied?
  'MsgBox "Color " & Color
  'In previous codes, there was  a bug here.

  Select Case Color
    Case 2   'the cell is empty, is it disruptive?
         If Not (Disruptive(m, A)) Then Call MoveOn(m, A)
    Case Else 'do nothing: the cell is occupied.

   End Select
End Sub

Private Sub AssessingFitness()
Dim A, B, m, Champ As Integer
Dim Max, Man As Integer
Dim Fitness(1 To 20) As Integer

'The fitness measures the vertical distance from
'actual position to the 100th row.

 For A = 1 To Nasps
 Max = 1
  For m = 1 To 12
   Man = asp(A).LinkAddress(m).y
   'MsgBox "vertical" & Man
   If Man > Max Then Max = Man
  Next m
   Fitness(A) = 100 - Max
   'MsgBox "Fitness " & Fitness(A)
 Next A

'Now, we sort asps by fitness
'and select the best four,
'whose numbers are recorded in Order.
For A = 1 To 4
 Champ = 1
 For B = 1 To 20
  If Fitness(B) > Fitness(Champ) Then Champ = B
  Next B
  Order(A) = Champ
  Fitness(Champ) = 0
   'MsgBox "Next champ " & Champ
Next A
End Sub

Private Sub Mutation()
 Dim A, m, n, l As Integer
 'Let us try out blind point mutations
 'in specific exons,
 'in the order of links to be moved
 'and in the direction of movement.
 For A = 1 To Nasps
 'Select at random  which exon to mutate:
'throw a number from 1 to 24
 m = Int(24 * Rnd) + 1
'Select at random which link to move:
'throw a number from 1 to 12
 n = Int(24 * Rnd) + 1
  asp(A).Chromosome(m).Link = n
 'Select at random which direction to move on
 'Throw a number from 1 to 9
  asp(A).Chromosome(m).Move = Int(9 * Rnd) + 1
 Next A

End Sub

Private Sub Reproduction()
Dim A, m, l As Integer
'There are always 20 asps.
'Our reproductive scheme is as follows:
'10 offspring for the champion,
'5 offspring for the second,
'3 for the third,
'2 for the fourth.
'We place children on order.
'Children are clones of the fathers
'that are copied to the asp2 array.

'Call InitialPosition(8)
'MsgBox "var1 " & asp(8).LinkAddress(7).x
'MsgBox "var2 " & asp2(8).LinkAddress(7).x
 'asp2(8) = asp(8)
 'MsgBox "var2n " & asp2(8).LinkAddress(7).x

 For A = 1 To Nasps
 Select Case A
   Case 1, 2, 3, 4, 5, 6, 7, 8, 9, 10
     asp2(A) = asp(Order(1))
   Case 11, 12, 13, 14, 15
     asp2(A) = asp(Order(2))
   Case 16, 17, 18
     asp2(A) = asp(Order(3))
  Case 19, 20
     asp2(A) = asp(Order(4))
 End Select
Next A

'Clones are subjected to mutation
Call Mutation
End Sub
Private Sub Eraseasp(A As Integer)
Dim l, i, j, h, Delay
 'Call Display
 'Call InitialPosition
 'Paints the body
 'Flag = 1 for initialization at random
 'Flag = 0 for reinitialization by inheritance.

  For l = 1 To 12
  j = asp(A).LinkAddress(l).x
  i = asp(A).LinkAddress(l).y
  'MsgBox "l " & l
  'MsgBox "i " & i
  'MsgBox "j " & j
   Range(Cells(i, j), Cells(i, j)).Select
   With Selection.Interior
        .ColorIndex = 2
        .Pattern = xlSolid
     End With
  Next l
  'If A = 15 Then MsgBox "Color " & asp(A).Color
End Sub



Private Sub Death()
Dim A As Integer
'The asp array is overwritten with
'the asp2 array.
For A = 1 To Nasps
 Call Eraseasp(A)
 asp(A) = asp2(A)
Next A
End Sub

Private Sub Initialization()
Dim A As Integer

  For A = 1 To Nasps
   Call InitialPosition(A)
   Call DrawBody(A, Flag)
    Call InitialChromosome(A)
  Next A
End Sub

Private Sub Reinitialization()
Dim A As Integer

'Beginning from the bottom
'with new genes
  For A = 1 To Nasps
   Call InitialPosition(A)
   Call DrawBody(A, Flag)
  Next A
End Sub

Public Sub asp7()



 'Evolution on its first trial.
 'Body structure is automatically conserved.
 'The task is to climb over the display.

Dim m As Integer
Dim i, j, l, n
Dim A As Integer
Dim Gen, NGen As Integer


 Call Moves
 Call Display
 Call Initialization

'Nasps is the number of asps,
'Default value = 20.
'NGen is the number of generations
'of the whole generation.
'LifeSpan is the number of   days
'in a life.
'Life ends with reproduction and dead.

 Nasps = 20
 NGen = 10
 LifeSpan = 15

 'To generate new random sequences
 'change the number below.

 Randomize (3)

'Flag = 1 for initialization at random
 'Flag = 0 for reinitialization by inheritance.


 Flag = 1
 Call Initialization
 Flag = 0
'm is the number of the link
'l is the counter of days of life.
'A is the index of the asp.

 For Gen = 1 To NGen
  MsgBox "Generation " & Gen
  For l = 1 To LifeSpan
   For m = 1 To 24
    For A = 1 To Nasps
     Call Ribosome(m, A)
    Next A
   Next m
  Next l
  Call AssessingFitness
  Call Reproduction
  Call Mutation
  Call Death
   'Reports the Generation

  Call Reinitialization
 Next Gen
End Sub

\end{verbatim}

 \bigskip
\textbf{\thenum. }  \addtocounter{num}{1} Exercise. Run the code. Notice that   no asp    rises the tail at generation 4 but by the end of  generation 3 there are 4 asps than can rise the tail. This is explained by mutation else by a bug.  Moreover, what we call the first generation is very vivid while the next one rapidly arrive to stagnation.

 \bigskip
\textbf{\thenum. }  \addtocounter{num}{1} In the previous code, the first generation has been programmed to have mutations in each day of life and no inheritance: thus it is a time where randomness is all to the dynamics of the population. But since the second generation and later on, mutation and reproduction enter the stage. This could be a sort of hand given 'origin of life'.

 \bigskip
\textbf{\thenum. }  \addtocounter{num}{1} Let us carry a test to know whether or not mutation is the cause of lost of information or if that is due to a bug. To know that, we silence Mutation just on the third generation to look what happens. The code follows:

\begin{verbatim}

Option Explicit

Type LinkAddresses
 x As Integer
 y As Integer
End Type

Type Moves
 x As Integer
 y As Integer
End Type

Type Exons
 Link As Integer
 Move As Integer
End Type

Type asps
'Each chromosome has 24 instructions
 Chromosome(1 To 25) As Exons
 LinkAddress(1 To 12) As LinkAddresses
 Color As Integer
End Type

Dim Nasps As Integer 'Less than 20
Dim asp(1 To 20) As asps
Dim asp2(1 To 20) As asps
Dim VectMoves(1 To 9) As Moves
Dim LifeSpan As Integer
Dim InitialRow
Dim Order(1 To 20) As Integer
Dim Flag As Integer

Private Sub Moves()
VectMoves(1).x = 0
VectMoves(1).y = 1
VectMoves(2).x = 1
VectMoves(2).y = 1
VectMoves(3).x = 1
VectMoves(3).y = 0
VectMoves(4).x = 1
VectMoves(4).y = -1
VectMoves(5).x = 0
VectMoves(5).y = -1
VectMoves(6).x = -1
VectMoves(6).y = -1
VectMoves(7).x = -1
VectMoves(7).y = 0
VectMoves(8).x = -1
VectMoves(8).y = 1
VectMoves(9).x = 0
VectMoves(9).y = 0
End Sub


Private Sub Display()
Dim i, j As Integer
'Prepares the screen
'Excel already works with this "new" style!
'To be sure, Excel works with the OOP style!


 Workbooks.add
    Range(Cells(1, 1), Cells(103, 170)).Select
    Selection.RowHeight = 3
    Selection.ColumnWidth = 0.3
    ActiveWindow.WindowState = xlMaximized

  'Paint the background in white
  'This is necessary because It is not at all clear
  'what is the default background color.
  'The color of a cell will be taken as
  'the indicative of occupation.

  For i = 1 To 103
   For j = 1 To 170
    Range(Cells(i, j), Cells(i, j)).Select
      With Selection.Interior
       .ColorIndex = 2
      End With
   Next j
  Next i
    Range("G9:T9").Select
End Sub

Private Sub InitialPosition(A As Integer)
Dim l, k
 'The index A distinguishes each asp.
 'Each asp has 12 links,
 'each part has its own coordinates on the screen,
 'which are kept in the LinkAddress   array.
 'The asps are posited at the bottom of the screen.
 'Head are at row InitialRow
 InitialRow = 100
For l = 1 To 12
 k = InitialRow - 12
 asp(A).LinkAddress(l).x = 8 * A
 asp(A).LinkAddress(l).y = k + l
 Next l
End Sub
Private Function Fcolor(A As Integer)
 'Color is a number less than 40
 'but different than 2.
Rebegin:  Fcolor = Int(40 * Rnd)
 If Fcolor = 2 Then GoTo Rebegin
 'MsgBox "Color  " & Fcolor
End Function

Private Sub DrawBody(A As Integer, Flag As Integer)
Dim l, i, j, h, Delay
 'Call Display
 'Call InitialPosition
 'Paints the body
 'Flag = 1 for initialization at random
 'Flag = 0 for reinitialization by inheritance.
  If Flag = 1 Then asp(A).Color = Fcolor(A)
  For l = 1 To 12
  j = asp(A).LinkAddress(l).x
  i = asp(A).LinkAddress(l).y
  'MsgBox "l " & l
  'MsgBox "i " & i
  'MsgBox "j " & j
   Range(Cells(i, j), Cells(i, j)).Select
   With Selection.Interior
        .ColorIndex = asp(A).Color
        .Pattern = xlSolid
     End With
  Next l
  'If A = 15 Then MsgBox "Color " & asp(A).Color
End Sub

Private Sub InitialChromosome(A As Integer)
Dim m, n, t
Dim x, y As Integer
Dim x1, y1 As Integer

  'Select at random  which link to move
'Throw a number from 1 to 12
For m = 1 To 24
 n = Int(12 * Rnd) + 1
  asp(A).Chromosome(m).Link = n

 'Select at random which direction to move on
 'Throw a number from 1 to 9
  asp(A).Chromosome(m).Move = Int(9 * Rnd) + 1
Next m

End Sub

Private Sub MoveOn(m As Integer, A As Integer)
Dim n, i, j As Integer
Dim Delay, h As Integer
Dim Move As Integer
 'To move a link:
 'the cell occupied by  the link is cleared,
 'the cell to be occupied is painted.

 'The occupied cell is cleared
  n = asp(A).Chromosome(m).Link
  j = asp(A).LinkAddress(n).x
  i = asp(A).LinkAddress(n).y
  Range(Cells(i, j), Cells(i, j)).Select
  Selection.Clear

  'The target addres is memorized
  Move = asp(A).Chromosome(m).Move
  j = j + VectMoves(Move).x
  i = i - VectMoves(Move).y
  asp(A).LinkAddress(n).x = j
  asp(A).LinkAddress(n).y = i

  'The target cell is painted
   Range(Cells(i, j), Cells(i, j)).Select


   'If movement is too fast, nothing can be seen.
   'To attune the velocity of the asp,
   'augment or diminish the delay as   pleased

   Delay = 1
   For h = 1 To Delay
      With Selection.Interior
        .ColorIndex = asp(A).Color
        .Pattern = xlSolid
     End With
   Next h
End Sub


 Private Function Disruptive(m As Integer, A As Integer)
'A move is disruptive if the target address
'is farther than 1.5 units from the neighboring links.
Dim n As Integer
Dim Move As Integer
Dim x, y, x1, y1, x2, x3, y2, y3 As Integer
Dim d1, d2 As Single

  'Reads the link to be moved
  n = asp(A).Chromosome(m).Link
  'Reads the   address of the link
  y = asp(A).LinkAddress(n).x
  x = asp(A).LinkAddress(n).y
  'The (integer) move to be done is read
  Move = asp(A).Chromosome(m).Move

 'The target address is calculated
  y1 = y + VectMoves(Move).x
  x1 = x - VectMoves(Move).y
  'The addresses of the two neighboring links is read
  'And disruption is calculated
    'MsgBox "n " & n
  If ((1 < n) And (n < 12)) Then
    y2 = asp(A).LinkAddress(n - 1).x
    x2 = asp(A).LinkAddress(n - 1).y
    d1 = ((x2 - x1) ^ 2 + (y2 - y1) ^ 2) ^ 0.5
    y3 = asp(A).LinkAddress(n + 1).x
    x3 = asp(A).LinkAddress(n + 1).y
    d2 = ((x3 - x1) ^ 2 + (y3 - y1) ^ 2) ^ 0.5

    'The disruption bug is fixed here.

    If ((d1 > 1.5) Or (d2 > 1.5)) Then
       Disruptive = True
      Else
       Disruptive = False
     End If
  End If

    'Debugging messages
    'MsgBox "d1 " & d1
    'MsgBox "d2 " & d2

  If n = 1 Then
    y2 = asp(A).LinkAddress(2).x
    x2 = asp(A).LinkAddress(2).y
    d1 = ((x2 - x1) ^ 2 + (y2 - y1) ^ 2) ^ 0.5
If d1 > 1.1 Then Disruptive = True Else Disruptive = False
   End If

  If n = 12 Then
    y2 = asp(A).LinkAddress(11).x
    x2 = asp(A).LinkAddress(11).y
    d2 = ((x2 - x1) ^ 2 + (y2 - y1) ^ 2) ^ 0.5
If d2 > 1.1 Then Disruptive = True Else Disruptive = False
  End If

 End Function



Private Sub Ribosome(m As Integer, A As Integer)
Dim n, i, j As Integer
Dim Delay, h As Integer
Dim Move As Integer
Dim x, y, x1, y1 As Integer
Dim Color As Integer
  'Reads the link to be moved
  n = asp(A).Chromosome(m).Link
  'Reads the   address of the link
  y = asp(A).LinkAddress(n).x
  x = asp(A).LinkAddress(n).y
  'The (integer) move to be done is read
  Move = asp(A).Chromosome(m).Move

 'The target address is calculated
  y1 = y + VectMoves(Move).x
  x1 = x - VectMoves(Move).y

  'There are forbidden borders on all directions
  If ((x1 < 1) Or (y1 < 1) Or (x1 > 103) Or (y1 > 168)) Then
    Exit Sub
  End If
  'Get the  color of the target address
  Range(Cells(x1, y1), Cells(x1, y1)).Select
      With Selection.Interior
       Color = .ColorIndex
      End With

  'Is the target address occupied?
  'MsgBox "Color " & Color
  'In previous codes, there was  a bug here.

  Select Case Color
    Case 2   'the cell is empty, is it disruptive?
         If Not (Disruptive(m, A)) Then Call MoveOn(m, A)
    Case Else 'do nothing: the cell is occupied.

   End Select
End Sub

Private Sub AssessingFitness()
Dim A, B, m, Champ As Integer
Dim Max, Man As Integer
Dim Fitness(1 To 20) As Integer

'The fitness measures the vertical distance from
'actual position to the 100th row.

 For A = 1 To Nasps
 Max = 1
  For m = 1 To 12
   Man = asp(A).LinkAddress(m).y
   'MsgBox "vertical" & Man
   If Man > Max Then Max = Man
  Next m
   Fitness(A) = 100 - Max
   'MsgBox "Fitness " & Fitness(A)
 Next A

'Now, we sort asps by fitness
'and select the best four,
'whose numbers are recorded in Order.
For A = 1 To 4
 Champ = 1
 For B = 1 To 20
  If Fitness(B) > Fitness(Champ) Then Champ = B
  Next B
  Order(A) = Champ
  Fitness(Champ) = 0
   'MsgBox "Next champ " & Champ
Next A
 MsgBox "Great champ " & Order(1)
End Sub
Private Sub Mutation()
 Dim A, m, n, l As Integer
 'Let us try out blind point mutations
 'in specific exons,
 'in the  link to be moved
 'and in the direction of movement.
 For A = 1 To Nasps
 'Select at random  which exon to mutate:
'throw a number from 1 to 24
 m = Int(24 * Rnd) + 1
'Select at random which link to move:
'throw a number from 1 to 12.
'Bug: Here there was a 24.
 n = Int(12 * Rnd) + 1
  asp(A).Chromosome(m).Link = n
 'Select at random which direction to move on.
 'Throw a number from 1 to 9
  asp(A).Chromosome(m).Move = Int(9 * Rnd) + 1
 Next A

End Sub

Private Sub Reproduction()
Dim A, m, l As Integer
'There are always 20 asps.
'Our reproductive scheme is as follows:
'10 offspring for the champion,
'5 offspring for the second,
'3 for the third,
'2 for the fourth.
'We place children on order.
'Children are clones of the fathers
'that are copied to the asp2 array.

'Call InitialPosition(8)
'MsgBox "var1 " & asp(8).LinkAddress(7).x
'MsgBox "var2 " & asp2(8).LinkAddress(7).x
 'asp2(8) = asp(8)
 'MsgBox "var2n " & asp2(8).LinkAddress(7).x

 For A = 1 To Nasps
 Select Case A
   Case 1, 2, 3, 4, 5, 6, 7, 8, 9, 10
     asp2(A) = asp(Order(1))
   Case 11, 12, 13, 14, 15
     asp2(A) = asp(Order(2))
   Case 16, 17, 18
     asp2(A) = asp(Order(3))
  Case 19, 20
     asp2(A) = asp(Order(4))
 End Select
Next A

'Clones are subjected to mutation
Call Mutation
End Sub

Private Sub Eraseasp(A As Integer)
Dim l, i, j, h, Delay
'To erase an asp, it is painted in white=2
 'Call Display
 'Call InitialPosition
 'Paints the body
 'Flag = 1 for initialization at random
 'Flag = 0 for reinitialization by inheritance.

  For l = 1 To 12
  j = asp(A).LinkAddress(l).x
  i = asp(A).LinkAddress(l).y
  'MsgBox "l " & l
  'MsgBox "i " & i
  'MsgBox "j " & j
   Range(Cells(i, j), Cells(i, j)).Select
   With Selection.Interior
        .ColorIndex = 2
        .Pattern = xlSolid
     End With
  Next l
  'If A = 15 Then MsgBox "Color " & asp(A).Color
End Sub



Private Sub Death()
Dim A As Integer
'The asp array is overwritten with
'the asp2 array.
For A = 1 To Nasps
 Call Eraseasp(A)
 asp(A) = asp2(A)
Next A
End Sub

Private Sub Initialization()
Dim A As Integer

  For A = 1 To Nasps
   Call InitialPosition(A)
   Call DrawBody(A, Flag)
    Call InitialChromosome(A)
  Next A

End Sub

Private Sub Reinitialization()
Dim A As Integer

'Beginning from the bottom
'with new genes
  For A = 1 To Nasps
   Call InitialPosition(A)
   Call DrawBody(A, Flag)
  Next A
End Sub

Public Sub asp6()
'We try to fix a bug with fidelity of inheritance.


 'Evolution on its first trial.
 'Body structure is automatically conserved.
 'The task is to climb over the display.

Dim m As Integer
Dim i, j, l, n
Dim A As Integer
Dim Gen, NGen As Integer


 Call Moves
 Call Display
 'Call Initialization

'Nasps is the number of asps,
'Default value = 20.
'NGen is the number of generations
'of the whole generation.
'LifeSpan is the number of   days
'in a life.
'Life ends with reproduction and dead.

 Nasps = 20
 NGen = 10
 LifeSpan = 15

 'To generate new random sequences
 'change the number below.

 Randomize (3)

'Flag = 1 for initialization at random
 'Flag = 0 for reinitialization by inheritance.


 Flag = 1
 Call Initialization
 Flag = 0
'm is the number of the link
'l is the counter of days of life.
'A is the index of the asp.

 For Gen = 1 To NGen
  MsgBox "Generation " & Gen
  For l = 1 To LifeSpan
   For m = 1 To 24
    For A = 1 To Nasps
     Call Ribosome(m, A)
    Next A
   Next m
  Next l
  Call AssessingFitness
  Call Reproduction
  If Not (Gen = 3) Then Call Mutation
  If Not (Gen = 4) Then Call Mutation
  Call Death
   'Reports the Generation

  Call Reinitialization
 Next Gen
End Sub

\end{verbatim}

 \bigskip
\textbf{\thenum. }  \addtocounter{num}{1} Exercise. Run this new code in which mutation has been silenced during generation 3rd. Which is the effect?

 \bigskip
\textbf{\thenum. }  \addtocounter{num}{1} Having concluded that we have a bug, we need to find it. After a more careful reading of the code, we found that mutation has been called twice: from the main public sub and from reproduction. So, we erased it from the main sub. But nothing happened. This implies that we have carried on a very deep bug since ancient times. Where we could look at for ideas?

 \bigskip
\textbf{\thenum. }  \addtocounter{num}{1} Let us pay attention to the following fact: during the first generation, mutation  seems to get  tired. Something similar happens here: mutation gets tired at generation 4th. So, we  shall do something to establish the cause. The following measures were taken: the velocity of the changes in the  screen was slowed down. It was clear that the inheritance works well but that mutation does not work. To test the inoperativeness  of mutation, the number of mutations per asp was increased from one to five: there was no effect. We conclude that mutation is not working. The reason was that the offspring live  in the asp2 array where mutation was working in the asp array. Thus, asp was replaced by asp2 in the sub mutation. The result was encouraging, so the velocity of the display was normalized and the resultant code follows.

\begin{verbatim}

Option Explicit

Type LinkAddresses
 x As Integer
 y As Integer
End Type

Type Moves
 x As Integer
 y As Integer
End Type

Type Exons
 Link As Integer
 Move As Integer
End Type

Type asps
'Each chromosome has 24 instructions
 Chromosome(1 To 24) As Exons
 LinkAddress(1 To 12) As LinkAddresses
 Color As Integer
End Type

Dim Nasps As Integer 'Less than 20
Dim asp(1 To 20) As asps
Dim asp2(1 To 20) As asps
Dim VectMoves(1 To 9) As Moves
Dim LifeSpan As Integer
Dim InitialRow
Dim Order(1 To 20) As Integer
Dim Flag As Integer
Dim numbmutation As Integer
Dim Velocity As Integer









Private Sub Moves()
VectMoves(1).x = 0
VectMoves(1).y = 1
VectMoves(2).x = 1
VectMoves(2).y = 1
VectMoves(3).x = 1
VectMoves(3).y = 0
VectMoves(4).x = 1
VectMoves(4).y = -1
VectMoves(5).x = 0
VectMoves(5).y = -1
VectMoves(6).x = -1
VectMoves(6).y = -1
VectMoves(7).x = -1
VectMoves(7).y = 0
VectMoves(8).x = -1
VectMoves(8).y = 1
VectMoves(9).x = 0
VectMoves(9).y = 0
End Sub


Private Sub Display()
Dim i, j As Integer
'Prepares the screen
'Excel already works with this "new" style!
'To be sure, Excel works with the OOP style!


 Workbooks.add
    Range(Cells(1, 1), Cells(103, 170)).Select
    Selection.RowHeight = 3
    Selection.ColumnWidth = 0.3
    ActiveWindow.WindowState = xlMaximized

  'Paint the background in white
  'This is necessary because It is not at all clear
  'what is the default background color.
  'The color of a cell will be taken as
  'the indicative of occupation.

  For i = 1 To 103
   For j = 1 To 170
    Range(Cells(i, j), Cells(i, j)).Select
      With Selection.Interior
       .ColorIndex = 2
      End With
   Next j
  Next i
    Range("G9:T9").Select
End Sub

Private Sub InitialPosition(A As Integer)
Dim l, k
 'The index A distinguishes each asp.
 'Each asp has 12 links,
 'each part has its own coordinates on the screen,
 'which are kept in the LinkAddress   array.
 'The asps are posited at the bottom of the screen.
 'Head are at row InitialRow
 InitialRow = 100
For l = 1 To 12
 k = InitialRow - 12
 asp(A).LinkAddress(l).x = 8 * A
 asp(A).LinkAddress(l).y = k + l
 Next l
End Sub

Private Function Fcolor(A As Integer)
 'Color is a number less than 40
 'but different than 2.
Rebegin:  Fcolor = Int(40 * Rnd)
 If Fcolor = 2 Then GoTo Rebegin
 'MsgBox "Color  " & Fcolor
End Function

Private Sub DrawBody(A As Integer, Flag As Integer)
Dim l, i, j, h, Delay
 'Call Display
 'Call InitialPosition
 'Paints the body
 'Flag = 1 for initialization at random
 'Flag = 0 for reinitialization by inheritance.
  If Flag = 1 Then asp(A).Color = Fcolor(A)
  For l = 1 To 12
  j = asp(A).LinkAddress(l).x
  i = asp(A).LinkAddress(l).y
  'MsgBox "l " & l
  'MsgBox "i " & i
  'MsgBox "j " & j
   Range(Cells(i, j), Cells(i, j)).Select
   With Selection.Interior
        .ColorIndex = asp(A).Color
        .Pattern = xlSolid
     End With
  Next l
  'If A = 15 Then MsgBox "Color " & asp(A).Color
End Sub

Private Sub InitialChromosome(A As Integer)
Dim m, n, t
Dim x, y As Integer
Dim x1, y1 As Integer


For m = 1 To 24
'Select at random  which link to move
'Throw a number from 1 to 12
 n = Int(12 * Rnd) + 1
  asp(A).Chromosome(m).Link = n
 'MsgBox "n" & n
 'Select at random which direction to move on
 'Throw a number from 1 to 9
  asp(A).Chromosome(m).Move = Int(9 * Rnd) + 1
Next m

End Sub

Private Sub MoveOn(m As Integer, A As Integer)
Dim n, i, j As Integer
Dim Delay, h As Integer
Dim Move As Integer


 'To move a link:
 'the cell occupied by  the link is cleared,
 'the cell to be occupied is painted.

 'The occupied cell is cleared
  n = asp(A).Chromosome(m).Link
  j = asp(A).LinkAddress(n).x
  i = asp(A).LinkAddress(n).y
  Range(Cells(i, j), Cells(i, j)).Select
  Selection.Clear

  'The target addres is calculated
  Move = asp(A).Chromosome(m).Move
  j = j + VectMoves(Move).x
  i = i - VectMoves(Move).y
 'The target addres is memorized
   asp(A).LinkAddress(n).x = j
  asp(A).LinkAddress(n).y = i

  'The target cell is painted
   Range(Cells(i, j), Cells(i, j)).Select


   'If movement is too fast, nothing can be seen.
   'To attune the velocity of the asp,
   'augment or diminish the delay as   pleased

   Delay = 1001 - Velocity
   For h = 1 To Delay
      With Selection.Interior
        .ColorIndex = asp(A).Color
        .Pattern = xlSolid
     End With
   Next h
End Sub


 Private Function Disruptive(m As Integer, A As Integer)
'A move is disruptive if the target address
'is farther than 1.5 units from the neighboring links.
Dim n As Integer
Dim Move As Integer
Dim x, y, x1, y1, x2, x3, y2, y3 As Integer
Dim d1, d2 As Single

  'Reads the link to be moved
  n = asp(A).Chromosome(m).Link
  'Reads the   address of the link
  y = asp(A).LinkAddress(n).x
  x = asp(A).LinkAddress(n).y
  'The (integer) move to be done is read
  Move = asp(A).Chromosome(m).Move

 'The target address is calculated
  y1 = y + VectMoves(Move).x
  x1 = x - VectMoves(Move).y
  'The addresses of the two neighboring links is read
  'And disruption is calculated
    'MsgBox "n " & n
  If ((1 < n) And (n < 12)) Then
    y2 = asp(A).LinkAddress(n - 1).x
    x2 = asp(A).LinkAddress(n - 1).y
    d1 = ((x2 - x1) ^ 2 + (y2 - y1) ^ 2) ^ 0.5
    y3 = asp(A).LinkAddress(n + 1).x
    x3 = asp(A).LinkAddress(n + 1).y
    d2 = ((x3 - x1) ^ 2 + (y3 - y1) ^ 2) ^ 0.5

    'The disruption bug is fixed here.

    If ((d1 > 1.5) Or (d2 > 1.5)) Then
       Disruptive = True
      Else
       Disruptive = False
     End If
  End If

    'Debugging messages
    'MsgBox "d1 " & d1
    'MsgBox "d2 " & d2

  If n = 1 Then
    y2 = asp(A).LinkAddress(2).x
    x2 = asp(A).LinkAddress(2).y
    d1 = ((x2 - x1) ^ 2 + (y2 - y1) ^ 2) ^ 0.5
If d1 > 1.1 Then Disruptive = True Else Disruptive = False
   End If

  If n = 12 Then
    y2 = asp(A).LinkAddress(11).x
    x2 = asp(A).LinkAddress(11).y
    d2 = ((x2 - x1) ^ 2 + (y2 - y1) ^ 2) ^ 0.5
If d2 > 1.1 Then Disruptive = True Else Disruptive = False
  End If

 End Function



Private Sub Ribosome(m As Integer, A As Integer)
Dim n, i, j As Integer
Dim Delay, h As Integer
Dim Move As Integer
Dim x, y, x1, y1 As Integer
Dim Color As Integer
  'Reads the link to be moved
  n = asp(A).Chromosome(m).Link
  'Reads the   address of the link
  y = asp(A).LinkAddress(n).x
  x = asp(A).LinkAddress(n).y
  'The (integer) move to be done is read
  Move = asp(A).Chromosome(m).Move

 'The target address is calculated
  y1 = y + VectMoves(Move).x
  x1 = x - VectMoves(Move).y

  'There are forbidden borders on all directions
  If ((x1 < 1) Or (y1 < 1) Or (x1 > 103) Or (y1 > 168)) Then
   Exit Sub
  End If
  'Get the  color of the target address
  Range(Cells(x1, y1), Cells(x1, y1)).Select
      With Selection.Interior
       Color = .ColorIndex
      End With

  'Is the target address occupied?
  'MsgBox "Color " & Color
  'In previous codes, there was  a bug here.

  Select Case Color
    Case 2   'the cell is empty, is it disruptive?
         If Not (Disruptive(m, A)) Then Call MoveOn(m, A)
    Case Else 'do nothing: the cell is occupied.

   End Select
End Sub

Private Sub AssessingFitness()
Dim A, B, m, Champ As Integer
Dim Max, Man As Integer
Dim Fitness(1 To 20) As Integer

'The fitness measures the vertical distance from
'actual position to the 100th row.

 For A = 1 To Nasps
 Max = 1
  For m = 1 To 12
   Man = asp(A).LinkAddress(m).y
   'MsgBox "vertical" & Man
   If Man > Max Then Max = Man
  Next m
   Fitness(A) = 100 - Max
   'MsgBox "Fitness " & Fitness(A)
 Next A

'Now, we sort asps by fitness
'and select the best four,
'whose numbers are recorded in Order.
For A = 1 To 4
 Champ = 1
 For B = 1 To 20
  If Fitness(B) > Fitness(Champ) Then Champ = B
  Next B
  Order(A) = Champ
  Fitness(Champ) = 0
 'MsgBox "Next champ " & Champ
Next A
 'MsgBox "Great champ " & Order(1)
End Sub

Private Sub Mutation()
 Dim A, m, n, l, k  As Integer
 'Let us try out blind point mutations
 'in specific exons,
 'in the  link to be moved
 'and in the direction of movement.
 For A = 1 To Nasps
 'Each offsprings receives numbmutation mutations
  For k = 1 To numbmutation
  'Select at random  which exon to mutate:
  'throw a number from 1 to 24
   m = Int(24 * Rnd) + 1
  'Select at random which link to move:
  'throw a number from 1 to 12.
  'Bug: there was a 24 in here.
  n = Int(12 * Rnd) + 1
  asp2(A).Chromosome(m).Link = n
  'Select at random which direction to move on.
  'Throw a number from 1 to 9
  asp2(A).Chromosome(m).Move = Int(9 * Rnd) + 1
  Next k
 Next A

End Sub

Private Sub Reproduction()
Dim A, m, l As Integer
'There are always 20 asps.
'Our reproductive scheme is as follows:
'10 offspring for the champion,
'5 offspring for the second,
'3 for the third,
'2 for the fourth.
'We place children on order.
'Children are clones of the fathers
'that are copied to the asp2 array.

'Call InitialPosition(8)
'MsgBox "var1 " & asp(8).LinkAddress(7).x
'MsgBox "var2 " & asp2(8).LinkAddress(7).x
 'asp2(8) = asp(8)
 'MsgBox "var2n " & asp2(8).LinkAddress(7).x

 For A = 1 To Nasps
 Select Case A
   Case 1, 2, 3, 4, 5, 6, 7, 8, 9, 10
     asp2(A) = asp(Order(1))
   Case 11, 12, 13, 14, 15
     asp2(A) = asp(Order(2))
   Case 16, 17, 18
     asp2(A) = asp(Order(3))
  Case 19, 20
     asp2(A) = asp(Order(4))
 End Select
Next A



End Sub
Private Sub Eraseasp(A As Integer)
Dim l, i, j, h, Delay
'To erase an asp, it is painted in white=2
 'Call Display
 'Call InitialPosition
 'Paints the body
 'Flag = 1 for initialization at random
 'Flag = 0 for reinitialization by inheritance.

  For l = 1 To 12
  j = asp(A).LinkAddress(l).x
  i = asp(A).LinkAddress(l).y
  'MsgBox "l " & l
  'MsgBox "i " & i
  'MsgBox "j " & j
   Range(Cells(i, j), Cells(i, j)).Select
   With Selection.Interior
        .ColorIndex = 2
        .Pattern = xlSolid
     End With
  Next l
  'If A = 15 Then MsgBox "Color " & asp(A).Color
End Sub



Private Sub Death()
Dim A As Integer
'The asp array is overwritten with
'the asp2 array.
For A = 1 To Nasps
 Call Eraseasp(A)
 asp(A) = asp2(A)
Next A
End Sub

Private Sub Initialization()
Dim A As Integer

  For A = 1 To Nasps
   Call InitialPosition(A)
   Call DrawBody(A, Flag)
    Call InitialChromosome(A)
  Next A

End Sub

Private Sub Reinitialization()
Dim A As Integer

'Beginning from the bottom
'with new genes
  For A = 1 To Nasps
   Call InitialPosition(A)
   Call DrawBody(A, Flag)
  Next A
End Sub

Public Sub asp5()
'We try to fix a bug with mutation.


 'Evolution on its first trial.
 'Body structure is automatically conserved.
 'The task is to climb over the display.

Dim m As Integer
Dim i, j, l, n
Dim A As Integer
Dim Gen, NGen As Integer


 Call Moves
 Call Display
 'Call Initialization

'Nasps is the number of asps,
'Default value = 20.
'NGen is the number of generations
'of the whole generation.
'LifeSpan is the number of   days
'in a life.
'Life ends with reproduction and dead.
'numbmutation is the number of mutations
'of the offspring
'Velocity is among 0 and 1000,
'it regulate sthe velocity of the display.

 Nasps = 20
 NGen = 100
 LifeSpan = 15
 numbmutation = 5
 Velocity = 1000
 'To generate new random sequences
 'change the number below.

 Randomize (3)

'Flag = 1 for initialization at random
 'Flag = 0 for reinitialization by inheritance.


 Flag = 1
 Call Initialization
 Flag = 0
'm is the number of the link
'l is the counter of days of life.
'A is the index of the asp.

 For Gen = 1 To NGen
  'MsgBox "Generation " & Gen
  For l = 1 To LifeSpan
   For m = 1 To 24
    For A = 1 To Nasps
     Call Ribosome(m, A)
    Next A
   Next m
  Next l
  If Not (Gen = NGen) Then
   Call AssessingFitness
   Call Reproduction
   Call Mutation
   Call Death
   Call Reinitialization
  End If
 Next Gen
End Sub

\end{verbatim}

 \bigskip
\textbf{\thenum. }  \addtocounter{num}{1} Exercise. Run the code and verify that the evolutionary process begins to gather information (instruction that help the goal of rising upon the screen), but then mutation erases it and one gets bored before the appearance of a new reacquisition of the information that once was lost.

  \bigskip
\textbf{\thenum. }  \addtocounter{num}{1} Let us  implement now another form of mutation, which consists in   copying information from one place of the chromosome to another. This copying procedure could be invoked at random by mutation. In genetics this mechanism is called gene duplication and it is considered to be transcendental for adaptive processes: if an innovation serves for certain function, it is not excluded that more of the same serves for more of the function.

\begin{verbatim}

Option Explicit

Type LinkAddresses
 x As Integer
 y As Integer
End Type

Type Moves
 x As Integer
 y As Integer
End Type

Type Exons
 Link As Integer
 Move As Integer
End Type

Type asps
'Each chromosome has 24 instructions
 Chromosome(1 To 24) As Exons
 LinkAddress(1 To 12) As LinkAddresses
 Color As Integer
End Type

Dim Nasps As Integer 'Less than 20
Dim asp(1 To 20) As asps
Dim asp2(1 To 20) As asps
Dim VectMoves(1 To 9) As Moves
Dim LifeSpan As Integer
Dim InitialRow
Dim Order(1 To 20) As Integer
Dim Flag As Integer
Dim NumbPointMutation As Integer
Dim Velocity As Integer
Dim NOriginalexons, Ncopies As Integer

Private Sub Moves()
VectMoves(1).x = 0
VectMoves(1).y = 1
VectMoves(2).x = 1
VectMoves(2).y = 1
VectMoves(3).x = 1
VectMoves(3).y = 0
VectMoves(4).x = 1
VectMoves(4).y = -1
VectMoves(5).x = 0
VectMoves(5).y = -1
VectMoves(6).x = -1
VectMoves(6).y = -1
VectMoves(7).x = -1
VectMoves(7).y = 0
VectMoves(8).x = -1
VectMoves(8).y = 1
VectMoves(9).x = 0
VectMoves(9).y = 0
End Sub


Private Sub Display()
Dim i, j As Integer
'Prepares the screen
'Excel already works with this "new" style!
'To be sure, Excel works with the OOP style!


 Workbooks.add
    Range(Cells(1, 1), Cells(103, 170)).Select
    Selection.RowHeight = 3
    Selection.ColumnWidth = 0.3
    ActiveWindow.WindowState = xlMaximized

  'Paint the background in white
  'This is necessary because It is not at all clear
  'what is the default background color.
  'The color of a cell will be taken as
  'the indicative of occupation.

  For i = 1 To 103
   For j = 1 To 170
    Range(Cells(i, j), Cells(i, j)).Select
      With Selection.Interior
       .ColorIndex = 2
      End With
   Next j
  Next i
    Range("G9:T9").Select
End Sub

Private Sub InitialPosition(A As Integer)
Dim l, k
 'The index A distinguishes each asp.
 'Each asp has 12 links,
 'each part has its own coordinates on the screen,
 'which are kept in the LinkAddress   array.
 'The asps are posited at the bottom of the screen.
 'Head are at row InitialRow
 InitialRow = 100
For l = 1 To 12
 k = InitialRow - 12
 asp(A).LinkAddress(l).x = 8 * A
 asp(A).LinkAddress(l).y = k + l
 Next l
End Sub

Private Function Fcolor(A As Integer)
 'Color is a number less than 40
 'but different than 2.
Rebegin:  Fcolor = Int(40 * Rnd)
 If Fcolor = 2 Then GoTo Rebegin
 'MsgBox "Color  " & Fcolor
End Function

Private Sub DrawBody(A As Integer, Flag As Integer)
Dim l, i, j, h, Delay
 'Call Display
 'Call InitialPosition
 'Paints the body
 'Flag = 1 for initialization at random
 'Flag = 0 for reinitialization by inheritance.
  If Flag = 1 Then asp(A).Color = Fcolor(A)
  For l = 1 To 12
  j = asp(A).LinkAddress(l).x
  i = asp(A).LinkAddress(l).y
  'MsgBox "l " & l
  'MsgBox "i " & i
  'MsgBox "j " & j
   Range(Cells(i, j), Cells(i, j)).Select
   With Selection.Interior
        .ColorIndex = asp(A).Color
        .Pattern = xlSolid
     End With
  Next l
  'If A = 15 Then MsgBox "Color " & asp(A).Color
End Sub

Private Sub InitialChromosome(A As Integer)
Dim m, n, t
Dim x, y As Integer
Dim x1, y1 As Integer


For m = 1 To 24
'Select at random  which link to move
'Throw a number from 1 to 12
 n = Int(12 * Rnd) + 1
  asp(A).Chromosome(m).Link = n
 'MsgBox "n" & n
 'Select at random which direction to move on
 'Throw a number from 1 to 9
  asp(A).Chromosome(m).Move = Int(9 * Rnd) + 1
Next m

End Sub

Private Sub MoveOn(m As Integer, A As Integer)
Dim n, i, j As Integer
Dim Delay, h As Integer
Dim Move As Integer


 'To move a link:
 'the cell occupied by  the link is cleared,
 'the cell to be occupied is painted.

 'The occupied cell is cleared
  n = asp(A).Chromosome(m).Link
  j = asp(A).LinkAddress(n).x
  i = asp(A).LinkAddress(n).y
  Range(Cells(i, j), Cells(i, j)).Select
  Selection.Clear

  'The target addres is calculated
  Move = asp(A).Chromosome(m).Move
  j = j + VectMoves(Move).x
  i = i - VectMoves(Move).y
 'The target addres is memorized
   asp(A).LinkAddress(n).x = j
  asp(A).LinkAddress(n).y = i

  'The target cell is painted
   Range(Cells(i, j), Cells(i, j)).Select


   'If movement is too fast, nothing can be seen.
   'To attune the velocity of the asp,
   'augment or diminish the delay as   pleased

   Delay = 1001 - Velocity
   For h = 1 To Delay
      With Selection.Interior
        .ColorIndex = asp(A).Color
        .Pattern = xlSolid
     End With
   Next h
End Sub


 Private Function Disruptive(m As Integer, A As Integer)
'A move is disruptive if the target address
'is farther than 1.5 units from the neighboring links.
Dim n As Integer
Dim Move As Integer
Dim x, y, x1, y1, x2, x3, y2, y3 As Integer
Dim d1, d2 As Single

  'Reads the link to be moved
  n = asp(A).Chromosome(m).Link
  'Reads the   address of the link
  y = asp(A).LinkAddress(n).x
  x = asp(A).LinkAddress(n).y
  'The (integer) move to be done is read
  Move = asp(A).Chromosome(m).Move

 'The target address is calculated
  y1 = y + VectMoves(Move).x
  x1 = x - VectMoves(Move).y
  'The addresses of the two neighboring links is read
  'And disruption is calculated
    'MsgBox "n " & n
  If ((1 < n) And (n < 12)) Then
    y2 = asp(A).LinkAddress(n - 1).x
    x2 = asp(A).LinkAddress(n - 1).y
    d1 = ((x2 - x1) ^ 2 + (y2 - y1) ^ 2) ^ 0.5
    y3 = asp(A).LinkAddress(n + 1).x
    x3 = asp(A).LinkAddress(n + 1).y
    d2 = ((x3 - x1) ^ 2 + (y3 - y1) ^ 2) ^ 0.5

    'The disruption bug is fixed here.

    If ((d1 > 1.5) Or (d2 > 1.5)) Then
       Disruptive = True
      Else
       Disruptive = False
     End If
  End If

    'Debugging messages
    'MsgBox "d1 " & d1
    'MsgBox "d2 " & d2

  If n = 1 Then
    y2 = asp(A).LinkAddress(2).x
    x2 = asp(A).LinkAddress(2).y
    d1 = ((x2 - x1) ^ 2 + (y2 - y1) ^ 2) ^ 0.5
If d1 > 1.1 Then Disruptive = True Else Disruptive = False
   End If

  If n = 12 Then
    y2 = asp(A).LinkAddress(11).x
    x2 = asp(A).LinkAddress(11).y
    d2 = ((x2 - x1) ^ 2 + (y2 - y1) ^ 2) ^ 0.5
If d2 > 1.1 Then Disruptive = True Else Disruptive = False
  End If

 End Function



Private Sub Ribosome(m As Integer, A As Integer)
Dim n, i, j As Integer
Dim Delay, h As Integer
Dim Move As Integer
Dim x, y, x1, y1 As Integer
Dim Color As Integer
  'Reads the link to be moved
  n = asp(A).Chromosome(m).Link
  'Reads the   address of the link
  y = asp(A).LinkAddress(n).x
  x = asp(A).LinkAddress(n).y
  'The (integer) move to be done is read
  Move = asp(A).Chromosome(m).Move

 'The target address is calculated
  y1 = y + VectMoves(Move).x
  x1 = x - VectMoves(Move).y

  'There are forbidden borders on all directions
  If ((x1 < 1) Or (y1 < 1) Or (x1 > 103) Or (y1 > 168)) Then
   Exit Sub
  End If

  'Get the  color of the target address
  Range(Cells(x1, y1), Cells(x1, y1)).Select
      With Selection.Interior
       Color = .ColorIndex
      End With

  'Is the target address occupied?
  'MsgBox "Color " & Color
  'In previous codes, there was  a bug here.

  Select Case Color
    Case 2   'the cell is empty, is it disruptive?
         If Not (Disruptive(m, A)) Then Call MoveOn(m, A)
    Case Else 'do nothing: the cell is occupied.

   End Select
End Sub

Private Sub AssessingFitness()
Dim A, B, m, Champ As Integer
Dim Max, Man As Integer
Dim Fitness(1 To 20) As Integer

'The fitness measures the vertical distance from
'actual position to the 100th row.

 For A = 1 To Nasps
 Max = 1
  For m = 1 To 12
   Man = asp(A).LinkAddress(m).y
   'MsgBox "vertical" & Man
   If Man > Max Then Max = Man
  Next m
   Fitness(A) = 100 - Max
   'MsgBox "Fitness " & Fitness(A)
 Next A

'Now, we sort asps by fitness
'and select the best four,
'whose numbers are recorded in Order.
For A = 1 To 4
 Champ = 1
 For B = 1 To 20
  If Fitness(B) > Fitness(Champ) Then Champ = B
  Next B
  Order(A) = Champ
  Fitness(Champ) = 0
 'MsgBox "Next champ " & Champ
Next A
 'MsgBox "Great champ " & Order(1)
End Sub

Private Sub PointMutation()
 Dim A, m, n, l, k  As Integer
 'Let us try out blind point mutations
 'in specific exons,
 'in the  link to be moved
 'and in the direction of movement.
 For A = 1 To Nasps
 'Each offsprings receives numbmutation mutations
  For k = 1 To NumbPointMutation
  'Select at random  which exon to mutate:
  'throw a number from 1 to 24
   m = Int(24 * Rnd) + 1
  'Select at random which link to move:
  'throw a number from 1 to 12.
  'Bug: there was a 24 in here.
  n = Int(12 * Rnd) + 1
  asp2(A).Chromosome(m).Link = n
  'Select at random which direction to move on.
  'Throw a number from 1 to 9
  asp2(A).Chromosome(m).Move = Int(9 * Rnd) + 1
  Next k
 Next A
End Sub

Private Sub GeneAmplification()
 'Some exons are overwriten with others.
Dim A, m, e, n, Move, k, ee As Integer

 For A = 1 To 20
  For m = 1 To NOriginalexons
    'Select  which exon to be amplified:
     'throw a number from 1 to 24
     e = Int(24 * Rnd) + 1
     n = asp2(A).Chromosome(e).Link
     Move = asp2(A).Chromosome(e).Move
     'The exon is amplified
     For k = 1 To Ncopies
     'Select at random where to be copied:
     'throw a number from 1 to 24.
     ee = Int(24 * Rnd) + 1
     asp2(A).Chromosome(ee).Link = n
     asp2(A).Chromosome(ee).Move = Move
    Next k
  Next m
 Next A
End Sub

Private Sub Mutation()
'Gene amplification dominates over
'point mutation.
  Call PointMutation
  Call GeneAmplification
End Sub



Private Sub Reproduction()
Dim A, m, l As Integer
'There are always 20 asps.
'Our reproductive scheme is as follows:
'10 offspring for the champion,
'5 offspring for the second,
'3 for the third,
'2 for the fourth.
'We place children on order.
'Children are clones of the fathers
'that are copied to the asp2 array.

'Call InitialPosition(8)
'MsgBox "var1 " & asp(8).LinkAddress(7).x
'MsgBox "var2 " & asp2(8).LinkAddress(7).x
 'asp2(8) = asp(8)
 'MsgBox "var2n " & asp2(8).LinkAddress(7).x

 For A = 1 To Nasps
 Select Case A
   Case 1, 2, 3, 4, 5, 6, 7, 8, 9, 10
     asp2(A) = asp(Order(1))
   Case 11, 12, 13, 14, 15
     asp2(A) = asp(Order(2))
   Case 16, 17, 18
     asp2(A) = asp(Order(3))
  Case 19, 20
     asp2(A) = asp(Order(4))
 End Select
Next A

End Sub

Private Sub Eraseasp(A As Integer)
Dim l, i, j, h, Delay
'To erase an asp, it is painted in white=2
 'Call Display
 'Call InitialPosition
 'Paints the body
 'Flag = 1 for initialization at random
 'Flag = 0 for reinitialization by inheritance.

  For l = 1 To 12
  j = asp(A).LinkAddress(l).x
  i = asp(A).LinkAddress(l).y
  'MsgBox "l " & l
  'MsgBox "i " & i
  'MsgBox "j " & j
   Range(Cells(i, j), Cells(i, j)).Select
   With Selection.Interior
        .ColorIndex = 2
        .Pattern = xlSolid
     End With
  Next l
  'If A = 15 Then MsgBox "Color " & asp(A).Color
End Sub

Private Sub Death()
Dim A As Integer
'The asp array is overwritten with
'the asp2 array.
For A = 1 To Nasps
 Call Eraseasp(A)
 asp(A) = asp2(A)
Next A
End Sub

Private Sub Initialization()
Dim A As Integer

  For A = 1 To Nasps
   Call InitialPosition(A)
   Call DrawBody(A, Flag)
    Call InitialChromosome(A)
  Next A

End Sub

Private Sub Reinitialization()
Dim A As Integer

'Beginning from the bottom
'with new genes
  For A = 1 To Nasps
   Call InitialPosition(A)
   Call DrawBody(A, Flag)
  Next A
End Sub

Public Sub asp4()
'We implement gene amplification


 'Evolution on its first trial.
 'Body structure is automatically conserved.
 'The task is to climb over the display.

Dim m As Integer
Dim i, j, l, n
Dim A As Integer
Dim Gen, NGen As Integer


 Call Moves
 Call Display
 'Call Initialization

'Nasps is the number of asps,
'Default value = 20.
'NGen is the number of generations
'of the whole generation.
'LifeSpan is the number of   days
'in a life.
'Life ends with reproduction and dead.
'numbmutation is the number of mutations
'of the offspring
'Velocity is among 0 and 1000,
'it regulate sthe velocity of the display.

 Nasps = 20
 NGen = 100
 LifeSpan = 15
 NumbPointMutation = 5
 'Gene amplification follows:
 'NOriginalexons is the number of different exons
 'to be copied.
 NOriginalexons = 2
 'Ncopies is the number of new copies
 'of each copied exon.
 Ncopies = 3
 Velocity = 1000
 'To generate new random sequences
 'change the number below.

 Randomize (3)

'Flag = 1 for initialization at random
 'Flag = 0 for reinitialization by inheritance.


 Flag = 1
 Call Initialization
 Flag = 0
'm is the number of the link
'l is the counter of days of life.
'A is the index of the asp.

 For Gen = 1 To NGen
  'MsgBox "Generation " & Gen
  For l = 1 To LifeSpan
   For m = 1 To 24
    For A = 1 To Nasps
     Call Ribosome(m, A)
    Next A
   Next m
  Next l
  If Not (Gen = NGen) Then
   Call AssessingFitness
   Call Reproduction
   'Gene amplification is included.
   Call Mutation
   Call Death
   Call Reinitialization
  End If
 Next Gen
End Sub

\end{verbatim}

 \bigskip
\textbf{\thenum. }  \addtocounter{num}{1} Exercise. Run the code and verify that the asp cannot climb  over the screen. Seemingly, mutation destroys any form of information and spoils everything. Thus, play with parameters in such a way as to diminish mutation to see what happens.

 \bigskip
\textbf{\thenum. }  \addtocounter{num}{1} We, seemingly,  have observed that a life span has two periods: in the first one, the information  of the chromosomes is expressed and in the second period things remain static. This means that the probability of generation of recursive behavior by the genetic information is nil and void.

 \bigskip
\textbf{\thenum. }  \addtocounter{num}{1} Exercise. Our failure could be due to the fact that climbing demands an extremely coordinated movements of 12 links. Make an estimation of the probability of assembling the correct information just by randomness. Explain, moreover, why we have failed.

 \bigskip
\textbf{\thenum. }  \addtocounter{num}{1} Coordination of 12 links could be too much. Let us shorten the length of the asp. The resultant code follows.

\begin{verbatim}

Option Explicit

Type LinkAddresses
 x As Integer
 y As Integer
End Type

Type Moves
 x As Integer
 y As Integer
End Type

Type Exons
 Link As Integer
 Move As Integer
End Type

Type asps
'Each chromosome has NExons instructions
 Chromosome(1 To 40) As Exons
 LinkAddress(1 To 20) As LinkAddresses
 Color As Integer
End Type

Dim Nasps As Integer 'Less than 20
Dim asp(1 To 20) As asps
Dim asp2(1 To 20) As asps
Dim VectMoves(1 To 9) As Moves
Dim LifeSpan As Integer
Dim InitialRow
Dim Order(1 To 20) As Integer
Dim Flag As Integer
Dim NumbPointMutations As Integer
Dim Velocity As Integer
Dim NOriginalexons, Ncopies As Integer
Dim NLinks, NExons As Integer

Sub Moves()
VectMoves(1).x = 0
VectMoves(1).y = 1
VectMoves(2).x = 1
VectMoves(2).y = 1
VectMoves(3).x = 1
VectMoves(3).y = 0
VectMoves(4).x = 1
VectMoves(4).y = -1
VectMoves(5).x = 0
VectMoves(5).y = -1
VectMoves(6).x = -1
VectMoves(6).y = -1
VectMoves(7).x = -1
VectMoves(7).y = 0
VectMoves(8).x = -1
VectMoves(8).y = 1
VectMoves(9).x = 0
VectMoves(9).y = 0
End Sub


Private Sub Display()
Dim i, j As Integer
'Prepares the screen
'Excel already works with this "new" style!
'To be sure, Excel works with the OOP style!


 Workbooks.add
    Range(Cells(1, 1), Cells(103, 170)).Select
    Selection.RowHeight = 3
    Selection.ColumnWidth = 0.3
    ActiveWindow.WindowState = xlMaximized

  'Paint the background in white
  'This is necessary because It is not at all clear
  'what is the default background color.
  'The color of a cell will be taken as
  'the indicative of occupation.

  For i = 1 To 103
   For j = 1 To 170
    Range(Cells(i, j), Cells(i, j)).Select
      With Selection.Interior
       .ColorIndex = 2
      End With
   Next j
  Next i
    Range("G9:T9").Select
End Sub

Private Sub InitialPosition(A As Integer)
Dim l, k
 'The index A distinguishes each asp.
 'Each asp has NLinks links,
 'each part has its own coordinates on the screen,
 'which are kept in the LinkAddress   array.
 'The asps are posited at the bottom of the screen.
 'Head are at row InitialRow

For l = 1 To NLinks
 k = InitialRow - NLinks
 asp(A).LinkAddress(l).x = 8 * A
 asp(A).LinkAddress(l).y = k + l
 Next l
End Sub

Private Function Fcolor(A As Integer)
 'Color is a number less than 40
 'but different than 2.
Rebegin:  Fcolor = Int(40 * Rnd)
 If Fcolor = 2 Then GoTo Rebegin
 'MsgBox "Color  " & Fcolor
End Function

Private Sub DrawBody(A As Integer, Flag As Integer)
Dim l, i, j, h, Delay
 'Call Display
 'Call InitialPosition
 'Paints the body
 'Flag = 1 for initialization at random
 'Flag = 0 for reinitialization by inheritance.
  If Flag = 1 Then asp(A).Color = Fcolor(A)
  For l = 1 To NLinks
  j = asp(A).LinkAddress(l).x
  i = asp(A).LinkAddress(l).y
  'MsgBox "l " & l
  'MsgBox "i " & i
  'MsgBox "j " & j
   Range(Cells(i, j), Cells(i, j)).Select
   With Selection.Interior
        .ColorIndex = asp(A).Color
        .Pattern = xlSolid
     End With
  Next l
  'If A = 15 Then MsgBox "Color " & asp(A).Color
End Sub

Private Sub InitialChromosome(A As Integer)
Dim m, n, t
Dim x, y As Integer
Dim x1, y1 As Integer


For m = 1 To NExons
'Select at random  which link to move
'Throw a number from 1 to NLinks
 n = Int(NLinks * Rnd) + 1
 'MsgBox "n " & n
 asp(A).Chromosome(m).Link = n

 'Select at random which direction to move on
 'Throw a number from 1 to 9
  asp(A).Chromosome(m).Move = Int(9 * Rnd) + 1
Next m

End Sub

Private Sub MoveOn(m As Integer, A As Integer)
Dim n, i, j As Integer
Dim Delay, h As Integer
Dim Move As Integer


 'To move a link:
 'the cell occupied by  the link is cleared,
 'the cell to be occupied is painted.

 'The occupied cell is cleared
  n = asp(A).Chromosome(m).Link
  j = asp(A).LinkAddress(n).x
  i = asp(A).LinkAddress(n).y
  Range(Cells(i, j), Cells(i, j)).Select
  Selection.Clear

  'The target addres is calculated
  Move = asp(A).Chromosome(m).Move
  j = j + VectMoves(Move).x
  i = i - VectMoves(Move).y
 'The target addres is memorized
   asp(A).LinkAddress(n).x = j
  asp(A).LinkAddress(n).y = i

  'The target cell is painted
   Range(Cells(i, j), Cells(i, j)).Select


   'If movement is too fast, nothing can be seen.
   'To attune the velocity of the asp,
   'augment or diminish the delay as   pleased

   Delay = 1001 - Velocity
   For h = 1 To Delay
      With Selection.Interior
        .ColorIndex = asp(A).Color
        .Pattern = xlSolid
     End With
   Next h
End Sub


 Private Function Disruptive(m As Integer, A As Integer)
'A move is disruptive if the target address
'is farther than 1.5 units from the neighboring links.
Dim n As Integer
Dim Move As Integer
Dim x, y, x1, y1, x2, x3, y2, y3 As Integer
Dim d1, d2 As Single

  'Reads the link to be moved
  n = asp(A).Chromosome(m).Link
  'Reads the   address of the link
  y = asp(A).LinkAddress(n).x
  x = asp(A).LinkAddress(n).y
  'The (integer) move to be done is read
  Move = asp(A).Chromosome(m).Move

 'The target address is calculated
  y1 = y + VectMoves(Move).x
  x1 = x - VectMoves(Move).y
  'The addresses of the two neighboring links is read
  'And disruption is calculated
    'MsgBox "n " & n
  If ((1 < n) And (n < NLinks)) Then
    y2 = asp(A).LinkAddress(n - 1).x
    x2 = asp(A).LinkAddress(n - 1).y
    d1 = ((x2 - x1) ^ 2 + (y2 - y1) ^ 2) ^ 0.5
    y3 = asp(A).LinkAddress(n + 1).x
    x3 = asp(A).LinkAddress(n + 1).y
    d2 = ((x3 - x1) ^ 2 + (y3 - y1) ^ 2) ^ 0.5

    'The disruption bug is fixed here.

   If ((d1 > 1.5) Or (d2 > 1.5)) Then
       Disruptive = True
      Else
       Disruptive = False
     End If
  End If

    'Debugging messages
    'MsgBox "d1 " & d1
    'MsgBox "d2 " & d2

  If n = 1 Then
    y2 = asp(A).LinkAddress(2).x
    x2 = asp(A).LinkAddress(2).y
    d1 = ((x2 - x1) ^ 2 + (y2 - y1) ^ 2) ^ 0.5
If d1 > 1.1 Then Disruptive = True Else Disruptive = False
   End If

  If n = NLinks Then
    y2 = asp(A).LinkAddress(NLinks - 1).x
    x2 = asp(A).LinkAddress(NLinks - 1).y
    d2 = ((x2 - x1) ^ 2 + (y2 - y1) ^ 2) ^ 0.5
If d2 > 1.1 Then Disruptive = True Else Disruptive = False
  End If

 End Function



Private Sub Ribosome(m As Integer, A As Integer)
Dim n, i, j As Integer
Dim Delay, h As Integer
Dim Move As Integer
Dim x, y, x1, y1 As Integer
Dim Color As Integer
  'Reads the link to be moved
  n = asp(A).Chromosome(m).Link
  'Reads the   address of the link
  y = asp(A).LinkAddress(n).x
  x = asp(A).LinkAddress(n).y
  'The (integer) move to be done is read
  Move = asp(A).Chromosome(m).Move

 'The target address is calculated
  y1 = y + VectMoves(Move).x
  x1 = x - VectMoves(Move).y

  'There are forbidden borders on all directions
  If ((x1 < 1) Or (y1 < 1) Or (x1 > 103) Or (y1 > 168)) Then
   Exit Sub
 End If

  'Get the  color of the target address
  Range(Cells(x1, y1), Cells(x1, y1)).Select
      With Selection.Interior
       Color = .ColorIndex
      End With

  'Is the target address occupied?
  'MsgBox "Color " & Color
  'In previous codes, there was  a bug here.

  Select Case Color
    Case 2   'the cell is empty, is it disruptive?
         If Not (Disruptive(m, A)) Then Call MoveOn(m, A)
    Case Else 'do nothing: the cell is occupied.

   End Select
End Sub

Private Sub AssessingFitness()
Dim A, B, m, Champ As Integer
Dim Max, Man As Integer
Dim Fitness(1 To 20) As Integer

'The fitness measures the vertical distance from
'actual position to the 100th row.

 For A = 1 To Nasps
 Max = 1
  For m = 1 To NLinks
   Man = asp(A).LinkAddress(m).y
   'MsgBox "vertical" & Man
   If Man > Max Then Max = Man
  Next m
   Fitness(A) = 100 - Max
   'MsgBox "Fitness " & Fitness(A)
 Next A

'Now, we sort asps by fitness
'and select the best four,
'whose numbers are recorded in Order.
For A = 1 To 4
 Champ = 1
 For B = 1 To 20
  If Fitness(B) > Fitness(Champ) Then Champ = B
  Next B
  Order(A) = Champ
  Fitness(Champ) = 0
 'MsgBox "Next champ " & Champ
Next A
 'MsgBox "Great champ " & Order(1)
End Sub

Private Sub PointMutation()
 Dim A, m, n, l, k  As Integer
 'Let us try out blind point mutations
 'in specific exons,
 'in the  link to be moved
 'and in the direction of movement.
 For A = 1 To Nasps
 'Each offsprings receives numbmutation mutations
  For k = 1 To NumbPointMutations
  'Select at random  which exon to mutate:
  'throw a number from 1 to NExons
   m = Int(NExons * Rnd) + 1
  'Select at random which link to move:
  'throw a number from 1 to NLinks.
  n = Int(NLinks * Rnd) + 1
  asp2(A).Chromosome(m).Link = n
  'Select at random which direction to move on.
  'Throw a number from 1 to 9
  asp2(A).Chromosome(m).Move = Int(9 * Rnd) + 1
  Next k
 Next A
End Sub

Private Sub GeneAmplification()
 'Some exons are overwriten with others.
Dim A, m, e, n, Move, k, ee As Integer

 For A = 1 To 20
  For m = 1 To NOriginalexons
    'Select  which exon to be amplified:
     'throw a number from 1 to NExons
     e = Int(NExons * Rnd) + 1
     n = asp2(A).Chromosome(e).Link
     Move = asp2(A).Chromosome(e).Move
     'The exon is amplified
     For k = 1 To Ncopies
     'Select at random where to be copied:
     'throw a number from 1 to NExons.
     ee = Int(NExons * Rnd) + 1
     asp2(A).Chromosome(ee).Link = n
     asp2(A).Chromosome(ee).Move = Move
    Next k
  Next m
 Next A
End Sub

Private Sub Mutation()
'Point mutation dominates over
'gene amplification.
  Call GeneAmplification
  Call PointMutation
End Sub



Private Sub Reproduction()
Dim A, m, l As Integer
'There are always 20 asps.
'Our reproductive scheme is as follows:
'10 offspring for the champion,
'5 offspring for the second,
'3 for the third,
'2 for the fourth.
'We place children on order.
'Children are clones of the fathers
'that are copied to the asp2 array.

'Call InitialPosition(8)
'MsgBox "var1 " & asp(8).LinkAddress(7).x
'MsgBox "var2 " & asp2(8).LinkAddress(7).x
 'asp2(8) = asp(8)
 'MsgBox "var2n " & asp2(8).LinkAddress(7).x

 For A = 1 To Nasps
 Select Case A
   Case 1, 2, 3, 4, 5, 6, 7, 8, 9, 10
     asp2(A) = asp(Order(1))
   Case 11, 12, 13, 14, 15
     asp2(A) = asp(Order(2))
   Case 16, 17, 18
     asp2(A) = asp(Order(3))
  Case 19, 20
     asp2(A) = asp(Order(4))
 End Select
Next A

End Sub

Private Sub Eraseasp(A As Integer)
Dim l, i, j, h, Delay
'To erase an asp, it is painted in white=2
 'Call Display
 'Call InitialPosition
 'Paints the body
 'Flag = 1 for initialization at random
 'Flag = 0 for reinitialization by inheritance.

  For l = 1 To NLinks
  j = asp(A).LinkAddress(l).x
  i = asp(A).LinkAddress(l).y
  'MsgBox "l " & l
  'MsgBox "i " & i
  'MsgBox "j " & j
   Range(Cells(i, j), Cells(i, j)).Select
   With Selection.Interior
        .ColorIndex = 2
        .Pattern = xlSolid
     End With
  Next l
  'If A = 15 Then MsgBox "Color " & asp(A).Color
End Sub

Private Sub Death()
Dim A As Integer
'The asp array is overwritten with
'the asp2 array.
For A = 1 To Nasps
 Call Eraseasp(A)
 asp(A) = asp2(A)
Next A
End Sub

Private Sub Initialization()
Dim A As Integer

  For A = 1 To Nasps
   Call InitialPosition(A)
   Call DrawBody(A, Flag)
    Call InitialChromosome(A)
  Next A

End Sub

Private Sub Reinitialization()
Dim A As Integer

'Beginning from the bottom
'with new genes
  For A = 1 To Nasps
   Call InitialPosition(A)
   Call DrawBody(A, Flag)
  Next A
End Sub

Public Sub asp3()
'We allow the lenght of the asp
'to be regulable so
'that it could be shortened.
'The number of exons is also released.


Dim m As Integer
Dim i, j, l, n
Dim A As Integer
Dim Gen, NGen As Integer


 Call Moves

 'InitialRow is the place in the screen
 'where the race begins
 InitialRow = 50
 Call Display


'Nasps is the number of asps,
'Default value = 20. This is the allowed max .
 Nasps = 20

'NGen is the number of generations
 NGen = 1

'LifeSpan is the number of   days
'in a life.
'Life ends with reproduction and dead.
 LifeSpan = 50

 'Each asp has NLinks number of links.
 'NLinks shall be smaller than 20
 NLinks = 2
 'Each chromosome has Nexons
 'number of instructions.
 'Nexons shall be smaller than 40
 NExons = 2 * NLinks


 'NumbPointMutations  is the number of
 'point mutations of the offspring
 NumbPointMutations = 1
 'Gene amplification follows:
 'NOriginalexons is the number of different exons
 'to be copied.
 NOriginalexons = 1
 'Ncopies is the number of new copies
 'of each copied exon.
 Ncopies = 1
 'Velocity is among 0 and 1000,
'it regulates  the velocity of the display.
 Velocity = 1000

 'To generate new random sequences
 'change the number below.
 Randomize (3)

'Flag = 1 for initialization at random
 'Flag = 0 for reinitialization by inheritance.
 Flag = 1
 Call Initialization
 Flag = 0

'm is the number of the link
'l is the counter of days of life.
'A is the index of the asp.

 For Gen = 1 To NGen
  MsgBox "Generation " & Gen
  For l = 1 To LifeSpan
   For m = 1 To NExons
    For A = 1 To Nasps
     Call Ribosome(m, A)
    Next A
   Next m
  Next l
  If Not (Gen = NGen) Then
   Call AssessingFitness
   Call Reproduction
   'Gene amplification is included.
   Call Mutation
   Call Death
   Call Reinitialization
  End If
 Next Gen
End Sub

\end{verbatim}

 \bigskip
\textbf{\thenum. }  \addtocounter{num}{1} Exercise. Run the code with the established parameters. We have shortened the length of the asp to 2 links and we only run for  one generation -hence, there was no reproduction, inheritance neither evolution.

 \bigskip
\textbf{\thenum. }  \addtocounter{num}{1} You can run the code various times to verify that  under this simplest setting,  recursive behavior that solves a specific task could eventually appear  as  encoded by information assembled at random.

 \bigskip
\textbf{\thenum. }  \addtocounter{num}{1} Thus far so good. But, where is our asp 12 links long? And, where is inheritance and its essential evolutionary role?

 \bigskip
\textbf{\thenum. }  \addtocounter{num}{1} Exercise. Set the number of generations equal to, say, 5. Verify that the  asps that multiply after the first generation are precisely descendant of those that tend to rise in the first generation. But their movement is slant.  Notice also that the traces left by the paths, although nice, are annoying for the development of the dynamics, so that it is a bug that we need to fix. Nevertheless, keep playing as much as necessary to  notice that this bug represents an unexpected selecting factor: a mutant may arise at the edge of the screen that   climbs directly upwards and so feels no thread from the border of the screen neither form the tracks of its companions.  Very surprising indeed is the arising of other mutants that   advance in a slant form and  bend when shock against the track left by a previous trajectory. This is very astonishing.

 \bigskip
\textbf{\thenum. }  \addtocounter{num}{1} The results show that evolution is more surprising that anyone can imagine. In the usual jargon it is said like this:  evolution is plenty of emergent features.  That is a happy result because it is completely clear than evolution is much more difficult to implement than 'robotic' life, and so one shall ask : how does  evolution have survived in nature until today ? In the light of our observations, the reason is that alive beings change the milieu in such unexpected ways that there is no provision for that. So adaptation shall be produced by a vivid, malleable process: Is there something better than evolution?

 \bigskip
\textbf{\thenum. }  \addtocounter{num}{1} Exercise. Run the program with asps with three links (NLinks= 3) and during some 20 generations. Verify that successful mutants arise rather late after some 13 generations. For asps with 4 links, use the following setting: Nlinks=4; LifeSpan = 8; NGen = 500. For another run, augment the number of exons from 8 to 12. Next, augment the number of point mutations from 1 to 2 (per chromosome, per generation). Make some scans with asps of 5 , 6 or 7 links to see what happens. If you have more patience than anybody, may be you will succeed in seen the arising of winners.

 \bigskip
\textbf{\thenum. }  \addtocounter{num}{1}  We have cherished the illusion of seeing an asp, synthesized by evolution and 12 links long, climbing over the screen. But we did not have found the way to get  moving asps with more than 4 links. Most possibly, by keeping running for hours and enhancing the power of the simulation, we will solve this and any other problem.

But to be sincere, we are saying with very soft words that evolution does not function: who will believe us that our problem will be solved by just keep our code running?  If someone does, let him or her know that he or she is insulting us: we are are not dreamers. We are simulators. We are like fishers: if we have a fish in our hands, we can cook and eat otherwise we must continue fishing. So, we need evolution functioning right now in front of us.  Yes, only in that way we will be clean of our shame of seeing evolution defeated. The evolution that we have created with our own hands cannot be defeated. This is a simple axiom of work. Thus, we agree that the evolutionary platform that we currently use does not function but we will create just now a new evolutionary  platform to see what its gives from itself.

 \bigskip
\textbf{\thenum. }  \addtocounter{num}{1} Important comment: each program of simulation defines an evolutionary platform.  Each platform comes with its own span of evolvable possibilities. For given restrictions of time and  memory, a given platform may solve a problem or may be not. Thus, an evolutionary platform may fail or may succeed.

 \bigskip
\textbf{\thenum. }  \addtocounter{num}{1} At the mean time, let us develop a program, an evolutionary platform,   in which an asp maybe enlarged from 2 links to 3 to 4 .. .. to 12. This experiment explores the possibility that complexity is built from extant complexity more easily than ab initio.

 \bigskip
\textbf{\thenum. }  \addtocounter{num}{1} Exercise. Run  asp3 but change the number of links from 5 to 2. Play with it until you see an asp that can repetitively change the direction of movement if an obstacle is found in its way. Let us notice that in the chromosomes of these new asps there is more place for information because they are longer than before.

 \bigskip
\textbf{\thenum. }  \addtocounter{num}{1} The code that allow asps  to get enlarged follows:

\begin{verbatim}


Option Explicit

Type LinkAddresses
 x As Integer
 y As Integer
End Type

Type Moves
 x As Integer
 y As Integer
End Type

Type Exons
 Link As Integer
 Move As Integer
End Type

Type asps
'Each chromosome has NExons instructions
 Chromosome(1 To 40) As Exons
 LinkAddress(1 To 20) As LinkAddresses
 Color As Integer
End Type

Dim Nasps As Integer 'Less than 20
Dim asp(1 To 20) As asps
Dim aspTwo(1 To 20) As asps
Dim VectMoves(1 To 9) As Moves
Dim LifeSpan As Integer
Dim InitialRow
Dim Order(1 To 20) As Integer
Dim Flag As Integer
Dim NumbPointMutations As Integer
Dim Velocity As Integer
Dim NOriginalexons, Ncopies As Integer
Dim NLinks, NExons As Integer
Dim BasalColor As Integer
Dim NGen As Integer

Private Sub Moves()
VectMoves(1).x = 0
VectMoves(1).y = 1
VectMoves(2).x = 1
VectMoves(2).y = 1
VectMoves(3).x = 1
VectMoves(3).y = 0
VectMoves(4).x = 1
VectMoves(4).y = -1
VectMoves(5).x = 0
VectMoves(5).y = -1
VectMoves(6).x = -1
VectMoves(6).y = -1
VectMoves(7).x = -1
VectMoves(7).y = 0
VectMoves(8).x = -1
VectMoves(8).y = 1
VectMoves(9).x = 0
VectMoves(9).y = 0
End Sub


Private Sub Display()
Dim i, j As Integer
'Prepares the screen
'Excel already works with this "new" style!
'To be sure, Excel works with the OOP style!


 Workbooks.Add
    Range(Cells(1, 1), Cells(103, 170)).Select
    Selection.RowHeight = 3
    Selection.ColumnWidth = 0.3
    ActiveWindow.WindowState = xlMaximized

  'Paint the background in white
  'This is necessary because It is not at all clear
  'what is the default background color.
  'The color of a cell will be taken as
  'the indicative of occupation.

  BasalColor = 2
  For i = 1 To 103
   For j = 1 To 170
    Range(Cells(i, j), Cells(i, j)).Select
      With Selection.Interior
       .ColorIndex = BasalColor
      End With
   Next j
  Next i

    Range("G9:T9").Select
    'With Selection.Interior
      'BasalColor = .ColorIndex
    'End With

End Sub

Private Sub InitialPosition(A As Integer)
Dim l, k
 'The index A distinguishes each asp.
 'Each asp has NLinks links,
 'each part has its own coordinates on the screen,
 'which are kept in the LinkAddress   array.
 'The asps are posited at the bottom of the screen.
 'Head are at row InitialRow

For l = 1 To NLinks
 k = InitialRow - NLinks
 asp(A).LinkAddress(l).x = 8 * A
 asp(A).LinkAddress(l).y = k + l
 Next l
End Sub

Private Function Fcolor(A As Integer)
 'Color is a number less than 40
 'but different than BasalColor.
Rebegin:  Fcolor = Int(40 * Rnd)
 If Fcolor = BasalColor Then GoTo Rebegin
 'MsgBox "Color  " & Fcolor
End Function

Private Sub DrawBody(A As Integer, Flag As Integer)
Dim l, i, j, h, Delay
 'Call Display
 'Call InitialPosition
 'Paints the body
 'Flag = 1 for initialization at random
 'Flag = 0 for reinitialization by inheritance.
  If Flag = 1 Then asp(A).Color = Fcolor(A)
  For l = 1 To NLinks
  j = asp(A).LinkAddress(l).x
  i = asp(A).LinkAddress(l).y
  'MsgBox "l " & l
  'MsgBox "i " & i
  'MsgBox "j " & j
   Range(Cells(i, j), Cells(i, j)).Select
   With Selection.Interior
        .ColorIndex = asp(A).Color
        .Pattern = xlSolid
     End With
  Next l
  'If A = 15 Then MsgBox "Color " & asp(A).Color
End Sub

Private Sub InitialChromosome(A As Integer)
Dim m, n, t
Dim x, y As Integer
Dim x1, y1 As Integer


For m = 1 To NExons
'Select at random  which link to move
'Throw a number from 1 to NLinks
 n = Int(NLinks * Rnd) + 1
 'MsgBox "n " & n
 asp(A).Chromosome(m).Link = n

 'Select at random which direction to move on
 'Throw a number from 1 to 9
  asp(A).Chromosome(m).Move = Int(9 * Rnd) + 1
Next m

End Sub

Private Sub MoveOn(m As Integer, A As Integer)
Dim n, i, j As Integer
Dim Delay, h As Integer
Dim Move As Integer


 'To move a link:
 'the cell occupied by  the link is cleared,
 'the cell to be occupied is painted.

 'The occupied cell is cleared
  n = asp(A).Chromosome(m).Link
  j = asp(A).LinkAddress(n).x
  i = asp(A).LinkAddress(n).y
  Range(Cells(i, j), Cells(i, j)).Select
  Selection.Clear

  'The target addres is calculated
  Move = asp(A).Chromosome(m).Move
  j = j + VectMoves(Move).x
  i = i - VectMoves(Move).y
 'The target addres is memorized
   asp(A).LinkAddress(n).x = j
  asp(A).LinkAddress(n).y = i

  'The target cell is painted
   Range(Cells(i, j), Cells(i, j)).Select


   'If movement is too fast, nothing can be seen.
   'To attune the velocity of the asp,
   'augment or diminish the delay as   pleased

   Delay = 1001 - Velocity
   For h = 1 To Delay
      With Selection.Interior
        .ColorIndex = asp(A).Color
        .Pattern = xlSolid
     End With
   Next h
End Sub


 Private Function Disruptive(m As Integer, A As Integer)
'A move is disruptive if the target address
'is farther than 1.5 units from the neighboring links.
Dim n As Integer
Dim Move As Integer
Dim x, y, x1, y1, x2, x3, y2, y3 As Integer
Dim d1, d2 As Single

  'Reads the link to be moved
  n = asp(A).Chromosome(m).Link
  'Reads the   address of the link
  y = asp(A).LinkAddress(n).x
  x = asp(A).LinkAddress(n).y
  'The (integer) move to be done is read
  Move = asp(A).Chromosome(m).Move

 'The target address is calculated
  y1 = y + VectMoves(Move).x
  x1 = x - VectMoves(Move).y
  'The addresses of the two neighboring links is read
  'And disruption is calculated
    'MsgBox "n " & n
  If ((1 < n) And (n < NLinks)) Then
    y2 = asp(A).LinkAddress(n - 1).x
    x2 = asp(A).LinkAddress(n - 1).y
    d1 = ((x2 - x1) ^ 2 + (y2 - y1) ^ 2) ^ 0.5
    y3 = asp(A).LinkAddress(n + 1).x
    x3 = asp(A).LinkAddress(n + 1).y
    d2 = ((x3 - x1) ^ 2 + (y3 - y1) ^ 2) ^ 0.5

    'The disruption bug is fixed here.

    If ((d1 > 1.5) Or (d2 > 1.5)) Then
       Disruptive = True
      Else
       Disruptive = False
     End If
  End If

    'Debugging messages
    'MsgBox "d1 " & d1
    'MsgBox "d2 " & d2

  If n = 1 Then
    y2 = asp(A).LinkAddress(2).x
    x2 = asp(A).LinkAddress(2).y
    d1 = ((x2 - x1) ^ 2 + (y2 - y1) ^ 2) ^ 0.5
If d1 > 1.1 Then Disruptive = True Else Disruptive = False
   End If

  If n = NLinks Then
    y2 = asp(A).LinkAddress(NLinks - 1).x
    x2 = asp(A).LinkAddress(NLinks - 1).y
    d2 = ((x2 - x1) ^ 2 + (y2 - y1) ^ 2) ^ 0.5
If d2 > 1.1 Then Disruptive = True Else Disruptive = False
  End If

 End Function



Private Sub Ribosome(m As Integer, A As Integer)
Dim n, i, j As Integer
Dim Delay, h As Integer
Dim Move As Integer
Dim x, y, x1, y1 As Integer
Dim Color As Integer
  'Reads the link to be moved
  n = asp(A).Chromosome(m).Link
  'Reads the   address of the link
  y = asp(A).LinkAddress(n).x
  x = asp(A).LinkAddress(n).y
  'The (integer) move to be done is read
  Move = asp(A).Chromosome(m).Move

 'The target address is calculated
  y1 = y + VectMoves(Move).x
  x1 = x - VectMoves(Move).y

  'There are forbidden borders on all directions
  If ((x1 < 1) Or (y1 < 1) Or (x1 > 103) Or (y1 > 168)) Then
   Exit Sub
  En If
  'Get the  color of the target address
  Range(Cells(x1, y1), Cells(x1, y1)).Select
      With Selection.Interior
       Color = .ColorIndex
      End With

  'Is the target address occupied?
  'MsgBox "Color " & Color
  'In previous codes, there was  a bug here.

  Select Case Color
    Case BasalColor   'the cell is empty, is it disruptive?
         If Not (Disruptive(m, A)) Then Call MoveOn(m, A)
    Case Else 'do nothing: the cell is occupied.

   End Select
End Sub

Private Sub AssessingFitness()
Dim A, B, m, Champ As Integer
Dim Max, Man As Integer
Dim Fitness(1 To 20) As Integer

'The fitness measures the vertical distance from
'actual position to the 100th row.

 For A = 1 To Nasps
 Max = 1
  For m = 1 To NLinks
   Man = asp(A).LinkAddress(m).y
   'MsgBox "vertical" & Man
   If Man > Max Then Max = Man
  Next m
   Fitness(A) = 100 - Max
   'MsgBox "Fitness " & Fitness(A)
 Next A

'Now, we sort asps by fitness
'and select the best four,
'whose numbers are recorded in Order.
For A = 1 To 4
 Champ = 1
 For B = 1 To 20
  If Fitness(B) > Fitness(Champ) Then Champ = B
  Next B
  Order(A) = Champ
  Fitness(Champ) = 0
 'MsgBox "Next champ " & Champ
Next A
 'MsgBox "Great champ " & Order(1)
End Sub

Private Sub PointMutation()
 Dim A, m, n, l, k  As Integer
 'Let us try out blind point mutations
 'in specific exons,
 'in the  link to be moved
 'and in the direction of movement.
 For A = 1 To Nasps
 'Each offsprings receives numbmutation mutations
  For k = 1 To NumbPointMutations
  'Select at random  which exon to mutate:
  'throw a number from 1 to NExons
   m = Int(NExons * Rnd) + 1
  'Select at random which link to move:
  'throw a number from 1 to NLinks.
  n = Int(NLinks * Rnd) + 1
  asp(A).Chromosome(m).Link = n
  'Select at random which direction to move on.
  'Throw a number from 1 to 9
  asp(A).Chromosome(m).Move = Int(9 * Rnd) + 1
  Next k
 Next A
End Sub

Private Sub GeneAmplification()
 'Some exons are overwriten with others.
Dim A, m, e, n, Move, k, ee As Integer

 For A = 1 To 20
  For m = 1 To NOriginalexons
    'Select  which exon to be amplified:
     'throw a number from 1 to NExons
     e = Int(NExons * Rnd) + 1
     n = aspTwo(A).Chromosome(e).Link
     Move = asp(A).Chromosome(e).Move
     'The exon is amplified
     For k = 1 To Ncopies
     'Select at random where to be copied:
     'throw a number from 1 to NExons.
     ee = Int(NExons * Rnd) + 1
     asp(A).Chromosome(ee).Link = n
     asp(A).Chromosome(ee).Move = Move
    Next k
  Next m
 Next A
End Sub

Private Sub OneMoreLink()
Dim A As Integer
Dim m, n, lim1 As Integer
'A new link is added to all asps.
'The link is added at the tail
lim1 = NExons + 1
NLinks = NLinks + 1
NExons = 3 * NLinks
For A = 1 To Nasps
 'New exons are filled in information
 For m = lim1 To NExons
 'Select at random  which link to move
 'Throw a number from 1 to NLinks
 n = Int(NLinks * Rnd) + 1
 'MsgBox "n " & n
 asp(A).Chromosome(m).Link = n

 'Select at random which direction to move on
 'Throw a number from 1 to 9
  asp(A).Chromosome(m).Move = Int(9 * Rnd) + 1
  Next m
Next A
End Sub

Private Sub Mutation()
'Point mutation dominates over
'gene amplification.
  Call GeneAmplification
  Call PointMutation
  Call OneMoreLink
End Sub



Private Sub Reproduction()
Dim A, m, l As Integer
'There are always 20 asps.
'Our reproductive scheme is as follows:
'10 offspring for the champion,
'5 offspring for the second,
'3 for the third,
'2 for the fourth.
'We place children on order.
'Children are clones of the fathers
'that are copied to the aspTwo array.

'Call InitialPosition(8)
'MsgBox "var1 " & asp(8).LinkAddress(7).x
'MsgBox "var2 " & aspTwo(8).LinkAddress(7).x
 'aspTwo(8) = asp(8)
 'MsgBox "var2n " & aspTwo(8).LinkAddress(7).x

 For A = 1 To Nasps
 Select Case A
   Case 1, 2, 3, 4, 5, 6, 7, 8, 9, 10
     aspTwo(A) = asp(Order(1))
   Case 11, 12, 13, 14, 15
     aspTwo(A) = asp(Order(2))
   Case 16, 17, 18
     aspTwo(A) = asp(Order(3))
  Case 19, 20
     aspTwo(A) = asp(Order(4))
 End Select
Next A

End Sub

Private Sub Eraseasp(A As Integer)
Dim l, i, j, h, Delay
'To erase an asp, it is painted in white=BasalCOlor
 'Call Display
 'Call InitialPosition
 'Paints the body
 'Flag = 1 for initialization at random
 'Flag = 0 for reinitialization by inheritance.

  For l = 1 To NLinks
  j = asp(A).LinkAddress(l).x
  i = asp(A).LinkAddress(l).y
  'MsgBox "l " & l
  'MsgBox "i " & i
  'MsgBox "j " & j
   Range(Cells(i, j), Cells(i, j)).Select
   With Selection.Interior
        .ColorIndex = BasalColor
        .Pattern = xlSolid
     End With
  Next l
  'If A = 15 Then MsgBox "Color " & asp(A).Color
End Sub

Private Sub Death()
Dim A As Integer
'The asp array is overwritten with
'the aspTwo array.
For A = 1 To Nasps
 Call Eraseasp(A)
 asp(A) = aspTwo(A)
Next A
End Sub

Private Sub Initialization()
Dim A As Integer

  For A = 1 To Nasps
   Call InitialPosition(A)
   Call DrawBody(A, Flag)
    Call InitialChromosome(A)
  Next A

End Sub

Private Sub Reinitialization()
Dim A As Integer

'Beginning from the bottom
'with new genes
  For A = 1 To Nasps
   Call InitialPosition(A)
   Call DrawBody(A, Flag)
  Next A
End Sub

Private Sub Parameters()
 'InitialRow is the place in the screen
 'where the race begins
 InitialRow = 50
 Call Display


'Nasps is the number of asps,
'Default value = 20. This is the allowed max .
 Nasps = 20

'NGen is the number of generations

 NGen = 500

'Each asp has NLinks number of links.
 'NLinks shall be smaller than 20.
 'Initial number of links
 NLinks = 2

 'Each chromosome has Nexons
 'number of instructions.
 'Nexons shall be smaller than 40
 NExons = 3 * NLinks

'LifeSpan is the number of   days
'in a life.
'Life ends with reproduction and dead.
 LifeSpan = 5 * NExons

 'NumbPointMutations  is the number of
 'point mutations of the offspring
 NumbPointMutations = 2
 'Gene amplification follows:
 'NOriginalexons is the number of different exons
 'to be copied.
 NOriginalexons = 2
 'Ncopies is the number of new copies
 'of each copied exon.
 Ncopies = 2
 'Velocity is among 0 and 1000,
'it regulates  the velocity of the display.
 Velocity = 1000

 'To generate new random sequences
 'change the number below.
 Randomize (4)

'Flag = 1 for initialization at random
 'Flag = 0 for reinitialization by inheritance.
 Flag = 1
 Call Initialization
 Flag = 0

'm is the number of the link
'l is the counter of days of life.
'A is the index of the asp.
End Sub

Private Sub Dynamics()
 'We allow the length of the asp
'to be regulated so
'that it could be lengthened.
'The number of exons is also released.


Dim m As Integer
Dim i, j, l, n
Dim A As Integer
Dim Gen  As Integer

 For Gen = 1 To NGen
  'MsgBox "Generation " & Gen
  For l = 1 To LifeSpan
   For m = 1 To NExons
    For A = 1 To Nasps
     Call Ribosome(m, A)
    Next A
   Next m
  Next l
  If Not (Gen = NGen) Then
   Call AssessingFitness
   Call Reproduction
   'Gene amplification is included.
   'Body enlargement is allowed.
   Call Death
   Call Mutation
   Call Reinitialization
  End If
 Next Gen
End Sub

Public Sub asp2()
  Call Moves
 Call Parameters
 Call Dynamics
End Sub

\end{verbatim}

 \bigskip
\textbf{\thenum. }  \addtocounter{num}{1} Exercise. Run the code and keep playing. Verify that asps with 4 links or lesser are mobile but we continue to be stopped by  a barrier at NLinks = 5. The program is aborted by the compiler because larger asps demand larger memory to be simulated and it that has not been dimensioned. If you consider necessary, solve the problem and continue playing.

\section{The final version}

 \bigskip
\textbf{\thenum. }  \addtocounter{num}{1} We have done so many things but nevertheless we continue without achieving our goal. What shall we do? Shall we recognize that we are in a hurry and it is because of it that we cannot fish anything? Shall we learn a lesson from ancient Greeks? They believed that there are some facts that only can be explained if time is allowed to run to infinite. Precisely, they set Chronos (Time) as the father of all gods. Unfortunately, we cannot: there have elapsed 2500 years from those Greeks to us, and  on our desks we have machines that, because of their velocity and power,    could surpass those one million dollars machines of around 1970. Moreover, we have access to VBE for Excel, which is a real wonder.

 \bigskip
\textbf{\thenum. }  \addtocounter{num}{1} We have so much that our one and only option is to  keep working and to drag with us the shame of having caught too small a fish. So, before passing to a new theme, let us make a last attempt. Let us implement two reforms. The first is that we will not pass to enlarging the body of the asp if it is motionless. The second is that we will erase the traces left by the previous paths because they could interfere with the arising of successful mutants at their very birth. The code follows:

\begin{verbatim}

Option Explicit

Type LinkAddresses
 x As Integer
 y As Integer
End Type

Type Moves
 x As Integer
 y As Integer
End Type

Type Exons
 Link As Integer
 Move As Integer
End Type

Type asps
'Each chromosome has NExons instructions
 Chromosome(1 To 40) As Exons
 LinkAddress(1 To 20) As LinkAddresses
 Color As Integer
End Type

Dim Nasps As Integer 'Less than 20
Dim asp(1 To 20) As asps
Dim aspTwo(1 To 20) As asps
Dim VectMoves(1 To 9) As Moves
Dim LifeSpan As Integer
Dim InitialRow
Dim Order(1 To 20) As Integer
Dim Flag As Integer
Dim NumbPointMutations As Integer
Dim Velocity As Integer
Dim NOriginalexons, Ncopies As Integer
Dim NLinks, NExons As Integer
Dim BasalColor As Integer
Dim NGen As Integer
Dim ThereWasAdvance As Integer


Private Sub Moves()
VectMoves(1).x = 0
VectMoves(1).y = 1
VectMoves(2).x = 1
VectMoves(2).y = 1
VectMoves(3).x = 1
VectMoves(3).y = 0
VectMoves(4).x = 1
VectMoves(4).y = -1
VectMoves(5).x = 0
VectMoves(5).y = -1
VectMoves(6).x = -1
VectMoves(6).y = -1
VectMoves(7).x = -1
VectMoves(7).y = 0
VectMoves(8).x = -1
VectMoves(8).y = 1
VectMoves(9).x = 0
VectMoves(9).y = 0
End Sub


Private Sub Display()
Dim i, j As Integer
'Prepares the screen
'Excel already works with this "new" style!
'To be sure, Excel works with the OOP style!


 Workbooks.Add
    Range(Cells(1, 1), Cells(103, 170)).Select
    Selection.RowHeight = 3
    Selection.ColumnWidth = 0.3
    ActiveWindow.WindowState = xlMaximized

  'Paint the background in white
  'This is necessary because It is not at all clear
  'what is the default background color.
  'The color of a cell will be taken as
  'the indicative of occupation.

  BasalColor = 2
  For i = 1 To 103
   For j = 1 To 170
    Range(Cells(i, j), Cells(i, j)).Select
      With Selection.Interior
       .ColorIndex = BasalColor
      End With
   Next j
  Next i

    'Range("G9:T9").Select
    'With Selection.Interior
      'BasalColor = .ColorIndex
    'End With

End Sub
Private Sub CleanScreen()
Dim i, j As Integer
  For i = InitialRow - NExons To InitialRow + 4
   For j = 1 To 170
    Range(Cells(i, j), Cells(i, j)).Select
      With Selection.Interior
       .ColorIndex = BasalColor
      End With
   Next j
  Next i
End Sub

Private Sub InitialPosition(A As Integer)
Dim l, k
 'The index A distinguishes each asp.
 'Each asp has NLinks links,
 'each part has its own coordinates on the screen,
 'which are kept in the LinkAddress   array.
 'The asps are posited at the bottom of the screen.
 'Head are at row InitialRow

For l = 1 To NLinks
 k = InitialRow - NLinks
 asp(A).LinkAddress(l).x = 8 * A
 asp(A).LinkAddress(l).y = k + l
 Next l
End Sub

Private Function Fcolor(A As Integer)
 'Color is a number less than 40
 'but different than BasalColor.
Rebegin:  Fcolor = Int(40 * Rnd)
 If Fcolor = BasalColor Then GoTo Rebegin
 'MsgBox "Color  " & Fcolor
End Function

Private Sub DrawBody(A As Integer, Flag As Integer)
Dim l, i, j, h, Delay
 'Call Display
 'Call InitialPosition
 'Paints the body
 'Flag = 1 for initialization at random
 'Flag = 0 for reinitialization by inheritance.
  If Flag = 1 Then asp(A).Color = Fcolor(A)
  For l = 1 To NLinks
  j = asp(A).LinkAddress(l).x
  i = asp(A).LinkAddress(l).y
  'MsgBox "l " & l
  'MsgBox "i " & i
  'MsgBox "j " & j
   Range(Cells(i, j), Cells(i, j)).Select
   With Selection.Interior
        .ColorIndex = asp(A).Color
        .Pattern = xlSolid
     End With
  Next l
  'If A = 15 Then MsgBox "Color " & asp(A).Color
End Sub

Private Sub InitialChromosome(A As Integer)
Dim m, n, t
Dim x, y As Integer
Dim x1, y1 As Integer


For m = 1 To NExons
'Select at random  which link to move
'Throw a number from 1 to NLinks
 n = Int(NLinks * Rnd) + 1
 'MsgBox "n " & n
 asp(A).Chromosome(m).Link = n

 'Select at random which direction to move on
 'Throw a number from 1 to 9
  asp(A).Chromosome(m).Move = Int(9 * Rnd) + 1
Next m

End Sub

Private Sub MoveOn(m As Integer, A As Integer)
Dim n, i, j As Integer
Dim Delay, h As Integer
Dim Move As Integer


 'To move a link:
 'the cell occupied by  the link is cleared,
 'the cell to be occupied is painted.

 'The occupied cell is cleared
  n = asp(A).Chromosome(m).Link
  j = asp(A).LinkAddress(n).x
  i = asp(A).LinkAddress(n).y
  Range(Cells(i, j), Cells(i, j)).Select
  Selection.Clear

  'The target addres is calculated
  Move = asp(A).Chromosome(m).Move
  j = j + VectMoves(Move).x
  i = i - VectMoves(Move).y
 'The target addres is memorized
   asp(A).LinkAddress(n).x = j
  asp(A).LinkAddress(n).y = i

  'The target cell is painted
   Range(Cells(i, j), Cells(i, j)).Select


   'If movement is too fast, nothing can be seen.
   'To attune the velocity of the asp,
   'augment or diminish the delay as   pleased

   Delay = 1001 - Velocity
   For h = 1 To Delay
      With Selection.Interior
        .ColorIndex = asp(A).Color
        .Pattern = xlSolid
     End With
   Next h
End Sub


 Private Function Disruptive(m As Integer, A As Integer)
'A move is disruptive if the target address
'is farther than 1.5 units from the neighboring links.
Dim n As Integer
Dim Move As Integer
Dim x, y, x1, y1, x2, x3, y2, y3 As Integer
Dim d1, d2 As Single

  'Reads the link to be moved
  n = asp(A).Chromosome(m).Link
  'Reads the   address of the link
  y = asp(A).LinkAddress(n).x
  x = asp(A).LinkAddress(n).y
  'The (integer) move to be done is read
  Move = asp(A).Chromosome(m).Move

 'The target address is calculated
  y1 = y + VectMoves(Move).x
  x1 = x - VectMoves(Move).y
  'The addresses of the two neighboring links is read
  'And disruption is calculated
    'MsgBox "n " & n
  If ((1 < n) And (n < NLinks)) Then
    y2 = asp(A).LinkAddress(n - 1).x
    x2 = asp(A).LinkAddress(n - 1).y
    d1 = ((x2 - x1) ^ 2 + (y2 - y1) ^ 2) ^ 0.5
    y3 = asp(A).LinkAddress(n + 1).x
    x3 = asp(A).LinkAddress(n + 1).y
    d2 = ((x3 - x1) ^ 2 + (y3 - y1) ^ 2) ^ 0.5

    'The disruption bug is fixed here.

    If ((d1 > 1.5) Or (d2 > 1.5)) Then
       Disruptive = True
      Else
       Disruptive = False
     End If
  End If

    'Debugging messages
    'MsgBox "d1 " & d1
    'MsgBox "d2 " & d2

  If n = 1 Then
    y2 = asp(A).LinkAddress(2).x
    x2 = asp(A).LinkAddress(2).y
    d1 = ((x2 - x1) ^ 2 + (y2 - y1) ^ 2) ^ 0.5
If d1 > 1.1 Then Disruptive = True Else Disruptive = False
   End If

  If n = NLinks Then
    y2 = asp(A).LinkAddress(NLinks - 1).x
    x2 = asp(A).LinkAddress(NLinks - 1).y
    d2 = ((x2 - x1) ^ 2 + (y2 - y1) ^ 2) ^ 0.5
If d2 > 1.1 Then Disruptive = True Else Disruptive = False
  End If

 End Function



Private Sub Ribosome(m As Integer, A As Integer)
Dim n, i, j As Integer
Dim Delay, h As Integer
Dim Move As Integer
Dim x, y, x1, y1 As Integer
Dim Color As Integer
  'Reads the link to be moved
  n = asp(A).Chromosome(m).Link
  'Reads the   address of the link
  y = asp(A).LinkAddress(n).x
  x = asp(A).LinkAddress(n).y
  'The (integer) move to be done is read
  Move = asp(A).Chromosome(m).Move

 'The target address is calculated
  y1 = y + VectMoves(Move).x
  x1 = x - VectMoves(Move).y

  'There are forbidden borders on all directions
  If ((x1 < 1) Or (y1 < 1) Or (x1 > 103) Or (y1 > 168)) Then
   Exit Sub
  End If
  'Get the  color of the target address
  Range(Cells(x1, y1), Cells(x1, y1)).Select
      With Selection.Interior
       Color = .ColorIndex
      End With

  'Is the target address occupied?
  'MsgBox "Color " & Color
  'In previous codes, there was  a bug here.

  Select Case Color
    Case BasalColor   'the cell is empty, is it disruptive?
         If Not (Disruptive(m, A)) Then Call MoveOn(m, A)
    Case Else 'do nothing: the cell is occupied.

   End Select
End Sub

Private Sub AssessingFitness()
Dim A, B, m, Champ As Integer
Dim Max, Man As Integer
Dim Fitness(1 To 20) As Integer

'The fitness measures the vertical distance from
'actual position to the 100th row.

 For A = 1 To Nasps
 Max = 1
  For m = 1 To NLinks
   Man = asp(A).LinkAddress(m).y
   'MsgBox "vertical" & Man
   If Man > Max Then Max = Man
  Next m
   Fitness(A) = 100 - Max
   'MsgBox "Fitness " & Fitness(A)
 Next A

'Now, we sort asps by fitness
'and select the best four,
'whose numbers are recorded in Order.
For A = 1 To 4
 Champ = 1
 For B = 1 To 20
  If Fitness(B) > Fitness(Champ) Then Champ = B
  Next B
  Order(A) = Champ
  Fitness(Champ) = 0
 'MsgBox "Next champ " & Champ
Next A
 'MsgBox "Great champ " & Order(1)
End Sub

Private Sub AssessAdvance()
 'Our criterion for advance is
 'that the head has surpassed
 'its original vertical position.
 Dim A, HeadPosition As Integer
  ThereWasAdvance = 0
  For A = 1 To Nasps
   HeadPosition = asp(A).LinkAddress(1).y
   If HeadPosition < InitialRow - NLinks Then
       ThereWasAdvance = 1
       Exit Sub
     End If
  Next A
End Sub

Private Sub PointMutation()
 Dim A, m, n, l, k  As Integer
 'Let us try out blind point mutations
 'in specific exons,
 'in the  link to be moved
 'and in the direction of movement.
 For A = 1 To Nasps
 'Each offsprings receives numbmutation mutations
  For k = 1 To NumbPointMutations
  'Select at random  which exon to mutate:
  'throw a number from 1 to NExons
   m = Int(NExons * Rnd) + 1
  'Select at random which link to move:
  'throw a number from 1 to NLinks.
  n = Int(NLinks * Rnd) + 1
  asp(A).Chromosome(m).Link = n
  'Select at random which direction to move on.
  'Throw a number from 1 to 9
  asp(A).Chromosome(m).Move = Int(9 * Rnd) + 1
  Next k
 Next A
End Sub

Private Sub GeneAmplification()
 'Some exons are overwriten with others.
Dim A, m, e, n, Move, k, ee As Integer

 For A = 1 To 20
  For m = 1 To NOriginalexons
    'Select  which exon to be amplified:
     'throw a number from 1 to NExons
     e = Int(NExons * Rnd) + 1
     n = aspTwo(A).Chromosome(e).Link
     Move = asp(A).Chromosome(e).Move
     'The exon is amplified
     For k = 1 To Ncopies
     'Select at random where to be copied:
     'throw a number from 1 to NExons.
     ee = Int(NExons * Rnd) + 1
     asp(A).Chromosome(ee).Link = n
     asp(A).Chromosome(ee).Move = Move
    Next k
  Next m
 Next A
End Sub

Private Sub OneMoreLink()
Dim A As Integer
Dim m, n, lim1 As Integer
'A new link is added to all asps.
'The link is added at the tail
lim1 = NExons + 1
NLinks = NLinks + 1
NExons = 3 * NLinks
For A = 1 To Nasps
 'New exons are filled in information
 For m = lim1 To NExons
 'Select at random  which link to move
 'Throw a number from 1 to NLinks
 n = Int(NLinks * Rnd) + 1
 'MsgBox "n " & n
 asp(A).Chromosome(m).Link = n

 'Select at random which direction to move on
 'Throw a number from 1 to 9
  asp(A).Chromosome(m).Move = Int(9 * Rnd) + 1
  Next m
Next A
End Sub


Private Sub Mutation()
'Point mutation dominates over
'gene amplification.
  Call GeneAmplification
  Call PointMutation
'If at least one asp   advanced even a bit,
'the next population is enlarged by one link.

 If ThereWasAdvance Then Call OneMoreLink
End Sub



Private Sub Reproduction()
Dim A, m, l As Integer
'There are always 20 asps.
'Our reproductive scheme is as follows:
'10 offspring for the champion,
'5 offspring for the second,
'3 for the third,
'2 for the fourth.
'We place children on order.
'Children are clones of the fathers
'that are copied to the aspTwo array.

'Call InitialPosition(8)
'MsgBox "var1 " & asp(8).LinkAddress(7).x
'MsgBox "var2 " & aspTwo(8).LinkAddress(7).x
 'aspTwo(8) = asp(8)
 'MsgBox "var2n " & aspTwo(8).LinkAddress(7).x

 For A = 1 To Nasps
 Select Case A
   Case 1, 2, 3, 4, 5, 6, 7, 8, 9, 10
     aspTwo(A) = asp(Order(1))
   Case 11, 12, 13, 14, 15
     aspTwo(A) = asp(Order(2))
   Case 16, 17, 18
     aspTwo(A) = asp(Order(3))
  Case 19, 20
     aspTwo(A) = asp(Order(4))
 End Select
Next A

End Sub

Private Sub Eraseasp(A As Integer)
Dim l, i, j, h, Delay
'To erase an asp, it is painted in white=BasalCOlor
 'Call Display
 'Call InitialPosition
 'Paints the body
 'Flag = 1 for initialization at random
 'Flag = 0 for reinitialization by inheritance.

  For l = 1 To NLinks
  j = asp(A).LinkAddress(l).x
  i = asp(A).LinkAddress(l).y
  'MsgBox "l " & l
  'MsgBox "i " & i
  'MsgBox "j " & j
   Range(Cells(i, j), Cells(i, j)).Select
   With Selection.Interior
        .ColorIndex = BasalColor
        .Pattern = xlSolid
     End With
  Next l
  'If A = 15 Then MsgBox "Color " & asp(A).Color
End Sub

Private Sub Death()
Dim A As Integer
'The asp array is overwritten with
'the aspTwo array.
For A = 1 To Nasps
 Call Eraseasp(A)
 asp(A) = aspTwo(A)
Next A
End Sub

Private Sub Initialization()
Dim A As Integer

  For A = 1 To Nasps
   Call InitialPosition(A)
   Call DrawBody(A, Flag)
    Call InitialChromosome(A)
  Next A

End Sub

Private Sub Reinitialization()
Dim A As Integer

'Beginning from the bottom
'with new genes
 Call CleanScreen
  For A = 1 To Nasps
   Call InitialPosition(A)
   Call DrawBody(A, Flag)
  Next A
End Sub

Private Sub Parameters()
 'InitialRow is the place in the screen
 'where the race begins
 InitialRow = 50
 Call Display


'Nasps is the number of asps,
'Default value = 20. This is the allowed max .
 Nasps = 20

'NGen is the number of generations

 NGen = 500

'Each asp has NLinks number of links.
 'NLinks shall be smaller than 20.
 'Initial number of links
 NLinks = 2

 'Each chromosome has Nexons
 'number of instructions.
 'Nexons shall be smaller than 40
 NExons = 1 * NLinks

'LifeSpan is the number of   days
'in a life.
'Life ends with reproduction and dead.
 LifeSpan = 5 * NExons

 'NumbPointMutations  is the number of
 'point mutations of the offspring
 NumbPointMutations = 2
 'Gene amplification follows:
 'NOriginalexons is the number of different exons
 'to be copied.
 NOriginalexons = 2
 'Ncopies is the number of new copies
 'of each copied exon.
 Ncopies = 2
 'Velocity is among 0 and 1000,
'it regulates  the velocity of the display.
 Velocity = 1000

 'To generate new random sequences
 'change the number below.
 Randomize (4)

'Flag = 1 for initialization at random
 'Flag = 0 for reinitialization by inheritance.
 Flag = 1
 Call Initialization
 Flag = 0

'm is the number of the link
'l is the counter of days of life.
'A is the index of the asp.
End Sub

Private Sub Dynamics()
 'Movile asps are allowed
 'to get enlarged.


Dim m As Integer
Dim i, j, l, n
Dim A As Integer
Dim Gen  As Integer

 For Gen = 1 To NGen
  'MsgBox "Generation " & Gen
  For l = 1 To LifeSpan
   For m = 1 To NExons
    For A = 1 To Nasps
     Call Ribosome(m, A)
    Next A
   Next m
  Next l
  If Not (Gen = NGen) Then
   Call AssessingFitness
   Call AssessAdvance
   Call Reproduction
   'Gene amplification is included.
   'Body enlargement is allowed.
   Call Death
   Call Mutation
   Call Reinitialization
  End If
 Next Gen
End Sub

Public Sub asp1()
  Call Moves
 Call Parameters
 Call Dynamics
End Sub

\end{verbatim}

 \bigskip
\textbf{\thenum. }  \addtocounter{num}{1} Exercise. Run the code and enjoy it.

 \bigskip
\textbf{\thenum. }  \addtocounter{num}{1} We have witness the evolutionary synthesis of functional code. In fact, the information written in the chromosomes is decoded  and converted in movement by the ribosome. Thus, evolution has been promoted to a software developer.

 \bigskip
\textbf{\thenum. }  \addtocounter{num}{1} Graduation.

You have the opportunity to graduate as a pedagogue else as a researcher.

As a pedagogue:

We have refined during 20 generations a   module that may be converted  into a laboratory of digital evolution. Add a  good interface and  appropriate documentation, test your material with your students and make a report of your experience.

As a Researcher:

The advantage of simulation is that everything is measurable. Indeed, this is more than an advantage: this is a responsibility. So,  devise detectors to catch interesting mutants  to see what do they have in their chromosomes. Deal carefully with those mutants that can bend over the obstacles. Explain in which way logical decisions have emerged from a set of instructions  that had none. This new work corresponds to that of a molecular biologist that  sequences various genomes.

As a multidisciplinary researcher:

Study the 20 files that were elaborated to achieve our simulation to see how is the evolution towards complexity along a line developed by an intelligent being. Then, you will have the opportunity to compare this type of evolution with the evolution towards complexity along the evolutionary line established by elongating asps.

 \bigskip
\textbf{\thenum. }  \addtocounter{num}{1} Dear Reader: if you  have accompanied me until here, working so hard, let me to express you my gratitude. You warm my heart. Thanks a lot.

\chapter{Conclusions}

Simulation consists in the building in the memory of a computer of a world with the specific properties we want and with   specific instruments of measurement  and appropriate procedures of reporting what is happening there.

The  applications of simulation to biology seem to be restricted only by the imagination. Very intriguing is the duplex method to analyze experiments with incomplete data. And, of course, the simulation of evolution is an essential elementary right, which is more difficult to exercise than one imagines. In general, simulation could become very complex and very difficult. This must not come as  a surprise because simulation is not intended to destroy the complexity of the nature that we study.  Nevertheless, some very important simulations are rather  simple while the corresponding mathematical models include hypercomplex items  such as stochastic differential equations.  This means that simulation has a great future as a pedagogical tool.

Evolution is rooted on mutations. Those  that occur in nature in asexual beings correspond exactly to the operations of a word processor. We have simulated many popular mutations, i.e., we have created a virtual world populated   by strings that can evolve thanks to mutations of the same sort as those happening in nature.

In simulated evolution, one can see   the emergence  of surprising features such as the emergence of behavior that seems to include logical decisions from a set of instructions that contained none.

Simulated evolution can develop software, which  is composed of a set of strings together with a  code of interpretation, an interpreter and a executer that can associate to each string a specific action.

The genome is the natural example of software: the code of interpretation is the genetic code,  the interpreter is the ensemble of t-RNA's, the primary executer is the ribosome.

Since our  simulated evolution is a transparent copy of some parts of natural evolution, we conclude that natural evolution is also a  natural software developer.

We have every reason to suspect that evolutionless life could  exits. Moreover, the genome of those alive beings would be orders of magnitude simpler than those that we see in nature, where evolution exists. Then, we ask: why shall life be so complex? Or,   Why did not evolution disappear long ago? Or, What does sustain evolution to exist until today?  Our simulations allow us to argue that alive beings add with their presence too much complexity to the environment and that that complexity could be best dealt with  evolution than with any given set of predefined strategies.

Our works allow us to pose the following intrigue: Software development is a very hard enterprise for human beings. Is there any equivalence for natural evolution? In particular, software made by humans is always plagued with bugs, errors of programming. Can we speak of evolutionary bugs? Can we say, for instance,  that the evolutionary bugs cause species to disappear and that dinosaurs were more bugged than mammals?

Can we compare the software developed by simulated evolution with that developed by humans, say, by skilled experts? Can we involve in this discussion the software of the natural genomes?

Our work allowed us to pose those questions, but it was too simple to shed some  light into the answers. One must venture to design simple platforms where these questions could be discussed. The advantage of simulation is that one knows exactly which are the conditions fulfilled by the background upon which one bold to draw conclusions. Thus, one might choose simulation as a profession for a life.

Simulation generates its own culture with its own ethics. The culture is so strong that is has the right to dare: you cannot understand something unless you submit it to simulation. Or, How can you say that you understand evolution if you have not simulated it? This complaint is very actual: with the popularization of genetic engineering, many people  seem to imagine that  mutation, directed mutation and undirected mutations are synonymous. No one, of course, would recognize that, but when they begin to simulate evolution they cannot avoid developing directed evolution as the solution to all problems.

The ethics of simulation is based on the historical responsibility that we have: any modern desktop is in the eyes of the history a million dollars   computer   and, moreover, present computers  are enlivened  by a software that might cost  no less.  Thus, the opportunities are so huge that simulation invites everybody to claim only what he or she has caught in his or her simulations. The challenge is   crude: who will stand it?

\backmatter

References

 The original sources from which one could learn about  Visual Basic for Excel are:

\bigskip
\textbf{\thenum. }  \addtocounter{num}{1} Microsoft Corporation (1994)  Microsoft Excel : version 5.0 : automating, customizing and programming in Microsoft Excel with the Microsoft Visual Basic Programming System, applications edition.    [Redmond, Washington] : Microsoft Corporation.

\bigskip
\textbf{\thenum. }  \addtocounter{num}{1} De Levie, Robert (2004): Advanced Excel for scientific data analysis.
 Oxford, England ; New York : Oxford University Press.

\bigskip
\textbf{\thenum. }  \addtocounter{num}{1} Steve Potts ... [et al.].(1995) Visual Basic 4 expert solutions. Indianapolis, Indiana : Que Corporation, c.

\bigskip

\end{document}